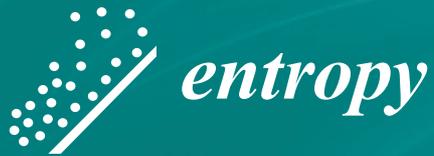

# Particle Theory and Theoretical Cosmology

Dedicated to Professor Paul Howard Frampton
on the Occasion of His 80th Birthday

Edited by
Thomas W. Kephart and Paul Howard Frampton

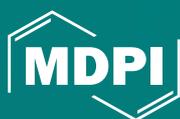

**Particle Theory and Theoretical Cosmology—Dedicated to Professor Paul Howard Frampton on the Occasion of His 80th Birthday**

# Particle Theory and Theoretical Cosmology—Dedicated to Professor Paul Howard Frampton on the Occasion of His 80th Birthday

Guest Editors

**Thomas W. Kephart**
**Paul Howard Frampton**

*Guest Editors*

Thomas W. Kephart
Vanderbilt University
Nashville, TN
USA

Paul Howard Frampton
Università del Salento and
INFN-Lecce
Lecce
Italy

This is a reprint of the Special Issue, published open access by the journal *Entropy* (ISSN 1099-4300), freely accessible at: https://www.mdpi.com/journal/entropy/special_issues/2M6D1ZZ85Y.

For citation purposes, cite each article independently as indicated on the article page online and as indicated below:

Lastname, A.A.; Lastname, B.B. Article Title. *Journal Name* **Year**, *Volume Number*, Page Range.

# Contents

# Preface

This Special Issue is a Festschrift containing articles written to commemorate the 80th anniversary of Frampton, who is encountering his second Festschrift after the first one on his 60th anniversary, organised by the University of Miami at the 2003 Coral Gables Meeting. This foreword can act as an introduction. We shall present a biography of Frampton which includes his early education, including how he became interested in particle theory. We then discuss selected accomplishments from his almost six decades of research productivity. Paul Frampton was born on October 31st, 1943 in Kidderminster, Worcestershire, UK, as the son of the late Harold Albert Frampton, a civil engineer, and the late Grace Elizabeth Frampton, née Howard, a housewife. In Kidderminster, he attended Foley Park Primary School 1948–54, then King Charles School 1954–62, where he became intellectually very excited at age 14 by the discovery of parity violation in weak interactions, which he could understand without quantum mechanics and which sparked his life-long interest in particle theory. That discovery resulted in the award of a Nobel prize to Lee and Yang in 1957. In the same year his headmaster, John Drake, very unusually noted on a school report that he considered it possible that Frampton might go on to win a Nobel prize in physics. Drake had a strong personality and was a larger than life figure both as a headmaster and as a physics teacher. Before Drake retired in 1958, he taught Frampton relativity and provided other intellectual inspiration. George Oxendale took over as Frampton's physics teacher and inspiration for the years 1958–62. In 1961, Frampton sat his A-level examinations; in 2006 he was invited back to present a hundred prizes to students at King Charles School, and his introducer was somehow able to quote Frampton's A-level results from 45 years earlier and stated that they had never been equalled. Also in 1961, he sat for the Oxford Entrance Examinations and was awarded an Open Hulme Scholarship to Brasenose College, Oxford, the same college of which his Latin teacher was an alumnus. Frampton studied physics in Oxford and was there educated in a time-honoured tutorial system where, with only one other student (the late Adrian Harford), he spent one hour a week in term time with a professor, including the late Simon Altmann. In the first-year examination, Frampton achieved First-Class Honours. At the end of the third and final undergraduate year, he again received First-Class Honours, thus completing a Double First bachelor degree. He stayed on at Brasenose College, now as a Senior Hulme Scholar, for a further three years 1965–68 to complete a doctoral DPhil research degree. His research supervisor was John Clayton Taylor and his thesis title was Strong Interactions of Elementary Particles: Regge Theory and Sum Rules. In it, he proved that certain sum rules which had previously been derived by assuming current algebra could be derived without that assumption using only more general considerations of analyticity and locality.

After his DPhil in 1968, Frampton spent the next twelve years until 1980 in a sequence of six postdoctoral positions. The first was in Chicago 1968–70 in a particle theory group headed by the late Yoichiro Nambu. Both of the senior Oxford particle theorists, J.C. Taylor and the late Richard Dalitz, recommended applying for a postdoctoral post in the USA because the quality of research was allegedly better there. Frampton received two good offers from Princeton and Chicago. The Proctor fellowship at Princeton seemed more prestigious than a regular postdoctoral position at the University of Chicago, but Dalitz was familiar with the exceptional gifts of Nambu as a former colleague and strongly urged accepting the Chicago offer instead. This advice immediately paid dividends. In late August 1968, just before leaving to Chicago, J.C. Taylor told Frampton about the Veneziano Model which provided an unexpectedly simple closed form solution of sum rules in his DPhil thesis. Upon arriving in Chicago and being greeted by Nambu, he discussed the Veneziano model then to answer a question posed by Nambu, and, through two weeks of very hard work, he determined how to factorise it by using string variables, perhaps his most fruitful piece of work in terms, not of agreeing with Nature, but of influencing many papers in the future. If the string invention had been published in 1968, it would by now have been cited thousands of times. He did publish three other papers with Nambu. Frampton's second postdoc was at CERN as a fellow1970–72 where he continued to work on dual resonance models and collaborated with Brink, Ellis, Goddard, and Nielsen. At age 28, after two postdocs, Frampton was ready for a permanent job back in England. He was a leading candidate for such a job at Imperial College London, but committed career suicide by saying in an interview that he preferred to teach as little as possible. He was offered instead a long-term postdoc there but oddly accepted fixed term jobs at Syracuse University 1972–75, UCLA 1975-77, Ohio State, 1977–78, and finally Harvard University 1978–80, where he began a long and fruitful collaboration with Sheldon Glashow. While at UCLA, when teaching postgraduate quantum field theory, Frampton solved the important problem of vacuum decay using instantons, resulting in a publication in 1976 Physical Review Letters, which a colleague, Gerard 't Hooft, named the "Frampton Disaster."

There was still the issue of a permanent job. However, this problem was solved by attending a meeting at UCLA in 1978 to celebrate the 60th birthday of Julian Schwinger. At the meeting, Frampton was approached by Henk Van Dam, a professor at UNC-Chapel Hill, who raised the possibility of moving permanently to his university. An offer of a tenure-track assistant professorship was made on Frampton's 35th birthday, October 31st 1978. He started as a UNC assistant professor on January 1st, 1981, sharing a DOE grant with Van Dam and Jack Ng. The first postdoc hired for 1981–83, and the best of all 23 was Thomas Kephart, who went on in 1985 to a professorship at Vanderbilt University. In 1983, Frampton and Kephart published a calculation of the hexagon chiral anomaly in ten spacetime dimensions, which was an important precursor of the first superstring revolution. They continued to collaborate with Kephart at Vanderbilt, creating a world-renowned collaboration which produced 45 joint papers of which five have at least 100 citations. In 1995, they introduced the successful flavour symmetry called T-prime. In 1984. Frampton was awarded an advanced DSc degree by Oxford University for the quality of his research papers. In the academic year 1986–87, Frampton was a visiting professor at Boston University where he and Glashow published a model called chiral colour, which predicted an axigluon particle. Although experimental searches revealed that the axigluon does not exist in Nature, the work was important to Frampton in terms of overcoming the psychological barrier to predicting a new particle. In 1992, Frampton published his most famous paper, which predicted the bilepton as an accessible resonance in like-sign leptons, This was the paper most likely to fulfil his headmaster's 1958 prediction about a Nobel prize. In 2002, Frampton published a well-known paper with Glashow and Yanagida about a "FGY" model in which the CP violation in leptogenesis is equal to that in neutrino oscillations. It has been argued that, if oscillation experiments must relate to the conundrum of matter–antimatter asymmetry, Nature must choose the FGY model. This will likely not be known quickly without information about right-handed neutrinos. In 2009-10, Frampton was a distinguished visiting professor at the University of Tokyo. While there, he published two popular papers about an entropic explanation for accelerating cosmic expansion with George Smoot, his third Nobelist coauthor (after Nambu and Glashow).

After retiring from UNC and returning to Oxford at the beginning of 2015, Frampton published a significant paper on the Origin and Nature of Dark Matter, arguing against microscopic constituents in favour of Primordial Intermediate Mass Black Holes (PIMBHs) inside the Milky Way. These PIMBHs could be discovered by microlensing of stars in the Magellanic Clouds. Frampton's 1992 prediction of the bilepton received a welcome boost in January 2021 when, after watching a video of a retirement colloquium at BU by Glashow in which he said his biggest career disappointment had been the failure of SU(5) theory, Frampton suggested to him that it could be explained by the bilepton. Glashow immediately contacted CERN to encourage a bilepton search; this search was started but requires a higher luminosity, which will be available at the High Luminosity HL-LHC in 2030. In 2022, Frampton began research on a topic as different as possible from the bilepton by studying whether the accelerated expansion of the universe may be caused by electromagnetic repulsion between same-sign Primordial Extremely Massive Naked Singularities (PEMNSs) in an EAU-model where the dominant force at the largest cosmological distances is electromagnetic rather than gravitational. Here, EAU means Electromagnetically Accelerating Universe. Preliminary supporting evidence that the EAU model is on the right track might, in the foreseeable future, be provided by successful discovery of PIMBHs in the Milky Way. It will be interesting to see whether the EAU model survives the test of time. Currently, Frampton has published more than 500 papers on particle theory and related topics.

Finally we must thank all the contributors to this Festschrift which commemorates Frampron's 80th birthday.

<div align="right">

**Thomas W. Kephart and Paul Howard Frampton**

*Guest Editors*

</div>

# Particle Physics and Cosmology Intertwined

**Pran Nath**

Department of Physics, Northeastern University, Boston, MA 02115-5000, USA; p.nath@northeastern.edu

**Abstract:** While the standard model accurately describes data at the electroweak scale without the inclusion of gravity, beyond the standard model, physics is increasingly intertwined with gravitational phenomena and cosmology. Thus, the gravity-mediated breaking of supersymmetry in supergravity models leads to sparticle masses, which are gravitational in origin, observable at TeV scales and testable at the LHC, and supergravity also provides a candidate for dark matter, a possible framework for inflationary models and for models of dark energy. Further, extended supergravity models and string and D-brane models contain hidden sectors, some of which may be feebly coupled to the visible sector, resulting in heat exchange between the visible and hidden sectors. Because of the couplings between the sectors, both particle physics and cosmology are affected. The above implies that particle physics and cosmology are intrinsically intertwined in the resolution of essentially all of the cosmological phenomena, such as dark matter and dark energy, and in the resolution of cosmological puzzles, such as the Hubble tension and the EDGES anomaly. Here, we give a brief overview of the intertwining and its implications for the discovery of sparticles, as well as the resolution of cosmological anomalies and the identification of dark matter and dark energy as major challenges for the coming decades.

**Keywords:** particle physics; cosmology

## 1. Introduction

This article is a contribution to Paul Frampton's 80th birthday volume, marking his over five decades of contributions as a prolific researcher to theoretical physics. He is one of the few theoretical physicists who recognized early on that there is no boundary between particle physics and cosmology and contributed freely to each in good measure. His prominent works include those in particle theory, such as those related to physics beyond the standard model and anomaly cancellations in higher dimensions, and in cosmology, such as those focusing on non-standard cosmological models and black-hole physics. Since particle physics and cosmology are the two major areas of his work, this paper elaborates on the progressive intertwining of the fields of particle physics and cosmology over the past several decades from the author's own perspective.

For a long period of time, up to and including the period of the emergence of the standard model [1–7] and its tests, it was largely accepted that gravity could be ignored in phenomena related to particle physics. The contrary, of course, was not true, as particle physics was already known to be central to a variety of astrophysical phenomena, such as the Chandrasekhar limit [8] and the synthesis of elements in the work of B²FH [9] and Peebles [10]. For particle physics, gravity became more relevant with the emergence of supersymmetry, supergravity, and strings. Further, supergravity models with the gravity-mediated breaking of supersymmetry lead to soft terms that allow for the radiative breaking of electroweak symmetry and predict sparticles observable at colliders. There is another aspect of supergravity and strings that has a direct impact on particle physics. In extended supergravity, string, and D-brane models, one finds hidden sectors that can couple feebly with the visible sector and affect particle physics phenomena observable at colliders and that also have implications for cosmology, as they can provide candidates for inflation, dark matter, and dark energy. Thus, with the emergence of supergravity and strings, a deeper connection between particle physics and cosmology has emerged. Of course, one hopes that particle physics and cosmology are parts of strings, and significant literature exists on the particle-physics–string connection (see, e.g., [11–15] and references therein) and on the cosmology–string connection (see, e.g., [16–18] and references therein).

In this paper, we will focus on the intertwining of particle physics and cosmology. As noted above, this intertwining has occurred on two fronts: first, in supergravity models with gravity-mediated breaking, the sparticle spectra are direct evidence that gravitational interactions are at work even at the scale of electroweak physics. Further, supergravity models with R-parity conservation lead to a candidate for dark matter, specifically a neutralino [19]. Most often, it turns out to be the lightest supersymmetric particle in the radiative breaking of electroweak symmetry [20]. The neutralino as dark matter importantly enters simulations of cosmological evolution. At the same time, supergravity provides models for the inflationary expansion of the universe. Second, as noted above, in extended supergravity and string models, one finds hidden sectors, some of which may be feebly coupled to the visible sector. Typically, the hidden sectors and the visible sector will have different temperatures, but they have heat exchange, which requires the synchronous evolution of the two sectors, thus intertwining the two and affecting both particle physics and cosmology. The outline of the rest of the paper is as follows. In Section 2, we will discuss the implications of the gravity-mediated breaking of supergravity at low energy, and Section 3 focuses on the intertwining of particle physics and cosmology via hidden sectors.

## 2. Gravitational Imprint on Particle Physics at the Electroweak Scale

As noted above, until the advent of Sugra, it was the prevalent view that gravity did not have much of a role in particle physics models. However, with the advent of supergravity grand unification [21,22], where supersymmetry is broken in the hidden sector and communicated to the visible sector through gravitational interactions, one finds that soft breaking terms are dependent on gravitational interactions [21,23,24]. Thus, the soft mass of scalars in the visible sector $m_s \propto \kappa m^2$, where $\kappa = \sqrt{8\pi G_N}$ and $G_N$ is Newton's constant, and $m$ is an intermediate hidden sector mass. Here, with $m \sim 10^{10}$ and $M_{Pl} = \kappa^{-1} = 2.43 \times 10^{18}$ GeV (in natural units: $\hbar = c = 1$), one finds $m_s$ to be of electroweak size. Since sparticle masses are controlled by the soft susy scale, the discovery of sparticles would be a signature indicating that gravity has a role in low-energy physics. This would very much be akin to the discovery that the $W$ and $Z$ bosons are a reflection of $SU(2)_L \times U(1)_Y$ unification. It is notable that the soft terms are also responsible for generating the spontaneous breaking of electroweak symmetry [21,25]. An indication that some of the sparticles may be low-lying comes from the $g - 2$ data from Fermilab [26], which point to a deviation from the standard-model prediction of about $4\sigma$. An attractive proposition is that the deviation arises from light-sparticle exchange, specifically light charginos and light sleptons (see, e.g., [27–29] and the references therein), a deviation that was predicted quite a while ago [30]. However, a word of caution is in order, in that the lattice analysis [31] for the hadronic vacuum polarization contribution gives a smaller deviation from the standard model than the conventional result, where the hadronic vacuum polarization contribution is computed using $e^+e^- \rightarrow \pi^+\pi^-$ data. Thus, further work is needed to reconcile the lattice analysis with the conventional result on the hadronic polarization contribution before drawing any definitive conclusions.

## 3. Hidden Sectors Intertwine Particle Physics and Cosmology

As already noted, in a variety of models beyond standard-model physics, which include extended supergravity models, string models, and extra-dimension models, one has hidden sectors. While these sectors are neutral under the standard-model gauge group, they may interact with the visible sectors via feeble interactions. Such feeble interactions can occur via a variety of portals, which include the Higgs portal [32], kinetic energy portal [33,34], Stueckelberg mass-mixing portal [35,36], kinetic-mass-mixing portal [37], and Stueckelberg–Higgs portal [38], as well as possible higher-dimensional operators. The hidden sectors could be endowed with gauge fields, as well as with matter. At the reheat temperature, the hidden sectors and the visible sector would, in general, lie in different heat baths. However, because of the feeble interactions between the sectors, there will be heat exchange between the visible and hidden sectors, and thus, their thermal evolution will be correlated. The evolution of the relative temperatures of the two sectors then depends on the initial conditions, specifically on the ratio $\xi(T) = T_h/T$ at the reheat temperature, where $T_h$ is the hidden-sector temperature, and $T$ is the visible-sector temperature. The Boltzmann equations governing the evolution of the visible and hidden sectors are coupled and involve the evolution equation for $\xi(T)$. Such an equation was derived in [39–41] and applied in a variety of settings in [42], consistent with all experimental constraints on hidden-sector matter from terrestrial and astrophysical data [43]. It is found that hidden sectors can affect observable phenomena in the visible sector, such as the density of thermal relics. Hidden sectors provide candidates for dark matter and dark energy and help resolve cosmological anomalies intertwining particle physics and cosmological phenomena. We discuss some of these topics in further detail below.

Green–Schwarz [44] found that in the low-energy limit of Type I strings, the kinetic energy of the two-tensor $B_{MN}$ of a 10D supergravity multiplet has Yang–Mills and Lorentz-group Chern–Simons terms (indicated by superscripts Y and L) so that $\partial_{[P}B_{MN]} \rightarrow \partial_{[P}B_{MN]} + \omega_{PMN}^{(Y)} - \omega_{PMN}^{(L)}$, where $M$, $N$, and $P$ are 10-dimensional indices. The inclusion of the Chern–Simons terms fully requires that one extend the 10D Sugra Lagrangian to order $O(\kappa)^2$. This was accomplished subsequent to Green-Schwarz's work in [45] (for related works, see [46–48]). Dimensional reduction to 4D with a vacuum expectation value for the internal-gauge-field strength, $\langle F_{ij} \rangle \neq 0$ (where the indices are for the six-dimensional compact manifold), leads to $\partial_\mu B_{ij} + A_\mu F_{ij} + \cdots \sim \partial_\mu \sigma + m A_\mu$ ($\mu$ in an index for four-dimensional Minkowskian space–time), where the internal components $B_{ij}$ give the pseudo-scalar $\sigma$, and $m$ arises from $< F_{ij} >$, which is a topological quantity, related to the Chern numbers of the gauge bundle. Thus, $A_\mu$ and $\sigma$ have a Stueckelberg coupling of the form $A_\mu\partial^\mu\sigma$. This provides the inspiration for building BSM theories with the Stueckelberg mechanism [35,36,49–51]. Specifically, this allows for the possibility of writing effective theories with gauge-invariant mass terms. For the case of a single $U(1)$ gauge field $A_\mu$, one may write a gauge-invariant mass term by letting $A_\mu \rightarrow A_\mu + \frac{1}{m}\partial_\mu\sigma$, where the gauge transformations are defined so that $\delta A_\mu = \partial_\mu \lambda$ and $\delta\sigma = -m\lambda$. In this case, $\sigma$'s role is akin to that of the longitudinal component of a massive vector. The above technique also allows one to generate invariant mass mixing between two $U(1)$ gauge fields. Thus, consider two gauge groups $U(1)_X$ and $U(1)_Y$ with gauge fields $A_\mu$ and $B_\mu$ and an axionic field $\sigma$. In this case, we can write a mass term $(m_1 A_\mu + m_2 B_\mu + \partial_\mu\sigma)^2$ that is invariant under $\delta_x A_\mu = \partial_\mu \lambda_x$, $\delta_x \sigma = -m_1 \lambda_x$ for $U(1)_X$, and $\delta_y B_\mu = \partial_\mu \lambda_y$, $\delta_y \sigma = -m_2 \lambda_y$ for $U(1)_y$. One of the interesting phenomena associated with effective gauge theories with gauge-invariant mass terms is that they generate millicharges when coupled to matter fields [35,37,49,52]. We will return to this feature of the Stueckelberg mass-mixing terms when we discuss the EDGES anomaly.

Hubble tension: Currently, there exists a discrepancy between the measured value of the Hubble parameter $H_0$ for low redshifts ($z < 1$) and high redshifts ($z > 1000$). Thus, for ($z < 1$), an analysis of data from Cepheids and SNIa gives [53] $H_0 = (73.04 \pm 1.04)$ km/s/Mpc. On the other hand, an analysis based on the $\Lambda CDM$ model by the SH0ES Collaboration [53] using data from the cosmic microwave background (CMB), Baryon Acoustic Oscillations (BAOs), and Big Bang Nucleosynthesis (BBN) determines the Hubble parameter at high $z$ to be [54] $H_0 = (67.4 \pm 0.5)$ km/s/Mpc. This indicates a $5\sigma$-level tension between the low-$z$ and the high-$z$ measurements. There is a significant amount of literature attempting to resolve this puzzle, at least partially, and recent reviews include [55,56]. One simple approach is introducing extra relativistic degrees of freedom during the period of recombination, which increases the magnitude of $H_0$, which helps alleviate the tension. Models using this idea introduce extra particles, such as the $Z'$ of an extra $U(1)$ gauge field that decays to neutrinos [57,58], or utilize other particles, such as the majoron [59,60]. The inclusion of extra degrees of freedom, however, must be consistent with the BBN constraints, which are sensitive to the addition of massless degrees of freedom. Thus, the standard-model prediction of $N_{\text{eff}}^{\text{SM}} \simeq 3.046$ [61] is consistent with the synthesis of light elements, and the introduction of new degrees of freedom must maintain this successful standard-model prediction. The above indicates that the extra degrees of freedom should emerge only beyond the BBN time and in the time frame of the recombination epoch. It is noted that new degrees of freedom are also constrained by the CMB data, as given by the Planck analysis [54].

A cosmologically consistent model based on the Stueckelberg extension of the SM with a hidden sector was proposed in [62] for alleviating the Hubble tension. The model is cosmologically consistent since the analysis is based on a consistent thermal evolution of the visible and hidden sectors, taking account of the thermal exchange between the two sectors. In addition to dark fermions and dark photons, the model also contains a massless pseudo-scalar particle field $\phi$ and a massive long-lived scalar field $s$. The fields $\phi$ and $s$ have interactions only in the dark sector, with no interactions with the standard-model fields. The decay of the scalar field occurs after BBN, close to the recombination time, via the decay $s \to \phi\phi$, which provides the extra degrees of freedom needed to alleviate the Hubble tension. It should be noted that the full resolution of the Hubble tension would require going beyond providing new degrees of freedom and would involve a fit to all of the CMB data that are consistent with all cosmological and particle physics constraints. For some recent related work on the Hubble tension, see [63–68].

EDGES anomaly: The 21 cm line plays an important role in the analysis of physics during the dark ages and the cosmic dawn in the evolution of the early universe. The 21 cm line arises from the spin transition from the triplet state to the singlet state and vice versa in the ground state of neutral hydrogen. The relative abundance of the triplet and singlet states defines the spin temperature $T_s$ (and $T_B = T_s$) of hydrogen gas and is given by $n_1/n_0 = 3e^{-T_*/T_s}$, where 3 is the ratio of the spin degrees of freedom for the triplet versus the singlet state, $T_*$ is defined by $\Delta E = kT_*$, where $\Delta E = 1420$ MHz is the energy difference at rest between the two spin states, and $T_* \equiv \frac{hc}{k\lambda_{21cm}} = 0.068$ K. EDGES (the Experiment to Detect the Global Epoch of Reionization Signature) reported an absorption profile centered at the frequency $\nu = 78$ MHz in the sky-averaged spectrum. The quantity of interest is the brightness temperature $T_{21}$ of the 21 cm line defined by $T_{21}(z) = (T_s - T_\gamma)(1 - e^{-\tau})/(1 + z)$, where $T_\gamma(z)$ is the photon temperature at redshift $z$, and $\tau$ is the optical depth for the transition. The analysis of Bowman et al. [69] finds (see, however, reference [70] on concerns regarding the modeling of data) that at redshift $z \sim 17$, $T_{21} = -500^{+200}_{-500}$ mK at 99% C.L. On the other hand, the analysis of [71] based on the $\Lambda$CDM model gives a $T_{21}$ of around $-230$ mK, which shows that the EDGES result is a $3.8\sigma$ deviation away from that of the standard cosmological paradigm.

The EDGES anomaly is not yet confirmed, but pending its possible confirmation, it is of interest to investigate what possible explanations there might be. In fact, several mechanisms have already been proposed to explain the $3.8\sigma$ anomaly [72–88]. A list of some of the prominent possibilities consists of the following: (1) astrophysical phenomena, such as radiation from stars and star remnants; (2) a hotter CMB background radiation temperature than expected; (3) cooler baryons than what $\Lambda$CDM predicts; (4) the modification of cosmological evolution: the inclusion of dark energy such as Chapligin gas. Of the above, there appears to be a leaning toward baryon cooling, and there is a substantial amount of work in this area following the earlier works of [89] and Barkana [78]. Specifically, it was pointed out in [78] that the observed anomaly could be explained if the baryons were cooled down by roughly 3 K. Here, one assumes that a small percentage of DM ($\sim$0.3%) is millicharged and that baryons become cooler through Rutherford scattering from the colder dark matter. As mentioned earlier, precisely such a possibility occurs via Stueckelberg mass mixing if we assume that one of the gauge fields $U(1)_Y$ is the hyper-charge gauge field, while $U(1)_X$ is a hidden-sector field, and the millicharged dark matter resides in the hidden sector, while the rest of the dark matter could be WIMPS. Within this framework, a cosmologically consistent analysis of a string-inspired millicharged model was proposed in [90], where a detailed fit to the data is possibly consistent within a high-scale model. For some recent work on the EDGES anomaly, see [91–94].

Inflation: As is well known, the problems associated with the Big Bang, such as the flatness, horizon, and monopole problems, are resolved in inflationary models. In models of this type, quantum fluctuations at the horizon exit encode information regarding the characteristics of the inflationary model that can be extracted from the cosmic microwave background (CMB) radiation anisotropy [95–99]. In fact, data from the Planck experiment [100–102] have already put stringent bounds on inflationary models, eliminating some. A model proposed in [103,104] is based on an axionic field with a potential of the form $V(a) = \Lambda^4 \left(1 + \cos\left(\frac{a}{f}\right)\right)$, where $a$ is the axion field and $f$ is the axion decay constant. However, for the simple model above to hold, the Planck data require $f > 10 M_{Pl}$, which is undesirable since string theory indicates that $f$ lies below $M_{Pl}$ [105,106]. However, a reduction in $f$ turns out to be a non-trivial issue. The techniques used to resolve this issue include the alignment mechanism [107,108], n-flation, coherent enhancement [109], and models using shift symmetry (for a review and more references, see [110,111]).

We mention another inflation model, which is based on an axion landscape with a $U(1)$ symmetry [112]. This model involves $m$ pairs of chiral fields, and the fields in each pair are oppositely charged under the same $U(1)$ symmetry. Our nomenclature is such that we label the pseudo-scalar component of each field as an axion and the corresponding real part as a saxion. Since the model has only $U(1)$ global symmetry, the breaking of the global symmetry leads to just one pseudo-Nambu–Goldstone boson (PNGB), and the remaining pseudo-scalars are not PNGBs. Thus, the superpotential of the model consists of a part that is invariant under the $U(1)$ global symmetry and a $U(1)$-symmetry-breaking part that simulates instanton effects. The analysis of this work shows that the potential contains a fast-roll–slow-roll-splitting mechanism, which splits the axion potential into fast-roll and slow-roll parts, where the fields entering the fast roll are eliminated early on, leaving the slow-roll part, which involves a single axion field that drives inflation. Here, under the constraints of stabilized saxions, one finds inflation models with $f < M_{Pl}$ to be consistent with the Planck data. Similar results are found in the Dirac–Born–Infeld-based models [113].

Dark energy: One of the most outstanding puzzles of both particle physics and of cosmology is dark energy, which constitutes about 70% of the energy budget of the universe and is responsible for the accelerated expansion of the universe. Dark energy is characterized by negative pressure such that $w$, defined by $w = p/\rho$, where $p$ is the pressure and $\rho$ is the energy density for dark energy, must satisfy $w < -1/3$. The CMB and the BAO data fit well with a cosmological constant $\Lambda$ that corresponds to $w = -1$. Thus, the Planck Collaboration [54] gives $w = -1.03 \pm 0.03$, consistent with the cosmological constant. There are two puzzles connected with dark energy. First, the use of the cosmological constant appears artificial, and it is desirable to replace it with a dynamical field, i.e., a so-called quintessence field (for a review, see [114]), which, at late times, can generate accelerated expansion similar to that given by $\Lambda$. The second problem relates to the very small size of the cosmological constant, which is not automatically resolved by simply replacing $\Lambda$ with a dynamical field. The extreme fine-tuning needed in a particle physics model to get to the size of $\Lambda$ requires a new idea, such as vacuum selection in a landscape with a large number of possible allowed vacua [115], for instance, those available in string theory. In any case, it is an example of the extreme intertwined nature of cosmology and particle physics. However, finding a quintessence solution that replaces $\Lambda$ and is consistent with all of the CMB data is itself progress. Regarding experimental measurement of $w = -1.03 \pm 0.03$, if more accurate data in the future give $w > -1$, it would point to something like quintessence, while $w < -1$ would indicate phantom energy and an entirely new sector.

## 4. Conclusions

In conclusion, it is clear that particle physics and cosmology are deeply intertwined, and in the future, models of physics beyond the standard model will be increasingly constrained by particle physics experiments as well as by astrophysical data. We congratulate

Paul for his notable contributions in the twin fields and wish him many productive years of contributions for the future.

**Funding:** PN is supported in part by the NSF Grant PHY-2209903.

**Data Availability Statement:** Data is contained within the article.

**Conflicts of Interest:** The author declares no conflicts of interest.

*Article*

# Bilocal Field Theory for Composite Scalar Bosons

Christopher T. Hill

Particle Theory Department, Fermi National Accelerator Laboratory, P. O. Box 500, Batavia, IL 60510, USA; hill@fnal.gov

**Abstract:** We give a bilocal field theory description of a composite scalar with an extended binding potential that reduces to the Nambu–Jona-Lasinio (NJL) model in the pointlike limit. This provides a description of the internal dynamics of the bound state and features a static internal wave function, $\phi(\vec{r})$, in the center-of-mass frame that satisfies a Schrödinger–Klein–Gordon equation with eigenvalues $m^2$. We analyze the "coloron" model (single perturbative massive gluon exchange) which yields a UV completion of the NJL model. This has a BCS-like enhancement of its interaction, $\propto N_c$ the number of colors, and is *classically critical* with $g_{critical}$ remarkably close to the NJL quantum critical coupling. Negative eigenvalues for $m^2$ lead to spontaneous symmetry breaking, and the Yukawa coupling of the bound state to constituent fermions is emergent.

**Keywords:** bound states; compositeness; coloron

## 1. Introduction

Many years ago, Yukawa proposed a multilocal field theory for the description of relativistic bound states [1–3]. For a composite scalar field, consisting of a pair of constituents, he introduced a complex bilocal field, $\Phi(x, y)$. This is factorized, $\Phi(x, y) \rightarrow \chi(X)\phi(r)$ where $X^\mu = (x^\mu + y^\mu)/2$ where $r^\mu = (x^\mu - y^\mu)/2$, and $\chi(X)$ describes the center-of-mass motion like any conventional point-like field. Then, $\phi(r)$ describes the internal structure of the bound state. The formalism preserves Lorentz covariance, though we typically "gauge fix" to the center-of-mass frame, and Lorentz covariance is then not manifested. Here, we must confront the issue of "relative time".

Each of the constituent particles in a relativistic bound state carries its own local clock, e.g., $x^0$ and $y^0$. These are, in principle, independent; so, the question "how can a description of a multi-particle bound state be given in a quantum theory with a single time variable, $X^0$?" arises. To answer this, Yukawa introduced an imaginary "relative time" $r^0 = (x^0 - y^0)/2$, but this did not seem to be effective and is an element of his construction we will abandon.

A bilocal field theory formalism can be constructed in an action by considering general properties of free field bilocal actions. However, we can "derive" the bilocal theory from a local constituent field theory by matching the conserved currents of the composite theory with those of the constituent theory. This leads to the removal of relative time, which then becomes associated with canonical normalization of the constituent fields $\chi$ and $\phi$. In the center-of-mass frame, the internal wave function reduces to a static field, $\phi(\vec{r})$, where $\vec{r} = (\vec{x} - \vec{y})/2$. The approach yields a fairly simple solution to the problem of relative time, matching the conclusions one obtains from the elegant Dirac Hamiltonian constraint theory [4–10]. The resulting $\phi(\vec{r})$ then appears as a straightforward result.

After first considering a bosonic construction, we apply this to a theory of chiral fermions with an extended interaction mediated by a perturbative massive gluon, i.e., the "coloron model" [11–16]. This provides a UV completion for the Nambu–Jona-Lasinio (NJL) model [17,18], which is recovered in the point-like limit, $\vec{r} \rightarrow 0$. This leads to an effective (mass)$^2$ Yukawa potential with coupling $g$. We form bound states with mass $m^2$, determined as the eigenvalue of a static Schrödinger–Klein–Gordon (SKG) equation for the internal wave function $\phi(\vec{r})$.

A key result of this analysis leads to a departure from the usual NJL model: the coloron model has a nontrivial *classical critical behavior*, $g > g_c$, leading to a bound state with a negative $m^2$. The classical interaction is analogous to the Fröhlich Hamiltonian interaction in a superconductor and has a BCS-like enhancement of the coupling by a factor of $N_c$ (number of colors) [19,20]. Remarkably, we find the classical $g_c$ is numerically close to the NJL critical coupling constant which arises in fermion loops.

The scalar bound state develops an effective Yukawa coupling to its constituent fermions, distinct from $g$, that is emergent in the theory. In the point-like limit this matches the NJL coupling when near criticality. However, in general, this depends in detail upon the internal wave function $\phi(\vec{r})$ and potential. In the point-like limit this is determined by $\phi(0)$ and we recover the NJL model. However, if we are far from the point-like limit in an extended wave function $\phi(\vec{r})$ might suppress the emergent Yukawa coupling, even though the coloron coupling $g$ is large.

The description of a relativistic bound state in the rest frame is similar to the eigenvalue problem of the nonrelativistic Schrödinger equation and some intuition carries over. However, the eigenvalue of the static Schrödinger–Klein–Gordon (SKG) equation is $m^2$ rather than energy. Hence, a bound state with positive $m^2$ is a resonance that can decay to its constituents and has a Lorentz line-shape in $m^2$, and thus has a large distance radiative component to its solution that represents incoming and outgoing open scattering states.

If the eigenvalue for $m^2$ is negative or tachyonic; contrary to the non-relativistic case, the bound state represents a chiral vacuum instability. This then requires consideration of a quartic interaction of the composite field, $\sim \lambda \Phi^4$, which is expected to be generated by the loops in the underlying theory. We treat this phenomenologically in the present paper. In the broken symmetry phase, the composite field $\Phi(x, y)$ acquires a vacuum expectation value (VEV), $\langle \Phi \rangle = v$. In the perturbative quartic coupling limit ($\lambda$), in the broken phase, $\phi(\vec{r})$ remains localized and the Nambu–Goldstone modes and Brout–Englert–Higgs (BEH) boson retain the common localized solution for their internal wave functions.

## 2. Constructing a Bilocal Composite Theory

### 2.1. Brief Review of the NJL Model

The Nambu–Jona-Lasinio model (NJL) [17,18] is the simplest field theory of a composite scalar boson, consisting of a pair of chiral fermions. A bound state emerges from an assumed point-like four-fermion interaction and is described by local effective field, $\Phi(x)$. The effective field arises as an auxiliary field from the factorization of the four-fermion interaction. In the usual formulation of the NJL model, chiral fermions induce loop effects in a leading large $N_c$ limit which, through the renormalization group, leads to interesting dynamic phenomena at low energies. We present a brief review of this.

We assume chiral fermions, each with $N_c$ "colors" labeled by $(a, b, \ldots)$. A non-confining, point-like chirally invariant $U(1)_L \times U(1)_R$ interaction then takes the form:

$$S_{NJL} = \int d^4x \left( i \overline{\psi}_L^a(x) \overline{\partial} \, \psi_{aL}(x) + i \overline{\psi}_R^a(x) \overline{\partial} \, \psi_{aR}(x) \right.$$
$$\left. + \frac{g^2}{M_0^2} \overline{\psi}_L^a(x) \psi_{aR}(x) \, \overline{\psi}_R^b(x) \psi_{bL}(x) \right) \tag{1}$$

This can be readily generalized to a $G_L \times G_R$ chiral symmetry. We then factorize Equation (1) by introducing the local auxiliary field $\Phi(x)$ and write for the interaction:

$$\int d^4x \left( g \overline{\psi}_L^a(x) \psi_{aR}(x) \Phi(x) + h.c. - M_0^2 \Phi^\dagger(x) \Phi(x) \right) \tag{2}$$

We view Equation (2) as the action defined at the high scale $\mu \sim M$. Then, following [21], we integrate out the fermions to obtain the effective action for the composite field $\Phi$ at a lower scale $\mu \ll M$.

The calculation in the large-$N_c$ limit and full renormalization group is discussed in detail in [22–24]. The leading $N_c$ fermion loop yields the result:

$$L_M \to L_\mu = g[\overline{\psi}_R \psi_L]\Phi + h.c + Z\partial_\mu \Phi^\dagger \partial^\mu \Phi$$
$$-m^2 \Phi^\dagger \Phi - \frac{\lambda}{2}(\Phi^\dagger \Phi)^2 \tag{3}$$

where

$$m^2 = M_0^2 - \frac{N_c g^2}{8\pi^2}(M_0^2 - \mu^2)$$
$$Z = \frac{N_c g^2}{8\pi^2}\ln(M_0/\mu), \quad \lambda = \frac{N_c g^4}{4\pi^2}\ln(M_0/\mu). \tag{4}$$

Here, $M_0^2$ is the UV loop momentum cut-off, and we include the induced kinetic and quartic interaction terms. The one-loop result can be improved by using the full renormalization group [22–24]. Hence, the NJL model is driven by fermion loops, which are $\propto \hbar$ intrinsically quantum effects.

Note the behavior of the composite scalar boson mass, $m^2$, of Equation (4) in the UV. The $-N_c g^2 M_0^2/8\pi^2$ term arises from the negative quadratic divergence in the loop involving the pair $(\psi_R, \psi_L)$ of Figure 1, with UV cut-off $M_0^2$. Therefore, the NJL model has a critical value of its coupling defined by the cancellation of the large $M_0^2$ terms for $\mu^2 = 0$

$$g_{c0}^2 = \frac{8\pi^2}{N_c} + \mathcal{O}\left(\frac{\mu^2}{M^2}\right) \tag{5}$$

Note that $\mu$ is the running RG mass and comes from the lower limit of the loop integrals and breaks scale invariance and can, in principle, be small. For super-critical coupling, $g > g'_c$, we see that $m^2 < 0$ and there will be a vacuum instability. The effective action, with a $\lambda|\Phi|^4$ term, is then the usual sombrero potential. The chiral symmetry is spontaneously broken, the chiral fermions acquire mass, and the theory generates Nambu–Goldstone bosons. Fine-tuning of $g^2 \approx g_c^2$ is possible if we want a theory with a hierarchy, $|m^2| \ll M_0^2$.

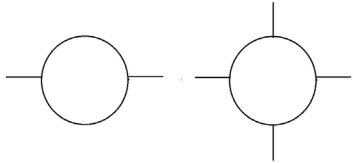

**Figure 1.** Diagrams contributing to the point-like NJL model effective Lagrangian, Equations (2) and (4). External lines are $\Phi$ and internal lines are fermions.

### 2.2. Construction of Bilocal Compositeness in a Local Scalar Field Theory

Presently, we obtain a theory of bound states by bilocal fields in a Lorentz invariant model, consisting of a point-like complex scalar field and an interaction mediated by a point-like real field (in Section 3, we extend this to a chiral fermion interaction via a massive gauge field, analogous to a heavy gluon, aka "coloron"; in Appendix A we give a summary of notation and formulas). Our present treatment will be semi-classical.

Consider local scalar fields $\varphi(x)$ (complex) and $A(x)$ (real) and action:

$$S = \int_x \left(|\partial\varphi|^2 + \frac{1}{2}(\partial A)^2 - \frac{1}{2}M^2 A^2 - gM|\varphi|^2 A - \frac{\lambda}{2}|\varphi|^4\right) \tag{6}$$

where we abbreviate $|\partial\varphi|^2 = \partial_\mu \varphi^\dagger \partial^\mu \varphi$ and $(\partial A)^2 = \partial_\mu A \partial^\mu A$. Here, $g$ is dimensionless and we refer all mass scales to the single scale $M$. We will discuss the quartic term separately below, and presently set it aside, $\lambda = 0$.

If we integrate out $A$, we obtain an effective, attractive, bilocal potential interaction term at leading order in $g^2$,

$$S = \int_x |\partial\varphi|^2 + \frac{g^2 M^2}{2} \int_{xy} \varphi^\dagger(y)\varphi(y) D_F(y-x)\varphi^\dagger(x)\varphi(x) \tag{7}$$

where the two-point function is given by $(i)\times$ the Feynman propagator,

$$D_F(x-y) = -\int \frac{e^{iq_\mu(x^\mu - y^\mu)}}{(q^2 - M^2)} \frac{d^4q}{(2\pi)^4} \tag{8}$$

The equation of motion of $\varphi$ is therefore

$$\partial^2 \varphi(x) - g^2 M^2 \int_y \varphi^\dagger(y)\varphi(x) D_F(x-y)\varphi(y) = 0 \tag{9}$$

(note we have transposed $\varphi^\dagger(y)$ and $\varphi(y)$ under the integral). In the action in Equation (7), the kinetic term is still local while the interaction is bilocal, and the theory is still classical in that this only involved a tree diagram that is $\mathcal{O}(\hbar^0)$.

We now define a bilocal field of mass dimension $d = 1$

$$\Phi(y,x) = M^{-1}\varphi(y)\varphi(x). \tag{10}$$

The free particle states described by the bilocal field trivially satisfy an equation of motion and a symmetry

$$\partial_x^2 \Phi(x,y) = 0 \qquad \Phi(x,y) = \Phi(y,x) \tag{11}$$

and this is generated by a bilocal action

$$S = M^4 \int_{xy} Z |\partial_x \Phi(x,y)|^2 \tag{12}$$

where we will specify the normalization, $Z$, and scale $M$ subsequently (we discuss the general properties of the bilocal fields and actions in Appendices B.2 and B.3). With the bilocal field the interaction of Equation (7), it becomes

$$\frac{g^2 M^4}{2} \int_{xy} \Phi^\dagger(x,y) D_F(x-y)\Phi(x,y) \tag{13}$$

We can therefore postulate a bilocalized action as a free particle part plus the interaction

$$S = \int_{xy} \left( Z M^4 |\partial_x \Phi(x,y)|^2 + \frac{1}{2} g^2 M^4 \Phi^\dagger(x,y) D_F(x-y)\Phi(x,y) \right) \tag{14}$$

In the limit $g = 0$, the field $\Phi(x,y)$ and the action faithfully represents two-particle kinematics, and we have the equation of motion

$$0 = Z \partial_x^2 \Phi(x,y) - \frac{1}{2} g^2 D_F(x-y)\Phi(x,y) \tag{15}$$

We see that a $U(1)$ conserved Noether current is generated by $\Phi(x,y) \to e^{i\theta(x)}\Phi(x,y)$

$$J_{\Phi\mu}(x) = iZ \int d^4y \left( \Phi^\dagger(x,y) \overset{\leftrightarrow}{\frac{\partial}{\partial x^\mu}} \Phi(x,y) \right) \tag{16}$$

where $A \overset{\leftrightarrow}{\partial} B = A\partial B - (\partial A)B$. This must match the conserved $U(1)$ current in the constituent theory

$$J_{\varphi\mu}(x) = i\varphi^\dagger(x) \overset{\leftrightarrow}{\frac{\partial}{\partial x^\mu}} \varphi(x) \tag{17}$$

Substituting Equation (10) into $J_{\Phi\mu}(x)$, we see that the matching requires

$$J_{\Phi\mu}(x) = J_{\varphi\mu}(x) \; ZM^2 \int d^4y |\varphi(y)|^2 \tag{18}$$

Hence

$$1 = ZM^2 \int d^4y \; |\varphi(y)|^2 \tag{19}$$

This is a required constraint for the bound state sector of the theory. Note that the square of the constraint is the four-normalization of $\Phi$

$$1 = Z^2M^4 \int d^4y \; d^4y \; |\varphi(y)|^2|\varphi(x)|^2$$
$$= Z^2M^6 \int d^4y \; d^4y \; |\Phi(x,y)|^2 \tag{20}$$

This implies that the presence of a correlation in the two-particle sector, $\Phi(x,y)$, acts as a constraint on the single particle action in that sector. We can now see how the underlying $\varphi$ action of Equation (7) leads to the $\Phi$ action by inserting the constraint of Equation (19) into the kinetic term of Equation (7) and rearranging to obtain

$$S = \int_{xy} \Bigg( ZM^2|\varphi(y)\partial_x\varphi(x)|^2$$
$$+ \frac{g^2M^2}{2} \varphi^\dagger(y)\varphi(x)D_F(x-y)\varphi^\dagger(x)\varphi(y) \Bigg) \tag{21}$$

and $S$ remains dimensionless. With the bilocal field of Equation (10) the bilocalized action Equation (21) becomes Equation (14).

Following Yukawa, we go to barycentric coordinates $(X, r)$

$$X = \frac{1}{2}(x+y), \qquad r = \frac{1}{2}(x-y). \tag{22}$$

where $r^\mu = (r^0, \vec{r})$, where $\vec{r}$ is the radius and $r^0$ is the relative time (Yukawa preferred to write things in terms of $\rho = 2r$, which has the advantage of a unit Jacobian, $\int d^4x d^4y = \int \int d^4X d^4\rho$ with $J = 1$. We find that the radius, $r$, is more convenient in loop calculations and derivatives are symmetrical, $\partial_{X,r} = (\partial_x \pm \partial_y)/2$ vs. $= \frac{1}{2}\partial_X + \partial_\rho$, but require the Jacobian. See Appendix A for a summary of notation).

Hence, we write

$$\Phi(x,y) = \Phi(X+r, X-r) \equiv \Phi(X, r) \tag{23}$$

Let $S = S_K + S_P$ and we can then rewrite the kinetic term, $S_K$, using the derivative $\partial_x = \frac{1}{2}(\partial_X + \partial_r)$

$$S_K = \frac{JM^4}{4} \int_{Xr} Z|(\partial_X + \partial_r)\Phi(X,r)|^2$$
$$= \frac{JM^4}{4} \int_{Xr} \Bigg( Z|\partial_X\Phi|^2 + Z|\partial_r\Phi|^2 + Z(\partial_X\Phi^\dagger\partial_r\Phi + h.c.) \Bigg) \tag{24}$$

Note the Jacobian $J = 16$

$$J^{-1} = \left| \frac{\partial(X,r)}{\partial(x,y)} \right| = \left( \frac{1}{2} \right)^4. \tag{25}$$

Likewise, the potential term is

$$S_P = \frac{JM^4}{2} \int_{Xr} g^2 D_F(2r)|\Phi(X,r)|^2. \tag{26}$$

We will treat the latter term in Equation (24) $Z(\partial_X \Phi^\dagger \partial_r \Phi + h.c.)$, as a constraint, with its contribution to the equation of motion

$$\frac{\partial}{\partial X^\mu} \frac{\partial}{\partial r_\mu} \Phi = 0 \tag{27}$$

We can redefine this term in the action as a Lagrange multiplier while preserving Lorentz invariance

$$\rightarrow \int_{Xr} \eta \left( \frac{\partial \Phi^\dagger}{\partial X^\mu} \frac{\partial \Phi}{\partial r_\mu} + h.c. \right)^2 \text{ hence, } \delta S/\delta \eta = 0 \tag{28}$$

which also enforces the constraint on a path integral in analogy to gauge fixing. In the following, we assume the constraint is present in the total action but not written explicitly. We therefore have the bilocal action with the constraint understood:

$$S = S_K + S_P =$$
$$\frac{JM^4}{4} \int_{Xr} (Z|\partial_X \Phi|^2 + Z|\partial_r \Phi|^2 + 2g^2 D_F(2r)|\Phi(X,r)|^2) \tag{29}$$

Following Yukawa, we factorize $\Phi$ (these factorized solutions form a complete set of basis functions)

$$\sqrt{J/4}\ \Phi(X,r) = \chi(X)\phi(r) \tag{30}$$

where $\phi$ is the internal wave function which we define to be dimensionless, $d = 0$, while $\chi$ is an ordinary local field with mass dimension $d = 1$. $\chi(X)$ determines the center-of-mass motion of the composite state. The full action for the factorized field takes the form

$$S = M^4 \int_{Xr} \Big( Z|\partial_X \chi|^2 |\phi^2|$$
$$+ |\chi|^2 (Z|\partial_r \phi|^2 + 2g^2 D_F(2r)|\phi(r)|^2) \Big) \tag{31}$$

The matching of the $U(1)$ current generated by $\chi \rightarrow e^{i\theta(X)}\chi$ (or to have a canonical normalization of $\chi(X)$), where we see that the normalization of the world-scalar four-integral is

$$1 = ZM^4 \int d^4r\ |\phi(r)|^2 \tag{32}$$

where $\phi$ replaces $\varphi$ in Equation (19).

We can then represent $S$ in terms of two "nested" actions. For the field $\chi$

$$S = \int_X \Big( |\partial_X \chi|^2 - m^2 |\chi|^2 \Big) \text{ where } m^2 = -S_\phi \tag{33}$$

and $S_\phi$ is an action for the internal wave function

$$S_\phi = M^4 \int_{r^0,\vec{r}} \Big( Z|\partial_r \phi(r^\mu)|^2 + 2g^2 D_F(2r^\mu)|\phi(r^\mu)|^2 \Big) \tag{34}$$

Equation (33) then implies

$$\partial_X^2 \chi = -m^2 \chi \text{ hence, } \chi \sim \exp(iP_\mu X^\mu) \tag{35}$$

$\chi(X)$ has free plane wave solutions with $P^2 = m^2$.

In the center-of-mass frame of the bound state, we can choose $\chi$ to have four-momentum $P_\mu = (m, 0, 0, 0)$ where we then have

$$\Phi(X,r) = \chi(X)\phi(r^\mu) \propto \exp(imX^0)\phi(r^\mu). \tag{36}$$

$\phi(r)$ must then satisfy the Lagrange multiplier constraint

$$P^{\mu} \frac{\partial}{\partial r^{\mu}} \phi(r^{\mu}) = 0 \tag{37}$$

and therefore becomes a *static function* of $r^{\mu} = (0, \vec{r})$.

While we have specified $Z$ in Equation (32), we still have the option of normalizing the internal wave function $\phi(\vec{r})$. This can be conveniently normalized in the center-of-mass frame as

$$M^3 \int d^3r \, |\phi(\vec{r})|^2 = 1 \tag{38}$$

Note that in Equation (38), we have implicitly defined the static internal wave function $\phi(\vec{r})$ to be dimensionless, $d = 0$.

We see that the relative time now emerges in the four-integral over $|\phi(r)|^2$ of Equation (32) together with Equation (38)

$$1 = ZM^4 \int d^4r \, |\phi(r)|^2 = ZM^4 \int dr^0 \int d^3r \, |\phi(\vec{r})|^2 = ZMT \tag{39}$$

where $\int dr^0 = \int dr^{\mu} P_{\mu}/m \equiv T$. Then, from Equation (39), we have

$$TZ = M^{-1} \tag{40}$$

With static $\phi(r) \to \phi(\vec{r})$, the internal action of Equation (34) becomes

$$S_{\phi} = M^4 \int_{r^0, \vec{r}} \left( -Z|\nabla_{\vec{r}}\phi(\vec{r})|^2 + 2g^2 D_F(2r^{\mu})|\phi(\vec{r})|^2 \right) \tag{41}$$

where $|\nabla_{\vec{r}}\phi(\vec{r})|^2 = \nabla_{\vec{r}}\phi^{\dagger} \cdot \nabla_{\vec{r}}\phi$. Note that $\nabla_{\vec{r}}\phi$ is spacelike, and the arguments of the constrained $\phi(\vec{r})$ are now three-vector; however, $D_F(2r^{\mu})$ still depends upon the four-vector $r^{\mu}$.

There remains the integral over relative time $r^0$ in the action. For the potential, we have the residues

$$-V(r) = 2 \int dr^0 D_F(2r) = \int \frac{e^{2i\vec{q}\cdot\vec{r}}}{\vec{q}^2 + M^2} \frac{d^3q}{(2\pi)^3} = \frac{e^{-2M|\vec{r}|}}{8\pi|\vec{r}|} \tag{42}$$

and the potential term in the action becomes the static Yukawa potential

$$S_P = -M^3 \int_{\vec{r}} g^2 M V(\vec{r})|\phi(\vec{r})|^2, \qquad V(\vec{r}) = -\frac{e^{-2M|\vec{r}|}}{8\pi|\vec{r}|} \tag{43}$$

The $\phi(\vec{r})$ kinetic term in Equation (41) becomes

$$S_K = -M^4 \int_{r^0, \vec{r}} Z|\nabla_{\vec{r}}\phi(\vec{r})|^2 = -M^4 Z T \int_{\vec{r}} |\nabla_{\vec{r}}\phi(\vec{r})|^2$$
$$= -M^3 \int_{\vec{r}} |\nabla_{\vec{r}}\phi(\vec{r})|^2 \tag{44}$$

where we use Equation (40). The action $S_{\phi}$ thus becomes

$$m^2 = -S_{\phi} = M^3 \int_{\vec{r}} \left( |\nabla_{\vec{r}}\phi|^2 + g^2 M V(r)|\phi(\vec{r})|^2 \right) \tag{45}$$

Note that $S_{\phi}$ has dimension $d = 2$, as it must for $m^2$. We thus see, as previously mentioned, that the combination $ZT$ occurs in the theory, and the relative time has disappeared into normalization constraints; see Equations (32) and (38).

The radicalization of $S_{\phi}$ leads to the Schrödinger–Klein–Gordon (SKG) equation in the center-of-mass frame

$$-\nabla_{\vec{r}}^2 \phi(r) - g^2 M \frac{e^{-2M|\vec{r}|}}{8\pi|\vec{r}|} \phi(r) = m^2 \phi(r). \tag{46}$$

where, for spherical symmetry in a ground state,

$$\nabla_r^2 = \partial_r^2 + \frac{2}{r}\partial_r \tag{47}$$

We see that the induced mass$^2$ of the bound state, $m^2$, is the eigenvalue of the SKG equation. We can compare this to a non-relativistic Schrödinger equation (NRSE)

$$-\frac{1}{2M}\nabla_r^2\phi(x) - g^2\frac{e^{-2Mr}}{16\pi r}\phi = E\phi(\vec{r}) \tag{48}$$

In the next section, we will obtain similar results for a bound state of chiral fermions and use the known results for the Yukawa potential in the NRSE to obtain the critical coupling. The negative eigenvalue of $E$ in the NRSE, which signals binding, presently implies a vacuum instability.

Integrating both parts, we then have, from Equation (45)

$$m^2 = M^3 \int_{\vec{r}} \left( \phi^\dagger(-\nabla_r^2\phi + g^2 MV(r)\phi(\vec{r})) \right) \tag{49}$$

Note the consistency, using Equation (48), and the normalization of the dimensionless field $\phi$ of Equation (38).

More generally, by promoting $\chi$ to a $(1+3)$ time-dependent field while maintaining a static $\phi$, we have the full joint action:

$$S = M^3 \int_{X\vec{r}} \left( |\phi|^2 \left| \frac{\partial\chi}{\partial X} \right|^2 - |\chi|^2 \left( |\nabla_r\phi|^2 + g^2 MV(r)|\phi(r)|^2 \right) \right) \tag{50}$$

In summary, we have constructed, by "bilocalization" of a local field theory, a bilocal field description $\Phi(x, y)$ for the dynamics of binding a pair of particles. The dynamics implies that, in barycentric coordinates, $\Phi(x, y) \sim \Phi(X, r) \sim \chi(X)\phi(\vec{r})$, where the internal wave function, $\phi(\vec{r})$, is a static function of $\vec{r}$ and satisfies an SKG equation with eigenvalue $m^2$, which determines the squared-mass of a bound state. This illustrates the removal of relative time in an action formalism, which is usually framed in the context of Dirac Hamiltonian constraints [4,5].

## 2.3. Simplified Normalization

The normalization system we have thus far used is awkward. We can facilitate this by defining a new integral over the internal wave function three-space $\vec{r}$:

$$\int_r' \equiv \int \frac{d^3r}{V} \quad \text{where} \ \ V = M^{-3} \tag{51}$$

We then have the key elements of the theory in this notation:

$$S = \int d^4X \left( |\partial_X \chi|^2 - m^2|\chi|^2 \right)$$

$$1 = \int_r' |\phi(\vec{r})|^2 = \int \frac{d^3r}{V} |\phi(\vec{r})|^2$$

$$m^2 = -S_\phi$$

$$S_\phi = \int_r' \left( -|\partial_r\phi|^2 - g^2 MV(r)|\phi(\vec{r})|^2 \right)$$

$$V(r) = -\frac{e^{-2M|\vec{r}|}}{8\pi|\vec{r}|} \tag{52}$$

Our general notation is summarized in Appendix A.

### 3. The Coloron Model

#### 3.1. Boundstate and $N_c$-Enhanced Coupling

The point-like NJL model can be viewed as the limit of a physical theory with a bilocal interaction. An example that motivates the origin of the NJL interaction is an analogue of QCD, with a massive and perturbatively coupled gluon. We call this a "coloron model", and it has been extensively deployed to describe chiral constituent and heavy–light quark models [13,25,26], the possibility of the BEH boson composed of top quarks, and as a generic model for experimental search strategies [11,12,14–16,27].

Consider a nonconfining $SU(N_c)$ gauge Theory with a broken global $SU(N_c)$, where the coloron gauge fields $A_\mu^A$ acquire mass $M$ and have a fixed coupling constant $g$. We assume chiral fermions, each with $N_c$ "colors" labeled by $(a, b, \ldots)$ with the local Dirac action

$$S_F = \int_x \left( i\overline{\psi}_L^a(x)D\!\!\!/\,\psi_{aL}(x) + i\overline{\psi}_R^a(x)D\!\!\!/\,\psi_{aR}(x) \right) \tag{53}$$

where the covariant derivative is

$$D_\mu = \partial_\mu - ig A_\mu^A(x) T^A \tag{54}$$

and $T^A$ are the adjoint representation generators of $SU(N_c)$. We assume the colorons have a common mass $M$.

The single coloron exchange interaction then takes a bilocal current-current form:

$$S_C = -g^2 \int_{xy} \overline{\psi}_L(x)\gamma_\mu T^A \psi_L(x) D^{\mu\nu}(x-y)\overline{\psi}_R(y)\gamma_\nu T^A \psi_R(y) \tag{55}$$

where $T^A$ are generators of $SU(N_c)$. The coloron propagator in a Feynman gauge yields:

$$D_{\mu\nu}(x-y) = \int \frac{-ig_{\mu\nu}}{q^2 - M^2} e^{iq(x-y)} \frac{d^4q}{(2\pi)^4} \tag{56}$$

A Fierz rearrangement of the interaction to leading order in $1/N_c$ leads to an attractive potential [11,12]:

$$S_C = g^2 \int_{xy} \overline{\psi}_L^a(x)\psi_{aR}(y)\, D_F(x-y)\, \overline{\psi}_R^b(y)\psi_{bL}(x) \tag{57}$$

where $D_F$ is defined in Equation (8). Note that if we suppress the $q^2$ term in the denominator of Equation (56)

$$D_F(x-y) \to \frac{1}{M^2}\delta^4(x-y) \tag{58}$$

and we immediately recover the point-like NJL model interaction.

Consider spin-0 fermion pairs of a given color $[\overline{a}b]$ $\overline{\psi}_R^a(x)\psi_{bL}(y)$. We will have free fermionic scattering states, $: \overline{\psi}_R^a(x)\psi_{bL}(y) :$ coexisting in the action with bound states $\sim \Phi(x, y)$

$$\overline{\psi}_R^a(x)\psi_{bL}(y) \to M^2 {:}\overline{\psi}_R^a(x)\psi_{bL}(y){:} + M^2\Phi_b^a(x, y), \tag{59}$$

The normal ordering $: \ldots :$ signifies that we have subtracted the bound state from the product. These will be eigenstates of the equation of motion and will be orthogonal wave functions.

We see that $\Phi_b^a(X, r)$ is an $N_c \times N_c$ complex matrix that transforms as a product of $SU(N_c)$ representations, $\overline{N}_c \times N_c$, and therefore decomposes into a singlet plus an adjoint representation of $SU(N_c)$. We write $\Phi_b^a$ it as a matrix $\widetilde{\Phi}$ by introducing the $N_c^2 - 1$ adjoint matrices, $T^A$, where $\text{Tr}(T^A T^B) = \frac{1}{2}\delta^{AB}$. The unit matrix is $T^0 \equiv \text{diag}(1,1,1,\ldots)/\sqrt{2N_c}$, and $\text{Tr}(T^0)^2 = 1/2$, hence we have

$$\bar{\Phi} = \sqrt{2}\left(T^0\Phi^0 + \sum_A T^A\Phi^A\right) \tag{60}$$

The $\sqrt{2}$ is present because $\Phi^0$ and $\Phi^A$ form complex representations since they also represent the $U(1)_L \times U(1)_R$ chiral symmetry.

For the bilocal fields, we have a bosonic kinetic term with the constraint

$$S_K = \frac{JZM^4}{2}\int_{Xr}\text{Tr}\left(|\partial_X\bar{\Phi}|^2 + |\partial_r\bar{\Phi}|^2 + \eta|\partial_X\bar{\Phi}^\dagger\partial_r\bar{\Phi}|^2\right) \tag{61}$$

Note the numerical factor differs from the scalar case by treating $(x,y)$ symmetrically as in Equation (A15). For the singlet representations this takes the form

$$S_K = \frac{JZM^4}{2}\int_{Xr}\left(|\partial_X\Phi^0|^2 + |\partial_r\Phi^0|^2 + \eta|\partial_X\Phi^{0\dagger}\partial_r\Phi^0|^2\right) \tag{62}$$

We assume the constraint in the barycentric frame, and integrate out relative time with $ZMT = 1$.

$$S_K = (J/2)\int_X\int_{\bar{r}}'\left(|\partial_X\Phi^0(X,\bar{r})|^2 - |\partial_r\Phi^0(X,\bar{r})|^2\right) \tag{63}$$

(where $\int_{\bar{r}}' = M^3\int d^3r$). Factorizing $\Phi^0$

$$\sqrt{J/2}\,\Phi^0(X,r) = \chi(X)\phi(r) \tag{64}$$

then the kinetic term action becomes identical to the bosonic case

$$S_K = \int_X\left(|\partial_X\chi(X)|^2 - |\chi(X)|^2\int_{\bar{r}}'|\partial_{\bar{r}}\phi(\bar{r})|^2\right) \tag{65}$$

with

$$\int_{\bar{r}}'|\phi(\bar{r})|^2 = 1 \tag{66}$$

If we include the free fermion scattering states, the full bound state interaction of Equation (57) becomes

$$\begin{aligned}
S_C \to g^2\int_{xy} &:\bar{\psi}_L^a(x)\psi_{aR}(y): D_F(x-y) :\bar{\psi}_R^b(y)\psi_{bL}(x):\\
&+g^2JM^2\sqrt{N_c}\int_{X,r} :\bar{\psi}_L^a(X-r)\psi_{aR}(X+r): D_F(2r)\,\Phi^0 + h.c.\\
&+g^2JM^4N_c\int_{X,r}\Phi^{0\dagger}(X,r)\,D_F(2r)\,\Phi^0(X,r)
\end{aligned} \tag{67}$$

where

$$D_F(2r) = -\int\frac{1}{(q^2-M^2)}e^{2iq_\mu r^\mu}\frac{d^4q}{(2\pi)^4} \tag{68}$$

The leading term $S_C$ of Equation (69) is just a free four-fermion scattering state interaction and has the structure of an NJL interaction in the limit of Equation (58). This identifies $g^2$ as the NJL coupling constant. This is best treated separately by the local interaction of Equation (57). We therefore omit this term in the discussion of the bound states.

The second term$\sim \text{Tr}(\psi^\dagger\psi)\Phi^0 + h.c.$ in Equation (69) determines the Yukawa interaction between the bound state $\Phi^0$ and the free fermion scattering states. We will treat this below.

Note that the third term is the binding interaction and it involves only the singlet, $\text{Tr}\,\tilde{\Phi} = \sqrt{N_c}\Phi^0$. It can then be written in Equation (42) as

$$S_C \to g^2 J M^4 N_c \int_{X,r} \Phi^{0\dagger}(X,r)\, D_F(2r)\, \Phi^0(X,r)$$

$$= g^2 N_c \int_X |\chi(X)|^2 \int_r' |\phi(\vec{r})|^2 M \frac{e^{-2M|\vec{r}|}}{8\pi|\vec{r}|} \tag{69}$$

We see that adjoint representation $\Phi^A$ is decoupled from the interaction and remain as two-body massless scattering states. Hence, they do not form bound states by the interaction.

We also see that the singlet $\Phi^0$ singlet field has an enhanced interaction by a factor of $N_c$. This is analogous to a BCS superconductor, where the $N_c$ color pairs are analogues of $N$ Cooper pairs and the weak four-fermion Fröhlich Hamiltonian interaction is enhanced by a factor of $N_{Cooper}$ [19,20]. The color enhancement also occurs in the NJL model, but at loop level. Here, we see that the color enhancement is occurring in the semi-classical (no loop) coloron theory by this coherent mechanism.

Hence, the removal of relative time is then the identical procedure as in the previous model (and absorbs away $Z$ and $T$ as in Equations (32), (39) and (40)), and leads to the same action, $S = S_K + S_C$, in the compact notation of Equation (52) with the interaction enhanced by $N_c$.

The radicalization of $\phi$ then leads to the SKG equation

$$-\nabla_r^2 \phi(\vec{r}) - g^2 N_c M \frac{e^{-2M|\vec{r}|}}{8\pi|\vec{r}|}\phi(\vec{r}) = m^2\phi(\vec{r}). \tag{70}$$

### 3.2. Classical Criticality of the Coloron Model

The coloron model furnishes a direct UV completion of the NJL model. However, in the coloron model, we do not need to invoke large-$N_c$ quantum loops to have a critical theory. Rather, it leads to an SKG potential of the Yukawa form which has a *classical critical coupling*, $g_c$. For $g < g_c$, the theory is subcritical and produces resonant bound states that decay into chiral fermions. For $g > g_c$, the theory produces a tachyonic bound state which implies a chiral instability and $\Phi$ must develop a VEV. This requires stabilization by, e.g., quartic interactions and a sombrero potential. All of this is treated bosonically in our present formalism.

The criticality of the Yukawa potential in the nonrelativistic Schrödinger equation is discussed in the literature in the context of "screening". The nonrelativistic Schrödinger equation $r = |\vec{r}|$ is:

$$-\nabla^2 \psi - 2m\alpha \frac{e^{-\mu r}}{r}\psi = 2mE \tag{71}$$

and criticality (eigenvalue $E = 0$) occurs for $\mu = \mu_c$ where a numerical analysis yields [28,29]

$$\mu_c = 1.19\alpha m \tag{72}$$

For us, the spherical SKG equation is now $r = |\vec{r}|$

$$-\nabla_r^2 \phi(r) - g^2 N_c M \frac{e^{-2Mr}}{8\pi r}\phi(r) = 0 \tag{73}$$

Comparing, gives us a critical value of the coupling constant, when $\mu_c \to 2M$, $m \to M/2$ and $\alpha \to g^2 N_c/8\pi$, then:

$$2M = (1.19)\left(\frac{M}{2}\right)\left(\frac{g^2 N_c}{8\pi}\right), \quad \text{hence: } g^2/4\pi = 6.72/N_c \tag{74}$$

We can compare the NJL critical value of Equation (4)

$$g_{cNJL}^2/4\pi = 2\pi/N_c = 6.28/N_c. \tag{75}$$

Hence, the NJL quantum criticality is a comparable effect, with a remarkably similar numerical value for the critical coupling.

Note that we can rewrite Equation (73) with dimensionless coordinates, $\vec{u} = M\vec{r}$, $u = M|\vec{r}|$

$$M^2\left(-\nabla_u^2 \phi(u) - g^2 N_c \frac{e^{-2u}}{8\pi u}\phi(u)\right) = 0 \tag{76}$$

and then $M^2$ only appears as an overall scale factor. Hence, we see that critical coupling is determined by Equation (76), and the scale $M$ cancels out at criticality. The mass scale $M$ is dictated in the exponential $e^{-2Mr}$ of the particular Yukawa potential, together with canonical normalization of $\phi$. In general, we can start with the dimensionless coordinate form of the SKG equation and infer the scale $M$ by matching it to the potential. In this way, solutions may exist where the scale in the potential is driven by the renormalization group. This will be investigated elsewhere.

However, it is important to realize that the NJL model involves Yukawa coupling, $g_{NJL}$, while the present criticality involves the coloron coupling constant. The NJL coupling is emergent in the coloron model, and we need to compute it.

### 3.3. Yukawa Interaction

The second term in Equation (67) is the induced Yukawa interaction $S_Y$, and can be written with the factorized field as:

$$S_Y = g^2(\sqrt{2JN_c})M^2 \times$$
$$\int_{Xr} : \overline{\psi}_L^a(X-r)\psi_{aR}(X+r) : D_F(2r)\chi(X)\phi(\vec{r}) + h.c. \tag{77}$$

This is the effective Yukawa interaction between the bound state $\Phi^0$ and the free fermion scattering states.

We cannot simply integrate the relative time here. However, we can first connect this to the point-like limit by suppressing the $q^2$ term in the denominator of $D(2r)$ with $z \to 0$ in:

$$D_F(2r) \to \int \frac{1}{M^2} e^{2iq_\mu r^\mu} \frac{d^4q}{(2\pi)^4} \to \frac{1}{JM^2}\delta^4(r) \tag{78}$$

where $\delta^4(2r) = J^{-1}\delta^4(r)$, hence with $J^{-1} = 1/16$

$$S_Y = g^2(\sqrt{N_c/8})\int_X \overline{\psi}_L^a \psi_{aR}(X)\chi(X)\phi(0) + h.c. \tag{79}$$

This gives a value of the Yukawa coupling

$$g_Y = g^2(\sqrt{N_c/8})\phi(0) \tag{80}$$

The wave function at the origin, $\phi(0)$, in the NJL limit is somewhat undefined. However, if we consider a spherical cavity of radius $R$, where $MR = \pi/2$, with a *confined* and dimensionless $\phi(r)$, then $\phi(0)$ is obtained (see [30] Equation (B58))

$$\phi(0) = \frac{1}{\pi}. \tag{81}$$

Plugging this into the expression for $g_Y$ in Equation (80) gives

$$g_Y = g^2\sqrt{N_c/8\pi^2} = g^2/g_{cNJL} \tag{82}$$

where $g_{cNJL}^2 = (8\pi^2/N_c)$ is the critical coupling of the NJL model, as seen in Equation (4).

Hence, if the coloron coupling constant, $g^2$, is critical, as in Equations (74) and (75), we have seen that $g^2 \approx g_{cNJL}^2$, and the induced Yukawa coupling from Equation (82) is then $g_Y \approx g_{cNJL}$. The coloron model is then consistent with the NJL model in the point-like limit where the NJL model coupling is the Yukawa coupling, as seen in Equation (2).

However, the induced Yukawa coupling in the bound state, $g_Y$, may be significantly different than the coloron coupling $g$ in realistic extended $\vec{r}$ models. The result we just obtained applies when we assume the strict point-like limit of $D_F(2r) \sim \delta^3(r)$, while in reality, as the potential becomes more extended, the $\int V(\vec{r}) \phi(\vec{r})$ may become smaller, even if the coloron coupling $g$ may be supercritical. We anticipate this could have implications for a composite Higgs model, which will be investigated elsewhere (together with loop effects including the extended wave function). There may also be additional new effects that occur at the loop level in extended potentials, such as the infall of zero modes, as suggested in [31].

### 3.4. Spontaneous Symmetry Breaking

For subcritical Coupling, there are resonance solutions with positive $m^2$ that have large distance tails of external incoming and outgoing radiation, representing a steady state of resonant production and decay. The portion of the wave function localized within the potential can be viewed as the resonant bound state for normalization purposes, while the large distance tail is non-normalizable radiation.

With super-critical coupling, $g > g_c$, the bilocal field $\Phi(X, r)$ has a negative squared mass eigenvalue (tachyonic), with a well-defined localized wave function. In the region external to the potential (forbidden zone), the field is exponentially damped. At exact criticality with $g = g_c - \epsilon$, there is a $1/r$ (quasi-radiative) tail that switches to exponential damping for $g = g_c + \epsilon$. The supercritical solutions are localized and normalizable over the entire space $\vec{r}$, but with $m^2 < 0$, they lead to exponential runaway in time of the field $\chi(X^0)$, and must be stabilized, typically with a $|\Phi|^4$ interaction.

We then treat the supercritical case as resulting in spontaneous symmetry breaking. In the point-like limit, $\Phi(X) \sim \Phi(X, 0)$, the theory has the "sombrero potential",

$$V(\Phi) = -|M^2 \Phi^2| + \frac{\lambda}{2} |\Phi|^4 \tag{83}$$

The point-like field develops a VEV, $\langle \Phi \rangle = |M|/\sqrt{\lambda}$. In this way, the bound state theory will drive the usual chiral symmetry breaking from the underlying dynamics of a potential induced by new physics.

A quartic potential generally exists in a local field theory, Equation (6), and would be induced by free loops in the coloron model. We can introduce a bilocalized quartic interaction as a model by presently introducing another world scalar factor with coefficient $Z'$

$$\frac{\lambda_0}{2} \int d^4x |\varphi|^4 \rightarrow \frac{\lambda_0}{2} \int d^4y \, Z' |\varphi(y)|^4 \int d^4x |\varphi|^4$$
$$= \frac{Z' M T \tilde{\lambda}}{2} \int_X \int_r' |\chi(X) \phi(\vec{r})|^4 \tag{84}$$

and $Z'MT = 1$ to absorb relative time.

The simplest sombrero potential can therefore be modeled as

$$S = \int_{Xr}' \left( |\varphi|^2 \left| \frac{\partial \chi}{\partial X} \right|^2 - |\chi|^2 (|\nabla_r \phi|^2 + g^2 N_c M V(r) |\phi(r)|^2) \right.$$
$$\left. - |\chi|^4 \frac{\tilde{\lambda}}{2} |\phi(\vec{r})|^4 \right) \tag{85}$$

In the case of a perturbatively small $\lambda$, we expect the eigensolution of $\phi$ to be essentially unaffected

$$\int_r' \left( |\nabla_{\vec{r}} \phi|^2 + g^2 N_c M V(r) |\phi(\vec{r})|^2 \right) \approx m^2 \tag{86}$$

The effective quartic coupling is then further renormalized by the internal wave function

$$\frac{\tilde{\lambda}}{2} |\chi|^4 \int_r' |\phi(\vec{r})|^4 = |\chi|^4 \frac{\tilde{\lambda}}{2} \tag{87}$$

In this case, we see that $\chi$ develops a VEV in the usual way:

$$\langle|\chi|^2\rangle = |m^2|/\tilde{\lambda} = v^2 \tag{88}$$

This is a consequence of $\phi(\vec{r})$ remaining localized in its potential.

The external scattering state fermions, $\psi^a(X)$, will then acquire mass through the emergent Yukawa interaction described in the previous section, $\sim g_Y\langle|\chi|\rangle$. However, an issue we have yet to resolve is whether the induced fermion masses back-react with the VEV solution itself. We segregated the free fermions from the bound state wave function, $\Phi$, by shifting, so we are presently arguing that $\Phi$ forms a VEV as described above, and the scattering state fermions independently acquire mass as spectators, but this may require a more detailed analysis, and for general large $\tilde{\lambda}$ (as in a nonlinear sigma model), the situation is potentially more complicated.

## 4. Summary and Conclusions

In the present paper, we have given a formulation of bilocal field theory, $\Phi(x, y)$, as a variation on Yukawa's original multilocal field theory of composite particles [1–3]. In particular, we focus on two-particle-bound states consisting of bosons or chiral fermions and scattering states. There are many foreseeable extensions of the present work. Here, we construct bilocal field theories from an underlying local interacting field theory via the introduction of "world-scalars". We then go to barycentric coordinates, and the bilocal field is "factorized"

$$X = \frac{1}{2}(x + y), \qquad r = \frac{1}{2}(x - y)$$
$$\Phi(x, y) \rightarrow \Phi(X, r) = \chi(X)\phi(r) \tag{89}$$

Here, $\chi(X)$ describes center-of-mass motion like any pointlike scalar field, while $\phi(r)$ is the internal wave function of the bound state.

This procedure enables the removal of the relative time, $r^0$, in the bilocalized theory, essentially by canonical renormalization. The bilocal kinetic term contains a constraint that leads to a static internal field, $\phi(r) \rightarrow \phi(\vec{r})$, in the center-of-mass frame. Hence, we obtain a static Schrödinger–Klein–Gordon (SKG) equation for the internal wave function. The eigenvalue of this equation is the $m^2$ of the bound state.

The SKG equation likewise contains a static potential that comes from the Feynman propagator of the exchanged particle in the parent theory. Typically we have a Yukawa potential, $\sim g^2 M e^{-2Mr}/8\pi r$, though the formalism can in principle accommodate any desired phenomenological potential. Here, $g$ is the exchanged particle coupling constant, such as the coloron coupling (massive perturbative gluon) for fermions. This is not the scalar–fermion Yukawa coupling, $g_Y$, which is subsequently emergent.

We find that the Yukawa potential is classically critical with coupling $g_c$. If the coupling is sub-critical, $g < g_c$, then $m^2$ is positive, and the bound state is therefore a resonance. It will decay to its constituents if kinematically allowed. $\phi(\vec{r})$ is then a localized "lump", with a radiative tail representing the two body decay and production by external free particles.

If the coupling is supercritical, $g > g_c$, then $m^2 < 0$ is tachyonic and $\Phi$ will acquire a VEV. We require an interaction, such as $\sim\lambda|\Phi|^4$, to stabilize the vacuum and we therefore have spontaneous symmetry breaking. $\phi(r)$ is expected to be localized in its potential and $\chi(X)$ acquires the VEV. If $\phi(r)$ becomes delocalized, both $\chi(X)$ and $\phi(r)$ acquire VEVs, which is an analogue of a Bose–Einstein condensate, e.g., in a slightly heated superconductor; however, we have not produced solutions to the SKG equation that demonstrates this behavior.

We consider a bound state of chiral fermions in the coloron model, where a coloron is a massive gluon with coupling $g$, such as in "topcolor" models [11,12] and chiral constituent quark models [13]. Fierz's rearrangement of the non-local, color-current-current, interaction yields a leading large $N_c$ interaction in $\bar{\psi}(x)_L\psi(y)_R \sim \Phi(x, y)$. For the color singlet, $\Phi$, the coupling $g^2$ is enhanced by $N_c$ in analogy to a BCS superconductor [19,20].

As our main result, we find that the coloron model can be classically critical. The critical coupling, $g_c^2$, extracted from [28,29], is astonishingly close to the critical value of the Yukawa coupling in the NJL model. While in the NJL model the critical behavior is $\mathcal{O}(\hbar)$ coming from fermion loops, in the bilocal model this is a semiclassical result, and the essential factor of $N_c$ comes from the coherent BCS-like enhancement of the four-fermion scattering amplitudes. It is therefore unclear what happens when we include the fermion loops in addition to the classical behavior in the large $N_c$ limit. Is criticality further enhanced by the additional loop contribution? Is there an additional $N_c$ enhancement of the underlying coupling due to the $\sqrt{N_c}$ factor in the emergent Yukawa interaction? These are interesting issues we will address elsewhere.

The induced Yukawa coupling of the bound state to fermions, $g_Y \overline{\psi}(x)_L \psi(y)_R \Phi(x, y)$, is extended and emergent in the composite models. We derive the coupling and find $g_Y \propto \int V(r)\phi(r) \sim g^2 \phi(0)$ in the point-like limit. For $\phi(0)$, in a tiny spherical cavity, we obtain $g_Y = g^2/g_{cNJL}$. Hence, the critical value of $g^2$ implies the critical value of $g_Y = g_{cNJL}$ in the point-like limit NJL model, which is consistent.

However, the $\int V(r)\phi(r)$ could in principle be reduced to an extended potential as the wave function spreads out, even if $g^2$ is critical. This is an intriguing possibility: the BEH–Yukawa coupling of the top quark is $g_{top} \sim 1$, which is perturbative and is insufficient to drive the formation of a composite, negative $m^2$, Brout–Englert–Higgs (BEH) boson at low energies in, e.g., a top condensation model. However, the present result suggests that perhaps $\int V(r)\phi(r)$ is small suppressing $g_{top}$, even though the underlying coloron coupling, $g^2 N_c$, is super-critical and leads to the composite BEH mechanism. In this picture, the BEH boson may be a large object, e.g., a "balloon" of size $\sim m_{top}^{-1}$ (see [31]).

While we have an eye to a composite BEH boson for the standard model, as in top condensation theories [22–24,32,33], our present analysis is more general, but does not yet include many details, e.g., gauge interactions and gravity. We think the emphasis on a bosonic field description, the treatment of the coloron model and its classical criticality, its linkage the NJL model as UV completion, and our treatment of relative time renormalization comprises a novel perspective.

**Funding:** This research received no external funding

**Data Availability Statement:** Data is contained within the article.

**Acknowledgments:** I thank W. Bardeen and Bogdan Dobrescu for critical comments.

**Conflicts of Interest:** The author declares no conflicts of interest.

## Appendix A. Summary of Notation

Two body kinematics:

$$p_1 x + p_2 y = (p_1 + p_2)X + (p_1 - p_2)r \tag{A1}$$
$$p_1^2 = p_2^2 = \mu^2 \qquad P = (p_1 + p_2) \qquad Q = (p_1 - p_2)$$
$$P^2 + Q^2 = 4\mu^2; \text{ rest frame: } P^0 = 2p_1^0; \ \vec{Q} = 2\vec{p}_1;$$

$P^2 = (p_1 + p_2)^2$, though commonly referred to as the "invariant mass" of a pair, is not a scale breaking mass in that it involves fields with a traceless stress tensor.

Barycentric coordinates:

$$X = \frac{1}{2}(x + y) \qquad \rho = (x - y) \qquad r = \frac{1}{2}(x - y)$$
$$\partial_x = \frac{1}{2}(\partial_X + \partial_r) = \frac{1}{2}\partial_X + \partial_\rho$$
$$\partial_y = \frac{1}{2}(\partial_X - \partial_r) = \frac{1}{2}\partial_X - \partial_\rho \tag{A2}$$

Two-body scattering states

$$\Phi(x,y) = \exp(ip_1 x + ip_2 y) = \exp(iP_\mu X^\mu + iQ_\mu r^\mu)$$

$$(\partial_x^2 + \partial_y^2)\Phi(x,y) = (p_1^2 + p_2^2)\Phi(x,y) = 2\mu^2 \Phi(x,y)$$

$$= \frac{1}{2}\left(\frac{\partial^2}{\partial X^\mu \partial X_\mu} + \frac{\partial^2}{\partial r^\mu \partial r_\mu}\right)\Phi(X,r) = 2\mu^2 \Phi(x,y)$$

$$\rightarrow \left(\frac{\partial^2}{\partial X^\mu \partial X_\mu} + \frac{\partial^2}{\partial r^\mu \partial r_\mu}\right)\Phi(X,r) = 4\mu^2$$

$$(\partial_x^2 - \partial_y^2)\Phi(x,y) = \frac{\partial^2}{\partial X^\mu \partial r_\mu}\Phi(X,r) = 0$$

$$dx^2 + dy^2 = 2dX^2 + \frac{1}{2}d\rho^2 = 2dX^2 + 2dr^2$$

$$\partial_x^2 + \partial_y^2 = \frac{1}{2}\partial_X^2 + \frac{1}{2}\partial_r^2 = \frac{1}{2}\partial_X^2 + 2\partial_\rho^2$$

$$\partial_x^2 - \partial_y^2 = \partial_X^\mu \partial_{\mu r} = 2\partial_X^\mu \partial_{\mu\rho} \qquad \text{(A3)}$$

Integration measures

$$\int_{u...v} = \int d^4 u...d^4 v \qquad \int_{\bar{x}...\bar{y}} = \int d^3 x...d^3 y$$

$$\int_{u...v;\bar{x}...\bar{y}} = \int d^4 u..d^4 v\, d^3 x...d^3 y$$

$$\int'_{u...v;\bar{x}...\bar{y}} = \int M^4 d^4 u...M^4 d^4 v\, M^3 d^3 x...M^3 d^3 y$$

$$\int d^4 x d^4 y = \int d^4 X d^4 \rho = J\int d^4 X d^4 r;$$

Jacobian $\ J = (2)^4$

$$\int d^4 x d^4 y\left(|\partial_x \phi|^2 + |\partial_y \phi|^2 - \mu^2 |\phi|^2\right)$$

$$= J\int d^4 X d^4 r\left(\frac{1}{2}|\partial_X \phi|^2 + \frac{1}{2}|\partial_r \phi|^2 - \mu^2 |\phi|^2\right)$$

$$= \int d^4 X d^4 \rho\left(\frac{1}{2}|\partial_X \phi|^2 + 2|\partial_\rho \phi|^2 - \mu^2 |\phi|^2\right)$$

Hermitian operator : $\ W + W^\dagger = W_{+h.c.}$ $\qquad$ (A4)

## Appendix B. Bilocal Field Theory

Here, we give a general discussion of our "revisited" bilocal field theory for a pair of particles, as inspired by Yukawa [1–3]. We begin with the "bilocalization" and subsequently construct the generic actions for bilocal fields containing free particles.

### *Appendix B.1. Free Fields*

Let us examine the bilocalization procedure in Section 2.2 for free particle states. Consider a pair of local scalar fields $\varphi_i(x)$ (complex):

$$S = \int_X\left(|\partial\varphi_1|^2 + |\partial\varphi_2|^2 - \mu^2(|\varphi_1|^2 + |\varphi_2|^2)\right) \qquad \text{(A5)}$$

with independent free particle equations of motion

$$\partial^2 \varphi_1 + \mu^2 \varphi_1 = 0 \qquad \partial^2 \varphi_2 + \mu^2 \varphi_2 = 0 \qquad \text{(A6)}$$

We want to describe a pair of particles by a bilocal field of mass dimension 1:

$$\Phi(x,y) = M^{-1}\varphi_1(x)\varphi_2(y) \qquad \text{(A7)}$$

We therefore have the two equations of motion for $\Phi$

$$(\partial_x^2 + \partial_y^2)\Phi(x,y) + 2\mu^2\Phi(x,y) = 0 \qquad (A8)$$

$$(\partial_x^2 - \partial_y^2)\Phi(x,y) = 0. \qquad (A9)$$

*Appendix B.2. Bilocalization of Scattering States*

We can obtain the action for $\Phi$ as follows. We multiply the kinetic and mass terms of Equation (A5) by world scalars:

$$W_i = ZM^2 \int d^4y \, |\varphi_i(y)|^2 \qquad (A10)$$

and hence,

$$\begin{aligned}
S &= \int_x \left( W_2|\partial\varphi_1|^2 + W_1|\partial\varphi_2|^2 - \mu^2(W_2|\varphi_1|^2 + W_1|\varphi_2|^2) \right) \\
&= ZM^2 \int_{xy} \Big( |\varphi_2(y)\partial_x\varphi_1(x)|^2 + |\varphi_1(y)\partial_x\varphi_2(x)|^2 \\
&\qquad\qquad -2\mu^2|\varphi_1(y)\varphi_2(x)|^2 \Big)
\end{aligned} \qquad (A11)$$

At this stage, we can still vary with respect to either $\varphi_1$ of $\varphi_2$, and the equations are modified. In terms of the bilocal field, we have

$$= ZM^4 \int_{xy} \Big( |\partial_x\Phi(x,y)|^2 + |\partial_y\Phi(x,y)|^2 - 2\mu^2|\Phi(x,y)|^2 \Big) \qquad (A12)$$

We go to barycentric coordinates and note the derivatives:

$$\begin{aligned}
X &= (x+y)/2 & r &= (x-y)/2 \\
\partial_x &= (\partial_X + \partial_r)/2 & \partial_y &= (\partial_X - \partial_r)/2 \\
\Phi(x,y) &\to \Phi(X,r)
\end{aligned} \qquad (A13)$$

The factor of $M$ is superfluous at this point, and in what follows we can set $M = 1$ (we'll restore it below). Hence:

$$\begin{aligned}
S = JZ \int_{Xr} \Big( \frac{1}{4}|(\partial_X + \partial_r)\Phi(X,r)|^2 + \frac{1}{4}|(\partial_X - \partial_r)\Phi(X,r)|^2 \\
-2\mu^2|\Phi(X,r)|^2 \Big)
\end{aligned} \qquad (A14)$$

which yields the action

$$S = \frac{JZ}{2} \int_{Xr} \Big( |\partial_X\Phi(X,r)|^2 + |\partial_r\Phi(X,r)|^2 - 4\mu^2|\Phi(X,r)|^2 \Big) \qquad (A15)$$

If we vary $\delta\Phi(x',y') = \delta^4(x-x')\delta^4(y-y')$, we obtain one equation of motion:

$$\partial_X^2\Phi(X,r) + \partial_r^2\Phi(X,r) + 4\mu^2\Phi(X,r) = 0 \qquad (A16)$$

We see that this is consistent with Equation (A8)

$$\partial_x^2 + \partial_y^2 = \frac{1}{2}(\partial_X^2 + \partial_r^2). \qquad (A17)$$

However, Equation (A9) is missing. We see using

$$(\partial_x^2 - \partial_y^2) = \frac{\partial}{\partial X^\mu}\frac{\partial}{\partial r^\mu} \qquad (A18)$$

that Equation (A9) takes the form

$$\frac{\partial}{\partial X^\mu}\frac{\partial}{\partial r^\mu}\Phi(X, r) = 0 \tag{A19}$$

As before, this can be viewed as a constraint, and we can treat it as a Lagrange multiplier as in Equation (28) to supply the second equation.

We factorize $\Phi$ in barycentric coordinates as

$$\sqrt{J/2}\,\Phi(X, r) = \chi(X)\phi(r) \tag{A20}$$

The action Equation (A15) becomes

$$S = Z\int_{X}\int_{r}\Big(|\phi(r)|^2|\partial_X\chi(X)|^2$$
$$+|\chi(X)|^2(|\partial_r\phi(r)|^2 - 4\mu^2|\phi(r)|^2)\Big) \tag{A21}$$

We then define

$$1 = Z\int d^4r|\phi(r)|^2 \tag{A22}$$

The Schrödinger–Klein–Gordon equation has eigenvalue $m^2$

$$\partial_r^2\phi(r) + 4\mu^2\phi(r) = m^2\phi(r)$$
$$m^2\int_{r}|\phi(r)|^2 = \int_{r}(|\partial_r\phi(r)|^2 + 4\mu^2|\phi(r)|^2) \tag{A23}$$

Then, the $\chi$ equation becomes

$$\partial_X^2\chi(X) + m^2\chi(X) = 0 \tag{A24}$$

The constraint then takes the form

$$\frac{\partial\chi(X)}{\partial X^\mu}\frac{\partial\phi(r)}{\partial r^\mu} = 0 \tag{A25}$$

In the barycentric (rest) frame, we can choose $\chi$ to have four-momentum $P_\mu = (m, 0, 0, 0)$. The constraint becomes $P_\mu\partial_r^\mu\phi(r) = 0$, Therefore, $\phi$ becomes a *static function* of $r^\mu = (0, \vec{r})$. The Schrödinger–Klein–Gordon equation is then a static equation

$$-\nabla_{\vec{r}}^2\phi(\vec{r}) + 4\mu^2\phi(\vec{r}) = m^2\chi(\vec{r}) \tag{A26}$$

We define the static internal wave function $\phi(\vec{r})$ to be dimensionless, $d = 0$, and we normalize the dimensionless static wave function (restoring $M$)

$$(M^3)\int d^3\vec{r}\,|\phi(\vec{r})|^2 = 1 \tag{A27}$$

We then have

$$1 = Z(M^4)\int_{r}|\phi(r)|^2 = Z(M^4)\int dr^0\int d^3\vec{r}|\phi(\vec{r})|^2 = ZT(M) \tag{A28}$$

where an integral over relative time has appeared leading to an overall factor $ZT(M) = 1$ in the action for $\chi$. The action becomes two nested parts

$$S = \int_{X}\Big(|\partial_X\chi(X)|^2 - m^2|\chi(X)|^2\Big)$$
$$m^2\int_{\vec{r}}d^3\vec{r}\,|\phi(\vec{r})|^2 = \int_{r}(|\nabla_{\vec{r}}\phi(r)|^2 + 4\mu^2|\phi(r)|^2) \tag{A29}$$

with Equation (A27).

*Appendix B.3. Kinematics of Scattering States*

We can more directly construct the bilocal fields, without appealing to world scalars, by considering simple free particle kinematics. For free particles of four-momenta $p_i$, we see that $\Phi$ describes a scattering state

$$\Phi(x, y) \sim \exp(ip_{1\mu}x^\mu + ip_{2\mu}y^\mu) = \exp(iP_\mu X^\mu + iQ_\mu r^\mu) \tag{A30}$$

where

$$P_\mu = p_{1\mu} + p_{2\mu} \qquad Q_\mu = p_{1\mu} - p_{2\mu} \tag{A31}$$

For massive particles, $p_1^2 = p_2^2 = \mu^2$, $\Phi$ satisfies

$$\left( \frac{\partial^2}{\partial X^\mu \partial X_\mu} + \frac{\partial^2}{\partial r^\mu \partial r_\mu} + 4\mu^2 \right) \Phi(X, r) = 0 \tag{A32}$$

The constraint is the second equation

$$(\partial_x^2 - \partial_y^2)\Phi(x, y) = \frac{\partial^2}{\partial X^\mu \partial r_\mu} \Phi(X, r) = 0 \tag{A33}$$

and forces $\Phi(X, \vec{r})$ to have $r^0 = 0$ in the center-of-mass frame; hence, $P_\mu Q^\mu = 0$ and $Q = (0, \vec{q})$.

**Constant Positive $m_0^2 = 4\mu^2$:**

In the center-of-mass frame, $p_1^0 = p_2^0$, $Q^0 = 0$,

$$Q^2 = (p_1 - p_2)^2 = -(\vec{p}_1 - \vec{p}_2)^2 = -\vec{q}^{\,2} \tag{A34}$$

Hence, from Equations (A32) and (A33)

$$P^2 = 4\mu^2 + \vec{q}^{\,2}$$
$$\frac{\partial^2}{\partial X^\mu \partial X_\mu} \Phi(X, \vec{r}) = (m_0^2 + \vec{q}^{\,2}) \Phi(X, \vec{r}) \tag{A35}$$

Therefore, if we were to try to interpret $\Phi$ as a "bound state", we see that is has continuum of invariant "masses" $m^2 = m_0^2 + \vec{q}^{\,2}$. This is evidently an "unparticle" [34]. This implies it is not localized and is a scattering state of asymptotic free particles each of mass $\mu = m_0/2$.

With the constraint of Equation (A33), we can choose $r^\mu = (0, \vec{r})$ and $X^\mu = (X^0, 0)$. We factorize $\Phi = \chi(X)\phi(r)$ and the factor field $\phi(r)$ then satisfies the static SKG equation with eigenvalue $M^2$

$$-\nabla_{\vec{r}}^2 \phi(\vec{r}) + m_0^2 \phi(\vec{r}) = m^2 \phi(\vec{r}) \tag{A36}$$

the solutions of the factorized field $\phi(r)$ are static, box normalized, plane waves, with eigenvalues $m^2 \equiv m_{\vec{q}}^2 = m_0^2 + \vec{q}^{\,2}$

$$\phi(\vec{r}) = \frac{1}{\sqrt{V}} \exp(i\vec{q} \cdot \vec{r}) \qquad m_{\vec{q}}^2 = \vec{q}^{\,2} + m_0^2 \tag{A37}$$

$\chi(X)$ then satisfies the KG equation with $X^0 = t$ and $\vec{X} = 0$

$$\partial_t^2 \chi(t) + m_{\vec{q}}^2 \chi(t) = 0 \qquad t = X^0 \tag{A38}$$

and the $\chi(X)$ solution becomes

$$\chi(X) \propto \exp(im_{\vec{q}} t) \tag{A39}$$

This is just the spectrum of the two body states of $\mu$-massive particles in the barycentric frame.

**Constant Negative $m_0^2$:**

If we suppose $m_0^2 < 0$ then the solutions for $\Phi(t, r)$ for small $\vec{q}$ are runaway exponentials, $\Phi \sim \exp(|m_0|t)$. The eigenvalues are

$$m_{\vec{q}}^2 = \vec{q}^{\,2} - |m_0|^2 \tag{A40}$$

This analogous to a "Dirac sea" of negative mass eigenvalues extending from $\vec{q}^{\,2} = 0$ to $\vec{q}^{\,2} = |m_0|^2$, where the lowest mass state has $\vec{q} = 0$ and eigenvalue $-|m_0|^2$.

This would therefore be a tachyonic instability of unparticles, and we then require higher field theoretic interactions, such as $\lambda |\Phi|^4$, to stabilize the vacuum. This represents a spontaneous breaking of the chiral symmetry, and the constituent fermions will dynamically acquire masses and the instability is halted.

However, a constant negative $m_0^2$ does not give a Higgs-like spontaneous symmetry breaking mechanism. If the potential is

$$V = -m_0^2 |\Phi|^2 + \frac{\lambda}{2} |\Phi|^4 \tag{A41}$$

where we have constant $-m_0^2$ rather than a localized potential. Then, indeed, the field develops a constant VEV:

$$\langle \Phi \rangle = V + \Phi^0 \qquad m_0^2 = v^2 \lambda \tag{A42}$$

Then, $\Phi^0$ indeed becomes a massive BEH mode, but with the continuous spectrum $m^2 = |m_0|^2 + \vec{q}^{\,2}$. To have an acceptable BEH mechanism, with a well-defined BEH boson, we therefore require a localized potential in $\vec{r}$ where $\langle \Phi \rangle = \langle \chi \rangle \phi(\vec{r})$ and $\phi(\vec{r})$ is a localized eigenmode.

*Appendix B.4. Removal of Relative Time and Generic Potential*

A point-like interaction in the underlying theory may produce an effective action with a potential term that is a function of $r^\mu$

$$S = \frac{JM^4}{2} \int_{X_r} (Z |\partial_X \Phi|^2 + Z |\partial_r \Phi|^2 - U(2r^\mu) |\Phi(X, r)|^2) \tag{A43}$$

and we assume the constraint:

$$S = \int_{X_r} \eta \left( \frac{\partial \Phi^\dagger}{\partial X^\mu} \frac{\partial \Phi}{\partial r_\mu} + h.c. \right)^2. \tag{A44}$$

We factorize $\Phi$

$$\sqrt{J/2} \, \Phi(X, r) = \chi(X) \phi(r) \tag{A45}$$

In the center-of-mass frame we impose the constraint, $\Phi(X, r) \to \chi(X^0) \phi(\vec{r})$. We can then integrate over relative time

$$S = M^3 \int_{X\vec{r}} \Big( ZTM |\partial_X \chi|^2 |\phi(\vec{r})|^2 - ZTM |\chi(X)|^2 |\nabla_{\vec{r}} \phi|^2$$
$$- V(\vec{r}) |\chi(X)|^2 |\phi(\vec{r})|^2 \Big) \tag{A46}$$

where

$$V(\vec{r}) = M \int dr^0 U(2r^\mu) \tag{A47}$$

and $V(\vec{r})$ has dimensions of $M^2$. We now define the normalizations,

$$ZTM = 1 \qquad 1 = M^3 \int d^3r |\phi(\vec{r})|^2. \tag{A48}$$

The purpose of these normalizations is to have a canonically normalized $\chi$ field in the center-of-mass frame, even in the limit $g = 0$. It also defines the conserved $\chi$ current to have unit charge (one boundstate pair) to match the underlying theory's conserved current.

Hence, we obtain the actions:

$$S = \int_X \left( |\partial_X \chi(X)| - m^2 |\chi(X)|^2 \right)$$
$$m^2 \int_{\vec{r}} d^3\vec{r} \, |\phi(\vec{r})|^2 = \int_{\vec{r}} \left( |\nabla_{\vec{r}} \phi(r)|^2 + V(r)|\phi(r)|^2 \right) \tag{A49}$$

with Equation (A48).

An attractive (repulsive) potential is $V < 0$ ($V > 0$). The main point is that the field $\chi(X)$ must have canonical normalization of its kinetic term, which dictates the introduction of $Z$. The kinetic term of $\chi$ is seen to be extensive in $T$, and this is absorbed by normalizing $Z$, as $ZMT = 1$. The potential is determined by the integral over relative time of the interaction and is not extensive in $T$.

From the factorization $\Phi = \chi(X)\phi(\vec{r})$ with normalized $\phi(r)$, as in Equation (38), the eigenvalue $m^2$ is computed in the rest frame

$$m^2 = m^2 \int_{\vec{r}} |\phi(\vec{r})|^2 = \int_{\vec{r}} \left( -\phi^\dagger \nabla^2 \phi + V(\vec{r})|\phi(\vec{r})|^2 \right) \tag{A50}$$

From this, we obtain a normalized effective action for $\chi$ in any frame

$$S_\chi = \int d^4 X \left( \frac{\partial \chi^\dagger}{\partial X^\mu} \frac{\partial \chi}{\partial X_\mu} - m^2 |\chi|^2 \right) \tag{A51}$$

The mass of the bound state is determined by the eigenvalue of the static Schrödinger–Klein–Gordan (SKG) equation in the center-of-mass frame:

$$-\nabla^2 \phi + V(\vec{r})\phi(\vec{r}) = m^2 \phi(\vec{r}). \tag{A52}$$

MDPI

*Article*

# Weak Scale Supersymmetry Emergent from the String Landscape

Howard Baer [1,*], Vernon Barger [2], Dakotah Martinez [1] and Shadman Salam [3]

1  Homer L. Dodge Department of Physics and Astronomy, University of Oklahoma, Norman, OK 73019, USA; dakotah.s.martinez-1@ou.edu
2  Department of Physics, University of Wisconsin, Madison, WI 53706, USA; barger@pheno.wisc.edu
3  Department of Mathematics and Natural Sciences, Brac University, Dhaka 1212, Bangladesh; ext.shadman.salam@bracu.ac.bd
*  Correspondence: baer@ou.edu

**Abstract:** Superstring flux compactifications can stabilize all moduli while leading to an enormous number of vacua solutions, each leading to different $4 - d$ laws of physics. While the string landscape provides at present the only plausible explanation for the size of the cosmological constant, it may also predict the form of weak scale supersymmetry which is expected to emerge. Rather general arguments suggest a power-law draw to large soft terms, but these are subject to an anthropic selection of a not-too-large value for the weak scale. The combined selection allows one to compute relative probabilities for the emergence of supersymmetric models from the landscape. Models with weak scale naturalness appear most likely to emerge since they have the largest parameter space on the landscape. For finetuned models such as high-scale SUSY or split SUSY, the required weak scale finetuning shrinks their parameter space to tiny volumes, making them much less likely to appear compared to natural models. Probability distributions for sparticle and Higgs masses from natural models show a preference for Higgs mass $m_h \sim 125$ GeV, with sparticles typically beyond the present LHC limits, in accord with data. From these considerations, we briefly describe how natural SUSY is expected to be revealed at future LHC upgrades. This article is a contribution to the Special Edition of the journal *Entropy*, honoring Paul Frampton on his 80th birthday.

**Keywords:** supersymmetry; string theory; landscape; lhc

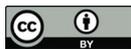

## 1. Introduction

Superstring theory provides the most promising avenue for unifying the Standard Model with gravity, but at the cost of requiring six or seven extra spatial dimensions [1–5]. The low energy limit $E < m_P$ (where $m_P$ is the reduced Planck mass) of string theory, once Kaluza–Klein modes are integrated out, is expected to be $10 - d$ supergravity (SUGRA). The $10 - d$ SUGRA theory is then assumed to be compactified to a tiny $6 - d$ space $K$ tensored with our usual $4 - d$ (approximately) Minkowski spacetime $M_4$: $M_{10} = M_4 \times K$. Originally, $K$ was taken to be a $6 - d$ compact Ricci-flat Kähler manifold with special holonomy [6]; such a Calabi–Yau manifold admits a conserved Killing spinor leading to a remnant $N = 1$ supersymmetry (SUSY) on $M_4$.

The cosmological constant (CC) problem remained a thorny issue until the early 2000s when it was realized that string flux compactifications could lead to an enormous number of vacuum states with different $4 - d$ laws of physics and, in particular, different $\Lambda_{CC}$ values [7]. Such large numbers of vacuum states ($N_{vac} \sim 10^{500}$ is an oft-quoted number [8]) provided a setting for Weinberg's anthropic solution to the CC problem [9]. But if the landscape [10] of string vacua provides a solution to the CC problem, might it also enter into other naturalness problems, such as the $m_{weak}/m_P$ (or related, $m_{SUSY}/m_P$) hierarchy problems (where $m_P \simeq 2.4 \times 10^{18}$ GeV)?

In this contribution to the volume of the journal *Entropy*, honoring Paul Frampton on his 80th birthday, we address this question. Here, we will put forward arguments in favor not only of weak scale SUSY as emergent from the string landscape, but indeed of a special form of weak scale SUSY (WSS)—SUSY with radiatively driven naturalness [11,12], or stringy-natural SUSY [13]. The specific form of WSS predicts at present that a light SUSY Higgs boson with mass $m_h \simeq 125$ GeV should emerge, whilst sparticles masses are at present somewhat or well beyond the reach of the CERN Large Hadron Collider (LHC) [14]. It also allows us to predict a variety of SUSY signatures, which may allow for SUSY discovery at LHC luminosity upgrades over the coming years. Perhaps most important of these are the soft isolated opposite-sign dileptons+MET which arise from light higgsino pair production [15] and which recoil against a hard initial state jet radiation [16–19]. At present, both ATLAS [20] and CMS [21] with 139 fb$^{-1}$ seem to have $2\sigma$ excesses in this channel, and an associated monojet signal may also be emerging [22].

## 2. Approximate Supersymmetric Vacua from String Theory

The main motivation for SUSY is that it provides a 'tHooft technical naturalness solution to the gauge hierarchy problem via the cancellation of quadratic divergences associated with the Higgs sector. This is true for SUSY breaking at any energy scale below $m_P$, since in the limit of $m_{SUSY} \to 0$, the model becomes more (super)symmetric. Thus, SUSY provides a technically natural solution to the so-called big hierarchy problem.

Specific motivation for weak scale SUSY comes from the little hierarchy problem and what we call *practical naturalness* [23,24]: an observable $\mathcal{O}$ is practically natural if all independent contributions to $\mathcal{O}$ are comparable to or less than $\mathcal{O}$. For the case of WSS, we can relate the weak scale $m_{weak} \sim m_{W,Z,h} \sim 100$ GeV to the weak scale soft SUSY-breaking terms and SUSY-conserving $\mu$ term via the scalar potential minimization conditions:

$$\frac{m_Z^2}{2} = \frac{m_{H_d}^2 + \Sigma_d^d - (m_{H_u}^2 + \Sigma_u^u)\tan^2\beta}{\tan^2\beta - 1} - \mu^2 \sim -m_{H_u}^2 - \Sigma_u^u(\bar{t}_{1,2}) - \mu^2 \qquad (1)$$

where $m_{H_{u,d}}^2$ are the soft SUSY-breaking Higgs masses and the $\Sigma_{u,d}^{u,d}$ contain an assortment of loop corrections to the scalar potential (explicit formulae are included in Ref's [12]). A measure of practical naturalness $\Delta_{EW}$ can be defined which compares the largest (absolute) contribution to the right-hand side of Equation (1) to $m_Z^2/2$. Requiring $\Delta_{EW} \lesssim 30$ fulfills the practical naturalness condition. From Equation (1), we see immediately that $m_{H_u}^2$ must be driven to *small* negative values at the weak scale while the $\mu$ term must also be $\mu \sim 100$–350 GeV. The latter condition means the higgsinos are usually the lightest SUSY particles, and the only ones required to be $\sim m_{weak}$. The other sparticles enter via the $\Sigma_{u,d}^{u,d}$ terms, and hence, are suppressed by loop factors, and so can live in the TeV or beyond range. We shall see shortly that practical naturalness is closely linked to the selection of SUSY models on the landscape.

On the theory side, we expect the $4 - d$ vacua emergent from the landscape to often contain some remnant SUSY.

- *Remnant spacetime SUSY:* In Ref. [25], Acharya argues that all stable, Ricci-flat manifolds in dimensions $< 11$ have special holonomy, and consequently a conserved Killing spinor. If so, then some remnant spacetime SUSY should exist in the $4 - d$ low-energy effective field theory (LE-EFT).
- *EW stability:* A problem for the Standard Model to be the low-energy effective field theory for $E < m_P$ is electroweak stability, in that the Higgs quartic term $\lambda$ may evolve to negative values at some intermediate scale, leading to a runaway scalar potential. For $m_t \sim 173.2$ GeV, the SM is just on the edge of metastability/runaway [26,27]. The Minimal Supersymmetric Standard Model (MSSM) has no such problem, since for the MSSM the quartic couplings involve the gauge couplings, which are always positive.
- *Landscape vacua stability:* In Ref's [28,29], Dine et al. ask what sort of conditions can stabilize landscape deSitter vacua against decay to AdS vacua, leading to a big crunch. The presence of SUSY leads to absolutely stable vacua, whilst the presence of approximate (broken) SUSY leads to (metastable) vacua decay rates $\Gamma \sim m_P e^{-m_P^2/m_{3/2}^2}$ far beyond the age of the universe.

- *Hierarchy of scales:* While a hierarchy of scales is typically hard to come by in many BSM models, SUSY models allow for dynamical SUSY breaking [30], where the SUSY-breaking scale $m_{hidden}$ is gained via dimensional transmutation $m_{hidden} \sim m_P e^{-8\pi^2/bg^2}$ and where the soft terms are developed as $m_{soft} \sim m_{hidden}^2/m_P$ under gravity-mediation. Here, we only consider gravity-mediation, since gauge mediation leads to tiny trilinear soft terms $A$, which then require unnatural top-squark contributions $\Sigma_u^u$ to gain $m_h \sim 125$ GeV [31].

- *Harmony:* Witten emphasizes that consistent QFTs exist for spin-0, 1/2, 1, 3/2, and 2. The graviton is the physical spin-2 particle and the spin-3/2 Rarita–Schwinger gravitino field would exist as the superpartner of the graviton, thus filling out all allowed spin states.

In addition, WSS is motivated experimentally by a variety of measurements.

- The measured values of the gauge couplings unify under MSSM RG evolution but do not under most other BSM extensions, including the SM itself [32].
- The measured top-quark mass is large enough to seed the required radiative breakdown of EW symmetry [33].
- The measured value of $m_h \simeq 125$ GeV falls squarely into the range allowed by the MSSM: $m_h \lesssim 130$ GeV [34].
- Precision EW corrections tend to prefer the (heavy spectra) MSSM over the SM [35].

A complaint often made, with good reason, is that gravity mediation has its own flavor and CP problems, the former arising from operators such as $\int d^4\theta S^\dagger S Q_i^\dagger Q_j/m_P^2$, where $S$ is a hidden sector superfield obtaining a SUSY-breaking vev $F_S \sim 10^{11}$ GeV and the $Q_i$ are visible-sector chiral superfields with generation index $i, j = 1 - 3$. Since no symmetry forbids such flavor mixing, then FCNCs are expected to be large in gravity-mediated SUSY breaking (historically, this strongly motivated the search for flavor-conserving models such as gauge-mediation and sometimes anomaly-mediation). It is pointed out in Ref. [36] that the landscape provides its own decoupling/quasi-degeneracy solution to the SUSY flavor and CP problems by pulling first-/second-generation matter scalars to a flavor-independent upper bound in the 20–40 TeV range.

For these reasons, we will assume a so-called "fertile patch" or friendly neighborhood [37] of the string landscape: those vacua which include the MSSM as the LE-EFT and where only the CC and the magnitude of the weak scale scan within the landscape. In this case, Yukawa couplings and gauge couplings are instead fixed by string dynamics rather than environmental selection. This leads to predictive landscape models [37]: if the CC is too large, then large scale structure will not form, which seems required for complexity to emerge (the structure principle, leading to Weinberg's successful prediction of $\Lambda_{CC}$). Only the magnitude of the weak scale scans. If $m_{weak}^{PU} \gtrsim 4m_{weak}^{OU}$, then the down–up quark mass difference becomes so large that neutrons are no longer stable in nuclei and the only atoms formed in the early universe are hydrogen. If $m_{weak}^{PU} \lesssim 0.5 m_{weak}^{OU}$, then we obtain a universe with only neutrons. This is the atomic principle [38], since complex nuclei are also apparently needed for complexity to emerge in any pocket universe (PU) within the greater multiverse (and where OU refers to $m_{weak}$ in our universe).

## 3. Natural SUSY from the Landscape

It is emphasized by Douglas that the CC scans independently of the SUSY-breaking scale in the landscape [39]. For the SUSY-breaking scale, we expect the vacua to be distributed as

$$dN_{vac} \sim f_{SUSY} \cdot f_{EWSB} \cdot dm_{SUSY}^2 \qquad (2)$$

where $m_{SUSY}$ is the overall hidden sector SUSY-breaking scale, expected to be $\sim 10^{11}$ GeV, such that the scale of soft terms is given by $m_{soft} \sim m_{SUSY}^2/m_P$.

### 3.1. Distribution of Soft Breaking Terms on the Landscape

How is $f_{SUSY}$ distributed? Douglas [39] emphasizes that there is nothing in string theory to favor any particular SUSY-breaking vev over another, and hence, $m_{soft}$ would be distributed as a power-law:

$$f_{SUSY} \sim m_{soft}^{2n_F + n_D - 1} \tag{3}$$

where $n_F$ is the total number of (complex-valued) $F$-breaking fields and $n_D$ is the total number of (real-valued) $D$-breaking fields contributing to the overall scale of SUSY breaking, $m_{SUSY}^4 = \sum_i |F_i|^2 + \sum_\alpha D_\alpha^2$. The prefactor of 2 in the exponent arises because the $F_i$ are distributed randomly as complex numbers. For the textbook case of SUSY breaking via a single $F$ term, then we expect $f_{SUSY} \sim m_{soft}^1$, i.e., a *linear* draw to large soft terms. If more hidden fields contribute to the overall SUSY-breaking scale, then the draw to large soft terms will be a stronger power-law.

While the overall SUSY-breaking scale is distributed as a power-law, the different functional dependence [40–42] of the soft terms on the hidden sector SUSY-breaking fields means that gaugino masses, the trilinear soft terms, and the various scalar masses will effectively scan independently on the landscape [43]. Now, it is an *advantage* that different scalar mass-squared terms scan independently (as expected in SUGRA), since the first-/second-generation scalars are pulled to much higher values than third-generation scalars, while the two Higgs soft masses are also non-universal and scan independently. This situation is borne out in Nilles et al.'s mini-landscape, where different fields gain different soft masses due to their different geographical locations on the compactification manifold [44]. In terms of gravity mediation, then the so-called *n*-extra-parameter non-universal Higgs model (NUHMn) with parameters [45,46]

$$m_0(i), \; m_{H_u}, \; m_{H_d}, \; m_{1/2}, A_0, \; \tan\beta \quad (NUHM4) \tag{4}$$

provides the proper template. Since the matter scalars fill out a complete spinor rep of $SO(10)$, we assume each generation $i = 1 - 3$ is unified to $m_0(i)$. Also, for convenience one may ultimately trade $m_{H_u}$ and $m_{H_d}$ for the more convenient weak scale parameters $m_A$ and $\mu$. One may also build (and scan separately) the natural anomaly-mediated SUSY-breaking model [47,48] (nAMSB) and the natural generalized mirage mediation model [49] (nGMM).

### 3.2. The ABDS Window

The anthropic selection on the landscape comes from $f_{EWSB}$. This involves a rather unheralded prediction of the MSSM: the value of the weak scale in terms of soft SUSY-breaking parameters and $\mu$, as displayed in Equation (1). However, in the multiverse, here we rely on the existence of a friendly neighborhood [37], wherein the LE-EFT contains the MSSM but where only dimensional quantities such as $\Lambda_{CC}$ and $v_u^2 + v_d^2$ scan, whilst dimensionless quantities like gauge and Yukawa couplings are determined by dynamics. This assumption leads to *predictivity* as we shall see.

Under these assumptions, then we ask what conditions lead to complex nuclei, atoms as we know them, and hence, the ability to generate complex lifeforms in a pocket universe? For different values of soft terms, frequently one is pushed into a weak scale scalar potential with charge-or-color-breaking minima (CCB) where one or more charged or colored scalar mass squared is driven tachyonic (i.e., $m^2 < 0$). Such CCB minima must be vetoed. Also, for values of $m_{H_u}^2$ that are too large, then its value is *not* driven to negative values and EW symmetry is generally not broken. These we label as "no EWSB" and veto them as well. In practice, we must check the boundedness of the scalar potential from below in the vacuum stability conditions and that the origin of field space has been destabilized at least at tree level.

At this point, we are left with (MS)SM vacua where the EW symmetry is properly broken, but where $m_{weak} \sim m_{W,Z,h}$ is at a different value from what we see in our universe. Here, we rely on the prescient analysis of Agrawal, Barr, Donoghue, and Seckel (ABDS) [38,50]. If the derived value of the weak scale is bigger than ours by a factor of 2–5, then the light quark mass difference $m_d - m_u$ becomes so large that neutrons are no longer stable in the nucleus and nuclei with $Z \gg N$ are not bound; such pocket universes would have nuclei of single protons only, and would be chemically inert. Likewise, if the PU value of the weak scale is a factor $\sim 0.5$ less than our measured value, then one obtains a universe with only neutrons—also chemically inert. The ABDS window of allowed values is that

$$0.5 m_{weak}^{OU} < m_{weak}^{PU} \lesssim 4 m_{weak}^{OU} \tag{5}$$

where we take the $(2 - 5) m_{weak}^{OU}$ to be $\sim 4 m_{weak}^{OU}$ for definiteness, which is probably a conservative value. This is very central to our analysis and so is displayed in Figure 1. Our anthropic condition $f_{EWSB}$ is then that the scalar potential leads to minima with not only the appropriate EWSB, but also that the derived value of the weak scale lies within the ABDS window. Vacua not fulfilling these conditions must be vetoed. Early papers on this topic used instead a naturalness "penalty" of $f_{EWSB} \sim m_{weak}^2 / m_{SUSY}^2$; this condition would allow for many of the vacua which are forbidden by our approach.

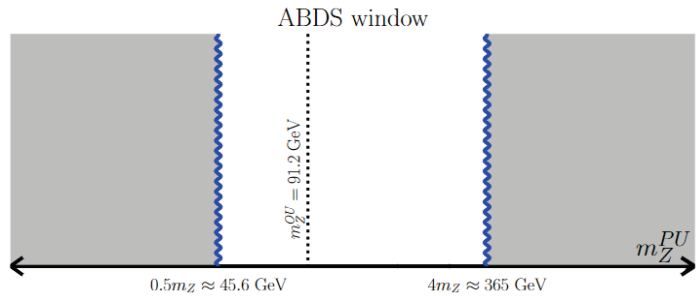

**Figure 1.** The ABDS-allowed window within the range of $m_Z^{PU}$ values.

### 3.3. EW-Natural SUSY Emergent from the Landscape

The next goal is to build a toy simulation of our friendly neighborhood of the string landscape. We can generate the soft terms of Equation (4) according to a power-law selection, usually taken to be $n = 2n_F + n_D - 1 = 1$ (linear draw). While Equation (3) favors the largest possible soft terms, the anthropic veto $f_{EWSB}$ places an upper bound on such terms because usually large soft terms lead to too large a value of $m_{weak}^{PU}$ beyond the ABDS window. The trick is to take the upper bound on scan limits beyond the upper bound posed by $f_{EWSB}$. However, in some cases larger soft terms are *more* apt to generate vacua within the ABDS window. A case in point is $m_{H_u}^2$: the smaller its value, the more negative it runs to unnatural values at the weak scale, while as it gets larger, then it barely runs negative: EW symmetry is barely broken. As its high scale value becomes even larger, it does not run negative by $m_{weak}$, and EW symmetry is typically not properly broken—such vacua failing to break the EW symmetry are vetoed. Also, for small $A_0$, the $\Sigma_u^u(\tilde{t}_{1,2})$ terms can be large. When $A_0$ becomes significantly negative, then cancellations occur in $\Sigma_u^u(\tilde{t}_{1,2})$ such that these loop corrections then lie within the ABDS window: large negative weak scale $A$ terms make $\Sigma_u^u(\tilde{t}_{1,2})$ more natural while raising the light Higgs mass to $m_h \sim 125$ GeV.

A plot of the weak scale values of $m_{H_u}$ and $\mu$ is shown in Figure 2 (taken from Ref. [51]) for the case where all radiative corrections—some negative and some positive [52]—lie within the ABDS window. The ABDS window lies between the red and green curves. Imagine playing darts with this target, trying to land your dart within the ABDS window. There is a large region to the lower left where both $m_{H_u}$ and $\mu$ are $\lesssim 350$ GeV, which leads to PUs with complexity. Alternatively, if you want to land your dart at a point with $\mu \sim 1000$ GeV, then the target space has pinched off to a tiny volume: the target space is finetuned and your dart will almost never land there. The EW-natural SUSY models live in the lower-left ABDS window while finetuned SUSY models with large $\Delta_{EW}$ lie within the extremely small volume between the red and green curves in the upper-right plane.

This tightly constrained region is labeled by split SUSY [53], high-scale SUSY [54] and mini-split [55].

It is often said that landscape selection offers an alternative to naturalness and allows for finetuned SUSY models. After all, is the CC not finetuned? However, from Figure 2 we see that models with EW naturalness (low $\Delta_{EW}$) have a far greater relative probability to emerge from landscape selection than finetuned SUSY models.

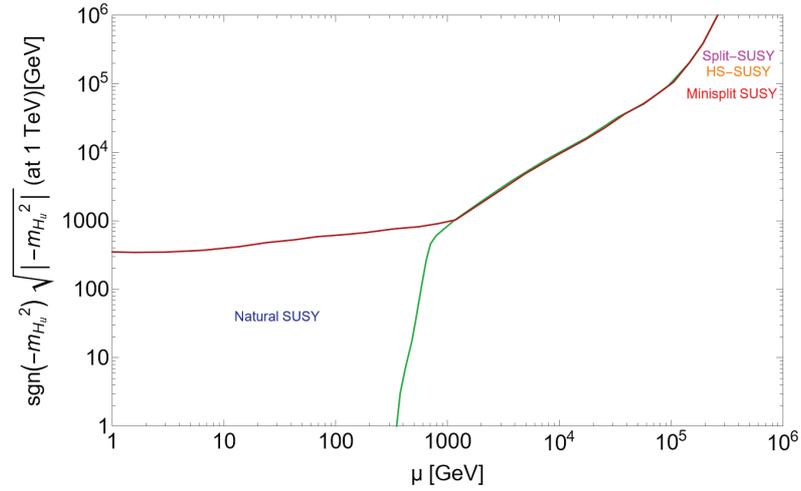

**Figure 2.** The $\mu^{PU}$ vs. $\sqrt{-m_{H_u}^2(weak)}$ parameter space in a toy model ignoring radiative corrections to the Higgs potential. The region between the red and green curves leads to $m_{weak}^{PU} < 4m_{weak}^{OU}$ so that the atomic principle is satisfied.

In Figure 3 (from Ref. [51]), we perform a numerical exercise to generate high-scale SUSY soft terms in accord with an $n = 1$ draw in Equation (3). The green dots are viable vacua states with appropriate EWSB and $m_{weak}^{PU}$ within the ABDS window. While some dots land in the finetuned region, the bulk of the points lie within the EW-natural SUSY parameter space.

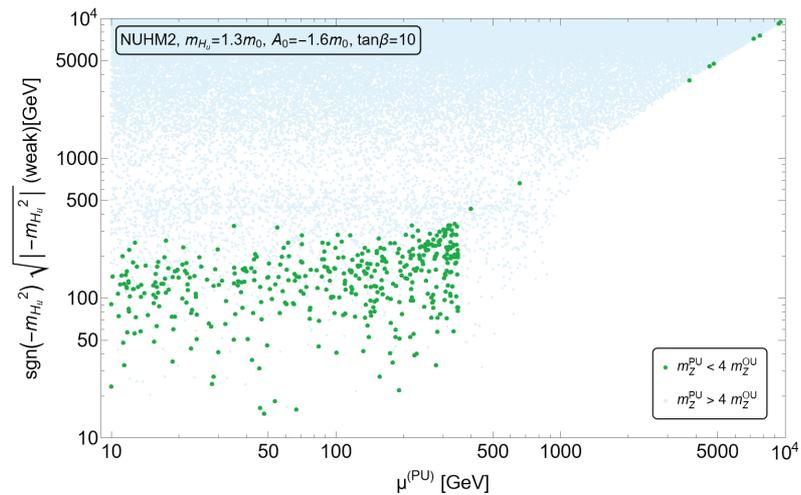

**Figure 3.** The value of $m_{H_u}(weak)$ vs. $\mu^{PU}$ The green points denote vacua with appropriate EWSB and with $m_{weak}^{PU} < 4m_{weak}^{OU}$ so that the atomic principle is satisfied. Blue points have $m_{weak}^{PU} > 4m_{weak}^{OU}$.

An alternative view is gained from Figure 4 from Ref. [56]. Here, we compute contributions to the scalar potential within a variety of SUSY models including RNS (radiatively driven natural SUSY [11]), CMSSM [57], G₂MSSM [58], high-scale SUSY [59], spread SUSY [60], mini-split [55], split SUSY [53], and the SM with cutoff $\Lambda = 10^{13}$ TeV, indicative of the neutrino see-saw scale [61]. The $x$-axis is either the SM $\mu$ parameter or the SUSY $\mu$ parameter while the $y$-axis is the calculated value of $m_Z$ within the PU. The ABDS window is the horizontal blue-shaded region. For $\mu$ distributed as equally likely at all scales (the distribution's probability density integrates to a log), then the length of the $x$-axis interval leading to $m_Z^{PU}$ within the ABDS window can be regarded as a relative probability measure $P_\mu$ for the model to emerge from the landscape. There is a substantial interval for the RNS model, but for finetuned SUSY models, the interval is typically much more narrow than the width of the printed curves. We can see that finetuned models have only a tiny range of $\mu$ values which allow habitation within the ABDS window.

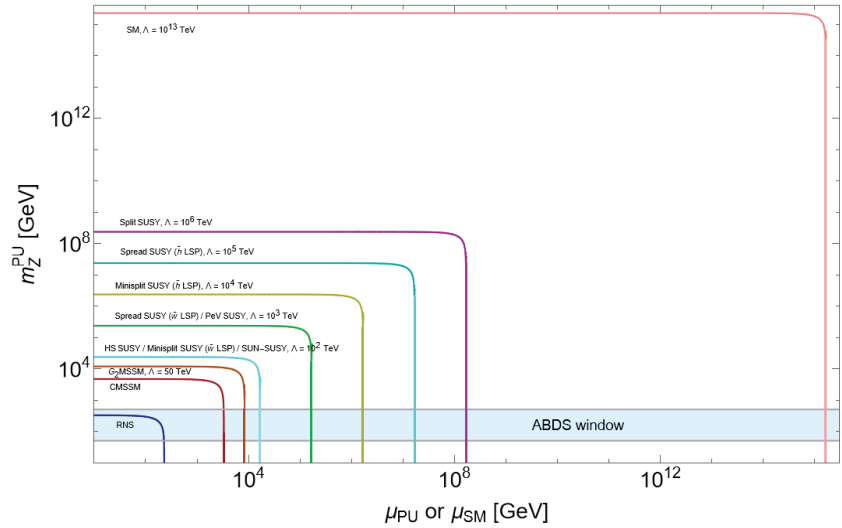

**Figure 4.** Values of $m_Z^{PU}$ vs. $\mu_{PU}$ or $\mu_{SM}$ for various natural (RNS) and unnatural SUSY models and the SM. The ABDS window extends here from $m_Z^{PU} \sim 50$ to 500 GeV.

Using the magic of algebra, the width of the $\mu$ intervals can be computed, and the results are shown in Table 1. Here, $P_\mu$ is to be considered as a *relative* probability. From the table, we see that the SM is about $10^{-27}$ times less likely to emerge as compared to RNS. Mini-split is $10^{-4}$–$10^{-8}$ times less likely to emerge (depending on the version of mini-split). Even the once-popular CMSSM model is $\sim 10^{-3}$ times less likely than RNS to emerge from the landscape.

**Table 1.** A survey of some unnatural and natural SUSY models along with general expectations for sparticle and Higgs mass spectra in TeV units. We also show the relative probability measure $P_\mu$ for the model to emerge from the landscape. For RNS, we take $\mu_{min} = 10$ GeV.

| Model | $\tilde{m}(1,2)$ | $\tilde{m}(3)$ | Gauginos | Higgsinos | $m_h$ | $P_\mu$ |
|---|---|---|---|---|---|---|
| SM | - | - | - | - | - | $7 \times 10^{-27}$ |
| CMSSM ($\Delta_{EW} = 2641$) | $\sim 1$ | $\sim 1$ | $\sim 1$ | $\sim 1$ | 0.1–0.13 | $5 \times 10^{-3}$ |
| PeV SUSY | $\sim 10^3$ | $\sim 10^3$ | $\sim 1$ | $1 - 10^3$ | 0.125–0.155 | $5 \times 10^{-6}$ |
| Split SUSY | $\sim 10^6$ | $\sim 10^6$ | $\sim 1$ | $\sim 1$ | 0.13–0.155 | $7 \times 10^{-12}$ |
| HS-SUSY | $\gtrsim 10^2$ | $\gtrsim 10^2$ | $\gtrsim 10^2$ | $\gtrsim 10^2$ | 0.125–0.16 | $6 \times 10^{-4}$ |
| Spread ($\tilde{h}$LSP) | $10^5$ | $10^5$ | $10^2$ | $\sim 1$ | 0.125–0.15 | $9 \times 10^{-10}$ |

**Table 1.** *Cont.*

| Model | $\tilde{m}(1,2)$ | $\tilde{m}(3)$ | Gauginos | Higgsinos | $m_h$ | $P_\mu$ |
|---|---|---|---|---|---|---|
| Spread ($\tilde{w}$LSP) | $10^3$ | $10^3$ | $\sim 1$ | $\sim 10^2$ | 0.125–0.14 | $5 \times 10^{-6}$ |
| Mini-Split ($\tilde{h}$LSP) | $\sim 10^4$ | $\sim 10^4$ | $\sim 10^2$ | $\sim 1$ | 0.125–0.14 | $8 \times 10^{-8}$ |
| Mini-Split ($\tilde{w}$LSP) | $\sim 10^2$ | $\sim 10^2$ | $\sim 1$ | $\sim 10^2$ | 0.11–0.13 | $4 \times 10^{-4}$ |
| SUN-SUSY | $\sim 10^2$ | $\sim 10^2$ | $\sim 1$ | $\sim 10^2$ | 0.125 | $4 \times 10^{-4}$ |
| $G_2$MSSM | 30–100 | 30–100 | $\sim 1$ | $\sim 1$ | 0.11–0.13 | $2 \times 10^{-3}$ |
| RNS/landscape | 5–40 | 0.5–3 | $\sim 1$ | 0.1–0.35 | 0.123–0.126 | 1.4 |

## 4. Radiatively Driven Natural SUSY

Along with probability distributions for models to emerge from the landscape, one can compute probability distributions for sparticle and Higgs mass values from particular models given an assumed value of $n$ in $f_{SUSY}$. Here, we use a linear draw, $n = 1$, to large soft terms with the NUHM4 model as the LE-EFT. We capture non-finetuned models by requiring $\Delta_{EW} \lesssim 30$, i.e., that the largest independent contribution to $m_Z$ lies within the ABDS window. These models have radiatively driven naturalness (RNS) [11], where RG running drives various soft terms to natural values at the weak scale. The value of $m_{H_u}(m_{GUT})$ is selected so that $m_Z = 91.2$ GeV in our universe.

The distribution for the light Higgs mass is shown in Figure 5. We see for $n = 1$ that the blue distribution rises to a maximum at $m_h \sim 125$ GeV. This is where $A_t$ is large enough to yield cancellations in the $\Sigma_u^u(\tilde{t}_{1,2})$ terms, but also lifts $m_h$ up to $\sim 125$ GeV via maximal stop mixing [11]. For comparison, we also show the orange histogram for $n = -1$, where soft terms are equally favored at any mass scale. Here, the distribution peaks at $m_h \sim 118$ GeV, with hardly any probability at $m_h \sim 125$ GeV.

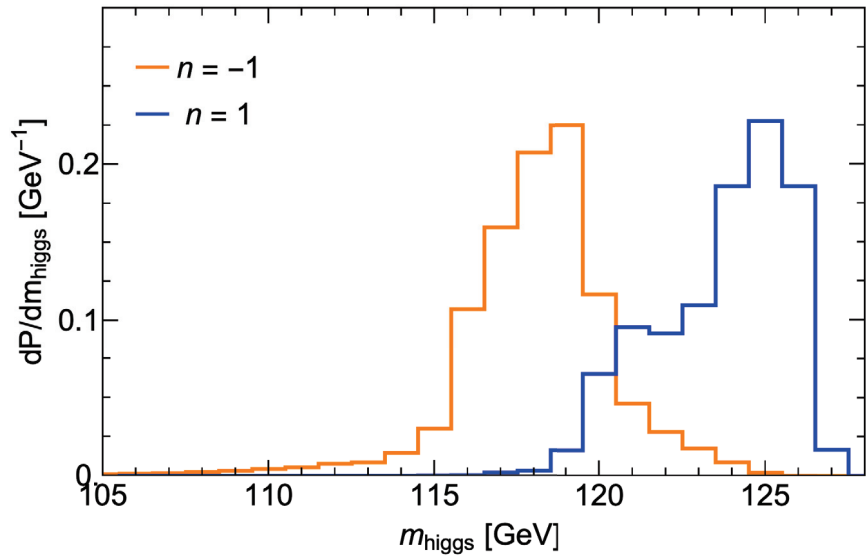

**Figure 5.** Probability distributions for the light Higgs scalar mass $m_h$ from the $f_{SUSY} = m_{soft}^{\pm 1}$ distributions of soft terms in the string landscape with $\mu = 150$ GeV.

In Figure 6, we show the corresponding probability distribution for the gluino mass. Here, for $n = 1$ the curve begins around $m_{\tilde{g}} \sim 1$ TeV and reaches a broad maximum around 3–4 TeV, while petering out beyond $m_{\tilde{g}} \sim 6$ TeV. The present LHC Run 2 limit from ATLAS/CMS [62,63] is $m_{\tilde{g}} \gtrsim 2.2$ TeV from searches within the simplified model context.

From the plot, we see that the LHC is only beginning to probe the expected range of $m_{\tilde{g}}$ values from the landscape.

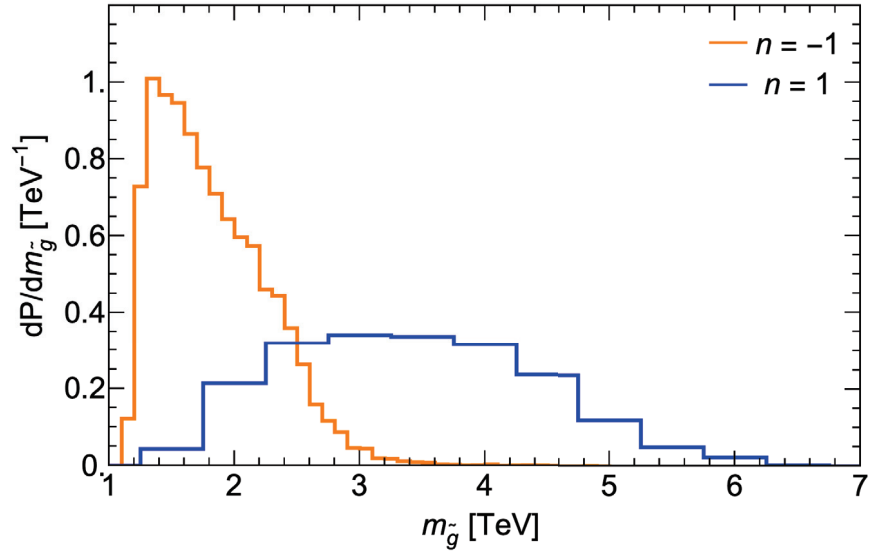

**Figure 6.** Probability distribution for $m_{\tilde{g}}$ from the $f_{SUSY} = m_{soft}^{\pm 1}$ distributions of soft terms in the string landscape with $\mu = 150$ GeV.

Other sparticle and Higgs mass distributions from the landscape are shown in Refs. [14,64], and they are typically beyond or even well beyond present LHC limits. For instance, light top-squarks are expected around $m_{\tilde{t}_1} \sim$1–2.5 TeV whilst first-/second-generation squarks and sleptons are expected near $m_{\tilde{q},\tilde{\ell}} \sim$10–30 TeV. From this point of view, LHC is at present seeing exactly what the string landscape predicts.

## 5. Conclusions

Theoretical arguments suggest many models which include a remnant spacetime SUSY to populate the string landscape of $4 - d$ vacua. We assume a friendly neighborhood of the landscape populated by the MSSM as the LE-EFT, but where the CC and also the soft SUSY-breaking gaugino masses, scalar masses, and $A$-terms scan via a power-law draw to large values. Landscape selection of soft terms then allows for a derived value of the weak scale, which must lie within the ABDS window in order for the atomic principle to be obeyed, leading to complex nuclei, and hence, atoms which are needed for complexity.

Under the landscape selection of soft SUSY-breaking terms, one expects radiative natural SUSY, or RNS, to be much more prevalent than finetuned SUSY models such as CMSSM, $G_2$MSSM, high-scale SUSY, split SUSY, or mini-split SUSY. This is evident because in RNS, where all contributions to the weak scale lie within the ABDS window, there is a much larger volume of scan space leading to $m_{weak} \in ABDS$. Alternatively, if even one contribution to the weak scale lies outside the ABDS window, then the remaining volume of parameter space leading to $m_{weak} \in ABDS$ shrinks to tiny values, and is relatively less likely. This is borne out by toy simulations of the string landscape and also allows for a relative probability measure $P_\mu$ for different models to emerge from the landscape. For instance, $P_\mu(RNS) \sim 1.4$, compared to, for instance, $P_\mu(HS\text{-}SUSY) \sim 6 \times 10^{-4}$. Finally, we show probability distributions of the light Higgs mass and gluinos, showing that the present LHC is seeing what one would expect from the string landscape. New SUSY signals, especially from higgsino pair production, could arise within the next few years at LHC.

With all these beautiful results, we anticipate that Paul will begin to work on landscape SUSY as well.

**Author Contributions:** Software, D.M. and S.S.; Writing—original draft, H.B.; Writing—review & editing, V.B. All authors have read and agreed to the published version of the manuscript.

**Funding:** This material is based upon work supported by the U.S. Department of Energy, Office of Science, Office of Basic Energy Sciences Energy Frontier Research Centers program under Award Number DE-SC-0009956 and DE-SC-0017647.

**Data Availability Statement:** No new data were created or analyzed in this study. Data sharing is not applicable to this article.

**Acknowledgments:** V.B. gratefully acknowledges support from the William F. Vilas estate.

**Conflicts of Interest:** The authors declare no conflicts of interest.

*Article*

# Dark Matter and Mirror World †

Rabindra N. Mohapatra

Maryland Center for Fundamental Physics, Department of Physics, University of Maryland, College Park, MD 20742, USA; rmohapat@umd.edu
† This paper is dedicated to Paul Frampton on his eightieth birthday with congratulations on his many contributions to particle theory and cosmology and best wishes for many more years of creativity.

**Abstract:** Overwhelming astronomical evidence for dark matter and absence of any laboratory evidence for it despite many dedicated searches have fueled speculation that dark matter may reside in a parallel universe interacting with the familiar universe only via gravitational interactions as well as possibly via some ultra-weak forces. In this scenario, we postulate that the visible universe co-exists with a mirror world consisting of an identical duplicate of forces and matter of our world, obeying a mirror symmetry. This picture, motivated by particle physics considerations, not only provides a natural candidate for dark matter but also has the potential to explain the matter dark matter coincidence problem, i.e., why the dark matter content of the universe is only a few times the visible matter content. One requirement for mirror models is that the mirror world must be colder than our world to maintain the success of big bang nucleosynthesis. After a review of the basic features of the model, we present several new results: first is that the consistency between the coldness of the mirror world and the explanation of the matter dark matter coincidence implies an upper bound on the inflation reheat temperature of the universe to be around $10^{6.5}$ GeV. We also argue that the coldness implies the mirror world consists mainly of mirror Helium and very little mirror hydrogen, which is the exact opposite of what we see in the visible world.

**Keywords:** mirror world; asymmetric inflation; matter–dark matter coincidence; helium dominated mirror sector

## 1. Introduction

There is now overwhelming evidence in favor of the existence of dark matter from many astrophysical observations such as the speed of galaxies in the Coma cluster, flat rotation curves of stars in galaxies, as well as the Chandra image of two galaxies crossing each other in the Bullet Cluster with dark matter moving ahead of the visible matter. This conclusion seems to have been further confirmed by the study of the cosmic microwave spectrum obtained by the NASA WMAP spacecraft followed by other space missions such as the Planck spacecraft of the European Space Agency, etc. This has granted urgency to the question of what dark matter is and if it is a collection of particles spread out over the universe, what particles are they and what kind of forces they experience other than gravity. The hope is that any understanding of dark matter will provide a glimpse into the nature of physics beyond the standard model.

Experiments in the laboratory set-ups deep underground as well as in colliders have been ongoing for the last thirty years to obtain the dark matter particle (or particles) with more and more sophisticated techniques, but they have all ended up with negative results (for a review, see [1]; see, however, the claims by the DAMA collaboration [2]). This has fueled speculation that dark matter could be residing in a parallel universe (or the mirror universe), in which case it only interaction with known matter (i.e., protons, neutrons or electrons) is via gravity forces or similar ultra-weak forces. This would explain why it seems to elude discovery by conventional detectors. It is the ramification of this idea that we discuss in this article. We call the familiar proton, neutron and electron the visible particles and their mirror partners the mirror protons ($p'$), mirror neutrons, etc.

The idea of a mirror universe first appeared in the famous parity violation paper of Lee and Yang in 1956 [3], where they noted that while parity is violated maximally in the beta decay process in our universe, it could be a good symmetry of nature if there was another sector to our universe (called mirror world here) with identical particle spectra and force content to our sector, with the opposite chirality fermions in the mirror sector participating in the mirror beta decay. Our world and the mirror world would transform to each other under mirror (or $Z_2$) symmetry. This picture is also motivated by a class of string theories based on the $E_8 \times E_8'$ group.

This picture provides a minimal extension of the standard model with very few additional parameters describing it. The phenomenological implications of this hypothesis were first discussed in a paper by Kovzarev, Okun and Pomeranchuk [4] in 1966. In recent years, these models have been the focus of many papers in the context of both particle physics and cosmology [5–12]. In particular, this model also provides a natural candidate for dark matter, which is the main motivation in this paper. The dark matter particle could either be mirror hydrogen or mirror neutron [13–17], whichever is the lightest baryonic particle. We choose the mirror neutron alternative here.

In this model, the dark matter displays self-interaction, which seems to be a useful attribute to explain several puzzles of the collisionless dark matter hypothesis [18,19].

After summarizing the basic ingredients of this model, the paper focuses on two of its salient features: (i) first is an important consistency requirement, which says that the mirror world must be cooler than ours. We outline how this can possibility be realized in concrete models; (ii) next, we present a scenario which provides a resolution [20] of the coincidence problem of matter and dark matter. For another recent proposal in this direction, see [21]. These scenarios require that the mirror fermions must have a higher mass than the fermions of our world. We then discuss the implications of these two ingredients for structure in the mirror universe.

The main new results of this paper are the following: (i) for the first time, we demonstrate the consistency between the colder mirror world and the scenario for matter–dark matter coincidence. (ii) This consistency requirement implies that the inflation reheat temperature of the universe must be less than $10^{6.5}$ GeV. (iii) A final interesting result is that the combined effect of lower temperature and higher mirror electroweak VEV implies that the mirror world consists mainly of mirror helium and very little mirror hydrogen, which is exactly the opposite of the situation in our universe.

## 2. The Mirror Model

As noted in the previous section, the mirror model consists of two sectors to our universe invariant under a discrete $Z_2$ symmetry, the mirror symmetry, which transforms all particles and forces of one sector to those of the other. The symmetry guarantees that the particles and forces in the mirror sector of the universe are duplicates of those in the visible sector with equal coupling strengths for the mirror-duplicated forces prior to symmetry breaking. In Table 1, we display the particle content of the model. The symmetry breaking may introduce differences between the two sectors. Depending on the way the gauge symmetries are broken, one can define two broad classes of mirror models. The first is called the symmetric mirror model, where the weak scale in both sectors are the same, whereas the second realization is one where the weak scales in the two sectors are different. In the symmetric mirror model, the visible particles in our world have the same or nearly the same mass as their mirror partners. This can lead to a new class of phenomena such as neutron–mirror-neutron oscillation [22] if there are interactions connecting $n$ to $n'$ such as $uddu'd'd'$. There are now several experiments searching for this process [23,24]. The Kaon oscillation can now involve four mesons $(K, \bar{K}, K', \bar{K}')$, if there are operators of type $\bar{d}s\bar{d}'s'$ in the theory. Note that such new oscillations are not generic to mirror models and need additional assumptions.

**Table 1.** Gauge quantum numbers of all the fields in the theory; $\eta$ is a mirror parity odd field.

| **Our World** | $SU(3)_c \times SU(2)_L \times U(1)_Y$ | **Mirror World** | $SU(3)'_c \times SU(2)'_L \times U(1)'_Y$ |
|---|---|---|---|
| Visible fermions | | mirror fermions | |
| $Q_L$ | $(3, 2, 1/3)$ | $Q'_L$ | $(3, 2, 1/3)$ |
| $u_R$ | $(3, 1, 4/3)$ | $u'_R$ | $(3, 1, 4/3)$ |
| $d_R$ | $(3, 1, -2/3)$ | $d'_R$ | $(3, 1, -2/3)$ |
| $\ell_L$ | $(1, 2, -1)$ | $\ell'_L$ | $(1, 2, -1)$ |
| $e_R$ | $(1, 1, -2)$ | $e'_R$ | $(1, 1, -2)$ |
| Gauge bosons | | Mirror Gauge bosons | |
| $W, Z, \gamma$, Gluons | | $W', Z', \gamma'$, mirror Gluons | |
| Scalar sector | | mirror scalar | |
| $H$ | $(1, 2, 1)$ | $H'$ | $(1, 2, +1)$ |
| $\eta$ | $(1, 1, 0)$ | $\eta$ | $(1, 1, 0)$ |

To implement other details such as $n'$ as dark matter, we must make sure that $n'$ is the lightest baryon in the mirror sector. A single Higgs doublet in each sector leads to relation $\frac{m_p}{m_n} = \frac{m_{p'}}{m_{n'}}$. To make $m_{n'}$ less than $m_{p'}$, we need to add another Higgs doublet to each sector. Furthermore, to understand neutrino masses, we add a $Y = 2$ triplet Higgs field to each sector as well [25]. Finally, we add three gauge singlet fermions $N_a$ connecting the mirror world and the visible one to explain the matter–dark matter coincidence puzzle.

The interactions of the mirror particles are identical duplicates of those in the visible sector, and we do not write them down here.

## 3. Consistency Requirements for the Mirror World Picture

The basic picture for the mirror world scenario is that at the big bang origin time, both ours and the mirror universe were present and started evolving in a completely identical manner. The next big event in the evolution of the universe was the inflation to explain the isotropy, homogeneity, causal connectedness and flatness of the universe. The question that now arises is the whether both the worlds inflate and reheat the in same way. It turns out that they do, but they must reheat to different temperatures after inflation with the mirror world reheating to a cooler temperature and remaining colder for the rest of its life. Thus, the two requirements for mirror models are: (i) asymmetric inflation [12,13,26] so that the reheat temperature in the mirror sector is lower and (ii) the absence of interactions connecting both worlds that will put them in equilibrium with each other after the reheat.

The reason why the mirror sector has to be colder is the fact that the mirror sector adds three extra neutrinos, the mirror electron and a mirror photon to the cosmic plasma of relativistic particles on top of the already known neutrinos, electron and the photon, thus doubling the relativistic degrees of freedom at the BBN epoch. This increased number is in sharp contradiction to the fact that the known neutrinos, the photon and the electrons are just enough to explain the observed helium, deuterium and lithium abundance in the visible universe. The extra degrees of freedom, if any, are collectively denoted by $\Delta N_{eff}$, which is restricted to be less than 0.3. Since the energy density of relativistic particles acts like $T^4$, a cooler mirror sector reduces the extra energy density contributed by the extra mirror particles to the desired level to restore consistency of the nucleosynthesis results. The necessary coldness of the mirror sector can be determined from this. If we denote $x = \frac{T'}{T}$, the present limits from BBN are satisfied for $x \leq 0.7$, assuming $\Delta N_{eff} \leq 0.3$. In the discussion below, we assume $x = 0.5$ for definiteness. The only way to avoid this requirement is to design the model in such a way that all mirror neutrinos, the mirror photon and the mirror electrons are much heavier so that they would have annihilated or decayed away by the BBN epoch. We do not make this assumption in what follows.

To obtain the cooler mirror sector, we resort to the mechanism of asymmetric inflation outlined in [12]. The $Z_2$ invariant Higgs potential for the model with a single Higgs doublet in each sector that implements asymmetric inflation is given by

$$V(\eta, H, H') = V(H, H') + m_\eta^2 \eta^2 + \lambda_\eta \eta^4 + \mu_\eta \eta(H^\dagger H - H'^\dagger H') + \lambda_{\eta H} \eta^2 (H^\dagger H + H'^\dagger H') \tag{1}$$

with

$$V(H, H') = \mu_H^2(H^\dagger H + H'^\dagger H') + \lambda_H[(H^\dagger H)^2 + (H'^\dagger H')^2] + \lambda'_H(H^\dagger H)(H'^\dagger H') \tag{2}$$

We note that the potential has no $\eta^3$ term since $\eta$ is a mirror $Z_2$ odd field. The $\eta$-field is the inflaton field which acquires VEV $< \eta > \neq 0$. This asymmetrises $\eta$ couplings to $H$ and $H'$ fields leading to $\Gamma(\eta \to HH) > \Gamma(\eta \to H'H')$ as we see later. After inflation ends, the inflaton field decays to the two sectors in an asymmetric way, leading to different reheat temperatures in the two sectors. The same asymmetric coupling of the $\eta$-field also leads to the electroweak VEVs in the two sectors being different. Thus, $\eta$ plays a dual role in the model, unifying two different aspects of it.

Another way to restore consistency of mirror models with BBN is to add heavy gauge singlet Majorana neutrinos $N, N'$ to the two sectors, respectively, and connect them via a mass term $MNN'$ + h.c. This can lead to $N$ eigenstates with different masses and different couplings to $\ell H$ and $\ell' H'$ states and a subsequent release of more relativistic particles from their decay to the visible sector compared to the mirror sector [27]. This, in turn, leads to $x < 1$ and solving the BBN problem. We do not follow this route here.

Thus, in our asymmetric mirror scenario, the two parameters that characterize the mirror sector of the universe are $x$, the ratio of the two temperatures and $\beta$, the ratio of mass scales, $\beta \equiv \frac{v_{wk}'}{v_{wk}}$. One immediate consequence of $\beta \ll 1$ is that the two strong couplings $\alpha_s$ and $\alpha_s'$, which start out being equal at very high energies due to mirror symmetry, become different at low scales, when we obtain $\alpha_s(\mu) \ll \alpha_s'(\mu)$. This results from the fact that the mirror top quark decouples much above the visible top quark from QCD running since $(m_{t'}/m_t) = (v'_{wk}/v_{wk}) \gg 1$. This leads to the following constraint on QCD and QCD' scales: $\Lambda'_{QCD} \gg \Lambda_{QCD}$, which makes the mirror sector particles, and in particular baryons, heavier.

## 4. Asymmetric Inflation, Weak Scale Asymmetry and Constraints on Model Parameters

As already noted, the $\eta$ field in Equation (1), which is odd under mirror symmetry, plays an important role in the model; (i) it is the inflaton field; (ii) its VEV asymmetrises the inflaton coupling to the Higgs and mirror Higgs fields, which leads to asymmetric reheating in the two sectors; and (iii) finally, its VEV also asymmetrises the particle spectrum in both sectors and becomes one of the keys to solving the matter–dark matter coincidence problem. We note, parenthetically, that ours is a model of chaotic inflation, which is slightly outside the Planck CMB data, but this problem can be cured by coupling $\eta$ non-minimally to gravity. We do not dwell on this aspect here. As an order of magnitude, we estimate that VEV of $\eta$ to be of order $|M_\eta|$, which is assumed in the discussion below.

We see that $\eta$ vev makes the $H$ and $H'$ masses different as follows [28]:

$$M_H^2 = \mu_H^2 + \lambda_{\eta H} v_\eta^2 + \mu_\eta v_\eta$$
$$M_{H'}^2 = \mu_H^2 + \lambda_{\eta H} v_\eta^2 - \mu_\eta v_\eta \tag{3}$$

The reheat temperatures are given by $T_{RH} \simeq \sqrt{\Gamma_\eta M_P}$. If we then use the width of $\eta \to HH$, etc., as (assuming $M_\eta \gg \mu_H$)

$$\Gamma(\eta \to HH) \simeq \frac{(\mu_\eta + \lambda_{\eta H} v_\eta)^2}{8\pi M_\eta} \tag{4}$$

and similarly for $\eta \to H'H'$ decay, we obtain

$$\frac{T'_{RH}}{T_{RH}} \sim \frac{\mu_\eta - \lambda_{\eta H} v_\eta}{\mu_\eta + \lambda_{\eta H} v_\eta} \approx 0.3 \qquad (5)$$

We now summarize the constraints on the parameters of the potential above that follow from weak scale asymmetry and asymmetric reheating. For the sake of illustration, we take $M_\eta \sim 10^8$ GeV as a benchmark parameter.

- First, we find that $< \eta > \sim M_\eta$ for $\lambda_\eta \sim 1$.
- Since $M^2_{H,H'}$ to break the electroweak symmetries in both the visible and mirror sectors, to obtain $v'_{wk} \sim 10^3 v_{wk}$, $|\mu^2_H + \lambda_{\eta H} v^2_\eta| \sim \mu_\eta v_\eta \sim 10^{10}$ GeV$^2$, $\mu_\eta v_\eta > 0$ and $\mu^2_H < 0$ is required, as well as $-|\mu^2_H| + \lambda_{\eta H} v^2 - \eta \mu_\eta v_\eta \approx 10^4$ GeV$^2$. This produces the desired parameter range for our model, i.e., $\beta \sim 10^{-3}$. This also implies that

$$\mu_\eta v_\eta \approx |\mu^2_H| \sim 10^{10} \text{GeV}^2 \qquad (6)$$

- This offers $\mu_\eta \sim 100$ GeV and $\lambda_\eta \sim 10^{-6}$ for our benchmark choice.
- We require $\frac{T'_{RH}}{T_{RH}} \sim x \sim 0.5$. This leads to $\mu_\eta \approx 2\lambda_{\eta H} v_\eta$.
- These results for $M_\eta \sim 10^8$ GeV produce $T_{RH} \sim 10^{6.5}$ GeV. Thus, we obtain an upper bound on the inflation reheat in the visible sector.
- The last constraint at this stage is that the mass of the $\eta$-field must be such that the $HH \to H'H'$ scattering via $\eta$ exchange does not thermalize the two sectors till the BBN epoch, i.e., $T_{BBN} \sim 10^{-3}$ GeV. The condition for this is that $\frac{\mu^4_\eta T}{8\pi M^4_\eta} \leq 10 \frac{T^2}{M_P}$. For $M_\eta \sim 10^8$ GeV, this implies that the temperature below which there is equilibrium is given by $T_* \sim \frac{\mu^4_\eta M_P}{100 M^4_\eta} \sim 10$ eV. This is acceptable since it does not affect BBN.
- In fact, by varying the value of $M_\eta$, we find that this condition implies an upper limit on $T_{RH}$ of order $10^{6.5}$ GeV. In Table 2, we present the value of $T_*$ for different choices of $M_\eta$ which helps us to obtain this upper limit on $T_{RH}$. This upper limit is important since it implies that the masses of the singlet fermions $N$ must be less than this if they have to be present in the universe to generate lepton asymmetry in both the visible and dark sectors (see the next section).

**Table 2.** Lowering $M_\eta$ brings the mirror and visible sectors to equilibrium and makes the theory unacceptable. Increasing $M_\eta$ keeps the theory acceptable but yields a lower $T_{RH}$. Thus, near about the value of $M_\eta \sim 10^8$, the maximum $T_{RH}$ is yielded. We choose this optimal value for our parameters.

| $M_\eta$ | $\mu_\eta$ | $T_{RH}$ | $T_*$ | Comment |
|---|---|---|---|---|
| $10^8$ GeV | 100 GeV | $10^{6.5}$ GeV | 10 eV | acceptable |
| $10^{10}$ GeV | 1 GeV | $10^3$ GeV | $10^{-25}$ GeV | acceptable |
| $10^6$ GeV | $10^4$ GeV | $10^{8.5}$ GeV | $10^9$ GeV | unacceptable |

## 5. Matter–Dark Matter Coincidence

As already noted, the lightest baryon of the mirror sector (in our case, $n'$) can be a dark matter of the universe and in the framework described below is an asymmetric dark matter [29,30]. After the end of inflation reheat, the universe undergoes usual Hubble expansion and processes leading to leptogenesis and Big Bang Nucleosynthesis start in both sectors. Due to asymmetric weak scales and colder mirror sector, the value of $g^*$, the number of degrees of freedom are not always same in both sectors, but for simple illustration of the phenomena we are interested in, we assume them to be same.

Let us first discuss how the mirror world explains the matter–dark matter coincidence puzzle via leptogenesis [12]. For this purpose, we add three SM singlet Majorana fermions $N_a$ portals which connect both sectors of the universe via the following couplings [12]:

$$\mathcal{L}_Y = M_N(NN) + hN(LH + L'H') + h.c. \tag{7}$$

where we drop the three flavor indices in coupling matrix $h$ and mass matrix $M_N$. We then use leptogenesis as the co-genesis mechanism for matter and dark matter following [12] (we note that there is no $\eta NN$ coupling in the theory, since $NN$ is $Z_2$ even when $\eta$ is $Z_2$ odd). We assume that the mass of the $N$ singlets is $10^6$ GeV so that after reheating is completed they exist in the cosmic fluid. They connect with the SM and mirror particles via their couplings in Equation (5). We further assume that they produce lepton asymmetry via leptogenesis in both the visible and mirror sectors. Due to mirror symmetry, they produce an equal amount of lepton asymmetry in both sectors, which is then converted to both visible and mirror baryons, producing $n_B = n_{B'}$ due to their respective sphaleron interactions. The lepton asymmetry is produced below $T \sim M_N \sim 10^6$ GeV, when despite a colder mirror sector, the mirror sphalerons are still active. This requirement also puts an upper limit on $v'_{wk} \leq M_N$.

Since the $N$ masses are low, the mechanism is the resonant leptogenesis mechanism [31], which requires that at least two of the portal right-handed neutrinos (RHN) are degenerate. The RHNs must exit the equilibrium at $T \simeq M_N$. The condition for that is

$$\frac{hh^\dagger M_N}{4\pi} \simeq 10 \frac{M_N^2}{M_P} \tag{8}$$

For $M_N \sim 10^6$ GeV, this implies $h \sim 10^{-5}$. In resonant leptogenesis, by adjusting the degree of degeneracy, one can produce an adequate amount of lepton asymmetry.

The addition of the cogenesis to the mirror model imposes these constraints on the model parameters:

- We must guarantee that the wash-out processes are out of equilibrium, which requires that $K = \frac{\Gamma}{H} \leq 10^6$, which is easily satisfied in the model.
- We must also ensure that the $N$-mediated $\ell H \to \ell' H'$ scatterings do not equilibrate the two worlds. This implies that

$$\frac{(hh^\dagger)^2 T^3}{4\pi M_N^2} < \frac{10 T^2}{M_P} \tag{9}$$

This condition is easily satisfied below the reheat temperature of $10^{6.5}$ GeV.

It follows from these considerations that $n_B = n_{B'}$. Therefore, if the mass of the mirror neutron dark matter is about 4–5 GeV, we have an understanding of the matter–dark matter coincidence puzzle.

## 6. Helium Universe in the Mirror Sector

In this section, we discuss some implications of the two outstanding features of the asymmetric mirror model, i.e., ($\beta \equiv \frac{v_{wk}}{v'_{wk}} < 1$) and a cold mirror sector ($x \equiv \frac{T'}{T} < 1$). As a first step, we discuss the consequences of nucleosynthesis in the mirror sector. We present a very simple approximate analysis, neglecting the difference in the degrees of freedom between the two sectors as the universe evolves and also subtleties associated with deuterium formation prior to helium synthesis.

First, we need to write down the expansion rate equation of the universe in terms of the temperature of the mirror sector, $T' = xT$, where $T$ is the temperature of the visible sector. We then have

$$M_P^2 H^2 = g^* T'^4 (1 + x^{-4}) \tag{10}$$

To discuss big bang nucleosynthesis (BBN) in the mirror sector [32], we first find out the value of $T'$ at which the mirror weak reactions involving mirror neutrinos, such as $\nu' + n' \to p' + e'$, that maintain the mirror neutron proton equality go out of equilibrium. The equation for that is

$$G_F^2 \beta^4 T'^5 \leq \frac{g^{*1/2} T'^2 (1 + x^{-4})^{1/2}}{M_P} \tag{11}$$

As just noted, we assume the number of degrees of freedom $g^*$ to be same in each sector for simplicity. Adding extra details on this would not significantly change our broad conclusion.

This leads to the $\nu'$ decoupling temperature $T'_*$ in the mirror sector to be

$$T'_* \simeq (g^*)^{1/6} G_F^{-2/3} \beta^{-4/3} (1 + x^{-4})^{1/6} M_P^{-1/3} \text{ GeV} \tag{12}$$

For our parameter choice, this $n'/p'$ freeze-out occurs at the mirror sector temperature equal to $\sim$70 GeV. It was shown in Ref. [20] that for a certain choice of the two Higgs doublet VEVs (or $\tan \beta'$) in the model, the mirror neutron can be lighter than the mirror proton and the mass difference $m_{p'} - m_{n'} \simeq 2$ GeV or so, which means that the number of mirror protons and mirror neutrons at their freeze-out epoch is about same, with the number of neutrons slightly exceeding that of protons. Since the nuclear forces in the visible and the mirror sectors are similar, we expect that all the protons ($p'$) combine with equal number of neutrons ($n'$) to form mirror Helium with very few mirror neutrons left over leading to a Helium dominated mirror universe as announced. Dark matter then consists of mirror helium and leftover mirror neutrons (for an interpretation of the DAMA results in the mirror model framework, see [33,34]).

## 7. Comments and Conclusions

In this brief note, we summarize the main points of the asymmetric mirror world model for dark matter, where the electroweak symmetry breaking in the mirror sector is higher than that of the visible sector. It turns out that the consistency between a colder mirror sector with electroweak VEV asymmetry implies an upper bound on the inflation reheat temperature of $10^{6.5}$ GeV, which is a new result of this paper. We outline the cogenesis for matter and dark matter in this set-up and show that the weak scale asymmetry, together with a colder mirror sector, leads to the mirror sector being helium-dominated. This has implications for structure formation in the mirror universe.

There are many relevant points about asymmetric mirror models that we do not discuss here. For example, in these models, there are other gauge-invariant interactions which can connect both sectors, e.g., photon mirror photon mixing coming from hypercharge gauge boson mixings $B_{\mu\nu} B'^{\mu\nu}$, Higgs mixings $H^\dagger H H'^\dagger H'$, etc. For the consistency of the model described here, these interactions must be highly suppressed. The other issue that we do not address is the formation of structure in a helium universe and mirror stellar evolution as weal as the possibility that familiar neutron stars could contain mirror dark matter in their core and how it can affect their evolution, the latter item discussed in [35–37].

**Funding:** This research received no external funding.

**Institutional Review Board Statement:** Not applicable.

**Data Availability Statement:** No new data were created or analyzed in this study. Data sharing is not applicable to this article.

**Acknowledgments:** I thank Yue Zhang for useful comments and discussions as well as reading through the manuscript.

**Conflicts of Interest:** The author declares no conflicts of interest.

*Review*

# Flavor's Delight

**Hans Peter Nilles ¹ and Saúl Ramos-Sánchez ²,***

¹ Bethe Center for Theoretical Physics, Physikalisches Institut der Universität Bonn, Nussallee 12, 53115 Bonn, Germany; nilles@th.physik.uni-bonn.de
² Instituto de Física, Universidad Nacional Autónoma de México, Ciudad de México C.P. 04510, Mexico
* Correspondence: ramos@fisica.unam.mx

**Abstract:** Discrete flavor symmetries provide a promising approach to understand the flavor sector of the standard model of particle physics. Top-down (TD) explanations from string theory reveal two different types of such flavor symmetries: traditional and modular flavor symmetries that combine to the eclectic flavor group. There have been many bottom-up (BU) constructions to fit experimental data within this scheme. We compare TD and BU constructions to identify the most promising groups and try to give a unified description. Although there is some progress in joining BU and TD approaches, we point out some gaps that have to be closed with future model building.

**Keywords:** flavor; string compactifications; eclectic symmetries

## 1. Introduction

The problem of flavor, the description of masses and mixing angles of quarks and leptons, remains one of the most important questions in elementary particle physics. A major approach to solve this problem is based on non-Abelian (discrete) flavor symmetries. In attempting to fit presently available data, many different symmetries and representations of flavor groups have been suggested and analysed. A comprehensive summary of these BU attempts can be found in the reviews [1–4]. In his book [5] with Jihn E. Kim, entitled *History of Particle Theory*, Paul Frampton (p. 172) mentions his preferred flavor group $T'$, the binary tetrahedral group. This choice is motivated through his early work on flavor symmetries: see ref. [6] and references therein.

Most attempts in the BU approach focus on the lepton sector to obtain solutions close to neutrino tribimaximal mixing [7]. Prominent examples have been $A_4$, $S_4$, $\Delta(27)$, $\Delta(54)$, $\Sigma(81)$, and $Q(24)$, among many others [8]. While they lead to acceptable solutions in the lepton sector, applications to the quark sector have been less frequent and usually less successful. Still, as there are many viable models it is difficult to draw a definite conclusion about the correct choice.

It seems that we need additional ingredients to select models from a more theoretical point of view. Such TD considerations draw their motivation from string theory model building, in particular orbifold compactifications of the heterotic string [9–14]. Early work [15] on the $\mathbb{Z}_3$ orbifold revealed the discrete flavor group $\Delta(54)$ with irreducible triplet representations to describe the three families of quarks and leptons. Even more earlier work, analyzing duality symmetries in string theory [16–19], provided an example of the discrete (modular) group $T'$. From this point of view, the predictions of the $\mathbb{Z}_3$ orbifold lead to the discrete groups $\Delta(54)$ and $T'$.

Fortunately, these groups allow many connections to models of the BU approach; in fact, $T'$ and $\Delta(54)$, as well as their "little sisters" $A_4$ and $\Delta(27)$, have played a major role (for an encyclopedia of discrete groups and technical details, we refer to ref. [20]). In the following, we want to analyze these specific constructions in detail. In Section 2, we start with the tetrahedral group $T$ (isomorphic to the group $A_4$ of even permutations of four objects), which played a major role in the discussion of neutrino tribimaximal mixing. We continue with its double cover $T'$ and potential applications to flavor physics. Section 3 introduces the motivation for the use of the group $\Delta(27)$ for leptonic mixing. It has 27 elements and is a discrete subgroup of SU(3). It is also a subgroup of $\Delta(54)$ that appeared in early discussions of flavor groups in string theory constructions [15]. Section 4 is devoted to TD considerations of flavor symmetries from string theory model building. There, we shall also introduce the concept of discrete modular symmetries that were discovered from an analysis of dualities in string theory [16–19]. The application of modular symmetries to flavor physics was pioneered in the BU approach by Feruglio [21] for the example of the discrete modular group $A_4$. We argue that the TD approach favors instead the modular flavor group $T'$, the double cover of $A_4$. Section 5 introduces the concept of the eclectic flavor group [22,23] that appears as a prediction in the string theory framework. It combines the traditional flavor symmetries (here, $\Delta(54)$) with the discrete modular flavor symmetries (here, $T'$). In Section 6, we shall try to make contact between the BU and TD approaches. Section 7 will give an outlook on strategies for further model building. The appendices will provide technical details of the properties of $A_4$, $T'$, $\Delta(27)$, and $\Delta(54)$.

## 2. The Tetrahedral Group and Its Double Cover

The symmetry group $T$ of the tetrahedron is one of the smallest non-Abelian discrete groups and found early applications in particle physics [24,25]. It has 12 elements and is isomorphic to $A_4$, the group of even permutations of four elements. There are three singlets ($\mathbf{1}, \mathbf{1'}, \mathbf{1''}$) and one irreducible triplet representation. Detailed properties of $T \cong A_4$ can be found in Appendix A.1. The presence of the triplet representation makes it attractive for flavor physics with three families of quarks and leptons. It became particularly relevant for the discussion of (nearly) tribimaximal mixing [26,27] in the lepton sector. An explicit discussion of this situation can be found in the reviews [1,8]. Tribimaximal mixing [7] is characterized (up to phases) through the PMNS structure

$$U_{\text{PMNS}} = \begin{pmatrix} \sqrt{\frac{2}{3}} & \frac{1}{\sqrt{3}} & 0 \\ -\frac{1}{\sqrt{6}} & \frac{1}{\sqrt{3}} & -\frac{1}{\sqrt{2}} \\ -\frac{1}{\sqrt{6}} & \frac{1}{\sqrt{3}} & \frac{1}{\sqrt{2}} \end{pmatrix}$$

and includes a $\mathbb{Z}_2 \times \mathbb{Z}_2$ symmetry acting (in the neutrino mass basis) as $U = \text{diag}(-1, -1, 1)$ and $V = \text{diag}(-1, 1, -1)$. This symmetry is a subgroup of $S_4$, the group of permutations of four elements. Tribimaximal mixing, however, is not exactly realized in nature as it would imply that the (reactor) angle $\theta_{13}$ vanishes. The $\mathbb{Z}_2$ transformation $V$ thus cannot be an exact symmetry. This brings $A_4$ into the game, a subgroup of $S_4$ that does not contain $V$. It allows satisfactory fits for the lepton sector, as reviewed in ref. [1]. These applications typically use the triplet representation for the left-handed lepton-SU(2)-doublets ($\nu_i, \ell_i$) and the representations ($\mathbf{1}, \mathbf{1'}$ and $\mathbf{1''}$) for the SU(2) singlets of the standard model of particle physics (SM). Various "flavon" fields have to be considered for the spontaneous breakdown of $A_4$, and this is subject to explicit model building, which we shall not discuss here in detail. In any case, $T \cong A_4$ is a very appealing discrete flavor symmetry for the description of the lepton sector.

A look at the quark sector reveals a completely different picture: there, all mixing angles are small and a fit similar to the lepton sector does not seem to work. One particular property of the quark sector is the fact that the top-quark is much heavier than the other quarks. This seems to indicate a special role of the third family, somewhat sequestered from the other two families. This could therefore imply that for quarks the third-family is a singlet under the discrete flavor group. Such a situation can be well described in the framework of $T'$, the double cover of $T \cong A_4$. This group has 24 elements with representations $\mathbf{1}, \mathbf{1'}, \mathbf{1''}, \mathbf{3}$ (as $A_4 \cong T$) and in addition doublet representations $\mathbf{2}, \mathbf{2'}, \mathbf{2''}$ (details of properties of $T'$ can be found in Appendix A.2). This double-cover is similar to the double-cover SU(2) of SO(3) when describing angular momentum. In fact, $T$ is a subgroup of SO(3) and $T'$ is a subgroup of SU(2). This implies that the dynamics and constraints associated with $T$ can equally stem from the larger group $T'$ (in analogy to the fact that one can also describe integer spin with SU(2)), while the doublet representations of $T'$ allow for more options [28,29].

This fact has been used in refs. [30,31] to obtain a simultaneous description of both, the lepton- and the quark-sector in the framework of $T'$ [6]. The lepton sector remains the same as in the $A_4$ case, while in the quark sector we do not use the irreducible triplet representation, but the representation $\mathbf{1} \oplus \mathbf{2'}$, to single out the third family. This seems to be a nice explanation of the difference of the quark and lepton sectors within the flavor group $T'$. As Paul Frampton says in his book with Jihn E. Kim [5] (page 172) "Clearly, it is better simultaneously to fit both the quark- and lepton-mixing matrices. This is possible using, for example, the binary tetrahedral group $T'$". There are, of course, many other attempts based on larger groups and representations, but $T'$ remains a very attractive option.

## 3. Towards Larger Groups: $\Delta(27)$ and $\Delta(54)$

Although small groups such as $A_4$ and $T'$ already lead to satisfactory fits, there are a lot of new parameters and ambiguities in explicit model building, and it is not evident whether this really gives the ultimate answer. In fact, there have been many more attempts with different groups and different representations, as can be seen in refs. [1–4]. Another attractive small group is $\Delta(27)$. It has 27 elements and 9 one-dimensional representations, as well as a triplet $\mathbf{3}$ and an anti-triplet $\bar{\mathbf{3}}$ representation. Technical details of the group are given in Appendix B.2. This is still a small group and is attractive in particular because of the $\mathbf{3}$ and $\bar{\mathbf{3}}$ representations, which are well suited for flavor model building with three families of quarks and leptons. As shown in the appendix, it can be constructed as a semi-direct product of $\mathbb{Z}_3 \times \mathbb{Z}_3$ and $\mathbb{Z}_3$ and is a subgroup of SU(3).

Early applications can be found in refs. [32–36], which exploit the presence of the $\mathbf{3}$ and $\bar{\mathbf{3}}$ representations. For more recent work and a detailed list of references, refer to refs. [37,38]. As in the case of $A_4$, $\Delta(27)$ is also well suited to accommodate near tribimaximal mixing. Again (as for $A_4$), the $\mathbb{Z}_2 \times \mathbb{Z}_2$ group of tribimaximal mixing is not a subgroup of $\Delta(27)$, but it appears approximately for specific alignments of the vacuum expectation values of flavon fields that appear naturally within $\Delta(27)$.

$\Delta(27)$ is the "little sister" and subgroup of $\Delta(54)$. This group has 54 elements, two singlet, four doublet, and two pairs of triplet and anti-triplet ($\mathbf{3} \oplus \bar{\mathbf{3}}$) representations. The properties of $\Delta(54)$ are collected in Appendix B.1. It is already quite a large group, somewhat unfamiliar to the BU flavor-community and found less applications than $\Delta(27)$. It became popular because of its appearance in string theory [15], which we shall discuss in Section 4 in detail. Explicit BU model building with $\Delta(54)$ was pioneered in ref. [20].

## 4. Top-Down Considerations: A Taste of Flavor from String Theory

In the BU approach, there are many successful models based on various groups and representations [1–4], too many to single out a "best" option. Such an answer might come from theoretical considerations and top-down model building. An attractive framework is provided by string theory. Here we shall concentrate on the orbifold compactifications of heterotic string theory, which provide many realistic models with gauge group SU(3)×SU(2)×U(1) and three families of quarks and leptons [10,13,39–44].

In these theories, discrete flavor symmetries arise as a result of the geometry of extra dimensions and the geography of fields localized in compact space. Strings are extended objects, and this reflects itself in generalized aspects of geometry that include the winding modes of strings. A full classification of the flavor symmetries of orbifold compactifications of the heterotic string is given by the outer automorphisms [45,46] of the Narain space group [47–50]. Here we shall not be able to give a full derivation of this fact, but only provide a glimpse of the general TD formalism and illustrate the results in simple examples based on a $D = 2$-dimensional torus and its orbifold.

In general, a string in $D$ dimensions has $D$ right-moving and $D$ left-moving degrees of freedom, encoded in $Y = (y_R, y_L)$. Compactifying the theory on a $D$-dimensional torus demands that the $2D$ degrees of freedom be subject to the toroidal boundary conditions

$$Y = \begin{pmatrix} y_R \\ y_L \end{pmatrix} \sim Y + E\,\hat{N} = \begin{pmatrix} y_R \\ y_L \end{pmatrix} + E\begin{pmatrix} n \\ m \end{pmatrix},$$

where the winding and the Kaluza–Klein (momentum) quantum numbers of the string, $n, m \in \mathbb{Z}^D$, define a $2D$-dimensional Narain lattice. $E$ denotes the so-called Narain vielbein and contains the moduli $M_i$ of the torus. In the Narain formulation, we achieve a $D$-dimensional orbifold by imposing the identifications

$$Y \sim \Theta^k\,Y + E\,\hat{N}, \quad \text{with the } \mathbb{Z}_K \text{ orbifold twist } \quad \Theta = \begin{pmatrix} \theta_R & 0 \\ 0 & \theta_L \end{pmatrix} \quad \text{satisfying} \quad \Theta^K = \mathbb{1}_{2D},$$

where $k = 0, \dots, K-1$ and the $SO(D)$ elements $\theta_L, \theta_R$ are set to be equal to obtain a symmetric orbifold. Excluding roto-translations, the Narain space group can then be generated by the twist $(\Theta, 0)$ and shifts $(\mathbb{1}, E_i)$ for $i = 1, \dots, 2D$.

It turns out that flavor symmetries correspond to the (rotational and translational) outer automorphisms of this Narain space group [45,46], which are transformations that map the group to itself but do not belong to the group.

Hence, it follows that the flavor symmetries of string theory come in two classes:

- Those symmetries that map momentum- to momentum- and winding- to winding-modes. These symmetries we call traditional flavor symmetries. They are the same type as those symmetries that would appear in a quantum field theory of point particles. In the Narain formulation, these can be understood as translational outer automorphisms of the Narain space group.
- Symmetries that exchange winding- and momentum-modes. They have their origin in the duality transformations of string theory. We call them modular flavor symmetries as (for the torus discussed here) they are connected to the modular group $SL(2, \mathbb{Z})$. These arise from rotational outer automorphisms of the Narain space group.

### 4.1. Traditional Flavor Symmetries

Here we concentrate on the two-dimensional cases $\mathbb{T}^2/\mathbb{Z}_K$ and $K = 2, 3, 4, 6$, which can be understood as the fundamental building blocks for the discussion of flavor symmetries. They have been discussed in detail in ref. [15]. Various groups can be obtained, prominently $D_8$ or $\Delta(54)$. As an illustrative example, we discuss here the case $\mathbb{T}^2/\mathbb{Z}_3$ with group $\Delta(54)$ because it has the nice property to provide irreducible triplet representations for three families of quarks and leptons [43,51].

The $\mathbb{Z}_3$ orbifold $\mathbb{T}^2/\mathbb{Z}_3$ is shown in Figure 1. Twisted fields are localized on the fixed points $X, Y, Z$ of the orbifold. This geometry leads to an $S_3$ symmetry from the interchange of the fixed points. String theory selection rules provide an additional $\mathbb{Z}_3 \times \mathbb{Z}_3$ flavor symmetry, as discussed in ref. [15]. The full traditional flavor symmetry is $\Delta(54)$, the multiplicative closure of these groups. The twisted states on the fixed points $X, Y, Z$ transform as (irreducible) triplets under $\Delta(54)$ (details can be found in Appendix B.1). $\Delta(54)$ has two independent triplet representations, $\mathbf{3}_1$ and $\mathbf{3}_2$. Both can be realized in string theory, depending on the presence or absence of twisted oscillator modes [23]. The untwisted states are in the trivial $\mathbf{1}$ representation in the absence and $\mathbf{1}'$ in the presence of oscillator modes. A nontrivial vacuum expectation value of a field in the $\mathbf{1}'$ representation will break $\Delta(54)$ to $\Delta(27)$. A discussion of the breakdown pattern of $\Delta(54)$ can be found in ref. [52]. Winding states transform as doublets under $\Delta(54)$. They are typically heavy and could play a prominent role in the discussion of $\mathcal{CP}$-violation in string theory as they provide a mechanism for baryogenesis through the decay of the heavy winding modes [53].

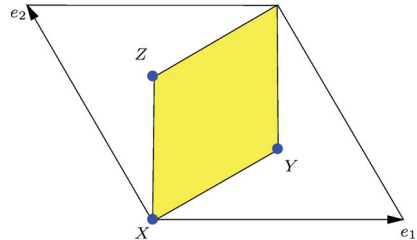

**Figure 1.** The $\mathbb{T}^2/\mathbb{Z}_3$ orbifold (yellow shaded region) with three fixed points, $X, Y, Z$. Twisted states are localized at theses fixed points. Figure taken from ref. [54].

### 4.2. Modular Flavor Symmetries

Modular flavor symmetries have their origin in the duality transformations of string theory. One example is *T*-duality, which exchanges winding and momentum modes. As a warm-up example, consider a string on a circle of radius *R*.

The masses of momentum modes are governed by $1/R$, while winding states become heavier as *R* grows. On the other hand, the T-duality of string theory is defined by the transformations

$$\text{winding modes} \longleftrightarrow \text{momentum modes} \qquad \text{and} \qquad R \longleftrightarrow \alpha'/R \,.$$

Hence, T-duality maps a theory to its T-dual, which coincides at the self-dual point $R^2 = \alpha' = 1/M_{\text{string}}^2$, where $1/\alpha'$ is the string tension. For a generic value of the modulus *R*, T-duality exchanges light and heavy states, which suggests that T-duality could be relevant to flavor physics. Since string theory demands the compactifications of more than one extra dimension, T-duality generalizes to large groups of nontrivial transformations of the moduli of higher-dimensional tori. For instance, in $D = 2$ the transformations on each of the (Kähler and complex structure) moduli build the modular group $\text{SL}(2, \mathbb{Z})^2$ of the $\mathbb{T}^2$ torus. The group $\text{SL}(2, \mathbb{Z})$ is generated by two elements,

$$\text{S and T,} \qquad \text{such that} \qquad \text{S}^4 = \mathbb{1}, \qquad \text{S}^2\text{T} = \text{TS}^2, \quad \text{and} \quad (\text{ST})^3 = \mathbb{1} \,.$$

For each modular group, $\text{SL}(2, \mathbb{Z})$, there exists an associated modulus, *M*, that transforms as

$$\text{S}: \quad M \mapsto -\frac{1}{M} \qquad \text{and} \qquad \text{T}: \quad M \mapsto M + 1 \,.$$

Further transformations include mirror symmetry (which exchanges Kähler and complex structure moduli) as well as the $\mathcal{CP}$-like transformation

$$\text{U}: \quad M \mapsto -\overline{M} \,,$$

where $\overline{M}$ denotes the complex conjugate of *M*. String dualities give important constraints on the action of the theory via the modular group $\text{SL}(2, \mathbb{Z})$ (or $\text{GL}(2, \mathbb{Z})$ when including U). A general $\text{SL}(2, \mathbb{Z})$ transformation of the modulus is given by

$$M \xmapsto{\;\gamma\;} \frac{a\,M + b}{c\,M + d}, \qquad \gamma = \begin{pmatrix} a & b \\ c & d \end{pmatrix} \in \text{SL}(2, \mathbb{Z}),$$

with $\det \gamma = 1$ and $a, b, c, d \in \mathbb{Z}$. The value of $M$ (originally in the upper complex half plane) is then restricted to the fundamental domain, as shown in (the dark shaded region of) Figure 2. Matter fields $\phi$ turn out to transform as

$$\phi \overset{\gamma}{\longmapsto} (c\,M + d)^k \rho(\gamma)\,\phi \qquad \text{for} \qquad \gamma \in \mathrm{SL}(2, \mathbb{Z}) \, ,$$

where $(c\,M + d)^k$ is known as automorphy factor, $k$ is a modular weight fixed by the compactification properties [55,56], and $\rho(\gamma)$ is a unitary representation of $\gamma$. Interestingly, $(\rho(\mathrm{T}))^N = \mathbb{1}$ even though $\mathrm{T}^N \neq \mathbb{1}$, such that $\rho(\gamma)$ generates a so-called finite modular group, as we shall shortly discuss. Among others, the modular weights, $k$, of the fields are important ingredients for flavor model building.

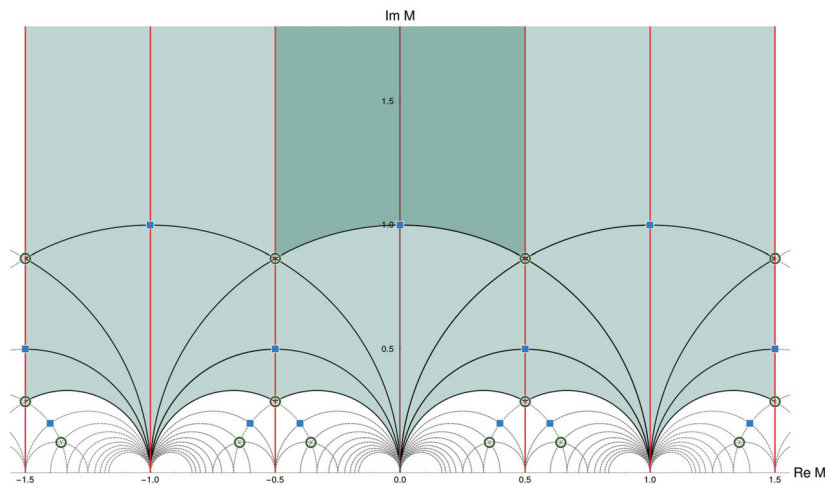

**Figure 2.** Fundamental domain of $\mathrm{SL}(2, \mathbb{Z})$ (dark shaded) and of its subgroup $\Gamma(3) \cong \mathrm{SL}(2, 3\mathbb{Z})$ (light shaded). Figure taken from ref. [54].

As in the one-dimensional case, duality maps one theory to its dual, and there remains the question of whether or not such transformations are relevant for the low-energy effective action of the massless fields. This has been discussed explicitly with the help of worldsheet conformal field theory methods [16–19]. It leads to field-dependent Yukawa couplings that transform as modular forms $Y^{(n_Y)}(M)$ of a weights $n_Y > 0$,

$$Y^{(n_Y)}(M) \overset{\gamma}{\longmapsto} (c\,M + d)^{n_Y} \rho_Y(\gamma)\, Y^{(n_Y)}(M) \quad \text{for} \quad \gamma \in \mathrm{SL}(2, \mathbb{Z}) \, ,$$

where, as for matter fields, $\rho_Y(\gamma)$ is also a unitary representation of $\gamma$ in a finite modular group. The description in terms of supergravity actions has been given in ref. [57]. From the transformation properties of matter fields and Yukawa couplings, it becomes clear that the action is subject to both invariance under the finite modular group and conditions on the modular weights, which are strongly restricted in the TD approach.

Let us illustrate the relevance to flavor physics in the case of the $\mathbb{Z}_3$ orbifold. We start with a two-torus and its two moduli: Kähler modulus $M$ and complex structure modulus $U$. On the orbifold, the $U$-modulus is frozen, such that the lengths of the lattice vectors $e_1$ and $e_2$ are equal, with an angle of 120 degrees (see Figure 1). This also gives restrictions on the modular transformations of the matter fields. The coefficients $a, b, c, d \in \mathbb{Z}$ of the modular transformation are defined only by modulo 3; hence, instead of the full modular group $\mathrm{SL}(2, \mathbb{Z})$, we have to deal with its so-called principal congruence subgroup (the principal congruence subgroup of level $N$ is denoted by $\Gamma(N)$ and built by all $\gamma \in \mathrm{SL}(2, \mathbb{Z})$, such that $\gamma = \mathbb{1} \mod N$), $\Gamma(3) \cong \mathrm{SL}(2, 3\mathbb{Z})$. Clearly, $\Gamma(3)$ still has infinitely many elements, but it is a normal subgroup of the finite index in $SL(2,\mathbb{Z})$. Hence, a finite discrete modular group can be obtained by the quotient $SL(2,\mathbb{Z})/\Gamma(3) = \Gamma'_3$. An explicit discussion is provided in ref. [58]. $\Gamma'_3$ is isomorphic to $T'$, the binary tetrahedral group. It is the double cover of $\Gamma_3 \cong A_4$, which one would obtain starting with $PSL(2,\mathbb{Z})$ instead of $SL(2,\mathbb{Z})$. In the first application of discrete modular flavor symmetry, Feruglio [21] used the group $\Gamma_3 \cong A_4$ with its representations $\mathbf{1}, \mathbf{1}', \mathbf{1}''$, and $\mathbf{3}$ to explain tribimaximal mixing in the standard way. Complications with flavon fields and many additional parameters could be avoided as the modular flavor symmetry is nonlinearly realized. This might lead to problems with the control of additional free parameters in the Kähler potential, which has been taken into account [59]. The modular flavor approach was picked up quickly [2,4,60–66] and led to many different BU constructions with various groups, representations of modular weights.

Unfortunately, the TD approach is much more restrictive and allows less freedom in model building. In our example, we obtain $T'$ and not $A_4$ (the double cover is necessary to obtain chiral fermions in the string construction). Moreover, the twisted states do not transform as irreducible triplets of $T'$ but as $\mathbf{1} \oplus \mathbf{2}'$, and the modular weights of the fields are correlated with the $T'$ representation (and thus cannot be chosen freely as done in the BU framework). Some details of $T'$ modular forms are provided in Appendix C.

## 5. Eclectic Flavor Groups

So far we have seen that string theory predicts the presence of both the traditional flavor group ($\Delta(54)$ in our example) and the modular flavor group ($T'$). You cannot have one of them without the other. This should be taken into account in flavor model building. The eclectic flavor group [22] is the multiplicative closure of $\Delta(54)$ and $T'$, here $\Omega(1) = [648, 533]$. (We have somewhat simplified the discussion here. In full string theory with six compact extra dimensions, we usually find additional $R$-symmetries that would increase the eclectic group, here the group $\Omega(1)$ to $\Omega(2) = [1944, 3448]$. A detailed discussion of these subtleties can be found in refs. [67,68].) Observe that this group has only 648 elements for the product of groups with 54 and 24 elements, respectively. There is one $\mathbb{Z}_2$-like element contained in both $\Delta(54)$ and $T'$. Incidentally, this is the same element that enhances $\Delta(27)$ to $\Delta(54)$. Thus, $\Delta(27)$ and $\Delta(54)$, together with $T'$, would lead to the same eclectic group [22].

The eclectic flavor group is nonlinearly realized. Part of it appears "spontaneously" broken through the vacuum expectation value of the modulus $M$. The modulus is confined to the fundamental domain of $\Gamma(3) = SL(2, 3\mathbb{Z})$, as displayed in Figure 2. This area is reduced by a factor two if we include the natural candidate for a $\mathcal{CP}$-symmetry that transforms $M$ to $-\overline{M}$. The $\mathcal{CP}$-symmetry extends $SL(2,\mathbb{Z})$ to $GL(2,\mathbb{Z})$, $T'$ to $GL(2,3)$ (a group with 48 elements), and the eclectic group $\Omega(1)$ to a group with 1296 elements. The fundamental domain includes fixed points and fixed lines with respect to the modular transformations S and T as well as the $\mathcal{CP}$-transformation $U : M \rightarrow -\overline{M}$, as shown in Figure 3.

For generic points in moduli space the traditional flavor symmetry, $\Delta(54)$ is linearly realized. At the fixed points and lines this symmetry is enhanced to larger groups, as illustrated in Figure 4.

We see that here the largest linearly realized group has 324 elements with GAP Id $[324, 39]$. (We use the group notation of the classification of GAP [69].) Thus, only part of the full eclectic flavor group with 1296 elements (including $\mathcal{CP}$) can be linearly realized. The enhancement of the traditional flavor symmetry at fixed loci (here points and lines) in the fundamental domain exhibits the phenomenon called "Local Flavor Unification" [45,46]. The flavor symmetry is non-universal in moduli space, and the spontaneous breakdown of modular flavor symmetry can be understood as a motion in moduli space. This has important consequences for flavor model building. At the loci of enhanced symmetry, some of the masses and mixing angles of the quark- and lepton-sector might vanish. The explanation of small parameters and hierarchies in flavor physics can thus find an explanation if the modulus is located close to the fixed points or lines [52,70–78]. The mechanism of moduli stabilization in string theory could therefore provide the ingredients to understand the mysteries of flavor [71,79–81].

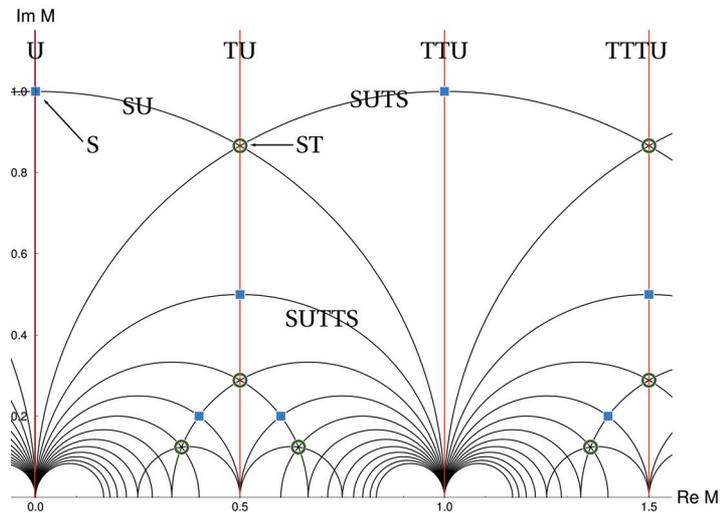

**Figure 3.** Unbroken modular symmetries at special curves in moduli space, including the $\mathcal{CP}$-like generator U, which maps $M \mapsto -\overline{M}$. Figure adapted from ref. [46].

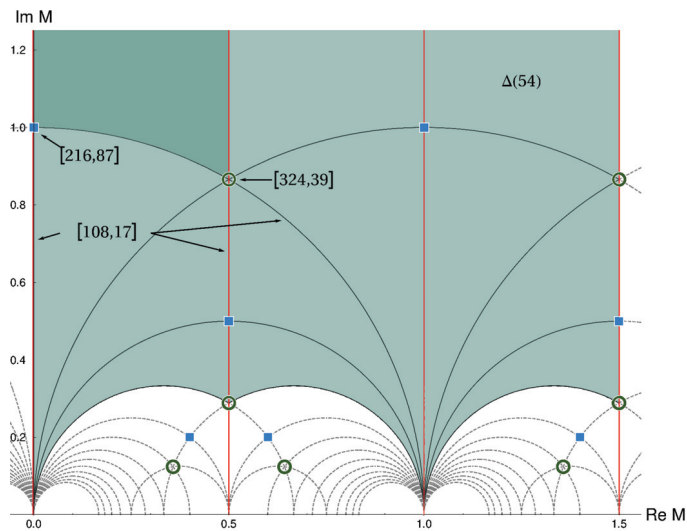

**Figure 4.** Local flavor unification at special points and curves in moduli space. The traditional flavor symmetry $\Delta(54)$, valid at generic points, is enhanced to two (different) groups with GAP Id [108,17] at the vertical lines and semi-circles, including $\mathcal{CP}$-like transformations. At the intersections of curves, the flavor symmetry is further enhanced to [216,87] and [324,39], also with $\mathcal{CP}$-like transformations. Figure adapted from ref. [45].

## 6. Top-Down Does Not Yet Meet Bottom-Up

There have been many BU constructions, but only a few take TD considerations into account [82–84]. From the presently available TD models, the groups $\Delta(54)$ for traditional and $T'$ for modular flavor symmetry seem to be the favorite choices. In fact, there is only one explicit model that incorporates the SM with gauge group $SU(3) \times SU(2) \times U(1)$ and three families of quarks and leptons [72]. We certainly need more work in the TD approach. Therefore, any conclusions about the connection between the two approaches is necessarily preliminary. Still, it is reassuring to see that the same groups $\Delta(54)$ and $T'$ and their "little sisters" $\Delta(27)$ and $A_4$ appear prominently in BU constructions. One could therefore try to make contact between the two approaches within this class of models.

Before we do that, we would like to stress some important properties of the TD approach that seem to be of more general validity and thus should have an influence on BU model building. The first of these is the prediction of string theory for the simultaneous presence of both traditional and modular flavor symmetry that combine to the eclectic flavor group. It is this eclectic group that is relevant, not one of the others in isolation. Up to now, many BU constructions only consider one of them. Therefore, a direct contact between the two approaches is very difficult at this point.

The TD approach is very restrictive. Apart from the limited type of groups that appear in the TD constructions, there are also severe restrictions on the choice of representations. Not everything is possible. In the case of modular symmetry $T'$, for example, the irreducible triplet representation does not appear in the spectrum, while many BU constructions exactly concentrate on this representation. Therefore, the TD approach cannot make contact with models based on modular $A_4$ flavor symmetry, where these triplets are generally used. For $T'$, we have the twisted fields in the $\mathbf{1} \oplus \mathbf{2'}$ representation. It seems to be more likely that irreducible triplet representations are found within the traditional flavor group, as seen in the example with $\Delta(54)$.

A second severe restriction concerns the choice of modular weights. In the TD approach we have essentially no choice. Once we know the representations of the eclectic group, the modular weights are fixed. This is an important restriction, as in the BU approach the choice of modular weights is an important ingredient of model building. With a careful choice of modular weights one can create additional "shaping symmetries", which are important for the success of the fit to the data. This is not possible in the TD approach. There the role of such symmetries could, however, be found in the traditional flavor symmetry.

As a result of these facts, there is presently still a crucial difference between the BU and TD approaches, and a direct comparison is not possible at this point. We are still at the very early stage of such investigations.

## 7. Outlook

Much more work in both approaches is needed in order to clarify the situation. In the BU approach, it would be desirable to consider models that fulfill the restrictions coming from TD. Traditional flavor symmetries and the eclectic framework should be taken into account. A toolkit for such a construction can be found in the consideration of a modular group that fits into the outer automorphism of the traditional flavor group, as explained in ref. [22]. A recent application of this connection for the traditional flavor group $\Delta(27)$ has been discussed in refs. [83,84]. Moreover, BU constructions should avoid the excessive use of modular weights in model building. A strict correlation between the representations and their modular weights might be the right way to proceed. Useful shaping symmetries might be found within the traditional flavor symmetry instead.

The TD approach needs to make serious attempts for the construction of more explicit models. In particular, it would be useful to increase the number of explicit string constructions that ressemble the SM with gauge group $SU(3) \times SU(2) \times U(1)$ and three families of quarks and leptons. This is important, as in generic string theory we might find huge classes of duality symmetries that might not survive in models with the properties of the SM. Of course, the size and nature of these large symmetry groups have to be explored. Modular invariance and its group $SL(2, \mathbb{Z})$ are closely related to torus compactifications, which can be realized in orbifold compactifications and more generally in Calabi–Yau compactifications with elliptic fibrations. This can be described by the basic building blocks $\mathbb{T}^2/\mathbb{Z}_K$ with $K = 2, 3, 4, 6$, which have been studied previously [68]. Explicit string model building shows that such situations are possible, but require particular constellations for the Wilson lines needed for realistic model building. Such Wilson lines and other background fields might otherwise break modular symmetries in various ways [85–87]. In some orbifolds, only a subgroup of $\mathrm{SL}(2,\mathbb{Z})$ is unbroken, even without background fields [88], which opens up the possibility of finite modular flavor symmetries beyond $\Gamma'_K$ [65,66,89]. However, a more general discussion has to go beyond $\mathrm{SL}(2,\mathbb{Z})$. A first step in the direction is the consideration of the Siegel modular group [78,90,91] or higher dimensional constructions [67,68,92,93]. Many exciting developments seem to be in front of us.

**Funding:** The work by SR-S was partly funded by UNAM-PAPIIT grant IN113223 and Marcos Moshinsky Foundation.

**Data Availability Statement:** No new data were created or analyzed in this study. Data sharing is not applicable to this article.

**Acknowledgments:** We acknowledge Alexander Baur, Mu-Chun Chen, Moritz Kade, Victoria Knapp-Pérez, Xiang-Gan Liu, Yesenia Olguín-Trejo, Ricardo Pérez-Martínez, Mario Ramos-Hamud, Michael Ratz, Andreas Trautner, and Patrick Vaudrevange for fruitful, interesting, and pleasant collaborations.

**Conflicts of Interest:** The authors declare no conflicts of interest.

## Abbreviations

The following abbreviations are used in this manuscript:

| | |
|---|---|
| TD | Top-Down |
| BU | Bottom-Up |
| SM | Standard Model |
| PMNS | Pontecorvo–Maki–Nakagawa–Sakata |

## Appendix A. The Group $A_4$ and its Double Cover $T'$

*Appendix A.1. $A_4$*

$A_4 \cong (\mathbb{Z}_2 \times \mathbb{Z}_2) \rtimes \mathbb{Z}_3$ (GAP Id [12,3]) is the alternating group of four elements and can also be understood as the rotational symmetry group of a regular tetrahedron. It contains 12 elements. $A_4$ has the irreducible representations $r \in \{\mathbf{1}, \mathbf{1}', \mathbf{1}'', \mathbf{3}\}$. With $\omega := e^{2\pi i/3}$, its character table reads

| class | $1C_1$ | $3C_2$ | $4C_3$ | $4C'_3$ |
|---|---|---|---|---|
| representative | $\mathbb{1}$ | S | T | $\mathrm{T}^2$ |
| $\mathbf{1}$ | 1 | 1 | 1 | 1 |
| $\mathbf{1}'$ | 1 | 1 | $\omega$ | $\omega^2$ |
| $\mathbf{1}''$ | 1 | 1 | $\omega^2$ | $\omega$ |
| $\mathbf{3}$ | 3 | $-1$ | 0 | 0 |

in terms of the generators S and T. They satisfy $\mathrm{S}^2 = \mathrm{T}^3 = (\mathrm{ST})^3 = \mathbb{1}$ and their representations $\rho_r$ can be expressed by

| $r$ | $\mathbf{1}$ | $\mathbf{1}'$ | $\mathbf{1}''$ | $\mathbf{3}$ |
|---|---|---|---|---|
| $\rho_r(\mathrm{S})$ | 1 | 1 | 1 | $\frac{1}{3}\begin{pmatrix} -1 & 2 & -2 \\ 2 & -1 & -2 \\ -2 & -2 & -1 \end{pmatrix}$ |
| $\rho_r(\mathrm{T})$ | 1 | $\omega$ | $\omega^2$ | $\mathrm{diag}(1, \omega, \omega^2)$ |

The $A_4$ product rules are

$$\mathbf{1}^a \otimes \mathbf{1}^b = \mathbf{1}^c \text{ with } c = a + b \mod 3, \quad \mathbf{1}^a \otimes \mathbf{3} = \mathbf{3}, \quad \mathbf{3} \otimes \mathbf{3} = \mathbf{1} \oplus \mathbf{1}' \oplus \mathbf{1}'' \oplus \mathbf{3} \oplus \mathbf{3},$$

where $a, b, c = 0, 1, 2$ correspond to the number of primes.

*Appendix A.2. $T'$*

$T'$ (GAP Id [24,3]) is the double cover of $A_4$ known also as the binary tetrahedral group. Its irreducible representations are $r \in \{\mathbf{1}, \mathbf{1}', \mathbf{1}'', \mathbf{2}, \mathbf{2}', \mathbf{2}'', \mathbf{3}\}$. This group can be generated by two generators, S and T, satisfying $S^4 = T^3 = (ST)^3 = \mathbb{1}$ and $S^2T = TS^2$. This leads to the character table

| class | $1C_1$ | $1C_2$ | $6C_4$ | $4C_3$ | $4C_3'$ | $4C_6$ | $4C_6'$ |
|---|---|---|---|---|---|---|---|
| representative | $\mathbb{1}$ | $S^2$ | $S$ | $T$ | $T^2$ | $S^2T$ | $S^2T^2$ |
| **1** | 1 | 1 | 1 | 1 | 1 | 1 | 1 |
| **1'** | 1 | 1 | 1 | $\omega$ | $\omega^2$ | $\omega$ | $\omega^2$ |
| **1''** | 1 | 1 | 1 | $\omega^2$ | $\omega$ | $\omega^2$ | $\omega$ |
| **2** | 2 | $-2$ | 0 | $-1$ | $-1$ | 1 | 1 |
| **2'** | 2 | $-2$ | 0 | $-\omega$ | $-\omega^2$ | $\omega$ | $\omega^2$ |
| **2''** | 2 | $-2$ | 0 | $-\omega^2$ | $-\omega$ | $\omega^2$ | $\omega$ |
| **3** | 3 | 3 | $-1$ | 0 | 0 | 0 | 0 |

Note that the triplet representation is unfaithful; it yields only $A_4 \cong T'/\mathbb{Z}_2$, where the normal $\mathbb{Z}_2$ subgroup is generated by $S^2$. The representations can be expressed as

| $r$ | **1** | **1'** | **1''** | **2** | **2'** | **2''** | **3** |
|---|---|---|---|---|---|---|---|
| $\rho_r(S)$ | 1 | 1 | 1 | $\Omega_S$ | $\Omega_S$ | $\Omega_S$ | $\rho_3(S)$ |
| $\rho_r(T)$ | 1 | $\omega$ | $\omega^2$ | $(\Omega_T\widetilde{\Omega}_T)^*$ | $\Omega_T$ | $\widetilde{\Omega}_T$ | $\rho_3(T)$ |

where we defined the two-dimensional matrices

$$\Omega_S = -\frac{i}{\sqrt{3}}\begin{pmatrix} 1 & \sqrt{2} \\ \sqrt{2} & -1 \end{pmatrix}, \qquad \Omega_T = \mathrm{diag}(1,\omega^2), \qquad \widetilde{\Omega}_T = \mathrm{diag}(\omega,1)$$

and the three-dimensional representation is given (as in $A_4$) by

$$\rho_3(S) = \frac{1}{3}\begin{pmatrix} -1 & 2 & -2 \\ 2 & -1 & -2 \\ -2 & -2 & -1 \end{pmatrix} \quad \text{and} \quad \rho_3(T) = \mathrm{diag}(1,\omega,\omega^2).$$

Finally, the tensor products of the $T'$ irreducible representations are given by

$$\mathbf{1}^a \otimes \mathbf{1}^b = \mathbf{1}^c, \quad \mathbf{1}^a \otimes \mathbf{2}^b = \mathbf{2}^c, \quad \mathbf{2}^a \otimes \mathbf{2}^b = \mathbf{1}^c \oplus \mathbf{3} \quad \text{with} \quad c = a+b \mod 3,$$

$$\mathbf{1}^a \otimes \mathbf{3} = \mathbf{3}, \quad \mathbf{2}^a \otimes \mathbf{3} = \mathbf{2} \oplus \mathbf{2}' \oplus \mathbf{2}'' \quad \text{and} \quad \mathbf{3} \otimes \mathbf{3} = \mathbf{1} \oplus \mathbf{1}' \oplus \mathbf{1}'' \oplus \mathbf{3} \oplus \mathbf{3},$$

where $a, b, c = 0, 1, 2$ correspond to the number of primes. The Clebsch–Gordan coefficients can be found, e.g., in [94].

## Appendix B. Group Theory Elements of Larger Groups

*Appendix B.1. $\Delta(54)$*

$\Delta(54) \cong (\mathbb{Z}_3 \times \mathbb{Z}_3) \rtimes S_3 \cong ((\mathbb{Z}_3 \times \mathbb{Z}_3) \rtimes \mathbb{Z}_3) \rtimes \mathbb{Z}_2$ (GAP Id [54,8]) has 54 elements, which can be generated by three generators, A, B, C, satisfying the presentation $A^3 = B^3 = C^2 = (AB)^3 = (AB^2)^3 = (AC)^2 = (BC)^2 = \mathbb{1}$. Its irreducible representations are two singlets, four doublets, and two triplets, plus their complex conjugates. Together, they lead to the character table

| class | $1C_1$ | $9C_2$ | $6C_3$ | $6C_3'$ | $6C_3''$ | $6C_3'''$ | $1C_3$ | $1C_3'$ | $9C_6$ | $9C_6'$ |
|---|---|---|---|---|---|---|---|---|---|---|
| repr. | $\mathbb{1}$ | C | A | B | AB | $AB^2$ | $(ABC)^2$ | $(ACB)^2$ | ABC | ACB |
| **1** | 1 | 1 | 1 | 1 | 1 | 1 | 1 | 1 | 1 | 1 |
| **1′** | 1 | −1 | 1 | 1 | 1 | 1 | 1 | 1 | −1 | −1 |
| **$2_1$** | 2 | 0 | 2 | −1 | −1 | −1 | 2 | 2 | 0 | 0 |
| **$2_2$** | 2 | 0 | −1 | 2 | −1 | −1 | 2 | 2 | 0 | 0 |
| **$2_3$** | 2 | 0 | −1 | −1 | −1 | 2 | 2 | 2 | 0 | 0 |
| **$2_4$** | 2 | 0 | −1 | −1 | 2 | −1 | 2 | 2 | 0 | 0 |
| **$3_1$** | 3 | 1 | 0 | 0 | 0 | 0 | $3\omega$ | $3\omega^2$ | $\omega^2$ | $\omega$ |
| **$3_2$** | 3 | −1 | 0 | 0 | 0 | 0 | $3\omega$ | $3\omega^2$ | $-\omega^2$ | $-\omega$ |
| **$\bar{3}_1$** | 3 | 1 | 0 | 0 | 0 | 0 | $3\omega^2$ | $3\omega$ | $\omega$ | $\omega^2$ |
| **$\bar{3}_2$** | 3 | −1 | 0 | 0 | 0 | 0 | $3\omega^2$ | $3\omega$ | $-\omega$ | $-\omega^2$ |

The doublets are unfaithful representations, which yield the quotient group $S_3 \cong \Delta(54)/\mathbb{Z}_3 \times \mathbb{Z}_3$, where the normal subgroup $\mathbb{Z}_3 \times \mathbb{Z}_3$ can be generated by A and $BAB^2A^2$. In the irreducible representations, the $\Delta(54)$ generators can be expressed as

| $r$ | **1** | **1′** | **$2_1$** | **$2_2$** | **$2_3$** | **$2_4$** | **$3_1$** | **$3_2$** | **$\bar{3}_1$** | **$\bar{3}_2$** |
|---|---|---|---|---|---|---|---|---|---|---|
| $\rho_r(A)$ | 1 | 1 | $\mathbb{1}_2$ | $\Omega_2$ | $\Omega_2$ | $\Omega_2$ | $\rho(A)$ | $\rho(A)$ | $\rho(A)$ | $\rho(A)$ |
| $\rho_r(B)$ | 1 | 1 | $\Omega_2$ | $\mathbb{1}_2$ | $\Omega_2$ | $\Omega_2^*$ | $\rho(B)$ | $\rho(B)$ | $\rho(B)^*$ | $\rho(B)^*$ |
| $\rho_r(C)$ | 1 | −1 | $S_2$ | $S_2$ | $S_2$ | $S_2$ | $\rho(C)$ | $-\rho(C)$ | $\rho(C)$ | $-\rho(C)$ |

where the doublet representations are generated by

$$\mathbb{1}_2 = \begin{pmatrix} 1 & 0 \\ 0 & 1 \end{pmatrix}, \quad \Omega_2 = \begin{pmatrix} \omega^2 & 0 \\ 0 & \omega \end{pmatrix}, \quad S_2 = \begin{pmatrix} 0 & 1 \\ 1 & 0 \end{pmatrix},$$

and the triplets by

$$\rho(A) = \begin{pmatrix} 0 & 1 & 0 \\ 0 & 0 & 1 \\ 1 & 0 & 0 \end{pmatrix}, \quad \rho(B) = \begin{pmatrix} 1 & 0 & 0 \\ 0 & \omega & 0 \\ 0 & 0 & \omega^2 \end{pmatrix}, \quad \rho(C) = \begin{pmatrix} 1 & 0 & 0 \\ 0 & 0 & 1 \\ 0 & 1 & 0 \end{pmatrix}.$$

It is useful to list the nontrivial tensor products of $\Delta(54)$ irreducible representations:

$$\mathbf{1}' \otimes \mathbf{1}' = \mathbf{1}, \quad \mathbf{1}' \otimes \mathbf{2}_k = \mathbf{2}_k, \quad \mathbf{1}' \otimes \mathbf{3}_1 = \mathbf{3}_2, \quad \mathbf{1}' \otimes \mathbf{3}_2 = \mathbf{3}_1, \quad \mathbf{1}' \otimes \bar{\mathbf{3}}_1 = \bar{\mathbf{3}}_2, \quad \mathbf{1}' \otimes \bar{\mathbf{3}}_1 = \bar{\mathbf{3}}_2,$$

$$\mathbf{2}_k \otimes \mathbf{2}_k = \mathbf{1} \oplus \mathbf{1}' \oplus \mathbf{2}_k, \quad \mathbf{2}_k \otimes \mathbf{2}_\ell = \mathbf{2}_m \oplus \mathbf{2}_n \quad \text{with } k \neq \ell \neq m \neq n, \, k, \ell, m, n = 1, \dots, 4,$$

$$\mathbf{2}_k \otimes \mathbf{3}_\ell = \mathbf{3}_1 \oplus \mathbf{3}_2, \quad \mathbf{2}_k \otimes \bar{\mathbf{3}}_\ell = \bar{\mathbf{3}}_1 \oplus \bar{\mathbf{3}}_2 \quad \text{for all } k = 1, \dots, 4, \, \ell = 1, 2,$$

$$\mathbf{3}_\ell \otimes \mathbf{3}_\ell = \bar{\mathbf{3}}_1 \oplus \bar{\mathbf{3}}_1 \oplus \bar{\mathbf{3}}_2, \quad \mathbf{3}_1 \otimes \mathbf{3}_2 = \bar{\mathbf{3}}_2 \oplus \bar{\mathbf{3}}_2 \oplus \bar{\mathbf{3}}_1, \quad \mathbf{3}_1 \otimes \bar{\mathbf{3}}_1 = \mathbf{1} \oplus \mathbf{2}_1 \oplus \mathbf{2}_2 \oplus \mathbf{2}_3 \oplus \mathbf{2}_4,$$

$$\mathbf{3}_1 \otimes \bar{\mathbf{3}}_2 = \mathbf{1}' \oplus \mathbf{2}_1 \oplus \mathbf{2}_2 \oplus \mathbf{2}_3 \oplus \mathbf{2}_4, \quad \mathbf{3}_2 \otimes \bar{\mathbf{3}}_1 = \mathbf{1}' \oplus \mathbf{2}_1 \oplus \mathbf{2}_2 \oplus \mathbf{2}_3 \oplus \mathbf{2}_4,$$

$$\mathbf{3}_2 \otimes \bar{\mathbf{3}}_2 = \mathbf{1} \oplus \mathbf{2}_1 \oplus \mathbf{2}_2 \oplus \mathbf{2}_3 \oplus \mathbf{2}_4, \quad \bar{\mathbf{3}}_\ell \otimes \bar{\mathbf{3}}_\ell = \mathbf{3}_1 \oplus \mathbf{3}_1 \oplus \mathbf{3}_2, \quad \bar{\mathbf{3}}_1 \otimes \bar{\mathbf{3}}_2 = \mathbf{3}_2 \oplus \mathbf{3}_2 \oplus \mathbf{3}_1 \, .$$

The explicit Clebsch–Gordan coefficients can be found, e.g., in [94].

*Appendix B.2. $\Delta(27)$*

The group $\Delta(27) \cong (\mathbb{Z}_3 \times \mathbb{Z}_3) \rtimes \mathbb{Z}_3$ (GAP Id [27,3]) can be obtained from $\Delta(54)$, excluding the $\mathbb{Z}_2$ generator C. Hence, the generators A and B yielding the 27 elements of the group are constrained to fulfill only the subset of conditions $A^3 = B^3 = (AB)^3 = (AB^2)^3 = \mathbb{1}$. By excluding C in the $\Delta(54)$ character table, we observe that the $\Delta(27)$ representations arise from the trivial singlet, the doublets, and combinations of triplets of $\Delta(54)$. One can show that they break into nine singlets and two triplets, which describe the character table

| class | $1C_1$ | $3C_3^a$ | $3C_3^b$ | $3C_3^c$ | $3C_3^d$ | $3C_3^e$ | $3C_3^f$ | $3C_3^g$ | $3C_3^h$ | $3C_3^i$ | $1C_3$ |
|---|---|---|---|---|---|---|---|---|---|---|---|
| repr. | $\mathbb{1}$ | A | B | $A^2$ | $B^2$ | AB | $AB^2$ | $(AB)^2$ | $BA^2$ | $A^2B^2AB$ | $A(AB)^2B$ |
| $\mathbf{1}_{0,0}$ | 1 | 1 | 1 | 1 | 1 | 1 | 1 | 1 | 1 | 1 | 1 |
| $\mathbf{1}_{0,1}$ | 1 | 1 | $\omega$ | 1 | $\omega^2$ | $\omega$ | $\omega^2$ | $\omega^2$ | $\omega$ | 1 | 1 |
| $\mathbf{1}_{0,2}$ | 1 | 1 | $\omega^2$ | 1 | $\omega$ | $\omega^2$ | $\omega$ | $\omega$ | $\omega^2$ | 1 | 1 |
| $\mathbf{1}_{1,0}$ | 1 | $\omega$ | 1 | $\omega^2$ | 1 | $\omega$ | $\omega$ | $\omega^2$ | $\omega^2$ | 1 | 1 |
| $\mathbf{1}_{1,1}$ | 1 | $\omega$ | $\omega$ | $\omega^2$ | $\omega^2$ | $\omega^2$ | 1 | $\omega$ | 1 | 1 | 1 |
| $\mathbf{1}_{1,2}$ | 1 | $\omega$ | $\omega^2$ | $\omega^2$ | $\omega^2$ | 1 | $\omega^2$ | 1 | $\omega$ | 1 | 1 |
| $\mathbf{1}_{2,0}$ | 1 | $\omega^2$ | 1 | $\omega$ | 1 | $\omega^2$ | $\omega^2$ | $\omega$ | $\omega$ | 1 | 1 |
| $\mathbf{1}_{2,1}$ | 1 | $\omega^2$ | $\omega^2$ | $\omega$ | $\omega$ | $\omega$ | 1 | $\omega^2$ | 1 | 1 | 1 |
| $\mathbf{1}_{2,2}$ | 1 | $\omega^2$ | $\omega$ | $\omega$ | $\omega^2$ | 1 | $\omega$ | 1 | $\omega^2$ | 1 | 1 |
| $\mathbf{3}$ | 3 | 0 | 0 | 0 | 0 | 0 | 0 | 0 | 0 | $3\omega$ | $3\omega^2$ |
| $\bar{\mathbf{3}}$ | 3 | 0 | 0 | 0 | 0 | 0 | 0 | 0 | 0 | $3\omega^2$ | $3\omega$ |

Here we immediately see that the singlets $\mathbf{1}_{r,s}$, $r,s = 0,1,2$, have the representations $\rho_{r,s}(A) = \omega^r$ and $\rho_{r,s}(B) = \omega^s$. Further, the triplet representations are given by $\rho_3(A) = \rho(A)$ and $\rho_3(B) = \rho_{\bar{3}}(B)^* = \rho(B)$, in terms of the $\Delta(54)$ matrices.

Finally, the tensor products of $\Delta(27)$ irreducible representations are given by

$$\mathbf{1}_{r,s} \otimes \mathbf{1}_{r',s'} = \mathbf{1}_{r'',s''} \quad \text{with} \quad r'' = r + r' \mod 3, \ s'' = s + s' \mod 3,$$
$$\mathbf{1}_{r,s} \otimes \mathbf{3} = \mathbf{3}, \ \ \mathbf{1}_{r,s} \otimes \bar{\mathbf{3}} = \bar{\mathbf{3}} \quad \text{for all} \ r,s = 0,1,2,$$
$$\mathbf{3} \otimes \mathbf{3} = \bar{\mathbf{3}} \oplus \bar{\mathbf{3}} \oplus \bar{\mathbf{3}}, \ \ \bar{\mathbf{3}} \otimes \bar{\mathbf{3}} = \mathbf{3} \oplus \mathbf{3} \oplus \mathbf{3} \quad \text{and} \quad \mathbf{3} \otimes \bar{\mathbf{3}} = \sum_{r,s} \mathbf{1}_{r,s}.$$

## Appendix C. $T'$ Modular Forms

The vector space of $SL(2,\mathbb{Z})$ modular forms of weight 1 associated with $T' \cong \Gamma_3' = SL(2,\mathbb{Z})/\Gamma(3)$ can be spanned by [63]

$$\hat{e}_1(M) := \frac{\eta^3(3M)}{\eta(M)} \quad \text{and} \quad \hat{e}_2(M) := \frac{\eta^3(M/3)}{\eta(M)},$$

in terms of the Dedekind $\eta$-function of the modulus $M$. One can show that the combinations

$$Y^{(1)}(M) = \begin{pmatrix} \hat{Y}_1(M) \\ \hat{Y}_2(M) \end{pmatrix} := \begin{pmatrix} -3\sqrt{2} & 0 \\ 3 & 1 \end{pmatrix} \begin{pmatrix} \hat{e}_1(M) \\ \hat{e}_2(M) \end{pmatrix}$$

transform under $\gamma \in SL(2,\mathbb{Z})$ as

$$Y^{(1)}(M) \xrightarrow{\gamma} (cM + d)\, \rho_{\mathbf{2}''}(\gamma)\, Y^{(1)}(M),$$

i.e., building a $\mathbf{2}''$ representation $\rho_{\mathbf{2}''}$ of $\Gamma_3' \cong T'$, which is given in Appendix A.2. Higher-weight modular forms of $T'$ are derived from $Y^{(1)}(M)$ by the products of this basic vector-valued modular form, such that $Y^{(n+m)}(M) = Y^{(n)}(M) \otimes Y^{(m)}(M)$. For instance, the modular forms of weight 2 are obtained from $Y^{(2)}(M) = Y^{(1)}(M) \otimes Y^{(1)}(M)$, which build the $T'$ (and $A_4$) triplet $(\hat{Y}_2(M)^2, \sqrt{2}\hat{Y}_1(M)\hat{Y}_2(M), \hat{Y}_1(M)^2)^{\mathsf{T}} =: (\hat{X}_1, \hat{X}_2, \hat{X}_3)^{\mathsf{T}}$ (other choices for the $T'$ Clebsch–Gordan coefficients lead to different but unimportant signs). The expected singlet from $\mathbf{2}'' \otimes \mathbf{2}'' = \mathbf{1}' \oplus \mathbf{3}$ vanishes, and we observe the known relation $\hat{X}_2^2 - 2\hat{X}_1\hat{X}_3 = 0$, which can lead to interesting consequences [95].

*Perspective*

# Quo Vadis Particula Physica?

**Xavier Calmet**

Department of Physics and Astronomy, University of Sussex, Brighton BN1 9QH, UK; x.calmet@sussex.ac.uk

**Abstract:** In this brief paper, I give a very personal account on the state of particle physics on the occasion of Paul Frampton's 80th birthday.

**Keywords:** standard model; quantum gravity

It is a pleasure to contribute to this Special Issue on the occasion of Paul Frampton's 80th birthday. I got to know Paul in 2004 when I moved from Caltech to UNC-Chapel Hill to take a postdoctoral position in his group. Paul very kindly came to knock at the door of my apartment on the day I arrived in Chapel Hill after an exhausting but amazing four and half days of driving from Pasadena to Chapel Hill. He took me to his place to have a chat and to show me the pictures of all the Nobel prize winners he had interacted with. I will never forget that Paul told me on that day that he had proposed so many extensions of the Standard Model of particle physics that he was bound to be awarded the Nobel Prize as one of his new particles would certainly be discovered at the CERN Large Hadron Collider (LHC). This was in 2004. The LHC discovered the Higgs boson in 2012. We are now in 2024 and sadly Paul is still waiting for the discovery of one of his particles and his prize. It is by now very unlikely that the LHC will be able to produce particles that are not part of the Standard Model, assuming they truly exist, at least not on-shell, see, e.g., [1].

To some, this lack of new physics beyond the Standard Model in the TeV region may not have been a surprise. My father, who was trained as a theoretical particle physicist in the 1960s, told me many times that his decision to move to computer science was strongly influenced by a discussion with Sheldon Glashow who was advocating in the 1970s the idea of a grand desert between the weak scale [2], i.e., the energy scale of the Standard Model, and the scale of unification at some $10^{16}$ GeV. My father, who was working on the heroic multi-loop calculation of the anomalous magnetic moment of the muon [3–5], became part of a small group of people who developed the field of computer algebra that was needed to perform these tedious calculations. He started a new research group in Grenoble in the mid-1980s, formally quitting physics and his original lab at Luminy in Marseille. He finally moved to the University of Karlsruhe, now KIT, in 1987, having accepted a professorship in computer algebra. There, he developed some contacts with the local particle physicists. In private, he would laugh at their effort to automatize Feynman diagram calculations in non-Abelian gauge theories including renormalization: he had done all these things in the 1970s and simply did not bother to publish them as they were trivial given his work on quantum electrodynamics. He had a very fulfilling career as a professor in computer science, never regretting having left physics.

What happened to physics since the 1970s? Well, first of all, people in the 1970s were too smart for the sake of my generation. The Higgs mechanism had been proposed by Peter Higgs [6,7] and embedded by Steven Weinberg in Sheldon Glashow's electroweak model [8]. Harald Fritzsch and Murray Gell-Mann had proposed what turned out to be the correct theory for strong interactions (see, e.g., [9], for a review). Gerardus 't Hooft and Martinus Veltman had shown [10] that these non-Abelian theories are renormalizable and thus mathematically consistent at all energy scales. This work enabled 't Hooft to show that Yang–Mills theories can be asymptotically safe, long before the minus sign controversy and the following dilemma of attribution [11]. Sadly, for theorists trying to extend the Standard

Model, experimentalists have found one particle predicted by the Standard Model after another, with the discovery of the Higgs boson being the last and final confirmation of the Standard Model.

Some will argue that neutrino masses are a clear sign of the breakdown of the electroweak Standard Model, but this is not something I find convincing. Neutrino masses are easily accounted for by the Standard Model if the Yukawa couplings between the left-handed neutrinos and right-handed neutrinos are not set to zero in full analogy to up-type quarks. There was never any real reason to set them to zero, besides the fact that their masses were compatible with zero given the experimental state of the art in the 1970s. I would see neutrino masses as a prediction of the Standard Model, because of the close analogy in the treatment of leptons and quarks in this model, rather than new physics. To a certain extent, this is a semantic question, but neutrino masses are not a theoretical challenge whichever point of view one takes.

As Glashow foresaw, it is thus conceivable that the Standard Model remains valid up to some very high energy scale, for example, the scale of grand unification, and threshold effects [12] could easily lead to the numerical unification of the gauge couplings of the Standard Model without the need for new physics between the weak scale and the grand unification scale (The issue of stability of the electroweak vacuum is an open one [13,14] as it depends on quantum gravity corrections.).

Despite Glashow's insight, a few generations of physicists worked on so-called beyond-the-Standard-Model physics between the early 1970s and the late 2010s. This program was motivated by different reasons related to the question of the spontaneous breaking of the electroweak symmetry. In particular, the Higgs mechanism implied the existence of a fundamental scalar boson, something that had never been observed until 2012. This was a strong motivation to consider alternatives to the Higgs mechanism using the idea of dynamical symmetry breaking, where the scalar would effectively be a condensate of fermions. Technicolor and other composite Higgs models were very attractive from a theoretical point of view; sadly, it quickly became clear that the simplest and most elegant models were not compatible with data accumulated at colliders. Another logical possibility was that there could be a lot of fundamental scalars and not just the Higgs boson. This is the case of supersymmetric extensions of the Standard Model, where there are a minimum of two scalar fields for each fermion field depending on the amount of supersymmetry envisaged. Supersymmetry came with its own model-building issues; namely, these new scalar fields had to be made heavy to explain why they had not been discovered yet and supersymmetry had to be broken as we do not observe it as an exact symmetry of nature, at low energies at least.

So why did this program of looking for physics beyond the Standard Model fail so badly? On the one hand, one could argue that physicists of the age of Paul have been very unlucky; indeed, nature picked a model that was proposed when they were finishing their studies. There were good reasons to doubt the Standard Model. On the other hand, applying Gell-Mann's criteria (one point for papers that are correct in the sense of being relevant to nature, minus one for papers that are not relevant to it) to evaluate particle physicists active in the last 50 years reveals that most of them are in the red and that they have been barking up the wrong tree.

Clearly one issue is that the main guiding principle to look for physics beyond the Standard Model was a red herring. Naturalness is the idea that the Higgs boson's mass should be stable under radiative corrections. Proponents of this idea argue that the bare mass of a scalar field receives corrections at the quantum level from self-interactions and interactions with other particles of the model. These corrections are argued to grow quadratically with a dimensionful cutoff that is introduced to regularize loop corrections. They argue that if the cutoff is taken of the order of the reduced Planck mass (i.e., the energy scale where quantum gravitational effects are expected to become important), there needs to be some unnatural adjustment between the quantum corrections and the bare mass to keep the Higgs boson's mass light. Most of the model building effort to go beyond the Standard Model has been motivated by this naturalness "problem". Four broad classes of solutions have been envisaged: models without fundamental scalars, supersymmetric models, models with a low scale of quantum gravity and models advocating the anthropic principle.

In my view, naturalness is absolutely meaningless in the context of a renormalizable quantum field theory, as masses and coupling constants cannot be calculated from first principles. They are renormalized parameters which are used to absorb divergent quantities appearing in the perturbative evaluation of quantum amplitudes. As such, they need to be measured at some energy scale and can be scaled up or down using renormalization group equations, but as we cannot calculate these parameters from first principles, it is meaningless to talk about large or small values. Furthermore, whether divergences are quadratic or logarithmic plays no role from a physical point of view. One could also argue that the problem is not even well posed from a mathematical point of view, as the nature of the divergences depends on the regularization schemed used. For example, in dimensional regularization, quadratic divergencies do not appear in four dimensions.

It is remarkable that this problem was indeed first introduced by proponents of String Theory, where it is indeed possibly an issue as they claim to be able to calculate all fundamental constants which appear as the expectation values of some moduli fields in their framework. But it is certainly not an issue for particle physics. From a particle physics point of view, it should be clear to any researcher that the naturalness problem was not a valid guiding principle. The discovery of a light Higgs boson without new physics to stabilize its mass is the final nail in the coffin for naturalness after the discovery of a cosmological constant that is small and again without any new physics to stabilize it (similar arguments to those for the Higgs boson's mass had been made for the cosmological constant). The particle physics community spent essentially 50 years trying to solve a problem which is not one.

It is fair to say that while experimental particle physics has been extremely successful for the last 50 years and found one particle of the Standard Model after the other, particle physics phenomenology has hit a wall and made very little progress partly because it has been guided by the wrong guiding principles.

Another issue that has affected theoretical particle physics overall is that because it has been increasingly disconnected from experimental physics, as it has been trying to solve a problem which is clearly not relevant to nature, it has become a beauty contest. An issue with beauty is that it obviously lies in the eye of the beholder and instead of applying Gell-Mann's principle to evaluate scientists, less objective criteria have been applied, resulting in high-energy theory groups at top universities being taken over by people convinced that the single most important problem was the naturalness problem. Young people had to follow their lead and research to hope to be able to obtain a job in academia. The problem we are describing here is not unique to particle phenomenology, but it also applies to String Theory for the same reason: this program is mostly completely disconnected from experiment or to a certain extent from physics which is an empirical science. Overall, theoretical physics has become extremely speculative and the "cutest" speculations get rewarded with prestigious faculty positions and academic prizes.

As we have argued, physicists of Paul's generation have been unlucky, but it is also clear that this generation decided to change the rules of the game when it became acceptable to invent new particles without being forced to do so by experiment or mathematical consistency of the theory.

How can my generation and younger theorists get out of this impasse? I can only offer a very personal opinion. We need to refocus research on what nature and mathematics are telling us. I see two clear problems, that while very difficult to solve, are certainly worth trying to address as they could guide us to an understanding of what lies beyond the Standard Model.

The first problem is obvious. It is dark matter, for which there is ample observational evidence. Unless all these uncorrelated observations are wrong, which would be very surprising, we know that Einstein's theory of gravitation with visible matter (which can be described by the Standard Model) is not able to explain, e.g., the galaxy rotation curves or the Cosmic Microwave Background power spectrum. These phenomena are clearly fully disconnected and related to physics at different energy scales. However, they both point towards physics beyond the Standard Model, as no particle of the Standard Model can account for these observations. While a modification of gravitational physics is a logical possibility, it is unclear whether this would be sufficient to explain all observation such as, e.g., bullet clusters. The most logical explanation is clearly that there is some hidden sector of dark matter particles that is weakly interacting with itself and extremely weakly interacting with Standard Model particles, possibly only gravitationally (There is a caveat here as primordial black holes could account for a least a good fraction if not all of dark matter. I personally like this scenario very much as it does not require physics beyond the Standard Model.).

Here again, the guiding principle described above has led people to consider mostly a limited class of models called wimps, which stands for Weakly Interacting Massive Particles. Wimps are common in supersymmetric extensions of the Standard Model. Wimps are now mostly excluded by searches at LEP, Tevatron and the LHC. Again, there was no real theoretical reason to expect wimps to be relevant to nature, but it did not stop the field from making an industry out of these models. From a theoretical point of view, very little is known about the masses of dark matter particles and their interactions with regular matter. Without any serious theoretical prejudice or guidance, it seems unrealistic and unreasonable to build a new collider to exclude a small fraction of the allowed parameter range for dark matter models. There is a recent effort that appears very promising to me which consists of using existing quantum sensors to probe for ultra-light dark matter, see, e.g., [15]. These are cheap experiments, which are mostly already operating, e.g., atomic clocks, for other reasons. While they may not find dark matter, these experiments clearly have other important outcomes in, e.g., the field of quantum metrology, and quantum sensors have important practical implications which are likely to benefit humanity. There is thus a no lose game argument to be made for these experiments.

My suggestion to young theorists is to make an effort to talk to the atomic, molecular, and optical physics (AMO) community and to learn their slang. Progress in quantum technology is fast and there are plenty of opportunities to propose tests of the Standard Model using these new technologies based on quantum physics.

The other direction I would like to mention is that of quantum gravity. While we are still far away from having a theory of quantum gravity which is ultra-violet finite, modern quantum field theoretical techniques can be used to derive an effective action for quantum gravity that enables one to perform calculations for any physical process taking place at energies below the reduced Planck scale. This approach is called the unique effective action [16–20]. The only required assumption is that general relativity is the correct low-energy limit of the theory of quantum gravity. The effective action enables some model-independent predictions of quantum gravity (see, e.g., [21]). While these effects are, as expected, very small, they demonstrate that calculations in quantum gravity are feasible and do not require any speculation.

While these quantum gravitational effects are small and unlikely to be relevant to currently conceivable experiments, they can provide us with some important insights into quantum gravity. My hope is that this program could give us some hints about the correct fundamental theory of quantum gravity, for example, by providing us with consistency conditions.

I would like to emphasize that this approach has already produced some important results. Indeed, it has enabled us to show that black holes have a quantum hair, which is the key feature to explain how information escapes an evaporating black hole, thereby resolving the famous Hawking paradox [22–26]. It has also enabled us to calculate the leading-order quantum gravitational corrections to the entropy of a Schwarzschild black hole, which forced us to introduce the notion of quantum pressure for black holes [27].

I strongly believe that this is not the end of the story for this approach to quantum gravity. I believe that connecting our results to some ideas coming from String Theory such as the AdS/CFT correspondence or the Swampland program could help us discover interesting results connecting gauge theories and quantum gravity.

In terms of probing quantum gravity experimentally, I think that one should again turn towards quantum technologies. Establishing that gravity can entangle macroscopic objects would be highly interesting [28,29] and proof, if one is needed, that gravity is a quantum force.

My feeling is that we are making some important progress and that while the way people have performed particle physics for the last 50 years has to change, there are plenty of interesting opportunities for bright young theorists if they are willing to take some risks, ignore famous people and try to follow their physical intuition, mathematical consistency and nature.

Finally, let me argue that while the approach followed by Paul and this generation did not lead to new discoveries, it was still valuable in the sense that it pushed experimentalists to keep an open mind about the type of physics that could supersede the Standard Model. Paul with his creativity and productivity has played a crucial role in this endeavour. On this note, I would like to use this opportunity to congratulate Paul on his 80th birthday.

**Funding:** The work of X.C. is supported in part by the Science and Technology Facilities Council (grants numbers ST/T006048/1 and ST/Y004418/1).

**Data Availability Statement:** This manuscript has no associated data. Data sharing not applicable to this article as no datasets were generated or analysed during the current study.

**Conflicts of Interest:** The author declares no conflict of interest.

*Review*

# The $SU(3)_C \times SU(3)_L \times U(1)_X$ (331) Model: Addressing the Fermion Families Problem within Horizontal Anomalies Cancellation

**Claudio Corianò * and Dario Melle**

Dipartimento di Matematica e Fisica, Università del Salento and INFN Sezione di Lecce, National Center for HPC, Big Data and Quantum Computing, Via Arnesano, 73100 Lecce, Italy; dario.melle@le.infn.it
* Correspondence: claudio.coriano@le.infn.it

**Abstract:** One of the most important and unanswered problems in particle physics is the origin of the three generations of quarks and leptons. The Standard Model does not provide any hint regarding its sequential charge assignments, which remain a fundamental mystery of Nature. One possible solution of the puzzle is to look for charge assignments, in a given gauge theory, that are inter-generational, by employing the cancellation of the gravitational and gauge anomalies horizontally. The 331 model, based on an $SU(3)_C \times SU(3)_L \times U(1)_X$ does this in an economical way and defines a possible extension of the Standard Model, where the number of families has necessarily to be three. We review the model in Pisano, Pleitez, and Frampton's formulation, which predicts the existence of bileptons. Another characteristics of the model is to unify the $SU(3)_C \times SU(2)_L \times U(1)_X$ into the 331 symmetry at a scale that is in the TeV range. Expressions of the scalar mass eigenstates and of the renormalization group equations of the model are also presented.

**Keywords:** particle theory; physics beyond the Standard Model; collider phenomenology

## 1. Introduction

In the quest to unveil new physics governing fundamental interactions at the Large Hadron Collider (LHC), resolving several crucial questions remains a challenge within the Standard Model (SM). These include the gauge hierarchy problem in the Higgs sector and the origin of light neutrino masses.

Addressing these issues often requires theories involving larger gauge groups and a broader spectrum of particles. Grand Unified Theories (GUTs) offer promising avenues, but their high energy scales (around $10^{12}$ to $10^{15}$ GeV) far exceed the electroweak scale probed by the LHC.

Bridging the gap between the GUT scale and the TeV scale, where the LHC operates, to identify signatures of symmetry breaking presents a significant challenge, due to the increased complexity of these extended theories.

However, specific scenarios exist where evidence for enlarged gauge symmetries might be discovered or excluded at the LHC scale, suggesting alternative exploration paths.

One such example is the 331 model ($SU(3)_C \times SU(3)_L \times U(1)_X$), where the constraint of real gauge couplings significantly restricts the parameter space for potential signal searches. This model was proposed as a potential extension to the SM, in order to address certain theoretical and experimental shortcomings, as well as to provide explanations for phenomena not accounted for within the SM.

The 331 model introduces a new gauge group, $SU(3)_L$, which is isomorphic to the color gauge group $SU(3)_C$. This implies that the strong force acting between quarks within hadrons is now governed by the $SU(3)_L$ symmetry, in addition to the color symmetry.

The fermion content of the 331 model differs from the SM, due to an extended gauge symmetry. Typically, in the 331 model, the quarks and leptons are organized into multiplets that transform under the representations of the $SU(3)_L$ gauge group. For example, quarks and leptons may be arranged in triplets or antitriplets of $SU(3)_L$, depending on their electric charge and other quantum numbers. One notable feature of the 331 model is the presence of new gauge bosons called "bileptons". These are bosons carrying both a lepton number and electric charge, with charges $Q = \pm 2$ and $L = \pm 2$. Bileptons arise due to the extended gauge symmetry and can have significant implications for various phenomena, including neutrino masses and decays of heavy particles.

Similar to the SM, the 331 model also involves spontaneous symmetry breaking, where the gauge symmetries are broken at a certain energy scale. This results in the generation of particle masses and the emergence of the familiar gauge bosons, such as the $W^\pm$ and $Z^0$ bosons.

The 331 model offers potential explanations for various phenomena beyond the scope of the SM, including neutrino masses and mixing, and the unification of fundamental forces at high energies.

Overall, the 331 model represents an intriguing extension of the SM, offering new avenues for exploring fundamental physics beyond the established framework. However, it remains subject to experimental scrutiny and theoretical refinement to fully ascertain its validity and implications for our understanding of the fundamental forces and particles in nature.

In the model, the constraint of real gauge couplings significantly restricts the parameter space for potential signal searches. This property establishes the vacuum expectation values (vevs) of the Higgs bosons, responsible for symmetry breaking from the 331 scale to the electroweak scale, around the TeV region. The model under consideration incorporates the presence of bileptons, denoted as gauge bosons $(Y^{--}, Y^{++})$, possessing a charge $Q = \pm 2$ and lepton number $L = \pm 2$. Consequently, we dub this framework the "bilepton model". Within the array of 331 models, the existence of bileptons within the spectrum only arises through specific embeddings of the $U(1)_X$ symmetry, as well as the charge $(Q)$ and hypercharge $(Y)$ generators within the local gauge structure.

An additional noteworthy aspect of this model is its departure from the conventions of the Standard Model or typical chiral models seen thus far. In contrast to merely extending the SM spectrum and symmetries, the determination of the (chiral) fermion generations hinges on the interfamily cancellation of gauge anomalies. Remarkably, gauge anomalies cancel across distinct fermion families, thereby pinpointing the number of generations as three. From this vantage point, the model emerges as distinctly singular. Furthermore, in the framework outlined by Frampton, which we adopt henceforth, the treatment of the third fermion family is asymmetrical in comparison to the initial two families.

Our work is organized as follows: We will first discuss the general structure of the model, starting with the anomaly constraints that crucially characterize the charge assignments of the spectrum. We then move on to characterize the gauge boson spectrum, turning afterwards to the Higgs sector. The structure of the potential is thoroughly examined, both in its triplet and sextet contributions. We discuss the energy bound present in the model, induced by the structure of the gauge coupling relation coming from the embedding of the Standard Model into the 331. This bound on the energy scale at which the model is characterized by real (as opposed to complex) values of the couplings is one of the salient features of this theory. A second important feature is the presence of a Landau pole in the renormalization group equations (RGEs) of the gauge couplings, which we briefly illustrate numerically. A list of results for the mass eigenstates of the Higgs sector is contained in Appendix A. Notice that an important feature of the model is the identification of the electric charge operator in terms of the diagonal generators of the fundamental gauge symmetry. In general this is given by

$$Q = T^3 + \beta\, T^8 + X\mathbb{1} \tag{1}$$

where $\beta$ is a parameter of the model, with $T^3$ and $T^8$ generators of $SU(3)_L$. Our discussion will focus on the choice $\beta = \sqrt{3}$, as in Pisano, Pleitez, and Frampton's original formulation.

This choice induces the presence of bileptons, which are doubly charged gauge bosons carrying lepton number $L = \pm 2$. We refer to this version of the model as to the minimal one.

## 2. The Particle Content of the Minimal 331 Model

One of the most important questions that arise in particle physics is why there are only three families of quarks and leptons. There are many observables in particle physics that depend on the family number and they all agree to constrain this number to three. In the Standard Model, there is no mechanism that prohibits the existence of more families than those observed. To answer this question P. H. Frampton, F. Pisano, and V. Pleitez [1,2] proposed a new model that extends the Standard Model and could provide an elegant answer to this fundamental question. Specifically, this is called the 331 model, and it is built on the gauge group

$$SU(3)_C \times SU(3)_L \times U(1)_X, \tag{2}$$

which enlarges the $SU(2)_L$ symmetry of the Standard Model. Three exotic quarks must be added to the particle content of the Standard Model to allow for $SU(3)_L$ symmetry in the quark sector. The 331 model democratically treats leptons in each of the three families, in fact color singlets are $SU(3)_L$ anti-triplets

$$\begin{pmatrix} e \\ -\nu_e \\ e^c \end{pmatrix}_L, \begin{pmatrix} \mu \\ -\nu_\mu \\ \mu^c \end{pmatrix}_L, \begin{pmatrix} \tau \\ -\nu_\tau \\ \tau^c \end{pmatrix}_L \to (\mathbf{1}, \bar{\mathbf{3}}, 0) \tag{3}$$

each with $X = 0$. Where with $e^c$, $\mu^c$, and $\tau^c$ we are denoting the left-handed Weyl spinors of the relative charge conjugate field. On the contrary, in the quark sector, the three families are treated differently, the first two generations are in triplets of $SU(3)_L$ with the corresponding left-handed field of exotic quark

$$\begin{pmatrix} u \\ d \\ D \end{pmatrix}_L, \begin{pmatrix} c \\ s \\ S \end{pmatrix}_L \to \left(\mathbf{3}, \mathbf{3}, -\frac{1}{3}\right), \tag{4}$$

both with $X = -\frac{1}{3}$. On the other hand, the third generation in the quark sector is embedded in anti-triplets of $SU(3)_L$

$$\begin{pmatrix} b \\ -t \\ T \end{pmatrix}_L \to \left(\mathbf{3}, \bar{\mathbf{3}}, \frac{2}{3}\right), \tag{5}$$

with $X = \frac{2}{3}$. To each left-handed field there is a corresponding right-handed field singlet under $SU(3)_L$

$$\begin{matrix} (u^c)_L \\ (c^c)_L \\ (t^c)_L \end{matrix} \to \left(\mathbf{1}, \mathbf{1}, -\frac{2}{3}\right), \tag{6}$$

$$\begin{matrix} (d^c)_L \\ (s^c)_L \\ (b^c)_L \end{matrix} \to \left(\mathbf{1}, \mathbf{1}, \frac{1}{3}\right), \tag{7}$$

$$\begin{matrix} (D^c)_L \\ (S^c)_L \end{matrix} \to \left(\mathbf{1}, \mathbf{1}, \frac{4}{3}\right), \tag{8}$$

$$(T^c)_L \to \left(\mathbf{1}, \mathbf{1}, -\frac{5}{3}\right), \tag{9}$$

The $U(1)_X$ charges are respectively $-\frac{2}{3}$, $\frac{1}{3}$ and $\frac{4}{3}$ for $u^c$, $d^c$, $D^c$ and the corresponding field of the second generation. The $X$ quantum numbers of the third family instead are respectively $\frac{1}{3}$, $-\frac{2}{3}$ and $-\frac{5}{3}$ for $b^c$, $t^c$, $T^c$. Three scalar triplets under $SU(3)_L$ are necessary to ensure the spontaneous symmetry breaking

$$\rho = \begin{pmatrix} \rho^{++} \\ \rho^{+} \\ \rho^{0} \end{pmatrix} \eta = \begin{pmatrix} \eta_1^+ \\ \eta^0 \\ \eta_2 \end{pmatrix} \chi = \begin{pmatrix} \chi^0 \\ \chi^- \\ \chi^{--} \end{pmatrix} \tag{10}$$

respectively with $X$ charge $X = 1, 0, -1$, and a scalar sextet

$$\sigma = \begin{pmatrix} \sigma_1^{++} & \frac{\sigma_1^+}{\sqrt{2}} & \frac{\sigma_1^0}{\sqrt{2}} \\ \frac{\sigma_1^+}{\sqrt{2}} & \sigma_2^0 & \frac{\sigma_2^-}{\sqrt{2}} \\ \frac{\sigma_1^0}{\sqrt{2}} & \frac{\sigma_2^-}{\sqrt{2}} & \sigma_2^{--} \end{pmatrix}, \tag{11}$$

in order to generate physical masses for leptons, as we will see in Section 6. All quantum numbers under $SU(3)_C \times SU(3)_L \times U(1)_X$ in the 331 model can be found in Table 1. This unconventional assignment of quantum numbers in the model ensures that gauge anomalies are not canceled vertically for each family, as in the Standard Model. It is necessary to add the contribution of each quark in the triangle anomaly to obtain the total cancellation. This is one of the most important and attractive features of the model, because it provides a possible explanation for the number of generations. This could provide a first step towards understanding the flavor puzzle and perhaps serve as a guide for new models inspired by it.

**Table 1.** Quantum numbers of the particle spectrum of the minimal 331 model.

| | $SU(3)_C$ | $SU(3)_L$ | $U(1)_X$ |
|---|---|---|---|
| $u^c \quad c^c$ | $\bar{3}$ | $1$ | $-\frac{2}{3}$ |
| $t^c$ | $\bar{3}$ | $1$ | $\frac{1}{3}$ |
| $d^c \quad s^c$ | $\bar{3}$ | $1$ | $\frac{1}{3}$ |
| $b^c$ | $\bar{3}$ | $1$ | $-\frac{2}{3}$ |
| $D^c \quad S^c$ | $\bar{3}$ | $1$ | $\frac{4}{3}$ |
| $T^c$ | $\bar{3}$ | $1$ | $-\frac{5}{3}$ |
| $\begin{pmatrix} u \\ d \\ D \end{pmatrix}_L \begin{pmatrix} c \\ s \\ S \end{pmatrix}_L$ | $3$ | $3$ | $-\frac{1}{3}$ |
| $\begin{pmatrix} b \\ t \\ T \end{pmatrix}_L$ | $3$ | $\bar{3}$ | $\frac{2}{3}$ |
| $\begin{pmatrix} e^- \\ \nu_e \\ e^+ \end{pmatrix}_L \begin{pmatrix} \mu^- \\ \nu_\mu \\ \mu^+ \end{pmatrix}_L \begin{pmatrix} \tau^- \\ \nu_\tau \\ \tau^+ \end{pmatrix}_L$ | $1$ | $\bar{3}$ | $0$ |
| $X_\mu$ | $1$ | $1$ | $0$ |
| $W_\mu^a$ | $1$ | $8$ | $0$ |
| $G_\mu^b$ | $8$ | $1$ | $0$ |
| $\sigma$ | $1$ | $3$ | $1$ |
| $\eta$ | $1$ | $3$ | $0$ |
| $\chi$ | $1$ | $3$ | $-1$ |
| $\sigma$ | $1$ | $6$ | $0$ |

### 3. Cross-Family Anomaly Cancellation and the Flavor Question

In this section, we analyze the non-trivial cancellation of anomalies in the 331 model and discuss why this method of eliminating them could represent an initial step towards addressing the flavor question.

There are six types of anomalies that occur in the 331 model

$$SU(3)_C^3, \qquad SU(3)_C^2 U(1)_X, \qquad SU(3)_L^3,$$
$$SU(3)_L^2 U(1)_X, \qquad U(1)_X^3, \qquad \text{grav}^2 U(1)_X, \tag{12}$$

where $\text{grav}^2 U(1)_X$ is the mixed chiral anomaly with two gravitons and a chiral $U(1)_X$ gauge current; i.e., the gravitational chiral anomaly.

Each vertex collects a factor

$$2tr(T_R^a\{T_R^b, T_R^c\}) = A(R)\, d^{abc}, \qquad \text{with } A(\text{fund}) = 1 \tag{13}$$

from the gauge current group generators, where $A(R)$ is the representation dependent anomaly coefficient, with $A(\text{fund})$ in the fundamental representation of $SU(3)$, and $d^{abc}$ is the totally symmetric invariant $SU(3)$ tensor.

The anomaly involving only gluons is obviously zero, as in the Standard Model. Another anomaly arises from the mixed $SU(3)_C^2\, U(1)_X$ vertex, which imposes the following constraint on the $X$ charges

$$SU(3)_C^2 U(1)_X \rightarrow 3X_{Q_i} + X_{u_i^c} + X_{d_i^c} + X_{J_i^c} = 0. \tag{14}$$

Here, the index $i$ ranges over families, but there is no summation over it, $Q_i$ stands for quark triplets, while $J_i^c$ denotes the complex conjugate of the exotic quark fields. Indeed, for this reason, the anomaly cancellation in this case occurs vertically between families, as in the Standard Model.

The same cannot be said for another anomaly, which involves only the gauge group $SU(3)_L$. In the Standard Model, the $SU(2)_L$ group possesses a vanishing anomaly coefficient, rendering it unnecessary. However, in the case of $SU(3)$ or, in general, $SU(N)$ with $N > 2$, it exhibits a non-zero anomaly coefficient. In the 331 model, such $SU(3)_L^3$ anomaly cancels due to the equal number of fermions in the **3** and **3̄** representations of $SU(3)_L$.

In the minimal 331 model, the $SU(3)_L^2 U(1)_X$ anomaly can only be canceled by accounting for contributions from all three families. This type of anomaly cancellation is referred to as a horizontal cancellation. If we consider anomaly cancellation on a generation-by-generation basis, it does not vanish, and requires summation over different families. The relative constraint is

$$SU(3)_L^2\, U(1)_X \rightarrow \sum_{i=1}^{3} X_{Q_i} = 0. \tag{15}$$

The same motivations lead to the cancellation of the cubic anomaly

$$U(1)_X^3 \rightarrow \sum_{i=1}^{3}\left(3X_{Q_i}^3 + X_{u_i^c}^3 + X_{d_i^c}^3 + X_{J_i}^3\right) = 0, \tag{16}$$

with the relative constraint.

The last anomaly that needs to be checked is the gravitational anomaly, but it is not difficult to show that it leads to the same constraint as for $SU(3)_C^2\, U(1)_X$

$$\text{grav}^2 U(1)_X \rightarrow 3X_{Q_i} + X_{u_i^c} + X_{d_i^c} + X_{J_i^c} = 0. \tag{17}$$

A summary of anomalies and relative constraints can be found in Table 2.

**Table 2.** Anomaly cancellation constraints on the fermion charges in the minimal 331 model.

| Anomaly | |
|---|---|
| $SU(3)_C^2 U(1)_X$ | $3X_Q + X_{u_i^c} + X_{d_i^c} + X_{J_i^c} = 0$ |
| $SU(3)_L^3$ | Equal number of $\mathbf{3}_L$ and $\bar{\mathbf{3}}_L$ representations |
| $SU(3)_L^2 U(1)_X$ | $\sum_{i=1}^{3} X_{Q_i} = 0$ |
| $U(1)_X^3$ | $\sum_{i=1}^{3} \left(3X_{Q_i}^3 + X_{u_i^c}^3 + X_{d_i^c}^3 + X_{J_i^c}^3\right) = 0$ |
| $grav^2 U(1)_X$ | $3X_{Q_i} + X_{u_i^c} + X_{d_i^c} + X_{J_i^c} = 0$ |

The non-trivial cancellation of anomalies in the model is arguably one of its most intriguing and distinctive features, which was first discussed in [3]. The horizontal approach, involving all three generations of quarks and leptons, inherently constrains the number of families to three. Unlike the Standard Model, which lacks a mechanism to limit the number of fermion generations, the 331 model provides such a constraint.

Notice that, in the Standard Model, the number of generations is fixed experimentally by the annihilation process $e^+e^- \rightarrow$ hadrons, which is sensitive to the number of families. Experimental constraints based on these observables restrict the number to three; nevertheless, the model lacks a theoretical guiding principle for predicting such a number. The horizontal approach to anomaly cancellation of the 331 model, on the other hand, as already mentioned, can serve as a guiding principle for investigating other UV-completions beyond the Standard Model (BSM) that aim to address the flavor question.

## 4. Spontaneous Symmetry Breaking

Below the electroweak scale, in the Standard Model, the gauge symmetry is $SU(3)_C \times U(1)_{em}$. Therefore, in the 331 model, spontaneous symmetry breaking (SSB) must also occur, in order to reduce the $SU(3)_C \times SU(3)_L \times U(1)_X$ gauge symmetry. The breaking can be divided into two stages. Initially, at energy scales greater than 246 GeV, the gauge symmetry of the 331 model can be broken down to that of the Standard Model, and subsequently to $SU(3)_C \times U(1)_{em}$. This can be represented as

$$SU(3)_C \times SU(3)_L \times U(1)_X \rightarrow SU(3)_C \times SU(2)_L \times U(1)_Y \rightarrow SU(3)_C \times U(1)_{em}. \quad (18)$$

Achieving this requires a more intricate Higgs sector comprising three scalar triplets and a scalar sextet of $SU(3)_L$. In the following two sections, we will analyze the pattern of SSB and explore how it predicts the existence of bileptons, specifically massive double-charged gauge bosons that carry lepton number of $L = \pm 2$, which can be classified as elementary bifermions (including also leptoquarks and biquarks) in the framework based on $SU(15)$ noted recently in [4].

### 4.1. The Breaking $SU(3)_C \times SU(3)_L \times U(1)_X \rightarrow SU(3)_C \times SU(2)_L \times U(1)_Y$

Spontaneous symmetry breaking to the Standard Model gauge group can be accomplished by means of a vacuum expectation value of a scalar triplet belonging to $SU(3)_L$, denoted as $\rho$,

$$\rho = \begin{pmatrix} \rho^{++} \\ \rho^+ \\ \rho^0 \end{pmatrix} \quad (19)$$

which carries charge under $U(1)_X$, namely $X = 1$

$$\langle \rho \rangle = \begin{pmatrix} 0 \\ 0 \\ v_\rho \end{pmatrix}. \quad (20)$$

The $SU(3)_L \times U(1)_X$ covariant derivative can be written as follows:

$$D_\mu = \partial_\mu - ig_1 XX_\mu - ig_2 \frac{\lambda^a}{2} W_\mu^a =$$

$$= \left( \partial_\mu - i\sqrt{\frac{2}{3}} g_1 XX_\mu \right) \begin{pmatrix} 1 & 0 & 0 \\ 0 & 1 & 0 \\ 0 & 0 & 1 \end{pmatrix} - ig_2 \begin{pmatrix} \frac{W_\mu^3}{2} + \frac{W_\mu^8}{2\sqrt{3}} & \frac{W_\mu^1}{2} - \frac{iW_\mu^2}{2} & \frac{W_\mu^4}{2} - \frac{iW_\mu^5}{2} \\ \frac{W_\mu^1}{2} + \frac{iW_\mu^2}{2} & -\frac{W_\mu^3}{2} + \frac{W_\mu^8}{2\sqrt{3}} & \frac{W_\mu^6}{2} - \frac{iW_\mu^7}{2} \\ \frac{W_\mu^4}{2} + \frac{iW_\mu^5}{2} & \frac{W_\mu^6}{2} + \frac{iW_\mu^7}{2} & -\frac{W_\mu^8}{\sqrt{3}} \end{pmatrix}, \tag{21}$$

where $X$ is the charge under $U(1)_X$ of the fermion, $X_\mu$ is the corresponding gauge boson, and $W_\mu^a$ are the generators of $SU(3)_L$. $\lambda^a$ are the Gell-Mann matrices normalized as $\mathrm{Tr}\left( \lambda^a \lambda^b \right) = 2\delta^{ab}$. Once the Higgs triplet $\rho$ acquires the vacuum expectation value, its kinetic term gives

$$\left( D_\mu \begin{pmatrix} 0 \\ 0 \\ v_\rho \end{pmatrix} \right)^\dagger \left( D_\mu \begin{pmatrix} 0 \\ 0 \\ v_\rho \end{pmatrix} \right) = -\frac{1}{6} v_\rho^2 \Big( -4\sqrt{2} g_1 g_2 X_\mu W^{\mu 8} + 4g_1^2 X^\mu X_\mu + 3g_2^2 Y^{++\mu} Y_\mu^{--}$$

$$+ 3g_2^2 V^{\mu+} V_\mu^- + 2g_2^2 W_\mu^8 W^{\mu 8} \Big). \tag{22}$$

From Equation (22), it is easy to observe that the mass terms obtained are expressed in a basis that is not completely diagonal. This implies that, in order to obtain the mass eigenstates of the bosons, it is necessary to perform an orthogonal rotation of the corresponding states. Before doing this, we mention that the $W^\pm$ bosons, given by

$$W_\mu^\pm = \frac{1}{\sqrt{2}} \left( W_\mu^1 \mp W_\mu^2 \right), \tag{23}$$

remain massless. This is due to the fact that the residual symmetry $SU(2)_L$ remains unbroken at this stage. We also recognize two correctly diagonalized kinetic energy contributions, given in terms of the charge operator eigenstates. As we will see, the charge operator in the 331 model is embedded as $Q = \frac{1}{2}\lambda^3 + \frac{\sqrt{3}}{2}\lambda^8 + X\mathbb{1}$. We have

$$Y_\mu^{\pm\pm} = \frac{1}{\sqrt{2}} \left( W_\mu^4 \mp iW_\mu^5 \right), \tag{24}$$

for the bileptons and

$$V_\mu^\pm = \frac{1}{\sqrt{2}} \left( W_\mu^6 \mp iW_\mu^7 \right), \tag{25}$$

for the exotic charged gauge bosons. The mass matrix that needs to be diagonalized in the $\{ X, W^8 \}$ bases is the following:

$$\begin{pmatrix} \frac{g_2^2 v_\rho^2}{3} & -\frac{\sqrt{2}}{3} g_1 g_2 v_\rho^2 \\ -\frac{\sqrt{2}}{3} g_1 g_2 v_\rho^2 & \frac{2}{3} g_1^2 v_\rho^2 \end{pmatrix}. \tag{26}$$

The diagonalization can be easily achieved through the orthogonal transformations

$$Z'_\mu = \frac{1}{\sqrt{g_2^2 + 2g_1^2}} \left( g_2 W_\mu^8 + \sqrt{2} g_1 X_\mu \right), \tag{27}$$

$$B_\mu = \frac{1}{\sqrt{g_2^2 + 2g_1^2}} \left( \sqrt{2} g_1 W_\mu^8 - g_2 X_\mu \right), \tag{28}$$

which can be thought as a rotation from the basis $\{X, W^8\}$ to $\{B, Z'\}$, with an angle

$$\sin\theta_{331} = \frac{g_2}{\sqrt{g_2^2 + \frac{g_1^2}{2}}},$$  (29)

that gives

$$\begin{pmatrix} 0 & 0 \\ 0 & \frac{1}{3}v_\rho^2(g_2^2 + 2g_1^2) \end{pmatrix}.$$  (30)

From the matrix in Equation (30), we can read off the squared masses of the mass eigenstates $\{B, Z'\}$

$$M_B^2 = 0 \qquad M_{Z'}^2 = \frac{1}{3}v_\rho^2\left(g_2^2 + 2g_1^2\right).$$  (31)

It is clear that we obtain a massless boson related to the $U(1)_Y$ symmetry of the Standard Model. It is not difficult to identify the embedding of the $Y$ charge operator in the 331 model, namely

$$\frac{Y}{2} = \sqrt{3}T^8 + X\mathbb{1},$$  (32)

where $T^8$ is the eighth generator of $SU(3)$. The matching condition between the $U(1)_X$ coupling and the Standard Model hypercharge can be easily computed, leading to the relation

$$\frac{1}{g_Y^2} = \frac{6}{g_1^2} + \frac{3}{g_2^2}.$$  (33)

### 4.2. The Breaking $SU(3)_C \times SU(2)_L \times U(1)_Y \rightarrow SU(3)_C \times U(1)_{em}$

Once the gauge symmetry has been decomposed into that of the Standard Model, another symmetry breaking is necessary to end up with the residual $SU(3)_C \times U(1)_{em}$ gauge symmetry. To realize the correct breaking scheme, we require two Higgs triplets, $\eta$ and $\chi$, which acquire the vacuum expectation values

$$\langle\eta\rangle = \begin{pmatrix} 0 \\ \frac{v_\eta}{\sqrt{2}} \\ 0 \end{pmatrix} \quad \text{with } X = 0,$$  (34)

and

$$\langle\chi\rangle = \begin{pmatrix} \frac{v_\chi}{\sqrt{2}} \\ 0 \\ 0 \end{pmatrix} \quad \text{with } X = -1,$$  (35)

and a sextet of $SU(3)_L$

$$\langle\sigma\rangle = \begin{pmatrix} 0 & 0 & \frac{v_\sigma}{2} \\ 0 & 0 & 0 \\ \frac{v_\sigma}{2} & 0 & 0 \end{pmatrix} \quad \text{with } X = 0.$$  (36)

After the first symmetry breaking, the covariant derivative can be written in terms of the mass eigenstate fields, namely in the basis $\{B, Z'\}$. Inserting the inverse of the Equations (27) and (28) into (21) gives

$$D_\mu = \begin{pmatrix} \partial_\mu - ig_2\frac{W_\mu^3}{2} + K_1 & -ig_2\left(\frac{W_\mu^1}{2} - \frac{iW_\mu^2}{2}\right) & -ig_2\left(\frac{W_\mu^4}{2} - \frac{iW_\mu^5}{2}\right) \\ -ig_2\left(\frac{W_\mu^1}{2} + \frac{iW_\mu^2}{2}\right) & \partial_\mu - ig_2\frac{W_\mu^3}{2} + K_1 & -ig_2\left(\frac{W_\mu^6}{2} - \frac{iW_\mu^7}{2}\right) \\ -ig_2\left(\frac{W_\mu^4}{2} + \frac{iW_\mu^5}{2}\right) & -ig_2\left(\frac{W_\mu^6}{2} + \frac{iW_\mu^7}{2}\right) & \partial_\mu + K_2 \end{pmatrix}$$  (37)

where $K_1$ and $K_2$ are given by

$$K_1 = \frac{i\left(\sqrt{2}B_\mu g_2 g_1 (2X-1) - Z'_\mu (g_2^2 + 4g_1^2 X)\right)}{2\sqrt{3}},\tag{38}$$

and

$$K_2 = \frac{i\left(\sqrt{2}B_\mu g_2 g_1 (X+1) + Z'_\mu (g_2^2 - 2g_1^2 X)\right)}{\sqrt{3}}.\tag{39}$$

Once all the scalar field has acquired a vacuum expectation value, the gauge fields $W^\pm$ and $Z$ become massive too, while $Y^{\pm\pm}$, $V^\pm$ and $Z'$ obtain more involved mass terms. The squared masses of $W^\pm$, $V^\pm$ and $Y^{\pm\pm}$ are given by

$$M_W^2 = \frac{g_2^2 v_\eta^2}{4} + \frac{g_2^2 v_\chi^2}{4} + \frac{g_2^2 v_\rho^2}{4},\tag{40}$$

$$M_V^2 = \frac{g_2^2 v_\rho^2}{4} + \frac{g_2^2 v_\eta^2}{4} + \frac{g_2^2 v_\sigma^2}{4},\tag{41}$$

$$M_Y^2 = \frac{g_2^2 v_\rho^2}{4} + \frac{g_2^2 v_\chi^2}{4} + g_2^2 v_\sigma^2,\tag{42}$$

where we recall that

$$W_\mu^\pm = \frac{1}{\sqrt{2}}\left(W_\mu^1 \mp i W_\mu^2\right), \qquad V_\mu^\pm = \frac{1}{\sqrt{2}}\left(W_\mu^6 \mp i W_\mu^7\right), \qquad Y_\mu^{\pm\pm} = \frac{1}{\sqrt{2}}\left(W_\mu^4 \mp i W_\mu^5\right).\tag{43}$$

On the other hand, the neutral gauge bosons also gain non-diagonal mass terms, which in the $\{W^3, W^8, X\}$ basis, are given by the following matrix:

$$\begin{pmatrix} g_1^2 v_\rho^2 + g_1^2 v_\chi^2 & -\frac{g_1 g_2 v_\rho^2}{\sqrt{3}} - \frac{g_1 g_2 v_\chi^2}{2\sqrt{3}} & -\frac{1}{2}g_1 g_2 v_\chi^2 \\ -\frac{g_1 g_2 v_\rho^2}{\sqrt{3}} - \frac{g_1 g_2 v_\chi^2}{2\sqrt{3}} & \frac{g_2^2 v_\rho^2}{3} + \frac{g_2^2 v_\eta^2}{12} + \frac{g_2^2 v_\chi^2}{12} + \frac{g_2^2 v_\sigma^2}{12} & -\frac{g_2^2 v_\eta^2}{4\sqrt{3}} + \frac{g_2^2 v_\chi^2}{4\sqrt{3}} - \frac{g_2^2 v_\sigma^2}{4\sqrt{3}} \\ -\frac{1}{2}g_1 g_2 v_\chi^2 & -\frac{g_2^2 v_\eta^2}{4\sqrt{3}} + \frac{g_2^2 v_\chi^2}{4\sqrt{3}} - \frac{g_2^2 v_\sigma^2}{4\sqrt{3}} & \frac{g_2^2 v_\eta^2}{4} + \frac{g_2^2 v_\chi^2}{4} + \frac{g_2^2 v_\sigma^2}{4} \end{pmatrix}\tag{44}$$

In a first step, the two neutral gauge bosons $W^8$ and $X$ mix, giving rise to the two bosons $B$ and $Z$. The mixing angle is denoted by $\theta_{331}$ and is given by

$$\sin\theta_{331} = \frac{g_2}{\sqrt{g_2^2 + \frac{g_1^2}{2}}},\tag{45}$$

which was the rotation discussed in the previous section. Then, we can proceed in complete analogy with the Standard Model, where $B$ mixes with $W^3$ through the Weinberg angle $\theta_W$, which in the minimal 331 model takes the form

$$\sin\theta_W = \frac{g_Y}{\sqrt{g^2 + g_Y^2}}.\tag{46}$$

Using (46), it is straightforward to show that we can express the $\theta_{331}$ angle in terms of the Weinberg angle, namely

$$\cos\theta_{331} = \sqrt{3}\tan\theta_W,\tag{47}$$

therefore the photon field $A$ and the two massive neutral gauge bosons $Z$ and $Z'$ are identified as

$$A = \sin\theta_W W_3 + \cos\theta_W\left(\sqrt{3}\tan\theta_W W_8 + \sqrt{1 - 3\tan^2\theta_W} X\right),\tag{48}$$

$$Z = \cos\theta_W W_3 - \sin\theta_W \left( \sqrt{3}\tan\theta_W W_8 + \sqrt{1 - 3\tan^2\theta_W}\, X \right), \tag{49}$$

$$Z' = -\sqrt{1 - 3\tan^2\theta_W}\, W_8 + \sqrt{3}\tan\theta_W X. \tag{50}$$

The Weinberg angle also relates the coupling of $SU(3)_L$, which through matching is equal to the $SU(2)_L$ of the Standard Model, to the $U(1)_X$ coupling, through

$$\frac{g_1^2}{g_2^2} = \frac{6\sin^2\theta_W}{1 - 4\sin^2\theta_W}, \tag{51}$$

which can be obtained from Equations (45) and (46). From Equations (48)–(50), it is evident that there is a residual mixing between the massive gauge boson, which is given by the matrix

$$\begin{pmatrix} C_{ZZ} & C_{ZZ'} \\ C_{ZZ'} & C_{Z'Z'} \end{pmatrix} \tag{52}$$

with entries

$$C_{ZZ} = g_2^2(2\theta_W)\left( v_\rho^2 + v_\eta^2 + v_\sigma^2 \right), \tag{53}$$

$$C_{Z'Z'} = \frac{g_2^2\sin^2\theta_W\left( \csc^2(\theta_W)\left( v_\rho^2 + v_\eta^2 + 4v_\chi^2 + v_\sigma^2 \right) + 9\sec^2(\theta_W)\left( v_\rho^2 + v_\eta^2 + v_\sigma^2 \right) - 4\left( v_\rho^2 + 4v_\eta^2 + v_\chi^2 + 4v_\sigma^2 \right) \right)}{24\cos(2\theta_W) - 12}, \tag{54}$$

$$C_{ZZ'} = \frac{g_2^2\sec^3(\theta_W)\left( -\cos(2\theta_W)\left( v_\rho^2 + 2\left( v_\eta^2 + v_\sigma^2 \right) \right) + 2v_\rho^2 + v_\eta^2 + v_\sigma^2 \right)}{4\sqrt{12 - 9\sec^2(\theta_W)}}. \tag{55}$$

Therefore, a further rotation is needed

$$\begin{pmatrix} Z_1 \\ Z_2 \end{pmatrix} = \begin{pmatrix} \cos\theta_Z & -\sin\theta_Z \\ \sin\theta_Z & \cos\theta_Z \end{pmatrix} \begin{pmatrix} Z \\ Z' \end{pmatrix} \tag{56}$$

in order to obtain the masses of propagating gauge bosons, namely

$$\begin{aligned} M_Z^2 = \frac{1}{6}\big(&3g_1^2(v_\rho^2 + v_\chi^2) - (9g_1^4(v_\rho^2 + v_\chi^2)^2 + 6g_1^2 g_2^2(v_\rho^4 - v_\rho^2(v_\eta^2 + v_\sigma^2) \\ &+ v_\chi^2(-v_\eta^2 + v_\chi^2 - v_\sigma^2)) + g_2^4(v_\rho^4 - v_\rho^2(v_\eta^2 + v_\chi^2 + v_\sigma^2) + v_\eta^4 + v_\sigma^2(2v_\eta^2 - v_\chi^2) \\ &- v_\eta^2 v_\chi^2 + v_\chi^4 + v_\sigma^4)^{\frac{1}{2}}) + g_2^2(v_\rho^2 + v_\eta^2 + v_\chi^2 + v_\sigma^2)), \end{aligned} \tag{57}$$

and

$$\begin{aligned} M_{Z'}^2 = \frac{1}{6}\big(&3g_1^2(v_\rho^2 + v_\chi^2) + (9g_1^4(v_\rho^2 + v_\chi^2)^2 + 6g_1^2 g_2^2(v_\rho^4 - v_\rho^2(v_\eta^2 + v_\sigma^2) \\ &+ v_\chi^2(-v_\eta^2 + v_\chi^2 - v_\sigma^2)) + g_2^4(v_\rho^4 - v_\rho^2(v_\eta^2 + v_\chi^2 + v_\sigma^2) + v_\eta^4 + v_\sigma^2(2v_\eta^2 - v_\chi^2) \\ &- v_\eta^2 v_\chi^2 + v_\chi^4 + v_\sigma^4))^{\frac{1}{2}} + g_2^2(v_\rho^2 + v_\eta^2 + v_\chi^2 + v_\sigma^2)). \end{aligned} \tag{58}$$

Once the orthogonal rotations have been performed and the Lagrangian has been written in terms of the mass eigenstates of bosonic fields, it is possible to extract the values of the couplings. In particular, it is possible to derive the expression of the electric charge in terms of the couplings of the minimal 331

$$e = \frac{g_1 g_2}{\sqrt{g_2^2 + 4g_1^2}}, \tag{59}$$

with the embedding of the charge operator given by

$$Q = T^3 + \sqrt{3}T^8 + X\mathbb{1}. \tag{60}$$

One of the most interesting features of the model is that the embedding of the Standard Model gauge group into the 331 model gauge group induces a bound on the UV completion of the model [5]. From (51), it is clear that we need to satisfy the following condition

$$\sin^2 \theta_W \leq \frac{1}{4},$$ (61)

in order to guarantee that the $g_1$ coupling of the minimal 331 model is finite. When $\sin^2 \theta_W(\mu) = 1/4$, the coupling constant $g_1(\mu)$ diverges, indicating a Landau pole in the renormalization group evolution of the model, which causes the theory to loose its perturbative character even at energy scales lower than $\mu$. With the particle content of the Standard Model, the condition $\sin^2 \theta_W(\mu) = 1/4$ is reached at an energy scale of around 4 TeV, and the presence of an additional particle at the TeV scale can make this behavior even faster, loosing a perturbative character before 4 TeV. The alternative scenario, where $g_2$ tends towards zero, is disregarded, as $g_2$ coincides with the Standard Model's $SU(2)_L$ coupling, $g_2$, due to the full embedding of $SU(2)_L$ into $SU(3)_L$. We show the running of the Standard Model couplings $g_2$ in Figure 1a and $g_y$ in Figure 1b, and in Figure 1c we show how the matching condition of the coupling $g_1$ evolves in terms of the matching scale. Here, we have defined

$$\overline{g_1} = \frac{g_2 g_y}{\sqrt{g_2^2 - 3g_y^2}},$$ (62)

which expresses the value of the $U(1)_X$ coupling in terms of the Standard Model. Finally, the evolution of $g_1$ in the context of the 331 Model is shown in Figure 1d, where we have plotted the quantity

$$\alpha_1 = \frac{g_1^2}{4\pi}$$ (63)

which makes it clear that when $\alpha_1 > 1$, the theory looses pertubativity. A more recent study was performed in [6].

The occurrence of a Landau-like pole in the minimal 331 model is not surprising, as many non-asymptotically free theories exhibit a similar behavior. What distinguishes some of these models is the possibility of encountering this behavior at energies as low as a few TeVs. Consequently, the cutoff scale, $\Lambda_{cutoff}$, cannot be removed by taking $\Lambda_{cutoff} \rightarrow \infty$, as in other renormalizable theories.

From a phenomenological perspective, this result is not overly concerning. The necessity of embedding QED within the electroweak theory at energies of a few hundred GeVs, along with the requirement to account for weak and strong corrections in calculations of physical observables, has already been acknowledged. Nevertheless, as a mathematical exercise, studying pure QED at infinitesimal distances proves intriguing. Lattice calculations suggest that chiral symmetry breaking within QED mitigates the Landau pole issue by shifting it above the cutoff scale. Interestingly, the potential existence of the Landau pole or the triviality of the theory arises even at low orders in perturbation theory, suggesting that this phenomenon is not merely a perturbative artifact.

The renormalization group offers qualitative insights into the asymptotic behavior of theories at very high energies, even when coupling constants at the relevant scale prohibit the use of perturbation theory. However, it is essential to remember that both QED and the Standard Model are effective, not fundamental, theories. Consequently, effective operators with dimensions higher than $d = 4$ must be considered for a realistic continuum limit in lattice calculations. Thus, employing the pure versions of these models remains inconclusive, and the renormalization group may provide valuable insights into this issue within the minimal 331 model.

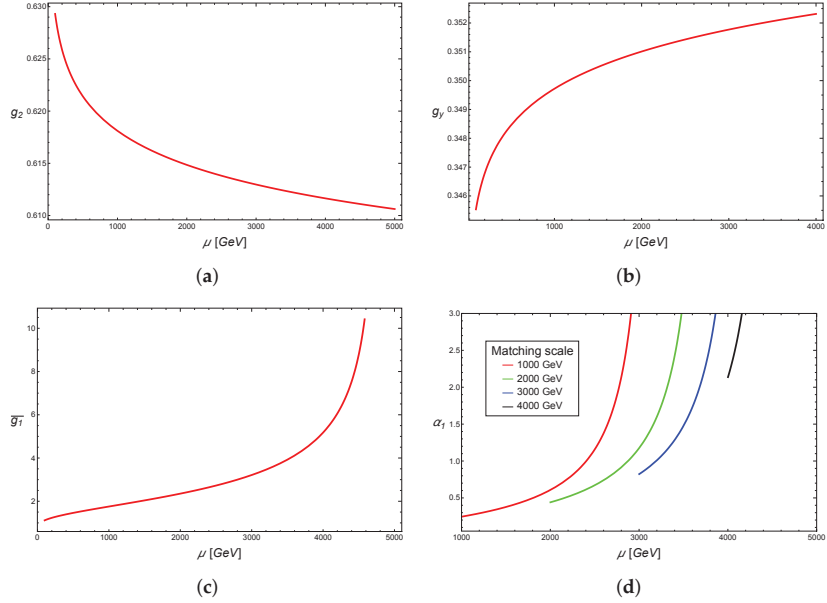

**Figure 1.** Behavior of the matching condition between the Standard Model and 331 model, which makes it clear how the Landau pole at the TeV scale emerges from the matching with the Standard Model. (**a**) Running of the coupling $g_2$ in the Standard Model. (**b**) Running of the coupling $g_y$ in the Standard Model. (**c**) Evolution of the matching condition of the coupling $g_1$ of the 331 model with the Standard Model coupling constants. (**d**) Running of the coupling $a_1$ in the minimal 331 model.

## 5. Higgs Sector

The inclusion of the sextet representation in the potential enriches the phenomenology of the model and enlarges the number of physical states in the spectrum. In fact, we now have, after electroweak symmetry breaking (EWSB) $SU(3)_L \times U(1)_X \to SU(2)_L \times U(1)_Y \to U(1)_{\text{em}}$, five scalar Higgses, three pseudoscalar Higgses, four charged Higgses, and three doubly-charged Higgses. The (lepton-number conserving) potential of the model is given by [7]

$$
\begin{aligned}
V =& m_1 \rho^\dagger \rho + m_2 \eta^\dagger \eta + m_3 \chi^\dagger \chi + \lambda_1 (\rho^\dagger \rho)^2 + \lambda_2 (\eta^\dagger \eta)^2 + \lambda_3 (\chi^\dagger \chi)^2 \\
& + \lambda_{12} \rho^\dagger \rho \eta^\dagger \eta + \lambda_{13} \rho^\dagger \rho \chi^\dagger \chi + \lambda_{23} \chi^\dagger \chi \eta^\dagger \eta + \zeta_{12} \rho^\dagger \eta \eta^\dagger \rho + \zeta_{13} \rho^\dagger \chi \chi^\dagger \rho + \zeta_{23} \eta^\dagger \chi \chi^\dagger \eta \\
& + m_4 \operatorname{Tr}\left(\sigma^\dagger \sigma\right) + \lambda_4 (\operatorname{Tr}\left(\sigma^\dagger \sigma\right))^2 + \lambda_{14} \rho^\dagger \rho \operatorname{Tr}\left(\sigma^\dagger \sigma\right) + \lambda_{24} \eta^\dagger \eta \operatorname{Tr}\left(\sigma^\dagger \sigma\right) + \lambda_{34} \chi^\dagger \chi \operatorname{Tr}\left(\sigma^\dagger \sigma\right) \\
& + \lambda_{44} \operatorname{Tr}\left(\sigma^\dagger \sigma \sigma^\dagger \sigma\right) + \zeta_{14} \rho^\dagger \sigma \sigma^\dagger \rho + \zeta_{24} \eta^\dagger \sigma \sigma^\dagger \eta + \zeta_{34} \chi^\dagger \sigma \sigma^\dagger \chi \\
& + (\sqrt{2} f_{\rho \eta \chi} \epsilon^{ijk} \rho_i \eta_j \chi_k + \sqrt{2} f_{\rho \sigma \chi} \rho^T \sigma^\dagger \chi \\
& + \zeta_{14} \epsilon^{ijk} \rho^{*l} \sigma_{li} \rho_j \eta_k + \zeta_{24} \epsilon^{ijk} \epsilon^{lmn} \eta_i \eta_l \sigma_{jm} \sigma_{kn} + \zeta_{34} \epsilon^{ijk} \chi^{*l} \sigma_{li} \chi_j \eta_k) + \text{h.c.} .
\end{aligned}
\tag{64}
$$

In principle, it is possible to extend the potential with additional lepton number violating terms that are singlet under 331 gauge group [7], this possibility has been extensively discussed in [8]. The EWSB mechanism will cause a mixing among the Higgs fields [9]. From Equation (64), it is possible to obtain the explicit expressions of the mass matrices of the scalar, pseudoscalar, charged, and doubly-charged Higgses by using standard procedures. In the broken Higgs phase, the minimization conditions

$$
\frac{\partial V}{\partial v_\phi} = 0, \quad \langle \phi^0 \rangle = v_\phi, \quad \phi = \rho, \eta, \chi, \sigma
\tag{65}
$$

will define the tree-level vacuum, one-loop contributions to the vacuum stability were recently analyzed in [10] for a simpler model version, namely the economical 331 model [11,12]. We remind that we are considering massless neutrinos by choosing the vev of the neutral field $\sigma_2^0$ as zero. This was the choice in Frampton's original formulation. This can be generalized in order to give a small Majorana neutrino mass to the neutrinos [13].

The explicit expressions of the minimization conditions are then given by

$$m_1 v_\rho + \lambda_1 v_\rho^3 + \frac{1}{2}\lambda_{12} v_\rho v_\eta^2 - f_{\rho\eta\chi} v_\eta v_\chi + \frac{1}{2}\lambda_{13} v_\rho v_\chi^2 - \frac{1}{\sqrt{2}}\xi_{14} v_\rho v_\eta v_\sigma + f_{\rho\sigma\chi} v_\chi v_\sigma$$
$$+ \frac{1}{2}\lambda_{14} v_\rho v_\sigma^2 + \frac{1}{4}\zeta_{14} v_\rho v_\sigma^2 = 0 \tag{66}$$

$$m_2 v_\eta + \frac{1}{2}\lambda_{12} v_\rho^2 v_\eta + \lambda_2 v_\eta^3 - f_{\rho\eta\chi} v_\rho v_\chi + \frac{1}{2}\lambda_{23} v_\eta v_\chi^2 - \frac{1}{2\sqrt{2}}\xi_{14} v_\rho^2 v_\sigma + \frac{1}{2\sqrt{2}} v_\chi^2 v_\sigma$$
$$+ \frac{1}{2}\lambda_{24} v_\eta v_\sigma^2 - \xi_{24} v_\eta v_\sigma^2 = 0 \tag{67}$$

$$m_3 v_\chi + \lambda_3 v_\chi^3 + \frac{1}{2}\lambda_{13} v_\rho^2 v_\chi - f_{\rho\eta\chi} v_\rho v_\eta + \frac{1}{2}\lambda_{23} v_\eta^2 v_\chi + \frac{1}{\sqrt{2}}\xi_{34} v_\eta v_\chi v_\sigma + f_{\rho\sigma\chi} v_\rho v_\sigma$$
$$+ \frac{1}{2}\lambda_{34} v_\chi v_\sigma^2 + \frac{1}{4}\zeta_{34} v_\chi v_\sigma^2 = 0 \tag{68}$$

$$m_4 v_\sigma + \frac{1}{2}\lambda_{14} v_\rho^2 v_\sigma + \lambda_{44} v_\sigma^3 + \frac{1}{2}\lambda_4 v_\sigma^3 + f_{\rho\sigma\chi} v_\rho v_\chi - \frac{1}{2\sqrt{2}}\xi_{14} v_\rho^2 v_\eta + \frac{1}{2\sqrt{2}}\xi_{34} v_\eta v_\chi^2$$
$$+ \frac{1}{2}\lambda_{14} v_\rho^2 v_\sigma + \frac{1}{4}\zeta_{14} v_\rho^2 v_\sigma + \frac{1}{2}\lambda_{24} v_\eta^2 v_\sigma - \xi_{24} v_\eta^2 v_\sigma + \frac{1}{2}\lambda_{34} v_\chi^2 v_\sigma + \frac{1}{4}\zeta_{34} v_\chi^2 v_\sigma = 0 \tag{69}$$

These conditions are inserted into the tree-level mass matrices of the CP-even and CP-odd Higgs sectors, derived from $M_{ij} = \partial^2 V / \partial\phi_i \partial\phi_j|_{vev}$, where $V$ is the potential in Equation (64). The mass eigenstates are defined as follows:

$$h = R^S \begin{pmatrix} \text{Re}\,\rho^0 \\ \text{Re}\,\eta^0 \\ \text{Re}\,\chi^0 \\ \text{Re}\,\sigma_1^0 \\ \sigma_2^0 \end{pmatrix} \quad Ah = R^P \begin{pmatrix} \text{Im}\,\rho^0 \\ \text{Im}\,\eta^0 \\ \text{Im}\,\chi^0 \\ \text{Im}\,\sigma_1^0 \end{pmatrix} \quad H^+ = R^C \begin{pmatrix} \rho^+ \\ \chi^+ \\ \eta_1^+ \\ \eta_2^+ \\ \sigma_1^+ \\ \sigma_2^+ \end{pmatrix} \quad H^{++} = R^{2C} \begin{pmatrix} \rho^{++} \\ \chi^{++} \\ \sigma_1^{++} \\ \sigma_2^{++} \end{pmatrix} \tag{70}$$

where the explicit expressions of the mass matrices are too cumbersome to be presented here, and are given in Appendix A.

In this case, we have five scalar Higgs bosons, and one of them will be the SM Higgs of mass about 125 GeV, along with four neutral pseudoscalar Higgs bosons, out of which, two are the Goldstones of the $Z$ and the $Z'$ massive vector bosons. In addition, there are six charged Higgses, two of which are the charged Goldstones, and three are doubly-charged Higgses, one of which is a Goldstone boson.

Hereafter, we shall give the schematic expression of the physical Higgs states, after EWSB, in terms of the gauge eigenstates, whose expressions contain only the vev of the various fields. In the following equations, $R_{ij}^K \equiv R_{ij}^K(m_1, m_2, m_3, \lambda_1, \lambda_2, \ldots)$ refers to the rotation matrix of each Higgs sector that depends on all the parameters of the potential in Equation (64). Starting from the scalar (CP-even) Higgs bosons, we have

$$H_i = R_{i1}^S \text{Re}\,\rho^0 + R_{i2}^S \text{Re}\,\eta^0 + R_{i3}^S \text{Re}\,\chi^0 + R_{i4}^S \text{Re}\,\sigma_1^0 + R_{i5}^S \text{Re}\,\sigma_2^0, \tag{71}$$

expressed in terms of the rotation matrix of the scalar components $R^S$. There are similar expressions for the pseudoscalars

$$Ah_i = R_{i1}^P \text{Im}\,\rho^0 + R_{i2}^P \text{Im}\,\eta^0 + R_{i3}^P \text{Im}\,\chi^0 + R_{i4}^P \text{Im}\,\sigma_1^0 + R_{i5}^P \text{Im}\,\sigma_2^0 \tag{72}$$

in terms of the rotation matrix of the pseudoscalar components $\mathrm{R}^P$. Here, however, we have two Goldstone bosons responsible for the generation of the masses of the neutral gauge bosons $Z$ and $Z'$ given by

$$A_0^1 = \frac{1}{N_1}\left(v_\rho \mathrm{Im}\,\rho^0 - v_\eta \mathrm{Im}\,\eta^0 + v_\sigma \mathrm{Im}\,\sigma_1^0\right), \qquad N_1 = \sqrt{v_\rho^2 + v_\eta^2 + v_\sigma^2}\,; \tag{73}$$

$$A_0^2 = \frac{1}{N_2}\left(-v_\rho \mathrm{Im}\,\rho^0 + v_\chi \mathrm{Im}\,\chi^0\right), \qquad N_2 = \sqrt{v_\rho^2 + v_\chi^2}. \tag{74}$$

For the charged Higgs bosons, the interaction eigenstates are

$$H_i^+ = \mathrm{R}_{i1}^C \rho^+ + \mathrm{R}_{i2}^C (\eta^-)^* + \mathrm{R}_{i3}^C \eta^+ + \mathrm{R}_{i4}^C (\chi^-)^* + \mathrm{R}_{i5}^C \sigma_1^+ + \mathrm{R}_{i6}^C (\sigma_2^-)^*, \tag{75}$$

with $\mathrm{R}^C$ being a rotation matrix of the charged sector. We recall that, even in this case, two $H_i^+$ are massless Goldstones bosons, because in the minimal 331 model there are $W^\pm$ and the $V^\pm$ gauge bosons that both become massive after EWSB. The explicit expressions of the Goldstones are

$$H_W^+ = \frac{1}{N_W}\left(-v_\eta \eta^+ + v_\chi (\chi^-)^* + v_\sigma (\sigma_2^-)^*\right), \qquad N_W = \sqrt{v_\eta^2 + v_\chi^2 + v_\sigma^2}\,; \tag{76}$$

$$H_V^+ = \frac{1}{N_V}\left(v_\rho \rho^+ - v_\eta (\eta^-)^* + v_\sigma \sigma_1^+\right), \qquad N_V = \sqrt{v_\rho^2 + v_\eta^2 + v_\sigma^2}. \tag{77}$$

In particular, we are interested in the doubly-charged Higgses, where the number of physical states, after EWSB, is three, whereas we would have had only one physical doubly-charged Higgs if we had not included the sextet. The physical doubly-charged Higgs states are expressed in terms of the gauge eigenstates and the elements of the rotation matrix $\mathrm{R}^C$ as

$$H_i^{++} = \mathrm{R}_{i1}^{2C} \rho^{++} + \mathrm{R}_{i2}^{2C} (\chi^{--})^* + \mathrm{R}_{i3}^{2C} \sigma_1^{++} + \mathrm{R}_{i4}^{2C} (\sigma_2^{--})^*. \tag{78}$$

In particular, the structure of the corresponding Goldstone boson is

$$H_0^{++} = \frac{1}{N}\left(-v_\rho \rho^{++} + v_\chi (\chi^{--})^* - \sqrt{2} v_\sigma \sigma_1^{++} + \sqrt{2} v_\sigma (\sigma_2^{--})^*\right) \tag{79}$$

where $N = \sqrt{v_\rho^2 + v_\chi^2 + 4v_\sigma^2}$ is a normalization factor.

## 6. The Yukawa Sector

The model presented in the previous section exhibits the interesting feature of having both scalar and vector doubly-charged bosons, which is a peculiarity of the minimal version of the 331 model. In fact, it is possible to consider various versions of the $SU(3)_c \times SU(3)_L \times U(1)_X$ gauge symmetry, usually parametrized by $\beta$ [14,15]. We discuss the case of $\beta = \sqrt{3}$, corresponding to the minimal version presented here [1,2], leading to vector bosons with an electric charge equal to $\pm 2$.

Doubly-charged states hold particular interest, due to their potential for unique characteristics in terms of permissible decay channels, such as the production of same-sign lepton pairs [16–20]. Within the framework of the minimal 331 model, an even more intriguing prospect arises. It becomes possible to discern whether a same-sign lepton pair originates from either a scalar or a vector boson. As we will elaborate, this distinction also offers insights into the existence of a higher representation within the $SU(3)_c \times SU(3)_L \times U(1)_X$ gauge group, notably the sextet.

### 6.1. The Triplet Sector

In the previous section, we saw that the EWSB mechanism is realized in the 331 model by giving a vev to the neutral component of the triplets $\rho$, $\eta$ and $\chi$. The Yukawa interactions for SM and exotic quarks are obtained by means of these scalar fields and are given by

$$
\begin{aligned}
\mathcal{L}_{q,\,triplet}^{Y} = & -\overline{Q}_m \left( Y_{ma}^d \eta^* d_{aR} + Y_{ma}^u \chi^* u_{ma} \right) - \overline{Q}_3 \left( Y_{3a}^d \chi d_{aR} + Y_{3a}^u \eta u_{ma} \right) + \\
& -\overline{Q}_m \left( Y_{mn}^J \chi J_{nR} \right) - \overline{Q}_3 Y_3^J \chi J_{3R} + \text{h.c.}
\end{aligned}
\tag{80}
$$

where $y_d^j$, $y_u^j$ and $y_L^j$ are the Yukawa couplings for down-, up-type, and exotic quarks, respectively. The masses of the exotic quarks are related to the vev of the neutral component of $\rho = (0, 0, v_\rho)$ via the invariants

$$
\begin{aligned}
Q_1\,\rho^* D_R^*, Q_1\,\rho^* S_R^* &\sim (3, 3, -1/3) \times (1, \bar{3}, -1) \times (\bar{3}, 1, 4/3) \\
Q_3\,\rho\, T_R^* &\sim (3, \bar{3}, 2/3) \times (1, 3, 1) \times (3, 1, -5/3),
\end{aligned}
\tag{81}
$$

responsible of the breaking $SU(3)_c \times SU(3)_L \times U(1)_X \to SU(3)_c \times SU(2)_L \times U(1)_Y$. It is clear that, being $v_\rho \gg v_{\eta,\chi}$, the masses of the exotic quarks are $\mathcal{O}(\text{TeV})$ whenever the relation $Y^J \sim \mathbb{1}$ is satisfied. As we will see, the model also needs a sextet.

### 6.2. The Sextet Yukawa Coupling

The need for introducing a sextet sector can be summarized as follows. A typical Dirac mass term for the leptons in the SM is associated with the operator $\bar{l}_L H e_R$, with $l_L = (v_{eL}, e_L)$ being the $SU(2)_L$ doublet, with the representation content $(2, 1/2) \times (2, 1/2) \times (1, -1)$ (for $l$, $H$ and $e_R$, respectively) in $SU(2)_L \times U(1)_Y$. In the 331 model, the $L$ and $R$ components of the lepton ($e$) belong to the same multiplet. Consequently, identifying an $SO(1, 3) \times SU(3)_L$ singlet requires two leptons in the same representation. This can be achieved (at least partially) with the operator

$$
\begin{aligned}
\mathcal{L}_{l,\,triplet}^{Yuk} &= G_{ab}^\eta (l_{a\alpha}^i \epsilon^{\alpha\beta} l_{b\beta}^j) \eta^{*k} \epsilon^{ijk} + \text{h.c.} \\
&= G_{ab}^\eta\, l_a^i \cdot l_b^j\, \eta^{*k} \epsilon^{ijk} + \text{h.c.}
\end{aligned}
\tag{82}
$$

where the indices $a$ and $b$ run over the three generations of flavor, $\alpha$ and $\beta$ are Weyl indices contracted in order to generate an $SO(1, 3)$ invariant ($l_a^i \cdot l_b^j \equiv l_{a\alpha}^i \epsilon^{\alpha\beta} l_{b\beta}^j$) from two Weyl fermions, and $i, j, k = 1, 2, 3$, are $SU(3)_L$ indices.

The use of $\eta$ as a Higgs field is mandatory, since the components of the multiplet $l^j$ are $U(1)_X$ singlets. The representation content of the operator $l_a^i l_b^j$ according to $SU(3)_L$ is given by $3 \times 3 = 6 + \bar{3}$, with the $\bar{3}$ extracted by an anti-symmetrization over $i$ and $j$ via $\epsilon^{ijk}$. This allows identifying $l_a^i l_b^j \eta^{*k} \epsilon^{ijk}$ as an $SU(3)_L$ singlet. Considering that the two leptons are anticommuting Weyl spinors, and that the $\epsilon^{\alpha\beta}$ (Lorentz) and $\epsilon^{ijk}$ ($SU(3)_L$) contractions introduce two sign flips under the $a \leftrightarrow b$ exchange, the combination

$$
M_{ab} = (l_a^i \cdot l_b^j) \eta^{*k} \epsilon^{ijk}
\tag{83}
$$

is therefore antisymmetric under the exchange of the two flavors, implying that even $G_{ab}$ has to be antisymmetric. However, an antisymmetric $G_{ab}^\eta$ is not sufficient to provide mass to all the leptons.

In fact, the diagonalization of $G^\eta$ by means of a unitary matrix $U$, namely $G^\eta = U\Lambda U^\dagger$, with $G^\eta$ antisymmetric in flavor space, implies that its three eigenvalues are given by $\Lambda = (0, \lambda_{22}, \lambda_{33})$, with $\lambda_{22} = -\lambda_{33}$; i.e., one eigenvalue is null and the other two are equal in magnitude. At the minimum of $\eta$, i.e., $\eta = (0, v_\eta, 0)$, one has

$$
G_{ab}^\eta M^{ab} = -Tr(\Lambda\, U M U^\dagger) = 2 v_\eta \lambda_{22}\, U_{2a}\, l_a^1 \cdot l_b^3\, U_{2b}^* + 2 v_\eta \lambda_{33}\, U_{3a}\, l_a^1 \cdot l_b^3\, U_{3b}^*,
\tag{84}
$$

with $l_a^1 = e_{aL}$ and $l_b^3 = e_{bR}^c$. Introducing the linear combinations

$$E_{2L} \equiv U_{2a} \, l_a^1 = U_{2a}' \, e_{aL} \qquad U_{2b}^* \, l_b^3 = U_{2b}^* \, e_{bR}^c = i\sigma_2 (U_{2b} \, e_{bR})^* \equiv E_{2R}^c, \tag{85}$$

then the antisymmetric contribution in flavor space becomes

$$\mathcal{L}_{l,triplet}^{Yuk} = 2v_\eta \lambda_{22} (E_{2L} E_{2R}^c - E_{3L} E_{3R}^c), \tag{86}$$

which is clearly insufficient to generate the lepton masses of three non-degenerate lepton families. We shall solve this problem by introducing a second invariant operator, with the inclusion of a sextet $\sigma$

$$\sigma = \begin{pmatrix} \sigma_1^{++} & \sigma_1^+ / \sqrt{2} & \sigma^0 / \sqrt{2} \\ \sigma_1^+ / \sqrt{2} & \sigma_1^0 & \sigma_2^- / \sqrt{2} \\ \sigma^0 / \sqrt{2} & \sigma_2^- / \sqrt{2} & \sigma_2^{--} \end{pmatrix} \in (1, 6, 0), \tag{87}$$

leading to the Yukawa term

$$\mathcal{L}_{l,sextet}^{Yuk.} = G_{ab}^\sigma \, l_a^i \cdot l_b^j \sigma_{i,j}^*, \tag{88}$$

which allows building a singlet out of the representation 6 of $SU(3)_L$, contained in $l_a^i \cdot l_b^j$, by combining it with the flavor-symmetric $\sigma^*$, i.e., $\bar{6}$. Notice that $G_{ab}^\sigma$ is symmetric in flavor space.

It is interesting to note that without considering the sextet, a doubly-charged scalar would not be able to decay into same-sign leptons. This is because, without the sextet, the interaction responsible for the leptons only involves the scalar triplet, denoted as $\eta$, which does not contain a doubly-charged state.

### 6.3. Lepton Mass Matrices

Let us now come to discuss the lepton mass matrices in the model. They are related to the Yukawa interactions by the Lagrangian

$$\mathcal{L}_l^{Yuk.} = \mathcal{L}_{l,sextet}^{Yuk.} + \mathcal{L}_{l,triplet}^{Yuk.} + \text{h.c.} \tag{89}$$

and are combinations of triplet and sextet contributions. The structure of the mass matrix that emerges from the vevs of the neutral components of $\eta$ and $\sigma$ is thus given by

$$\mathcal{L}_l^{Yuk.} = \left( \sqrt{2}\sigma_0 G_{a,b}^\sigma + 2v_\eta G_{ab}^\eta \right) (e_{aL} \cdot e_{bR}^c) + \sigma_1^0 G_{ab}^\sigma \left( \nu_L^T i\sigma_2 \nu_L \right) + \text{h.c.}, \tag{90}$$

which generates a Dirac mass matrix for the charged leptons $M_{ab}^l$ and a Majorana mass matrix for neutrinos $M_{ab}^{\nu_l}$

$$M_{ab}^l = \sqrt{2}\langle \sigma_0 \rangle \, G_{a,b}^\sigma + 2v_\eta \, G_{ab}^\eta \qquad , \qquad M_{ab}^{\nu_l} = \langle \sigma_1^0 \rangle \, G_{ab}^\sigma. \tag{91}$$

In the expression above, $\langle \sigma^0 \rangle$ and $\langle \sigma_1^0 \rangle$ are the vacuum expectation values of the neutral components of $\sigma$. For a vanishing $G^\sigma$, as we have already discussed, we will not be able to generate the lepton masses consistently, nor any mass for the neutrinos, i.e.,

$$M_{ab}^l = 2v_\eta \, G_{ab}^\eta \qquad , \qquad M^{\nu_l} = 0. \tag{92}$$

On the contrary, in the limit $G^\eta \to 0$, Equation (91) becomes

$$M_{ab}^l = \sqrt{2}\langle \sigma_0 \rangle \, G_{ab}^\sigma \qquad , \qquad M_{a,b}^{\nu_l} = \frac{\langle \sigma_1^0 \rangle}{\sqrt{2}} G_{ab}^\sigma, \tag{93}$$

which has some interesting consequences. Since the Yukawa couplings are the same for both leptons and neutrinos, we have to require $\langle \sigma_1^0 \rangle \ll \langle \sigma^0 \rangle$, in order to obtain small neutrino masses. For the goal of our analysis, we will assume that the vev of $\sigma_1^0$ vanishes; i.e., $\langle \sigma_1^0 \rangle \equiv 0$. Clearly, if the matrix $G^\sigma$ is diagonal in flavor space, from Equation (93), we will immediately conclude that the Yukawa coupling $G^\sigma$ has to be chosen to be proportional to the masses of the SM leptons. An interesting consequence of this is that the decay $H^{\pm\pm} \to l^\pm l^\pm$, which is also proportional to $G^\sigma$, and therefore to the lepton masses, will be enhanced for the heavier leptons, in particular for the $\tau$, as thoroughly discussed in [21]. This is an almost unique situation that is not encountered in other models with doubly-charged scalars decaying into same-sign leptons [22].

## 7. Flavor Physics in the Minimal 331 Model

One of the features of the minimal 331 model lies in its arrangement of fermions within triplets of $SU(3)_L$. However, to maintain anomaly cancellation, it becomes necessary to assign one of the quark families to a different representation than the other two, ensuring an equal number of triplets and anti-triplets in the fermion sector.

This introduces several complexities, particularly concerning the flavor physics within the model. First, to achieve the spontaneous symmetry breaking of the 331 gauge symmetry to the $SU(3)_C \times U(1)_{em}$ symmetry, at least three scalar triplets of $SU(3)_L$ must be introduced. While these are sufficient to impart masses to quarks in the quark sector, the flavor structure becomes intricate due to the differing group representations of the three quark families.

Conversely, in the lepton sector, a fundamentally different situation arises. Realistic masses cannot be obtained using only three triplets. As previously argued, the introduction of a scalar sextet belonging to $(\mathbf{1}, \mathbf{6}, 0)$ becomes necessary to generate appropriate masses for charged leptons.

A general feature of models of this kind, where mass terms arise from different scalar fields, is the introduction of flavor-changing neutral currents mediated by neutral scalars [23].

### 7.1. Quark Sector

Let us revisit the presence of three scalar triplets within the model. The first triplet,

$$\rho = \begin{pmatrix} \rho^{++} \\ \rho^+ \\ \rho^0 \end{pmatrix} \in (\mathbf{1}, \mathbf{3}, 1),$$ (94)

acquires a vacuum expectation value (vev) on the order of the spontaneous symmetry breaking of the 331 symmetry. The other two triplets,

$$\eta = \begin{pmatrix} \eta_1^+ \\ \eta^0 \\ \eta_2^- \end{pmatrix} \in (\mathbf{1}, \mathbf{3}, 0),$$ (95)

$$\chi = \begin{pmatrix} \chi^0 \\ \chi^- \\ \chi^{--} \end{pmatrix} \in (\mathbf{1}, \mathbf{3}, -1),$$ (96)

acquire non-zero vevs at the electroweak scale. In the quark sector, the Yukawa interactions are described by

$$\mathcal{L}_{q,\,triplet}^Y = -\overline{Q}_m \left( Y_{ma}^d \eta^* d_{aR} + Y_{ma}^u \chi^* u_{ma} \right) - \overline{Q}_{3a}(Y_{3a}^d \chi d_{aR} + Y_{3a}^u \eta u_{ma}) +$$
$$- \overline{Q}_m (Y_{mn}^J \chi J_{nR}) - \overline{Q}_3 Y_3^J \chi J_{3R} + \text{h.c.}.$$ (97)

The mass matrices for the up-type and down-type quarks arise when all scalar triplets acquire real vevs: $\rho \to v_\rho/\sqrt{2}$, $\eta \to v_\eta/\sqrt{2}$, and $\chi \to v_\chi/\sqrt{2}$. Both matrices involve contributions from the triplets $\eta$ and $\chi$. Exotic quark mass terms emerge following the initial spontaneous symmetry breaking.

Since there are two quarks, namely $D$ and $S$ with electric charge $Q = -4/3\ e$, they undergo mixing via a Cabibbo-like $2 \times 2$ matrix

$$\begin{pmatrix} D'_R \\ S'_R \end{pmatrix} = \tilde{V}_R^{-1} \begin{pmatrix} D_R \\ S_R \end{pmatrix}, \qquad \begin{pmatrix} D'_L \\ S'_L \end{pmatrix} = \tilde{V}_L^{-1} \begin{pmatrix} D_L \\ S_L \end{pmatrix}. \tag{98}$$

Notably, only the quark $T$ remains unmixed, being the sole quark with $Q = 5/3\ e$. The resulting mass matrices for ordinary quarks from the Lagrangian in Equation (97) are as follows:

$$M_u = \begin{pmatrix} v_\chi Y_{11}^u & v_\chi Y_{12}^u & v_\chi Y_{13}^u \\ v_\chi Y_{21}^u & v_\chi Y_{22}^u & v_\chi Y_{23}^u \\ v_\eta Y_{31}^u & v_\eta Y_{32}^u & v_\eta Y_{33}^u \end{pmatrix}, \tag{99}$$

$$M_d = \begin{pmatrix} v_\eta Y_{11}^d & v_\eta Y_{12}^d & v_\eta Y_{13}^d \\ v_\eta Y_{21}^d & v_\eta Y_{22}^d & v_\eta Y_{23}^d \\ v_\chi Y_{31}^d & v_\chi Y_{32}^d & v_\chi Y_{33}^d \end{pmatrix}. \tag{100}$$

In the minimal version, there is typically no reason to initially place one type of quark in the diagonal basis, unless some specific additional symmetry is introduced. Therefore, similar to the Standard Model, we proceed by independently rotating down-type and up-type quarks into their mass eigenstates

$$\begin{pmatrix} d'_L \\ s'_L \\ b'_L \end{pmatrix} = V_L^{-1} \begin{pmatrix} d_L \\ s_L \\ b_L \end{pmatrix} \qquad \begin{pmatrix} d'_R \\ s'_R \\ b'_R \end{pmatrix} = V_R^{-1} \begin{pmatrix} d_R \\ s_R \\ b_R \end{pmatrix} \tag{101}$$

$$\begin{pmatrix} u'_L \\ c'_L \\ t'_L \end{pmatrix} = U_L^{-1} \begin{pmatrix} u_L \\ c_L \\ t_L \end{pmatrix} \qquad \begin{pmatrix} u'_R \\ c'_R \\ t'_R \end{pmatrix} = U_R^{-1} \begin{pmatrix} u_R \\ c_R \\ t_R \end{pmatrix} \tag{102}$$

Here, the primed fields represent flavor eigenstates, while unprimed fields denote mass eigenstates. Matrices $U_L$, $U_R$, $V_L$, and $V_R$ are unitary matrices satisfying

$$V_L^\dagger V_L = V_R^\dagger V_R = U_L^\dagger U_L = U_R^\dagger U_R = \mathbb{1}, \tag{103}$$

which diagonalize $M_u$ and $M_d$ respectively via bi-unitary transformations

$$U_L^\dagger M_u U_R = \hat{M}_u \qquad V_L^\dagger M_d V_R = \hat{M}_d, \tag{104}$$

where $\hat{M}_u$ and $\hat{M}_d$ are diagonal $3 \times 3$ matrices containing quark masses.

In the Standard Model, all three families are placed within the same group representation of the gauge symmetry, and this results in no family distinction in the Lagrangian apart from the Yukawa sector. Here, non-diagonal terms between quark masses arise if the couplings are non-diagonal. The same field rotation procedure is implemented in the Standard Model to diagonalize Yukawa interactions, while leaving the Lagrangian mostly unchanged, and the only part sensitive to flavor rotation is the $W$ boson interactions, where the CKM matrix arises as

$$V_{CKM} = U_L^\dagger V_L. \tag{105}$$

Every other sector in the Lagrangian remains unaffected. Consequently, the GIM mechanism is naturally implemented, since there are no flavor-changing neutral currents (FCNCs) at tree-level, resulting in a natural suppression of these processes, which can only occur at loop-level. Up-type quark masses can be assumed as diagonal from the outset, while down-type quark flavor and mass eigenstates can be related through the CKM matrices, as any other rotation effects cancel out in the Lagrangian.

However, in the context of the minimal 331 model, the situation is fundamentally different. As previously emphasized, the appealing feature of constraining the number of families is that it causes every sector of the Lagrangian to be sensitive to flavor, owing to the different group representations to which the quarks are assigned. Consequently, flavor rotation matrices persist in various combinations in fermion interactions, either with scalars or gauge bosons. From the Yukawa Lagrangian in Equation (97), we can derive the fermion-scalar interactions after the field rotations discussed previously

$$\mathcal{L}_{scalar}^{cc} = -\overline{\mathbf{d}}_L V_L S_{du}^{cc} U_R^\dagger \mathbf{u}_R - \overline{\mathbf{u}}_L U_L S_{ud}^{cc} V_R^\dagger \mathbf{d}_R + \text{h.c.,} \tag{106}$$

where $\mathbf{d}$ and $\mathbf{u}$ represent the down-type and up-type flavor vectors in the mass basis, and

$$S_{du}^{cc} = \begin{pmatrix} \chi^- Y_{11}^u & \chi^- Y_{12}^u & \chi^- Y_{13}^u \\ \chi^- Y_{21}^u & \chi^- Y_{22}^u & \chi^- Y_{23}^u \\ \eta_1^- Y_{31}^u & \eta_1^- Y_{32}^u & \eta_1^- Y_{33}^u \end{pmatrix}, \tag{107}$$

$$S_{ud}^{cc} = \begin{pmatrix} \eta_1^+ Y_{11}^d & \eta_1^+ Y_{12}^d & \eta_1^+ Y_{13}^d \\ \eta_1^+ Y_{21}^d & \eta_1^+ Y_{22}^d & \eta_1^+ Y_{23}^d \\ \chi^+ Y_{31}^d & \chi^+ Y_{32}^d & \chi^+ Y_{33}^d \end{pmatrix}. \tag{108}$$

In the expression above, all the charged scalars are interaction eigenstates, and we need to perform a rotation in (111) and (110) in order to extract the corresponding mass eigenstates. It is evident that, in general, the combination of flavor rotation matrices cannot be further reduced, and they persist in a combination distinct from the CKM matrix. This behavior also manifests in the neutral currents mediated by scalar fields, where the Lagrangian becomes

$$\mathcal{L}_{scalar}^{nc} = -\overline{\mathbf{u}}_L U_L S_u^{nc} U_R^\dagger \mathbf{u}_R - \overline{\mathbf{d}}_L V_L S_d^{nc} V_R^\dagger \mathbf{d}_R + \text{h.c.,} \tag{109}$$

with

$$S_u^{nc} = \begin{pmatrix} \chi^0 Y_{11}^u & \chi^0 Y_{12}^u & \chi^0 Y_{13}^u \\ \chi^0 Y_{21}^u & \chi^0 Y_{22}^u & \chi^0 Y_{23}^u \\ \eta^0 Y_{31}^u & \eta^0 Y_{32}^u & \eta^0 Y_{33}^u \end{pmatrix} \tag{110}$$

and

$$S_d^{nc} = \begin{pmatrix} \eta^0 Y_{11}^d & \eta^0 Y_{12}^d & \eta^0 Y_{13}^d \\ \eta^0 Y_{21}^d & \eta^0 Y_{22}^d & \eta^0 Y_{23}^d \\ \chi^0 Y_{31}^d & \chi^0 Y_{32}^d & \chi^0 Y_{33}^d \end{pmatrix} \tag{111}$$

where, as in the case of charged scalars, here we also need a further rotation in order to extract the mass eigenstates of this sector.

Although the matrix combinations $U_L S_u^{nc} U_R^\dagger$ and $V_L S_d^{nc} V_R^\dagger$ resemble those in Equations (101) and (102), they are insufficient to diagonalize the interaction. Hence, flavor-changing neutral currents in the scalar sector are a prediction of the minimal 331 model. They can be controlled, but not entirely avoided.

In the minimal 331 model, eight bosons are present, including the neutral ones: the photon $A^\mu$, with straightforward interactions, $Z^\mu$ and $Z'^\mu$, and the charged ones, $W^\pm$, $V^\pm$, and the doubly charged bileptons $Y^{\pm\pm}$. However, rotations in flavor space do not leave the interactions unaffected. The d-type and u-type quarks interact with the $W^\pm$ boson as in the Standard Model

$$\mathcal{L}_{duW} = \frac{g_2}{\sqrt{2}} \overline{\mathbf{d}}_L \gamma^\mu V_{CKM} \mathbf{u}_L W_\mu^- + \text{h.c.} \tag{112}$$

Here, the CKM matrix appears as expected. Further interactions are possible between exotic quarks and ordinary quarks mediated by exotic gauge bosons. The interactions of u-type and d-type quarks with the exotic $T$ quark are given by

$$\mathcal{L}_{uTV} = \frac{g_2}{\sqrt{2}} \overline{T}_L \gamma^\mu (U_L)_{3j} \mathbf{u}_{jL} V_\mu^+ + \text{h.c.} \tag{113}$$

$$\mathcal{L}_{dTY} = \frac{g_2}{\sqrt{2}}\overline{T_L}\gamma^\mu(V_L)_{3j}\mathbf{d}_{jL}Y_\mu^{++} + \text{h.c.} \tag{114}$$

Here, only the third column of the rotation matrix appears due to the difference between the third quark generation and the other two. Similar interactions occur between up-type and down-type quarks with the other two exotic quarks $D$ and $S$

$$\mathcal{L}_{djY} = \frac{g_2}{\sqrt{2}}\overline{j_L}\gamma^\mu(O^\dagger V_L)_{mj}\mathbf{d}_{jL}V_\mu^- + \text{h.c.} \quad \text{with } j_L = (D_L,\, S_L) \tag{115}$$

$$\mathcal{L}_{ujY} = \frac{g_2}{\sqrt{2}}\overline{j_L}\gamma^\mu(O^\dagger U_L)_{mj}\mathbf{d}_{jL}Y_\mu^{++} + \text{h.c.} \quad \text{with } j_L = (D_L,\, S_L) \tag{116}$$

where the $O$ matrix in Equations (115) and (116) is a $2 \times 2$ Cabibbo-like matrix, which mixes the exotic quarks $D$ and $S$.

Finally, the interaction of ordinary quarks with the $Z'$ boson can be schematized as follows—omitting an overall coefficient-

$$\mathcal{L}_{qqZ'} = \left( \overline{\mathbf{u}_L}U_L^\dagger\gamma^\mu Y_L^u U_L\mathbf{u}_L + \overline{\mathbf{d}_L}V_L^\dagger\gamma^\mu Y_L^d V_L\mathbf{d}_L \right.$$
$$\left. + \overline{\mathbf{u}_R}U_R^\dagger\gamma^\mu Y_R^u U_R\mathbf{u}_R + \overline{\mathbf{d}_R}V_R^\dagger\gamma^\mu Y_R^d V_R\mathbf{d}_R \right) Z'_\mu \tag{117}$$

where we have defined the couplings with $Y_L^u$ and $Y_L^d$, which are proportional to the following matrix

$$Y_L^u = Y_L^d \propto \begin{pmatrix} 1 - 2\sin^2(\theta_W) & 0 & 0 \\ 0 & 1 - 2\sin^2(\theta_W) & 0 \\ 0 & 0 & -1 \end{pmatrix} \tag{118}$$

In this interaction, there are flavor-changing neutral currents (FCNCs) in the left-handed interactions, but the right-handed neutral currents mediated by the gauge boson $Z'$ are diagonal in flavor space, as $Y_R^u \propto \mathbb{1}$ and $Y_R^d \propto \mathbb{1}$ [24].

It is noteworthy that in the quark sector, the rotation matrices of the right-handed quarks cancel out from the Lagrangian, similarly to the case of the Standard Model. Conversely, the left-handed $U_L$, $V_L$ matrices not only survive in a combination analogous to the CKM matrix of the Standard Model, but also independently. From a practical standpoint, it is possible to redefine the fields in the interactions—using the unitary condition $U_L^\dagger U_L = \mathbb{1}$—to construct a Lagrangian for the quark sector in which only two matrices appear: the CKM matrix and $V_L$.

To obtain an appropriate parameterization for the matrix $V_L$, it is necessary to initially enumerate the additional parameters present within this matrix. Upon examining all conceivable interaction terms, it becomes apparent that, subsequently to employing phase transformations of the up and down-type quarks to simplify the CKM matrix, three more potential phases emerge from transformations in the $D$, $S$, and $T$ quarks. This results in a total of six supplementary parameters, encompassing three mixing angles and three phases. However, it is evident that only the $\tilde{V}_{3j}$ elements are essential when computing FCNCs, thus allowing for a parameterization that effectively diminishes the number of parameters involved. We obtain [25]

$$V_L = \begin{pmatrix} c_{12}c_{13} & s_{12}c_{23}e^{i\delta_3} - c_{12}s_{13}s_{23}e^{i(\delta_1-\delta_2)} & c_{12}c_{23}s_{13}e^{i\delta_1} + s_{12}s_{23}e^{i(\delta_2+\delta_3)} \\ -c_{13}s_{12}e^{-i\delta_3} & c_{12}c_{23} + s_{12}s_{13}s_{23}e^{i(\delta_1-\delta_2-\delta_3)} & -s_{12}c_{23}s_{13}e^{i(\delta_1-\delta_3)} - c_{12}s_{23}e^{i\delta_2} \\ -s_{13}e^{-i\delta_1} & -c_{13}s_{23}e^{-i\delta_2} & c_{13}c_{23} \end{pmatrix}. \tag{119}$$

where only two additional CP violating quantities $\delta_1$ and $\delta_2$ appear, which are responsible for the additional CP violating effects.

### 7.2. Lepton Sector

In the minimal 331 model, similarly to quarks, the leptonic sector exhibits a multitude of parameters. Despite the $Z$ and $Z'$ interactions being diagonal in flavor space (owing to the consistent transformation behavior of the three lepton generations under the electroweak symmetries $SU(3)_L \times U(1)_X$), flavor-changing neutral currents (FCNCs) occur in the scalar sector [26]. These parameters result from the diverse contributions to the mass matrices of charged leptons, mirroring the scenario observed in quarks.

The Yukawa interactions in the lepton sector must incorporate the triplet $\eta$ and the sextet, whose components are detailed as follows:

$$\sigma = \begin{pmatrix} \sigma_1^{++} & \frac{\sigma_1^+}{\sqrt{2}} & \frac{\sigma_1^0}{\sqrt{2}} \\ \frac{\sigma_1^+}{\sqrt{2}} & \sigma_2^0 & \frac{\sigma_2^-}{\sqrt{2}} \\ \frac{\sigma_1^0}{\sqrt{2}} & \frac{\sigma_2^-}{\sqrt{2}} & \sigma_2^{--} \end{pmatrix}. \tag{120}$$

In the minimal 331 model, including the sextet in the scalar sector is necessary to assign physical masses to charged leptons. This necessity arises from how the Yukawa interaction is constructed from group theory. When combining three triplets according to $\mathbf{3} \otimes \mathbf{3} \otimes \mathbf{3}$, the resulting invariant structure demands antisymmetry among the triplets. As a consequence, the Yukawa matrix must exhibit antisymmetry in flavor indices to allow for vanishing interactions. However, this interaction pattern leads to eigenvalues of $(0, m, -m)$ of this matrix, which is evidently an unphysical solution.

Once spontaneous symmetry breaking occurs, the scalars acquire vacuum expectation values as follows:

$$\eta = \frac{1}{\sqrt{2}} \begin{pmatrix} \frac{1}{\sqrt{2}} v_\eta + \frac{1}{\sqrt{2}} \Sigma_\eta + \frac{i}{\sqrt{2}} \zeta_\eta \\ \eta_1^+ \\ \eta_2^- \end{pmatrix} \tag{121}$$

$$\sigma = \frac{1}{\sqrt{2}} \begin{pmatrix} \sigma_1^{++} & \frac{\sigma_1^+}{\sqrt{2}} & \frac{v_{\sigma} + \Sigma_\sigma + i\xi_\sigma}{2} \\ \frac{\sigma_1^+}{\sqrt{2}} & \sigma_2^0 & \frac{\sigma_2^-}{\sqrt{2}} \\ \frac{v_{\sigma} + \Sigma_\sigma + i\xi_\sigma}{2} & \frac{\sigma_2^-}{\sqrt{2}} & \sigma_2^{--} \end{pmatrix}. \tag{122}$$

In the context of the minimal 331 model, as originally proposed, neutrinos are massless at the tree level. However, a mechanism to generate massive neutrinos can be obtained through the scalar sextet. Indeed, the component $\sigma_2^0$ can also acquire a vev, which can be used in the Yukawa interactions to construct a Majorana mass term for the neutrinos

$$\mathcal{L}_\nu^Y = -\overline{\nu_L^c} G^\sigma \nu_L \sigma + \text{h.c.} \,. \tag{123}$$

This interaction is not invariant under flavor rotation, therefore an additional rotation matrix appears in this sector, in order to diagonalize the neutrino mass terms

$$V_\nu^T M_\nu V_\nu = \hat{M}_\nu, \tag{124}$$

where $\hat{M}_\nu$ is a diagonal matrix containing physical neutrino masses.

The inclusion of this additional vev carries further implications, such as the mixing between the singly charged gauge bosons $W^\pm$ and $V^\pm$, which can be predicted to be a small value of $v_{\sigma_2}$ chosen to yield small neutrino masses. The interactions between charged leptons, neutrinos, and $W^\pm$ remain the same as in the Standard Model

$$\mathcal{L}_{l\nu W} = \frac{ig_2}{2\sqrt{2}} \left( \overline{\nu_L} V_{PMNS}^\dagger \gamma^\mu l_L + \overline{(l^c)_R} V_{PMNS} \gamma^\mu (\nu^c)_R \right) W_\mu^+ + \text{h.c.} \tag{125}$$

where $V_{PMNS}$ is the Pontecorvo–Maki–Nakagawa–Sakata matrix given by

$$V_{PMNS} = V_l^\dagger V^\nu. \tag{126}$$

The model also predicts that the interaction between charged leptons and neutrinos must include additional contributions from $V^\pm$

$$\mathcal{L}_{l\nu V} = \frac{ig_2}{2\sqrt{2}} \left( \overline{l_L^c} V_{l\nu}^* \gamma^\mu \nu_L + \overline{\nu_R^c} V_{l\nu} \gamma^\mu (l)_R \right) V_\mu^+ + \text{h.c.} \tag{127}$$

where the definition

$$V_{l\nu} = (V_L^\nu)^\dagger (V_R^l)^*. \tag{128}$$

In this model, there are also interactions between charged leptons and doubly charged vector bosons given by the Lagrangian

$$\mathcal{L}_{llY} = \frac{ig_2}{2\sqrt{2}} \left( \overline{l^c} \gamma^\mu \left( \tilde{V}_{l\nu} - \tilde{V}_{l\nu}^T \right) - \gamma^5 \gamma^\mu \left( \tilde{V}_{l\nu} + \tilde{V}_{l\nu}^T \right) \right) Y^{++} \tag{129}$$

with

$$\tilde{V}_{l\nu} = (V_R^l)^T V_L^l. \tag{130}$$

Finally, leptons couple universally to neutral vector bosons, since no distinction has been made between generations of leptons, where the $Z'$ boson also has the property of being Leptophobic [27].

Therefore, in the minimal 331 model, $V_\nu$ is always not equal to $V_{PMNS}$, and $V_L^l$ and $V_R^l$ both appear separately in the combination shown in Equations (128) and (130). Because the charged lepton mass matrix includes an anti-symmetric contribution, we cannot assume it is diagonal from the beginning. This means we cannot simply set $V_R^l$ as equal to $\mathbb{1}$ in the interactions in Equation (127). However, if interactions between leptons and the scalar triplet $\eta$ are forbidden by some discrete symmetry, then this simplification becomes possible.

In this scenario, the charged lepton mass matrix is diagonalized using the same unitary matrix as the neutrino mass matrix, making the PMNS matrix straightforward. Similarly, just like in the quark sector, determining the values of the entries in the matrices $V_\nu$, $V_L^l$, and $V_R^l$ is crucial for understanding the realistic behavior of this model.

## 8. Conclusions

The 331 model stands as a remarkable embodiment of the anomaly cancellation mechanism, encompassing all fermion generations. It represents a compelling avenue for extending the fermion sector of the Standard Model in an alternative direction. Within the framework of the 331 model, the significance of anomaly constraints cannot be overstated. They serve as pivotal guiding principles, steering us away from the conventional sequential structure of the Standard Model's fermion families. Unlike the traditional model, which lacks a clear rationale regarding the number of fermion families, the 331 model offers a fresh perspective, inviting exploration and deeper understanding of fundamental particle interactions.

**Author Contributions:** Writing—original draft, C.C. and D.M. All authors have read and agreed to the published version of the manuscript.

**Funding:** This work was partially supported by INFN within the Iniziativa Specifica QG-sky. The work of C.C. and S.L. was funded by the European Union, Next Generation EU, PNRR project "National Centre for HPC, Big Data and Quantum Computing", project code CN00000013. This work was partially supported by the the grant PRIN 2022BP52A MUR "The Holographic Universe for all Lambdas" Lecce-Naples.

**Acknowledgments:** This work is dedicated to Paul Frampton, whose exceptional and insightful contributions to physics have greatly enriched our understanding over the years. We extend our heartfelt gratitude to him for the invaluable knowledge shared during our time together at the University of Salento, in Lecce and Martignano, and in numerous gatherings around the globe. We fondly remember the enriching discussions and memorable moments. Wishing him many more years filled with joy, success, and continued inspiration. We thank Gennaro Corcella and Antonio Costantini for long-term collaborations and for sharing their insight on the analysis of the 331 model.

## Appendix A. Rotations for the Determination of the Mass Eigenstates

In this appendix, we summarize some results concerning the scalar/psesudoscalar sectors of the model [28,29].

- **Mass matrix of the scalar sector**, Basis: $\left( \rho_1, \rho_2, \rho_3, \rho_4, \sigma_2^{\,0} \right), \left( \rho_1, \rho_2, \rho_3, \rho_4, \sigma_2^{\,0,*} \right)$

$$m_h^2 = \begin{pmatrix} m_{\rho_1 \rho_1} & m_{\rho_2 \rho_1} & m_{\rho_3 \rho_1} & m_{\rho_4 \rho_1} & 0 \\ m_{\rho_1 \rho_2} & m_{\rho_2 \rho_2} & m_{\rho_3 \rho_2} & m_{\rho_4 \rho_2} & 0 \\ m_{\rho_1 \rho_3} & m_{\rho_2 \rho_3} & m_{\rho_3 \rho_3} & m_{\rho_4 \rho_3} & 0 \\ m_{\rho_1 \rho_4} & m_{\rho_2 \rho_4} & m_{\rho_3 \rho_4} & m_{\rho_4 \rho_4} & 0 \\ 0 & 0 & 0 & 0 & m_{\sigma_2^0 \sigma_2^{0,*}} \end{pmatrix} \tag{A1}$$

$$m_{\rho_1 \rho_1} = \frac{1}{4} \left( 12 \lambda_1 v_\rho^2 + 2 \left( \lambda_{12} v_\eta^2 + \lambda_{13} v_\chi^2 \right) - 2 \sqrt{2} v_\eta v_\sigma \xi_{14} + v_\sigma^2 \left( 2 \lambda_{14} + \zeta_{14} \right) \right) + \mu_1 \tag{A2}$$

$$m_{\rho_1 \rho_2} = f v_\chi + v_\rho \left( -\frac{1}{\sqrt{2}} v_\sigma \xi_{14} + \lambda_{12} v_\eta \right) \tag{A3}$$

$$m_{\rho_2 \rho_2} = \frac{1}{2} \left( 6 \lambda_2 v_\eta^2 + \lambda_{12} v_\rho^2 + \lambda_{23} v_\chi^2 + v_\sigma^2 \left( -2 \xi_{24} + \lambda_{24} \right) \right) + \mu_2 \tag{A4}$$

$$m_{\rho_1 \rho_3} = \frac{1}{\sqrt{2}} f_\sigma v_\sigma + f v_\eta + \lambda_{13} v_\rho v_\chi \tag{A5}$$

$$m_{\rho_2 \rho_3} = f v_\rho + v_\chi \left( \frac{1}{\sqrt{2}} v_\sigma \xi_{34} + \lambda_{23} v_\eta \right) \tag{A6}$$

$$m_{\rho_3 \rho_3} = \frac{1}{4} \left( 12 \lambda_3 v_\chi^2 + 2 \left( \lambda_{13} v_\rho^2 + \lambda_{23} v_\eta^2 + \sqrt{2} v_\eta v_\sigma \xi_{34} \right) + v_\sigma^2 \left( 2 \lambda_{34} + \zeta_{34} \right) \right) + \mu_3 \tag{A7}$$

$$m_{\rho_1 \rho_4} = \frac{1}{2} \left( \sqrt{2} f_\sigma v_\chi + v_\rho \left( -\sqrt{2} v_\eta \xi_{14} + v_\sigma \left( 2 \lambda_{14} + \zeta_{14} \right) \right) \right) \tag{A8}$$

$$m_{\rho_2 \rho_4} = \frac{1}{2} \frac{1}{\sqrt{2}} \left( v_\chi^2 \xi_{34} - v_\rho^2 \xi_{14} \right) + v_\eta v_\sigma \left( -2 \xi_{24} + \lambda_{24} \right) \tag{A9}$$

$$m_{\rho_3 \rho_4} = \frac{1}{2} \left( \sqrt{2} f_\sigma v_\rho + v_\chi \left( \sqrt{2} v_\eta \xi_{34} + v_\sigma \left( 2 \lambda_{34} + \zeta_{34} \right) \right) \right) \tag{A10}$$

$$m_{\rho_4 \rho_4} = \frac{1}{4} \left( 2 v_\eta^2 \left( -2 \xi_{24} + \lambda_{24} \right) + 6 \left( 2 \lambda_4 + \lambda_{44} \right) v_\sigma^2 + v_\chi^2 \left( 2 \lambda_{34} + \zeta_{34} \right) \right. \tag{A11}$$
$$\left. + v_\rho^2 \left( 2 \lambda_{14} + \zeta_{14} \right) \right) + \mu_4$$

$$m_{\sigma_2^0 \sigma_2^{0,*}} = \frac{1}{2} \left( 2 \lambda_4 v_\sigma^2 + \lambda_{14} v_\rho^2 + \lambda_{34} v_\chi^2 + v_\eta^2 \left( \lambda_{24} + \zeta_{24} \right) \right) + \mu_4 \tag{A12}$$

This matrix is diagonalized by $R^S$

$$R^S m_h^2 R^{S,\dagger} = m_{2,h}^{dia} \tag{A13}$$

with

$$\rho_1 = \sum_j R_{j1}^{S,*} h_j, \qquad \rho_2 = \sum_j R_{j2}^{S,*} h_j, \qquad \rho_3 = \sum_j R_{j3}^{S,*} h_j \tag{A14}$$

$$\rho_4 = \sum_j R_{j4}^{S,*} h_j, \qquad \sigma_2^{\,0} = \sum_j R_{j5}^{S,*} h_j \tag{A15}$$

- **Mass matrix of the pseudoscalar sector**, Basis: $\left( \sigma_1, \sigma_2, \sigma_3, \sigma_4 \right), \left( \sigma_1, \sigma_2, \sigma_3, \sigma_4 \right)$

$$m_{A^0}^2 = \begin{pmatrix} m_{\sigma_1 \sigma_1} & -f v_\chi & -\frac{1}{\sqrt{2}} f_\sigma v_\sigma - f v_\eta & -\frac{1}{\sqrt{2}} f_\sigma v_\chi \\ -f v_\chi & m_{\sigma_2 \sigma_2} & -f v_\rho & m_{\sigma_4 \sigma_2} \\ -\frac{1}{\sqrt{2}} f_\sigma v_\sigma - f v_\eta & -f v_\rho & m_{\sigma_3 \sigma_3} & -\frac{1}{\sqrt{2}} f_\sigma v_\rho \\ -\frac{1}{\sqrt{2}} f_\sigma v_\chi & m_{\sigma_2 \sigma_4} & -\frac{1}{\sqrt{2}} f_\sigma v_\rho & m_{\sigma_4 \sigma_4} \end{pmatrix} + \xi_Z m^2(Z) + \xi_{Z'} m^2(Z') \tag{A16}$$

$$m_{\sigma_1\sigma_1} = \frac{1}{4}\left(2\left(\lambda_{12}v_\eta^2 + \lambda_{13}v_\chi^2\right) - 2\sqrt{2}v_\eta v_\sigma \xi_{14} + 4\lambda_1 v_\rho^2 + v_\sigma^2\left(2\lambda_{14} + \zeta_{14}\right)\right) + \mu_1 \tag{A17}$$

$$m_{\sigma_2\sigma_2} = \frac{1}{2}\left(2\lambda_2 v_\rho^2 + \lambda_{12}v_\rho^2 + \lambda_{23}v_\chi^2 + v_\sigma^2\left(2\xi_{24} + \lambda_{24}\right)\right) + \mu_2 \tag{A18}$$

$$m_{\sigma_3\sigma_3} = \frac{1}{4}\left(2\left(\lambda_{13}v_\rho^2 + \lambda_{23}v_\eta^2 + \sqrt{2}v_\eta v_\sigma \xi_{34}\right) + 4\lambda_3 v_\chi^2 + v_\sigma^2\left(2\lambda_{34} + \zeta_{34}\right)\right) + \mu_3 \tag{A19}$$

$$m_{\sigma_2\sigma_4} = \frac{1}{4}\left(-8v_\eta v_\sigma \xi_{24} + \sqrt{2}\left(v_\chi^2 \xi_{34} - v_\rho^2 \xi_{14}\right)\right) \tag{A20}$$

$$m_{\sigma_4\sigma_4} = \frac{1}{4}\Big(2\big(\left(2\lambda_4 + \lambda_{44}\right)v_\sigma^2 + v_\eta^2\left(2\xi_{24} + \lambda_{24}\right)\big) + v_\chi^2\left(2\lambda_{34} + \zeta_{34}\right)$$
$$+ v_\rho^2\left(2\lambda_{14} + \zeta_{14}\right)\Big) + \mu_4 \tag{A21}$$

The gauge fixing contributions are

$$m^2(\xi_Z) = \begin{pmatrix} m_{\sigma_1\sigma_1} & m_{\sigma_2\sigma_1} & m_{\sigma_3\sigma_1} & m_{\sigma_4\sigma_1} \\ m_{\sigma_1\sigma_2} & m_{\sigma_2\sigma_2} & m_{\sigma_3\sigma_2} & m_{\sigma_4\sigma_2} \\ m_{\sigma_1\sigma_3} & m_{\sigma_2\sigma_3} & m_{\sigma_3\sigma_3} & m_{\sigma_4\sigma_3} \\ m_{\sigma_1\sigma_4} & m_{\sigma_2\sigma_4} & m_{\sigma_3\sigma_4} & m_{\sigma_4\sigma_4} \end{pmatrix} \tag{A22}$$

$$m_{\sigma_1\sigma_1} = \frac{1}{3}v_\rho^2\left(-2\sqrt{3}g_1 g_2 R_{12}^Z R_{22}^Z + 3g_1^2 R_{12}^{Z\,2} + g_2^2 R_{22}^{Z\,2}\right) \tag{A23}$$

$$m_{\sigma_1\sigma_2} = \frac{1}{6}g_2 v_\rho v_\eta\left(g_1 R_{12}^Z\left(-3R_{32}^Z + \sqrt{3}R_{22}^Z\right) + g_2 R_{22}^Z\left(-R_{22}^Z + \sqrt{3}R_{32}^Z\right)\right) \tag{A24}$$

$$m_{\sigma_2\sigma_2} = \frac{1}{12}g_2^2 v_\eta^2\left(-2\sqrt{3}R_{22}^Z R_{32}^Z + 3R_{32}^{Z\,2} + R_{22}^{Z\,2}\right) \tag{A25}$$

$$m_{\sigma_1\sigma_3} = -\frac{1}{6}v_\rho v_\chi\left(-3g_1 g_2 R_{12}^Z\left(\sqrt{3}R_{22}^Z + R_{32}^Z\right) + 6g_1^2 R_{12}^{Z\,2} + g_2^2 R_{22}^Z\left(\sqrt{3}R_{32}^Z + R_{22}^Z\right)\right) \tag{A26}$$

$$m_{\sigma_2\sigma_3} = \frac{1}{12}g_2 v_\eta v_\chi\left(-2g_1 R_{12}^Z\left(-3R_{32}^Z + \sqrt{3}R_{22}^Z\right) + g_2\left(-3R_{32}^{Z\,2} + R_{22}^{Z\,2}\right)\right) \tag{A27}$$

$$m_{\sigma_3\sigma_3} = \frac{1}{12}v_\chi^2\Big(12g_1^2 R_{12}^{Z\,2} - 4g_1 g_2 R_{12}^Z\left(3R_{32}^Z + \sqrt{3}R_{22}^Z\right)$$
$$+ g_2^2\left(2\sqrt{3}R_{22}^Z R_{32}^Z + 3R_{32}^{Z\,2} + R_{22}^{Z\,2}\right)\Big) \tag{A28}$$

$$m_{\sigma_1\sigma_4} = \frac{1}{6}g_2 v_\rho v_\sigma\left(g_1 R_{12}^Z\left(-3R_{32}^Z + \sqrt{3}R_{22}^Z\right) + g_2 R_{22}^Z\left(-R_{22}^Z + \sqrt{3}R_{32}^Z\right)\right) \tag{A29}$$

$$m_{\sigma_2\sigma_4} = \frac{1}{12}g_2^2 v_\eta v_\sigma\left(-2\sqrt{3}R_{22}^Z R_{32}^Z + 3R_{32}^{Z\,2} + R_{22}^{Z\,2}\right) \tag{A30}$$

$$m_{\sigma_3\sigma_4} = \frac{1}{12}g_2 v_\chi v_\sigma\left(-2g_1 R_{12}^Z\left(-3R_{32}^Z + \sqrt{3}R_{22}^Z\right) + g_2\left(-3R_{32}^{Z\,2} + R_{22}^{Z\,2}\right)\right) \tag{A31}$$

$$m_{\sigma_4\sigma_4} = \frac{1}{12}g_2^2 v_\sigma^2\left(-2\sqrt{3}R_{22}^Z R_{32}^Z + 3R_{32}^{Z\,2} + R_{22}^{Z\,2}\right) \tag{A32}$$

$$m^2(\xi_{Z'}) = \begin{pmatrix} m_{\sigma_1\sigma_1} & m_{\sigma_2\sigma_1} & m_{\sigma_3\sigma_1} & m_{\sigma_4\sigma_1} \\ m_{\sigma_1\sigma_2} & m_{\sigma_2\sigma_2} & m_{\sigma_3\sigma_2} & m_{\sigma_4\sigma_2} \\ m_{\sigma_1\sigma_3} & m_{\sigma_2\sigma_3} & m_{\sigma_3\sigma_3} & m_{\sigma_4\sigma_3} \\ m_{\sigma_1\sigma_4} & m_{\sigma_2\sigma_4} & m_{\sigma_3\sigma_4} & m_{\sigma_4\sigma_4} \end{pmatrix} \tag{A33}$$

$$m_{\sigma_1\sigma_1} = \frac{1}{3}v_\rho^2\left(-2\sqrt{3}g_1g_2R_{13}^Z R_{23}^Z + 3g_1^2 R_{13}^{Z,2} + g_2^2 R_{23}^{Z,2}\right) \tag{A34}$$

$$m_{\sigma_1\sigma_2} = \frac{1}{6}g_2 v_\rho v_\eta \left(g_1 R_{13}^Z\left(-3R_{33}^Z + \sqrt{3}R_{23}^Z\right) + g_2 R_{23}^Z\left(-R_{23}^Z + \sqrt{3}R_{33}^Z\right)\right) \tag{A35}$$

$$m_{\sigma_2\sigma_2} = \frac{1}{12}g_2^2 v_\eta^2\left(-2\sqrt{3}R_{23}^Z R_{33}^Z + 3R_{33}^{Z,2} + R_{23}^{Z,2}\right) \tag{A36}$$

$$m_{\sigma_1\sigma_3} = -\frac{1}{6}v_\rho v_\chi\left(-3g_1g_2R_{13}^Z\left(\sqrt{3}R_{23}^Z + R_{33}^Z\right) + 6g_1^2 R_{13}^{Z,2} + g_2^2 R_{23}^Z\left(\sqrt{3}R_{33}^Z + R_{23}^Z\right)\right) \tag{A37}$$

$$m_{\sigma_2\sigma_3} = \frac{1}{12}g_2 v_\eta v_\chi\left(-2g_1 R_{13}^Z\left(-3R_{33}^Z + \sqrt{3}R_{23}^Z\right) + g_2\left(-3R_{33}^{Z,2} + R_{23}^{Z,2}\right)\right) \tag{A38}$$

$$m_{\sigma_3\sigma_3} = \frac{1}{12}v_\chi^2\left(12g_1^2 R_{13}^{Z,2} - 4g_1g_2R_{13}^Z\left(3R_{33}^Z + \sqrt{3}R_{23}^Z\right)\right. \\ \left. + g_2^2\left(2\sqrt{3}R_{23}^Z R_{33}^Z + 3R_{33}^{Z,2} + R_{23}^{Z,2}\right)\right) \tag{A39}$$

$$m_{\sigma_1\sigma_4} = \frac{1}{6}g_2 v_\rho v_\sigma\left(g_1 R_{13}^Z\left(-3R_{33}^Z + \sqrt{3}R_{23}^Z\right) + g_2 R_{23}^Z\left(-R_{23}^Z + \sqrt{3}R_{33}^Z\right)\right) \tag{A40}$$

$$m_{\sigma_2\sigma_4} = \frac{1}{12}g_2^2 v_\eta v_\sigma\left(-2\sqrt{3}R_{23}^Z R_{33}^Z + 3R_{33}^{Z,2} + R_{23}^{Z,2}\right) \tag{A41}$$

$$m_{\sigma_3\sigma_4} = \frac{1}{12}g_2 v_\chi v_\sigma\left(-2g_1 R_{13}^Z\left(-3R_{33}^Z + \sqrt{3}R_{23}^Z\right) + g_2\left(-3R_{33}^{Z,2} + R_{23}^{Z,2}\right)\right) \tag{A42}$$

$$m_{\sigma_4\sigma_4} = \frac{1}{12}g_2^2 v_\sigma^2\left(-2\sqrt{3}R_{23}^Z R_{33}^Z + 3R_{33}^{Z,2} + R_{23}^{Z,2}\right) \tag{A43}$$

where $R^Z$ is the rotation matrix that diagonalize the mass of the neutral gauge boson components in the $\{W^3, W^8, X\}$ basis, which is given by the following matrix

$$\begin{pmatrix} g_1^2 v_\rho^2 + g_1^2 v_\chi^2 & -\frac{g_1g_2v_\rho^2}{\sqrt{3}} - \frac{g_1g_2v_\chi^2}{2\sqrt{3}} & -\frac{1}{2}g_1g_2v_\chi^2 \\ -\frac{g_1g_2v_\rho^2}{\sqrt{3}} - \frac{g_1g_2v_\chi^2}{2\sqrt{3}} & \frac{g_2^2v_\rho^2}{3} + \frac{g_2^2v_\eta^2}{12} + \frac{g_2^2v_\chi^2}{12} + \frac{g_2^2v_\sigma^2}{12} & -\frac{g_2^2v_\eta^2}{4\sqrt{3}} + \frac{g_2^2v_\chi^2}{4\sqrt{3}} - \frac{g_2^2v_\sigma^2}{4\sqrt{3}} \\ -\frac{1}{2}g_1g_2v_\chi^2 & -\frac{g_2^2v_\eta^2}{4\sqrt{3}} + \frac{g_2^2v_\chi^2}{4\sqrt{3}} - \frac{g_2^2v_\sigma^2}{4\sqrt{3}} & \frac{g_2^2v_\eta^2}{4} + \frac{g_2^2v_\chi^2}{4} + \frac{g_2^2v_\sigma^2}{4} \end{pmatrix} \tag{A44}$$

The matrix (A16) is diagonalized by the matrix $R^P$

$$R^P m_{A^0}^2 R^{P,\dagger} = m_{2,A^0}^{dia} \tag{A45}$$

with

$$\sigma_1 = \sum_j R_{j1}^{P,*} A_j^0, \qquad \sigma_2 = \sum_j R_{j2}^{P,*} A_j^0, \qquad \sigma_3 = \sum_j R_{j3}^{P,*} A_j^0 \tag{A46}$$

$$\sigma_4 = \sum_j R_{j4}^{P,*} A_j^0 \tag{A47}$$

- **Mass matrix for Charged Higgs**, Basis: $\left(\rho_+^*, \eta_1^{+,*}, \sigma_2^{P,*}, \eta_2^-, \chi_-, \sigma_1^-\right)$, $\left(\rho_+, \eta_1^+, \sigma_2^P, \eta_2^{-,*}, \chi_-^*, \sigma_1^{-,*}\right)$

$$m_{H^-}^2 = \begin{pmatrix} m_{\rho_+^*\rho_+} & 0 & 0 & m_{\eta_2^-\rho_+}^* & 0 & m_{\sigma_1^-\rho_+}^* \\ 0 & m_{\eta_1^{+,*}\eta_1^+} & m_{\sigma_2^{P,*}\eta_1^+}^* & 0 & m_{\chi_-\eta_1^+}^* & 0 \\ 0 & m_{\eta_1^{+,*}\sigma_2^P} & m_{\sigma_2^{P,*}\sigma_2^P} & 0 & m_{\chi_-\sigma_2^P}^* & 0 \\ m_{\rho_+^*\eta_2^{-,*}} & 0 & 0 & m_{\eta_2^-\eta_2^{-,*}} & 0 & m_{\sigma_1^-\eta_2^{-,*}}^* \\ 0 & m_{\eta_1^{+,*}\chi_-^*} & m_{\sigma_2^{P,*}\chi_-^*} & 0 & m_{\chi_-\chi_-^*} & 0 \\ m_{\rho_+^*\sigma_1^{-,*}} & 0 & 0 & m_{\eta_2^-\sigma_1^{-,*}} & 0 & m_{\sigma_1^-\sigma_1^{-,*}} \end{pmatrix} + \xi_{W^+}m^2(W^+) + \xi_{W'^+}m^2(W'^+) \tag{A48}$$

$$m_{\rho_+^* \rho_+} = \frac{1}{2}\left(2\lambda_1 v_\rho^2 + \left(\lambda_{12} + \zeta_{12}\right)v_\eta^2 + \lambda_{13}v_\chi^2 + \lambda_{14}v_\sigma^2\right) + \mu_1 \tag{A49}$$

$$m_{\eta_1^{+,*} \eta_1^+} = \frac{1}{4}\left(2\left(\lambda_{12}v_\rho^2 + \left(\lambda_{23} + \zeta_{23}\right)v_\chi^2\right) + 4\lambda_2 v_\eta^2 + v_\sigma^2\left(2\lambda_{24} + \zeta_{24}\right)\right) + \mu_2 \tag{A50}$$

$$m_{\eta_1^{+,*} \sigma_2^p} = \frac{1}{4}\left(\sqrt{2}v_\rho^2\xi_{14} + v_\eta v_\sigma\left(4\xi_{24} + \zeta_{24}\right)\right) \tag{A51}$$

$$m_{\sigma_2^{p,*} \sigma_2^p} = \frac{1}{4}\left(2\left(\left(2\lambda_4 + \lambda_{44}\right)v_\sigma^2 + \lambda_{34}v_\chi^2\right) + v_\eta^2\left(2\lambda_{24} + \zeta_{24}\right) + v_\rho^2\left(2\lambda_{14} + \zeta_{14}\right)\right) + \mu_4 \tag{A52}$$

$$m_{\rho_+^* \eta_2^{-,*}} = \frac{1}{4}\left(-4fv_\eta + v_\rho\left(2\zeta_{12}v_\eta + \sqrt{2}v_\sigma\xi_{14}\right)\right) \tag{A53}$$

$$m_{\eta_2^- \eta_2^{-,*}} = \frac{1}{4}\left(2\left(\left(\lambda_{12} + \zeta_{12}\right)v_\rho^2 + \lambda_{23}v_\chi^2\right) + 4\lambda_2 v_\eta^2 + v_\sigma^2\left(2\lambda_{24} + \zeta_{24}\right)\right) + \mu_2 \tag{A54}$$

$$m_{\eta_1^{+,*} \chi_-^*} = \frac{1}{4}\left(-4fv_\rho + v_\chi\left(2\zeta_{23}v_\eta - \sqrt{2}v_\sigma\zeta_{34}\right)\right) \tag{A55}$$

$$m_{\sigma_2^{p,*} \chi_-^*} = \frac{1}{4}\left(\sqrt{2}\left(2f_\sigma v_\rho + v_\eta v_\chi\xi_{34}\right) + v_\chi v_\sigma\zeta_{34}\right) \tag{A56}$$

$$m_{\chi_- \chi_-^*} = \frac{1}{2}\left(2\lambda_3 v_\chi^2 + \lambda_{13}v_\rho^2 + \left(\lambda_{23} + \zeta_{23}\right)v_\eta^2 + \lambda_{34}v_\sigma^2\right) + \mu_3 \tag{A57}$$

$$m_{\rho_+^* \sigma_1^{-,*}} = \frac{1}{4}\left(\sqrt{2}\left(2f_\sigma v_\chi - v_\rho v_\eta\xi_{14}\right) + v_\rho v_\sigma\zeta_{14}\right) \tag{A58}$$

$$m_{\eta_2^- \sigma_1^{-,*}} = \frac{1}{4}\left(-\sqrt{2}v_\chi^2\xi_{34} + v_\eta v_\sigma\left(4\xi_{24} + \zeta_{24}\right)\right) \tag{A59}$$

$$m_{\sigma_1^- \sigma_1^{-,*}} = \frac{1}{4}\left(2\left(\left(2\lambda_4 + \lambda_{44}\right)v_\sigma^2 + \lambda_{14}v_\rho^2\right) + v_\chi^2\left(2\lambda_{34} + \zeta_{34}\right) + v_\eta^2\left(2\lambda_{24} + \zeta_{24}\right)\right) + \mu_4 \tag{A60}$$

The gauge fixing contributions are

$$m^2(\xi_{W^+}) = \begin{pmatrix} 0 & 0 & 0 & 0 & 0 & 0 \\ 0 & \frac{1}{4}g_2^2 v_\eta^2 & -\frac{1}{4}g_2^2 v_\eta v_\sigma & 0 & -\frac{1}{4}g_2^2 v_\eta v_\chi & 0 \\ 0 & -\frac{1}{4}g_2^2 v_\eta v_\sigma & \frac{1}{4}g_2^2 v_\sigma^2 & 0 & \frac{1}{4}g_2^2 v_\chi v_\sigma & 0 \\ 0 & 0 & 0 & 0 & 0 & 0 \\ 0 & -\frac{1}{4}g_2^2 v_\eta v_\chi & \frac{1}{4}g_2^2 v_\chi v_\sigma & 0 & \frac{1}{4}g_2^2 v_\chi^2 & 0 \\ 0 & 0 & 0 & 0 & 0 & 0 \end{pmatrix} \tag{A61}$$

$$m^2(\xi_{W'^+}) = \begin{pmatrix} \frac{1}{4}g_2^2 v_\rho^2 & 0 & 0 & -\frac{1}{4}g_2^2 v_\rho v_\eta & 0 & \frac{1}{4}g_2^2 v_\rho v_\sigma \\ 0 & 0 & 0 & 0 & 0 & 0 \\ 0 & 0 & 0 & 0 & 0 & 0 \\ -\frac{1}{4}g_2^2 v_\rho v_\eta & 0 & 0 & \frac{1}{4}g_2^2 v_\eta^2 & 0 & -\frac{1}{4}g_2^2 v_\eta v_\sigma \\ 0 & 0 & 0 & 0 & 0 & 0 \\ \frac{1}{4}g_2^2 v_\rho v_\sigma & 0 & 0 & -\frac{1}{4}g_2^2 v_\eta v_\sigma & 0 & \frac{1}{4}g_2^2 v_\sigma^2 \end{pmatrix} \tag{A62}$$

This matrix is diagonalized by $R^C$

$$R^C m_{H^-}^2 R^{C,\dagger} = m_{2,H^-}^{dia} \tag{A63}$$

with

$$\rho_+ = \sum_j R_{j1}^C H_j^+, \qquad \eta_1^+ = \sum_j R_{j2}^C H_j^+, \qquad \sigma_2^p = \sum_j R_{j3}^C H_j^+ \tag{A64}$$

$$\eta_2^- = \sum_j R_{j4}^{C,*} H_j^-, \qquad \chi_- = \sum_j R_{j5}^C H_j^-, \qquad \sigma_1^- = \sum_j R_{j6}^{C,*} H_j^- \tag{A65}$$

- **Mass matrix for Doubly Charged Higgs**, Basis: $(\rho^{++,*}, \chi^{--}, \sigma_1^{--}, \sigma_2^{++,*}), (\rho^{++}, \chi^{--,*}, \sigma_1^{--,*}, \sigma_2^{++})$

$$m^2_{H^{--}} = \begin{pmatrix} m_{\rho^{++,*}\rho^{++}} & m^*_{\chi^{--}\rho^{++}} & m^*_{\sigma_1^{--}\rho^{++}} & m^*_{\sigma_2^{++,*}\rho^{++}} \\ m_{\rho^{++,*}\chi^{--,*}} & m_{\chi^{--}\chi^{--,*}} & m^*_{\sigma_1^{--}\chi^{--,*}} & m^*_{\sigma_2^{++,*}\chi^{--,*}} \\ m_{\rho^{++,*}\sigma_1^{--,*}} & m_{\chi^{--}\sigma_1^{--,*}} & m_{\sigma_1^{--}\sigma_1^{--,*}} & \frac{1}{2}\lambda_{44}v_\sigma^2 + v_\eta^2\xi_{24} \\ m_{\rho^{++,*}\sigma_2^{++}} & m_{\chi^{--}\sigma_2^{++}} & \frac{1}{2}\lambda_{44}v_\sigma^2 + v_\sigma^2\xi_{24} & m_{\sigma_2^{++,*}\sigma_2^{++}} \end{pmatrix} + \xi_{Y^{++}}m^2(Y^{++}) \quad (A66)$$

$$m_{\rho^{++,*}\rho^{++}} = \frac{1}{4}\Big(2\Big(\lambda_{12}v_\eta^2 + \big(\lambda_{13} + \xi_{13}\big)v_\chi^2 + \sqrt{2}v_\eta v_\sigma \xi_{14}\Big) + 4\lambda_1 v_\rho^2 \\ + v_\sigma^2\big(2\lambda_{14} + \xi_{14}\big)\Big) + \mu_1 \quad (A67)$$

$$m_{\rho^{++,*}\chi^{--,*}} = \frac{1}{2}\Big(-2fv_\eta + \sqrt{2}f_\sigma v_\sigma + \xi_{13}v_\rho v_\chi\Big) \quad (A68)$$

$$m_{\chi^{--}\chi^{--,*}} = \frac{1}{4}\Big(2\big(\big(\lambda_{13} + \xi_{13}\big)v_\rho^2 + \lambda_{23}v_\eta^2\big) - 2\sqrt{2}v_\eta v_\sigma\xi_{34} + 4\lambda_3 v_\chi^2 \\ + v_\sigma^2\big(2\lambda_{34} + \xi_{34}\big)\Big) + \mu_3 \quad (A69)$$

$$m_{\rho^{++,*}\sigma_1^{--,*}} = \frac{1}{4}\Big(4f_\sigma v_\chi + v_\rho\big(-2v_\eta\xi_{14} + \sqrt{2}v_\sigma\xi_{14}\big)\Big) \quad (A70)$$

$$m_{\chi^{--}\sigma_1^{--,*}} = \frac{1}{4}v_\chi\big(-2v_\eta\xi_{34} + \sqrt{2}v_\sigma\xi_{34}\big) \quad (A71)$$

$$m_{\sigma_1^{--}\sigma_1^{--,*}} = \frac{1}{2}\Big(2\big(\lambda_4 + \lambda_{44}\big)v_\sigma^2 + \lambda_{14}v_\rho^2 + \lambda_{24}v_\eta^2 + v_\chi^2\big(\lambda_{34} + \xi_{34}\big)\Big) + \mu_4 \quad (A72)$$

$$m_{\rho^{++,*}\sigma_2^{++}} = \frac{1}{4}v_\rho\big(2v_\eta\xi_{14} + \sqrt{2}v_\sigma\xi_{14}\big) \quad (A73)$$

$$m_{\chi^{--}\sigma_2^{++}} = \frac{1}{4}\Big(4f_\sigma v_\rho + v_\chi\big(2v_\eta\xi_{34} + \sqrt{2}v_\sigma\xi_{34}\big)\Big) \quad (A74)$$

$$m_{\sigma_2^{++,*}\sigma_2^{++}} = \frac{1}{2}\Big(2\big(\lambda_4 + \lambda_{44}\big)v_\sigma^2 + \lambda_{24}v_\eta^2 + \lambda_{34}v_\chi^2 + v_\rho^2\big(\lambda_{14} + \xi_{14}\big)\Big) + \mu_4 \quad (A75)$$

The gauge fixing contributions are

$$m^2(\xi_{Y^{++}}) = \begin{pmatrix} \frac{1}{4}g_2^2 v_\rho^2 & -\frac{1}{4}g_2^2 v_\rho v_\chi & \frac{1}{2}\frac{1}{\sqrt{2}}g_2^2 v_\rho v_\sigma & -\frac{1}{2}\frac{1}{\sqrt{2}}g_2^2 v_\rho v_\sigma \\ -\frac{1}{4}g_2^2 v_\rho v_\chi & \frac{1}{4}g_2^2 v_\chi^2 & \frac{1}{2}\frac{1}{\sqrt{2}}g_2^2 v_\chi v_\sigma & \frac{1}{2}\frac{1}{\sqrt{2}}g_2^2 v_\chi v_\sigma \\ \frac{1}{2}\frac{1}{\sqrt{2}}g_2^2 v_\rho v_\sigma & -\frac{1}{2}\frac{1}{\sqrt{2}}g_2^2 v_\chi v_\sigma & \frac{1}{2}g_2^2 v_\sigma^2 & -\frac{1}{2}g_2^2 v_\sigma^2 \\ -\frac{1}{2}\frac{1}{\sqrt{2}}g_2^2 v_\rho v_\sigma & \frac{1}{2}\frac{1}{\sqrt{2}}g_2^2 v_\chi v_\sigma & -\frac{1}{2}g_2^2 v_\sigma^2 & \frac{1}{2}g_2^2 v_\sigma^2 \end{pmatrix} \quad (A76)$$

This matrix is diagonalized by $R_{2C}$

$$R_{2C}m^2_{H^{--}}R^\dagger_{2C} = m^{dia}_{2,H^{--}} \quad (A77)$$

with

$$\rho^{++} = \sum_j R_{2C,j1}H^{++}_j, \qquad \chi^{--} = \sum_j R^*_{2C,j2}H^{--}_j, \qquad \sigma_1^{--} = \sum_j R^*_{2C,j3}H^{--}_j \quad (A78)$$

$$\sigma_2^{++} = \sum_j R_{2C,j4}H^{++}_j \quad (A79)$$

# Analytic Formulae for T Violation in Neutrino Oscillations

**Osamu Yasuda**

Department of Physics, Tokyo Metropolitan University, Tokyo 192-0397, Japan; yasuda@phys.se.tmu.ac.jp

**Abstract:** Recently, a concept known as $\mu$TRISTAN, which involves the acceleration of $\mu^+$, has been proposed. This initiative has led to considerations of a new design for a neutrino factory. Additionally, leveraging the polarization of $\mu^+$, measurements of T violation in neutrino oscillations are also being explored. In this paper, we present analytical expressions for T violation in neutrino oscillations within the framework of standard three-flavor neutrino oscillations, a scenario involving nonstandard interactions, and a case of unitarity violation. We point out that examining the energy spectrum of T violation may be useful for probing new physics effects.

**Keywords:** neutrino oscillation; T violation; $\mu$TRISTAN

## 1. Introduction

Results from various neutrino oscillation experiments have nearly determined the three mixing angles and the absolute values of the mass squared differences in the standard three-flavor mixing scenario within the lepton sector [1]. The remaining undetermined parameters, such as the mass ordering, the octant of the atmospheric neutrino oscillation mixing angle, and the CP phase, are expected to be resolved by the high-intensity neutrino long-baseline experiments currently under construction, such as T2HK and DUNE. Once the CP phase is established, the standard three-flavor lepton mixing scheme will be solidified, achieving the final goal of studies on standard three-flavor neutrino oscillations. To explore physics beyond this framework using neutrino oscillations, experiments in previously unexplored channels will be necessary.

Recently, a concept known as $\mu$TRISTAN [2] has been proposed, which involves creating a low-emittance $\mu^+$ beam using ultra-cold muon technology and accelerating it to energies suitable for a $\mu^+$ collider. The expected number of muons at $\mu$TRISTAN is on the order of $10^{13}$ to $10^{14}$ muons per second. This proposal has reignited interest [3] in the neutrino factory concept [4,5], which could be developed en route to achieving a muon collider. At such a neutrino factory, the decay of $\mu^+$ in the storage ring would produce $\bar{\nu}_\mu$ and $\nu_e$. Ref. [6] explored the potential to polarize the $\mu$ beam to reduce the flux of $\nu_\mu$ or $\bar{\nu}_\mu$, thereby enabling the measurement of $\nu_e \to \nu_\mu$ transitions. If this can be achieved, it would allow for the measurement of T violation in neutrino oscillations, i.e., the difference between the oscillation probabilities $P(\nu_\mu \to \nu_e)$ and $P(\nu_e \to \nu_\mu)$.

T violation in neutrino oscillations has been discussed by many researchers in the past [6–28]. T violation has attracted significant attention primarily because its structure is simpler than that of CP violation, which compares $P(\nu_\mu \to \nu_e)$ with $P(\bar{\nu}_\mu \to \bar{\nu}_e)$ and involves complications due to the presence of the matter effect. (To justify discussions of T violation on an equal footing with CP violation, CPT symmetry is necessary in the neutrino sector. Refs. [29–33] studied CPT symmetry in the neutrino sector and concluded that there are strong constraints on CPT violation.) In this paper, we derive the analytical forms of T violation in three scenarios: the standard three-flavor scheme, a scenario with nonstandard interactions, and a case of unitarity violation. We also briefly comment on the feasibility of probing new physics effects by examining the energy dependence of T violation.

In Section 2, we review the formalism by Kimura, Takamura, and Yokomakura [34,35] to derive analytical formulas for the oscillation probabilities. In Section 3, we derive the analytic forms for T violation in the three cases: the standard three-flavor mixing framework, a scenario involving flavor-dependent nonstandard interactions, and a case with unitarity violation. In Section 4, we summarize our conclusions.

## 2. Analytical Formula for Oscillation Probabilities

It is known [36] (see also earlier works [37–39]) that after eliminating the negative energy states by a Tani–Foldy–Wouthusen-type transformation, the Dirac equation for neutrinos propagating in matter is reduced to the familiar form:

$$i\frac{d\Psi}{dt} = \left(U\mathcal{E}U^{-1} + \mathcal{A}\right)\Psi,$$ (1)

where $U$ is the PMNS matrix,

$$\Psi \equiv \begin{pmatrix} \nu_e \\ \nu_\mu \\ \nu_\tau \end{pmatrix}$$

is the flavor eigenstate,

$$\mathcal{E} \equiv \text{diag}(E_1, E_2, E_3)$$ (2)

is the diagonal matrix of the energy eigenvalue $E_j \equiv \sqrt{m_j^2 + \vec{p}^2}$ $(j = 1, 2, 3)$ of each mass eigenstate with momentum $\vec{p}$, and the matrix

$$\mathcal{A} \equiv \sqrt{2}\,G_F \left\{\text{diag}\left(N_e - \frac{N_n}{2}, -\frac{N_n}{2}, -\frac{N_n}{2}\right)\right\}.$$

stands for the matter effect, which is characterized by the Fermi coupling constant $G_F$, the electron density $N_e$, and the neutron density $N_n$. Throughout this paper we assume for simplicity that the density of matter is constant. The $3 \times 3$ matrix on the right-hand side of Equation (1) is Hermitian and can be formally diagonalized by a unitary matrix $\widetilde{U}$ as

$$U\mathcal{E}U^{-1} + \mathcal{A} = \widetilde{U}\widetilde{\mathcal{E}}\widetilde{U}^{-1},$$ (3)

where

$$\widetilde{\mathcal{E}} \equiv \text{diag}\left(\widetilde{E}_1, \widetilde{E}_2, \widetilde{E}_3\right)$$

is a diagonal matrix with the energy eigenvalue $\widetilde{E}_j$ in the presence of the matter effect. Equation (1) can be easily solved, resulting in the flavor eigenstate at the distance $L$:

$$\Psi(L) = \widetilde{U}\exp\left(-i\widetilde{\mathcal{E}}L\right)\widetilde{U}^{-1}\Psi(0).$$ (4)

Thus, we have the probability amplitude $A(\nu_\beta \to \nu_\alpha)$ of the flavor transition $\nu_\beta \to \nu_\alpha$:

$$A(\nu_\beta \to \nu_\alpha) = \left[\widetilde{U}\exp\left(-i\widetilde{\mathcal{E}}L\right)\widetilde{U}^{-1}\right]_{\alpha\beta}.$$ (5)

From Equation (5) we observe that the shift $\mathcal{E} \to \mathcal{E} - \mathbf{1}E_1$, where $\mathbf{1}$ stands for the $3 \times 3$ identity matrix, changes only the overall phase of the probability amplitude $A(\nu_\beta \to \nu_\alpha)$, and this phase does not affect the value of the probability $P(\nu_\beta \to \nu_\alpha) = |A(\nu_\beta \to \nu_\alpha)|^2$ of the flavor transition $\nu_\beta \to \nu_\alpha$. In the following discussions, therefore, for simplicity, we define the diagonal energy matrix $\mathcal{E}$ and the potential one $\mathcal{A}$ as follows:

$$\mathcal{E} \equiv \mathrm{diag}(E_1, E_2, E_3) - E_1 \, \mathbf{1}$$
$$= \mathrm{diag}(0, \Delta E_{21}, \Delta E_{31}) \tag{6}$$
$$\mathcal{A} \equiv \sqrt{2} \, G_F \left\{ \mathrm{diag}\left( N_e - \frac{N_n}{2}, -\frac{N_n}{2}, -\frac{N_n}{2} \right) + \left( \frac{N_n}{2} \right) \mathbf{1} \right\}$$
$$= \mathrm{diag}(A, 0, 0) \tag{7}$$

with

$$\Delta E_{jk} \equiv E_j - E_k \simeq \frac{m_j^2 - m_k^2}{2|\vec{p}|} \equiv \frac{\Delta m_{jk}^2}{2|\vec{p}|} \equiv \frac{\Delta m_{jk}^2}{2E}$$
$$A \equiv \sqrt{2} \, G_F \, N_e \,. \tag{8}$$

Thus, the appearance of the oscillation probability $P(\nu_\beta \to \nu_\alpha)$ $(\alpha \neq \beta)$ is given by

$$P(\nu_\beta \to \nu_\alpha)$$
$$= \left| \left[ \tilde{U} \exp\left( -i \tilde{\mathcal{E}} L \right) \tilde{U}^{-1} \right]_{\alpha\beta} \right|^2$$
$$= \left| \sum_{j=1}^{3} \tilde{X}_j^{\alpha\beta} e^{-i\tilde{E}_j L} \right|^2$$
$$= \left| e^{-i\tilde{E}_1 L} \sum_{j=1}^{3} \tilde{X}_j^{\alpha\beta} e^{-i\Delta\tilde{E}_{j1} L} \right|^2$$
$$= \left| \sum_{j=1}^{3} \tilde{X}_j^{\alpha\beta} \left( e^{-i\Delta\tilde{E}_{j1} L} - 1 \right) \right|^2 \tag{9}$$
$$= \left| (-2i) \sum_{j=2}^{3} e^{-i\Delta\tilde{E}_{j1} L/2} \tilde{X}_j^{\alpha\beta} \sin\left( \frac{\Delta\tilde{E}_{j1} L}{2} \right) \right|^2$$
$$= 4 \left| e^{-i\Delta\tilde{E}_{31} L/2} \tilde{X}_3^{\alpha\beta} \sin\left( \frac{\Delta\tilde{E}_{31} L}{2} \right) + e^{-i\Delta\tilde{E}_{21} L/2} \tilde{X}_2^{\alpha\beta} \sin\left( \frac{\Delta\tilde{E}_{21} L}{2} \right) \right|^2$$
$$= 4 \left| \tilde{X}_3^{\alpha\beta} \sin\left( \frac{\Delta\tilde{E}_{31} L}{2} \right) + e^{i\Delta\tilde{E}_{32} L/2} \tilde{X}_2^{\alpha\beta} \sin\left( \frac{\Delta\tilde{E}_{21} L}{2} \right) \right|^2 \tag{10}$$

where

$$\tilde{X}_j^{\alpha\beta} \equiv \tilde{U}_{\alpha j} \tilde{U}_{\beta j}^*$$
$$\Delta\tilde{E}_{jk} \equiv \tilde{E}_j - \tilde{E}_k$$

have been defined,

$$\sum_{j=1}^{3} \tilde{X}_j^{\alpha\beta} = \delta_{\alpha\beta} = 0 \quad \text{for } \alpha \neq \beta \tag{11}$$

was subtracted in Equation (9) and throughout this paper the indices $\alpha, \beta = (e, \mu, \tau)$ and $j, k = (1, 2, 3)$ stand for those of the flavor and mass eigenstates, respectively. Once we know the eigenvalues $\tilde{E}_j$ and the quantity $\tilde{X}_j^{\alpha\beta}$, the oscillation probability can be expressed analytically. ( In the case of three neutrino flavors in matter, the energy eigenvalues, $\tilde{E}_j$, can in principle be analytically determined using the cubic equation root formula [40]. However, the analytic expression for $\tilde{E}_j$ involving the inverse cosine function is not practically useful. Therefore, below we will calculate $\tilde{E}_j$ using perturbation theory with small parameters, such as $\Delta E_{21}/\Delta E_{31} = \Delta m_{21}^2/\Delta m_{31}^2 \approx 1/30$ and those relevant to nonstandard scenarios).

So the only non-trivial problem in the standard case is to obtain the expression for $\widetilde{X}_j^{\alpha\beta}$, and this was achieved by Kimura, Takamura and Yokomakura [34,35]. Their arguments are based on the trivial identities. From the unitarity condition of the matrix $\widetilde{U}$, we have

$$\delta_{\alpha\beta} = \left[\widetilde{U}\widetilde{U}^{-1}\right]_{\alpha\beta} = \sum_j \widetilde{U}_{\alpha j}\widetilde{U}_{\beta j}^* = \sum_j \widetilde{X}_j^{\alpha\beta}. \tag{12}$$

Next, we take the $(\alpha, \beta)$ component of both sides in Equation (3):

$$\left[U\mathcal{E}U^{-1} + \mathcal{A}\right]_{\alpha\beta} = \left[\widetilde{U}\widetilde{\mathcal{E}}\widetilde{U}^{-1}\right]_{\alpha\beta} = \sum_j \widetilde{U}_{\alpha j}\widetilde{E}_j\widetilde{U}_{\beta j}^* = \sum_j \widetilde{E}_j\widetilde{X}_j^{\alpha\beta} \tag{13}$$

Furthermore, we take the $(\alpha, \beta)$ component of the square of Equation (3):

$$\left[\left(U\mathcal{E}U^{-1} + \mathcal{A}\right)^2\right]_{\alpha\beta} = \left[\widetilde{U}\widetilde{\mathcal{E}}^2\widetilde{U}^{-1}\right]_{\alpha\beta} = \sum_j \widetilde{U}_{\alpha j}\widetilde{E}_j^2\widetilde{U}_{\beta j}^* = \sum_j \widetilde{E}_j^2\widetilde{X}_j^{\alpha\beta} \tag{14}$$

Putting Equations (12)–(14) together, we have

$$\begin{pmatrix} 1 & 1 & 1 \\ \widetilde{E}_1 & \widetilde{E}_2 & \widetilde{E}_3 \\ \widetilde{E}_1^2 & \widetilde{E}_2^2 & \widetilde{E}_3^2 \end{pmatrix} \begin{pmatrix} \widetilde{X}_1^{\alpha\beta} \\ \widetilde{X}_2^{\alpha\beta} \\ \widetilde{X}_3^{\alpha\beta} \end{pmatrix} = \begin{pmatrix} Y_1^{\alpha\beta} \\ Y_2^{\alpha\beta} \\ Y_3^{\alpha\beta} \end{pmatrix} \tag{15}$$

with

$$Y_j^{\alpha\beta} \equiv \left[\left(U\mathcal{E}U^{-1} + \mathcal{A}\right)^{j-1}\right]_{\alpha\beta} \quad \text{for } j = 1, 2, 3,$$

which can be easily solved by inverting the Vandermonde matrix:

$$\begin{pmatrix} \widetilde{X}_1^{\alpha\beta} \\ \widetilde{X}_2^{\alpha\beta} \\ \widetilde{X}_3^{\alpha\beta} \end{pmatrix} = \begin{pmatrix} \dfrac{1}{\Delta\widetilde{E}_{21}\Delta\widetilde{E}_{31}}(\widetilde{E}_2\widetilde{E}_3, & -(\widetilde{E}_2 + \widetilde{E}_3), & 1) \\ \dfrac{-1}{\Delta\widetilde{E}_{21}\Delta\widetilde{E}_{32}}(\widetilde{E}_3\widetilde{E}_1, & -(\widetilde{E}_3 + \widetilde{E}_1), & 1) \\ \dfrac{1}{\Delta\widetilde{E}_{31}\Delta\widetilde{E}_{32}}(\widetilde{E}_1\widetilde{E}_2, & -(\widetilde{E}_1 + \widetilde{E}_2), & 1) \end{pmatrix} \begin{pmatrix} Y_1^{\alpha\beta} \\ Y_2^{\alpha\beta} \\ Y_3^{\alpha\beta} \end{pmatrix}. \tag{16}$$

## 3. Analytic Form of T Violation

In this section, we derive the analytic form of T violation in the cases with and without unitarity, using the formalism described in Section 2.

### 3.1. The Three-Flavor Case with Unitarity

First, let us discuss the case where time evolution is unitary. From Equation (10), we have

$$\begin{aligned} &P(\nu_\mu \to \nu_e) - P(\nu_e \to \nu_\mu) \\ &= 4\sin\left(\frac{\Delta\widetilde{E}_{31}L}{2}\right)\sin\left(\frac{\Delta\widetilde{E}_{21}L}{2}\right) \\ &\quad \times \left[e^{i\Delta\widetilde{E}_{32}L/2}\widetilde{X}_3^{e\mu*}\widetilde{X}_2^{e\mu} + e^{-i\Delta\widetilde{E}_{32}L/2}\widetilde{X}_3^{e\mu}\widetilde{X}_2^{e\mu*} \right. \\ &\quad\quad \left. -e^{i\Delta\widetilde{E}_{32}L/2}\widetilde{X}_3^{e\mu}\widetilde{X}_2^{e\mu*} - e^{-i\Delta\widetilde{E}_{32}L/2}\widetilde{X}_3^{e\mu*}\widetilde{X}_2^{e\mu}\right] \\ &= 16\,\mathrm{Im}\left[\widetilde{X}_2^{e\mu}\widetilde{X}_3^{e\mu*}\right]\sin\left(\frac{\Delta\widetilde{E}_{32}L}{2}\right)\sin\left(\frac{\Delta\widetilde{E}_{31}L}{2}\right)\sin\left(\frac{\Delta\widetilde{E}_{21}L}{2}\right), \end{aligned} \tag{17}$$

Furthermore, from Equation (16), the factor $\text{Im}\left[\widetilde{X}_2^{e\mu}\widetilde{X}_3^{e\mu*}\right]$ in Equation (17) can be rewritten as

$$
\begin{aligned}
\text{Im}\left[\widetilde{X}_2^{e\mu}\widetilde{X}_3^{e\mu*}\right] &= \frac{-1}{\Delta\widetilde{E}_{21}\Delta\widetilde{E}_{32}}\frac{1}{\Delta\widetilde{E}_{31}\Delta\widetilde{E}_{32}} \\
&\quad \times \text{Im}\left[\left\{Y_3^{e\mu}-(\widetilde{E}_3+\widetilde{E}_1)Y_2^{e\mu}\right\}\left\{Y_3^{e\mu*}-(\widetilde{E}_1+\widetilde{E}_2)Y_2^{e\mu*}\right\}\right] \\
&= \frac{1}{\Delta\widetilde{E}_{21}\Delta\widetilde{E}_{31}\Delta\widetilde{E}_{32}}\,\text{Im}\left[Y_2^{e\mu}Y_3^{e\mu*}\right]
\end{aligned}
\tag{18}
$$

Equations (17) and (18) are applicable for a generic case, as long as the unitarity relation (11) holds.

### 3.1.1. The Standard Three-Flavor Case

In the standard three-flavor case, $Y_{j+1}^{\alpha\beta}\equiv[(U\mathcal{E}U^{-1}+\mathcal{A})^j]_{\alpha\beta}$ $(j=1,2)$ can be expressed as follows:

$$
\begin{aligned}
Y_2^{\alpha\beta} &\equiv \left[U\mathcal{E}U^{-1}+\mathcal{A}\right]_{\alpha\beta} \\
&= \sum_{j=2}^{3}\Delta E_{j1}X_j^{\alpha\beta}+A\,\delta_{\alpha e}\delta_{\beta e}
\end{aligned}
\tag{19}
$$

$$
\begin{aligned}
Y_3^{\alpha\beta} &\equiv \left[\left(U\mathcal{E}U^{-1}+\mathcal{A}\right)^2\right]_{\alpha\beta} \\
&= \sum_{j=2}^{3}(\Delta E_{j1})^2X_j^{\alpha\beta}+A\sum_{j=2}^{3}\Delta E_{j1}\left(\delta_{\alpha e}X_j^{e\beta}+\delta_{\beta e}X_j^{\alpha e}\right)+A^2\,\delta_{\alpha e}\delta_{\beta e}\,,
\end{aligned}
\tag{20}
$$

where we have also defined the quantity in vacuum:

$$
X_j^{\alpha\beta}\equiv U_{\alpha j}U_{\beta j}^*\,.
\tag{21}
$$

From Equations (19) and (20), the factor $\text{Im}\left[Y_2^{e\mu}Y_3^{e\mu*}\right]$ in Equation (18) can be rewritten as

$$
\begin{aligned}
&\text{Im}\left[Y_2^{e\mu}Y_3^{e\mu*}\right] \\
&= \text{Im}\left[\left(\Delta E_{21}X_2^{e\mu}+\Delta E_{31}X_3^{e\mu}\right)\right. \\
&\qquad \left.\times\left\{\Delta E_{21}(\Delta E_{21}+A)X_2^{e\mu*}+\Delta E_{31}(\Delta E_{31}+A)X_3^{e\mu*}\right\}\right] \\
&= \text{Im}\left[X_2^{e\mu}X_3^{e\mu*}\right]\Delta E_{21}\Delta E_{31}\Delta E_{32}
\end{aligned}
\tag{22}
$$

$\text{Im}\left[X_2^{e\mu}X_3^{e\mu*}\right]$ in Equation (22) is the Jarlskog factor [41] for the lepton sector, and is given in the standard parametrization [1] with the three mixing angles $\theta_{jk}$ $(j,k=1,2,3)$ and the Dirac CP phase $\delta$ by

$$
\begin{aligned}
J &\equiv \text{Im}\left[X_2^{e\mu}X_3^{e\mu*}\right] \\
&= -\frac{1}{8}\sin\delta\cos\theta_{13}\sin 2\theta_{12}\sin 2\theta_{13}\sin 2\theta_{23}\,.
\end{aligned}
$$

Hence, we obtain

$$
\begin{aligned}
&P(\nu_\mu\to\nu_e)-P(\nu_e\to\nu_\mu) \\
&= 16\,J\,\frac{\Delta E_{21}\Delta E_{31}\Delta E_{32}}{\Delta\widetilde{E}_{21}\Delta\widetilde{E}_{31}\Delta\widetilde{E}_{32}}\,\sin\left(\frac{\Delta\widetilde{E}_{32}L}{2}\right)\sin\left(\frac{\Delta\widetilde{E}_{31}L}{2}\right)\sin\left(\frac{\Delta\widetilde{E}_{21}L}{2}\right)
\end{aligned}
\tag{23}
$$

Equation (23) is a well-known formula [10] for the standard three-flavor case. It is remarkable that in the standard three-flavor case, T violation in matter, when divided by the $\delta$-independent factor $16 \prod_{j>k}[\sin(\Delta \tilde{E}_{jk} L/2)\Delta E_{jk}/\Delta \tilde{E}_{jk}]$, coincides with the Jarlskog factor in vacuum. This implies that the only source of T violation is the CP phase $\delta$ in the standard three-flavor case, and it is the reason why T violation is simpler than CP violation in neutrino oscillations.

### 3.1.2. The Case with Nonstandard Interactions

As long as unitarity in the three-flavor framework is maintained, Equation (18) holds. In this subsection, let us consider the scenario with flavor-dependent nonstandard interactions [42,43] during neutrino propagation. This scenario has garnered significant attention due to its potential implications for phenomenology. In this case, the mass matrix is given by

$$U\mathcal{E}U^{-1} + \mathcal{A} + \mathcal{A}_{NP} \tag{24}$$

with

$$\mathcal{A}_{NP} \equiv A \begin{pmatrix} \epsilon_{ee} & \epsilon_{e\mu} & \epsilon_{e\tau} \\ \epsilon_{e\mu}^* & \epsilon_{\mu\mu} & \epsilon_{\mu\tau} \\ \epsilon_{e\tau}^* & \epsilon_{\mu\tau}^* & \epsilon_{\tau\tau} \end{pmatrix},$$

where $\mathcal{A}$ and $A$ are given by Equations (7) and (8), respectively. The dimensionless quantities $\epsilon_{\alpha\beta}$ stand for the ratio of the nonstandard Fermi coupling constant interaction to the standard one. Since the matrix (24) is Hermitian, time evolution is unitary and all the arguments up to Equation (18) hold also in this case. The oscillation probability is given by Equations (10) and (16), where the standard potential matrix $\mathcal{A}$ must be replaced by $\mathcal{A} + \mathcal{A}_{NP}$.

The extra complication compared to the standard case is calculations of the eigenvalues $\tilde{E}_j$ and the elements $[(U\mathcal{E}U^{-1} + \mathcal{A} + \mathcal{A}_{NP})^m]_{\alpha\beta}$ ($m = 1, 2$). Here, we work with perturbation theory with respect to the small parameters $\Delta E_{21}/\Delta E_{31} = \Delta m_{21}^2/\Delta m_{31}^2 \simeq 1/30$ and $\epsilon_{\alpha\beta}$, which we assume to be as small as $\Delta E_{21}/\Delta E_{31}$. Namely, throughout this paper we assume

$$|\Delta E_{31}| \sim A \gg |\Delta E_{21}| \gtrsim A\,|\epsilon_{\alpha\beta}|\,.$$

and take into consideration to first order in these small parameters. Then, to first order in them, we obtain

$$\begin{aligned} Y_2^{e\mu} &\equiv \left[U\mathcal{E}U^{-1} + \mathcal{A} + \mathcal{A}_{NP}\right]_{e\mu} \\ &= \Delta E_{31} X_3^{e\mu} + \Delta E_{21} X_2^{e\mu} + A\,\epsilon_{e\mu} \end{aligned} \tag{25}$$

$$\begin{aligned} Y_3^{e\mu} &\equiv \left[\left(U\mathcal{E}U^{-1} + \mathcal{A} + \mathcal{A}_{NP}\right)^2\right]_{e\mu} \\ &\simeq \left\{(\Delta E_{31})^2 + A\,\Delta E_{31}\right\}X_3^{e\mu} + A\,\Delta E_{21} X_2^{e\mu} \\ &\quad + A^2\,\epsilon_{e\mu} + \Delta E_{31}\{X_3, \mathcal{A}_{NP}\}_{e\mu} \\ &= (\Delta E_{31} + A)\,Y_2^{e\mu} - \Delta E_{31}\Delta E_{21} X_2^{e\mu} - A\,\Delta E_{31}\,\epsilon_{e\mu} + \Delta E_{31}\{X_3, \mathcal{A}_{NP}\}_{e\mu}\,, \end{aligned} \tag{26}$$

where the curly bracket stands for an anticommutator of matrices $P$ and $Q$: $\{P, Q\} \equiv PQ + QP$, and $X_3$ is a $3 \times 3$ matrix defined by

$$\begin{aligned} X_3 &\equiv U\,\mathrm{diag}(0, 0, 1)\,U^{-1} \\ (X_3)_{\alpha\beta} &= U_{\alpha 3}U_{\beta 3}^* = X_3^{\alpha\beta}\,. \end{aligned} \tag{27}$$

From this, the first term on the right-hand side of Equation (26) drops in the factor $\text{Im}\left[Y_2^{e\mu}Y_3^{e\mu*}\right]$, and we obtain

$$\text{Im}\left[Y_2^{e\mu}Y_3^{e\mu*}\right] = -\text{Im}\left[Y_2^{e\mu*}Y_3^{e\mu}\right]$$
$$\simeq \Delta E_{31}\,\text{Im}\left[Y_2^{e\mu*}\left(\Delta E_{21}X_2^{e\mu} + A\,\epsilon_{e\mu} - \{X_3, \mathcal{A}_{NP}\}_{e\mu}\right)\right]$$
$$\simeq (\Delta E_{31})^2\,\text{Im}\left[X_3^{e\mu*}\left(\Delta E_{21}X_2^{e\mu} + A\,\epsilon_{e\mu} - \{X_3, \mathcal{A}_{NP}\}_{e\mu}\right)\right]$$
$$= (\Delta E_{31})^2\,\text{Im}\left[X_3^{e\mu*}\left\{\Delta E_{21}X_2^{e\mu} + A\left(X_3^{e\tau}\epsilon_{e\mu} - X_3^{e\tau}\epsilon_{\tau\mu} - X_3^{\tau\mu}\epsilon_{e\tau}\right)\right\}\right],$$

where we have ignored terms of order $O((\Delta E_{21})^2)$, $O(A^2(\epsilon_{\alpha\beta})^2)$, and $O(A\Delta E_{21}\epsilon_{\alpha\beta})$. Thus, we finally obtain the form for T violation:

$$P(\nu_\mu \to \nu_e) - P(\nu_e \to \nu_\mu)$$
$$\simeq \frac{16\,(\Delta E_{31})^2}{\Delta\tilde{E}_{21}\Delta\tilde{E}_{31}\Delta\tilde{E}_{32}}\sin\left(\frac{\Delta\tilde{E}_{32}L}{2}\right)\sin\left(\frac{\Delta\tilde{E}_{31}L}{2}\right)\sin\left(\frac{\Delta\tilde{E}_{21}L}{2}\right)$$
$$\times\text{Im}\left[X_3^{e\mu*}\left\{\Delta E_{21}X_2^{e\mu} + A\left(X_3^{e\tau}\epsilon_{e\mu} - X_3^{e\tau}\epsilon_{\tau\mu} - X_3^{\tau\mu}\epsilon_{e\tau}\right)\right\}\right] \tag{28}$$

Note that the form of the standard contribution $(\Delta E_{31})^2\Delta E_{21}$, $\text{Im}\left[X_2^{e\mu}X_3^{e\mu*}\right]$ in Equation (28) differs slightly from that in Equation (22) because we are neglecting terms of order $O((\Delta E_{21})^2)$. The terms proportional to $A$ in the parenthesis in Equation (28) represent the additional contributions to T violation due to nonstandard interactions. These additional contributions are constant with respect to the neutrino energy $E$, and they exhibit a different energy dependence from that of the standard one, $\Delta E_{21}X_2^{e\mu} = (\Delta m_{21}^2/2E)U_{e2}U_{\mu2}^*$. Therefore, if the magnitude of the additional contributions from nonstandard interactions is significant enough, then their effects are expected to be observable in the energy spectrum of T violation. Constraints on the parameters $\epsilon_{\alpha\beta}$ have been provided in Refs. [44–46]. Depending on the sensitivity of each experiment, it may or may not be possible to detect the signal or to improve the existing bounds on $\epsilon_{\alpha\beta}$. The aim of this paper is to derive the analytic form of T violation; estimating experimental sensitivity is beyond its scope.

### 3.2. The Three-Flavor Case with Unitarity Violation

The discussions in Section 3.1 are based on the assumption that time evolution is unitary. In Ref. [47], the possibility to have a nonunitary leptonic mixing matrix was pointed out. In that case, the relation between the mass eigenstate $\nu_j$ and the flavor eigenstate $\nu_\alpha$ is given by a nonunitary matrix $N$:

$$\nu_\alpha = N_{\alpha j}\,\nu_j$$
$$N \equiv (\mathbf{1} + \eta)\,U$$

with

$$\eta^\dagger = \eta. \tag{29}$$

In the so-called minimal unitarity violation, which was discussed in Ref. [47], the constraint on the deviation matrix $\eta$ turned out to be strong. Here, we take phenomenologically the form of the nonunitary matrix $N$ and assume that the elements of the deviation matrix $\eta$ are of order $\Delta E_{21}/\Delta E_{31}$ or smaller, as in Section 3.1.2, namely,

$$|\Delta E_{31}| \sim A \gg |\Delta E_{21}| \gtrsim A\,|\eta_{\alpha\beta}|.$$

It was argued in Ref. [48] that time evolution in the case of a nonunitary mixing matrix can be discussed in terms of the mass eigenstate

$$\Psi_m \equiv \begin{pmatrix} \nu_1 \\ \nu_2 \\ \nu_3 \end{pmatrix},$$

and its time evolution is described by

$$i\frac{d\Psi_m}{dt} = \left\{ \mathcal{E} + N^T \mathcal{A} N^* - A_n \left( N^T N^* - \mathbf{1} \right) \right\} \Psi_m, \tag{30}$$

where $\mathcal{E}$ and $\mathcal{A}$ are defined by Equations (6) and (7),

$$A_n \equiv \frac{1}{\sqrt{2}} G_F \, N_n$$

stands for the absolute value of the contribution to the matter effect from the neutral current interaction, and the term $A_n \mathbf{1}$ was added to simplify the calculations without changing the absolute value of the probability amplitude. The $3 \times 3$ matrix on the right-hand side of Equation (30) can be diagonalized with a unitary matrix $W$:

$$\mathcal{E} + N^T \mathcal{A} N^* - A_n \left( N^T N^* - \mathbf{1} \right) = W \widetilde{\mathcal{E}} W^{-1}, \tag{31}$$

where

$$\widetilde{\mathcal{E}} \equiv \mathrm{diag}\left( \widetilde{E}_1, \widetilde{E}_2, \widetilde{E}_3 \right)$$

is the energy eigenvalue matrix in matter with unitarity violation. The mass eigenstate at distance $L$ can be solved as

$$\Psi_m(L) = W \exp(-i\widetilde{\mathcal{E}} L) W^{-1} \Psi_m(0). \tag{32}$$

In cases involving unitarity violation, due to the modified form of the charged current interaction [47], after computing the probability amplitude from Equation (32) we must multiply the probability amplitude by an additional factor of $(NN^\dagger)_{\beta\beta}^{-1/2}$ for the production process and $(NN^\dagger)_{\alpha\alpha}^{-1/2}$ for detection. Defining the modified amplitude

$$\hat{A}(\nu_\beta \to \nu_\alpha) \equiv A(\nu_\beta \to \nu_\alpha)(NN^\dagger)_{\alpha\alpha}^{1/2}(NN^\dagger)_{\beta\beta}^{1/2}$$
$$= [N^* W \exp(-i\widetilde{\mathcal{E}} L) W^{-1} N^T]_{\alpha\beta},$$

the modified probability

$$\hat{P}(\nu_\alpha \to \nu_\beta) \equiv |\hat{A}(\nu_\alpha \to \nu_\beta)|^2,$$

and the quantity

$$\widetilde{\mathcal{X}}_j^{\alpha\beta} \equiv (N^* W)_{\alpha j} (NW^*)_{\beta j},$$

we have the following expression for the appearance oscillation probability:

$$\hat{P}(\nu_\beta \to \nu_\alpha)$$

$$= \left| \left[ N^* W \exp\left(-i\tilde{\mathcal{E}}L\right) W^{-1} N^T \right]_{\alpha\beta} \right|^2$$

$$= \left| \sum_{j=1}^3 \tilde{\mathcal{X}}_j^{\alpha\beta} e^{-i\tilde{E}_j L} \right|^2$$

$$= \left| e^{-i\tilde{E}_1 L} \sum_{j=1}^3 \tilde{\mathcal{X}}_j^{\alpha\beta} e^{-i\Delta\tilde{E}_{j1}L} \right|^2$$

$$= \left| \sum_{j=1}^3 \tilde{\mathcal{X}}_j^{\alpha\beta} \left\{ 1 - \left( 1 - e^{-i\Delta\tilde{E}_{j1}L} \right) \right\} \right|^2$$

$$= \left| [N^* N^T]_{\alpha\beta} - 2i \sum_{j=2}^3 e^{-i\Delta\tilde{E}_{j1}L/2} \tilde{\mathcal{X}}_j^{\alpha\beta} \sin\left( \frac{\Delta\tilde{E}_{j1}L}{2} \right) \right|^2$$

$$= \left| [\{(1+\eta)^2\}^T]_{\alpha\beta} + 2e^{-i\Delta\tilde{E}_{31}L/2 - i\pi/2} \tilde{\mathcal{X}}_3^{\alpha\beta} \sin\left( \frac{\Delta\tilde{E}_{31}L}{2} \right) \right.$$
$$\left. + 2e^{-i\Delta\tilde{E}_{21}L/2 - i\pi/2} \tilde{\mathcal{X}}_2^{\alpha\beta} \sin\left( \frac{\Delta\tilde{E}_{21}L}{2} \right) \right|^2$$

$$\simeq 4 \left| \eta_{\beta\alpha} + e^{-i\Delta\tilde{E}_{31}L/2 - i\pi/2} \tilde{\mathcal{X}}_3^{\alpha\beta} \sin\left( \frac{\Delta\tilde{E}_{31}L}{2} \right) \right.$$
$$\left. + e^{-i\Delta\tilde{E}_{21}L/2 - i\pi/2} \tilde{\mathcal{X}}_2^{\alpha\beta} \sin\left( \frac{\Delta\tilde{E}_{21}L}{2} \right) \right|^2. \tag{33}$$

T violation $P(\nu_\mu \to \nu_e) - P(\nu_e \to \nu_\mu)$ is a small quantity, and the difference between the probability $P(\nu_\mu \to \nu_e)$ and the modified one $\hat{P}(\nu_\mu \to \nu_e)$ comes from the factor $(NN^\dagger)_{\alpha\alpha}(NN^\dagger)_{\beta\beta} = [(1+\eta)^2]_{\alpha\alpha}[(1+\eta)^2]_{\beta\beta} \simeq 1 + 2\eta_{\alpha\alpha} + 2\eta_{\beta\beta}$, which has a small deviation from 1. Therefore, T violation of the probability $P(\nu_\mu \to \nu_e) - P(\nu_e \to \nu_\mu)$ can be approximated by that of the modified probability $\hat{P}(\nu_\mu \to \nu_e) - \hat{P}(\nu_e \to \nu_\mu)$. Hence, T violation is given by

$$P(\nu_\mu \to \nu_e) - P(\nu_e \to \nu_\mu)$$
$$\simeq \hat{P}(\nu_\mu \to \nu_e) - \hat{P}(\nu_e \to \nu_\mu)$$

$$= 4 \left| \eta_{\mu e} + e^{-i\Delta\tilde{E}_{31}L/2 - i\pi/2} \tilde{\mathcal{X}}_3^{e\mu} \sin\left( \frac{\Delta\tilde{E}_{31}L}{2} \right) + e^{-i\Delta\tilde{E}_{21}L/2 - i\pi/2} \tilde{\mathcal{X}}_2^{e\mu} \sin\left( \frac{\Delta\tilde{E}_{21}L}{2} \right) \right|^2$$

$$- 4 \left| \eta_{e\mu} + e^{-i\Delta\tilde{E}_{31}L/2 - i\pi/2} \tilde{\mathcal{X}}_3^{\mu e} \sin\left( \frac{\Delta\tilde{E}_{31}L}{2} \right) + e^{-i\Delta\tilde{E}_{21}L/2 - i\pi/2} \tilde{\mathcal{X}}_2^{\mu e} \sin\left( \frac{\Delta\tilde{E}_{21}L}{2} \right) \right|^2$$

$$= 4 \sin\left( \frac{\Delta\tilde{E}_{31}L}{2} \right) \left( \eta_{\mu e} \tilde{\mathcal{X}}_3^{e\mu*} - \eta_{\mu e}^* \tilde{\mathcal{X}}_3^{e\mu} \right) (e^{i\Delta\tilde{E}_{31}L/2 + i\pi/2} - e^{-i\Delta\tilde{E}_{31}L/2 - i\pi/2})$$

$$+ 4 \sin\left( \frac{\Delta\tilde{E}_{21}L}{2} \right) \left( \eta_{\mu e} \tilde{\mathcal{X}}_2^{e\mu*} - \eta_{\mu e}^* \tilde{\mathcal{X}}_2^{e\mu} \right) (e^{i\Delta\tilde{E}_{21}L/2 + i\pi/2} - e^{-i\Delta\tilde{E}_{21}L/2 - i\pi/2})$$

$$- 4 \sin\left( \frac{\Delta\tilde{E}_{31}L}{2} \right) \sin\left( \frac{\Delta\tilde{E}_{21}L}{2} \right) \left( \tilde{\mathcal{X}}_3^{e\mu} \tilde{\mathcal{X}}_2^{e\mu*} - \tilde{\mathcal{X}}_3^{e\mu*} \tilde{\mathcal{X}}_2^{e\mu} \right)$$

$$\times (e^{i\Delta\tilde{E}_{31}L/2 + i\pi/2 - i\Delta\tilde{E}_{21}L/2 - i\pi/2} - e^{-i\Delta\tilde{E}_{31}L/2 - i\pi/2 + i\Delta\tilde{E}_{21}L/2 + i\pi/2})$$

$$= -16 \, \mathrm{Im}\left[ \tilde{\mathcal{X}}_2^{e\mu} \tilde{\mathcal{X}}_3^{e\mu*} \right] \sin\left( \frac{\Delta\tilde{E}_{31}L}{2} \right) \sin\left( \frac{\Delta\tilde{E}_{21}L}{2} \right) \sin\left( \frac{\Delta\tilde{E}_{32}L}{2} \right)$$

$$+ 8 \, \mathrm{Im}\left[ \eta_{\mu e} \tilde{\mathcal{X}}_3^{e\mu*} \right] \sin\left( \Delta\tilde{E}_{31}L \right) + 8 \, \mathrm{Im}\left[ \eta_{\mu e} \tilde{\mathcal{X}}_2^{e\mu*} \right] \sin\left( \Delta\tilde{E}_{21}L \right) \tag{34}$$

We observe that the energy dependence of T violation in this case is different from that with unitarity, since we have extra contributions which are proportional to $\sin\left(\Delta\widetilde{E}_{31}L\right)$ or $\sin\left(\Delta\widetilde{E}_{21}L\right)$. As in the case with unitarity, $\widetilde{\mathcal{X}}_j^{\alpha\beta}$ can be expressed in terms of the quantity $X_j^{\alpha\beta} \equiv U_{\alpha j}U_{\beta j}^*$ in vacuum, $\widetilde{E}_j$, and $\eta_{\alpha\beta}$. First of all, we note the following relations:

$$\sum_j (\widetilde{E}_j)^m \widetilde{\mathcal{X}}_j^{\alpha\beta}$$

$$= \sum_j (N^*W)_{\alpha j}(\widetilde{E}_j)^m (NW^*)_{\beta j}$$

$$= \left[ N^* \left\{ \mathcal{E} + N^T \mathcal{A}N^* - A_n\left(N^T N^* - \mathbf{1}\right) \right\}^m N^T \right]_{\alpha\beta}$$

$$= \left[ N \left\{ \mathcal{E} + N^\dagger \mathcal{A}N - A_n\left(N^\dagger N - \mathbf{1}\right) \right\}^m N^\dagger \right]_{\beta\alpha}$$

$$\equiv \mathcal{Y}_{m+1}^{\alpha\beta} \qquad \text{for } m = 0, 1, 2. \tag{35}$$

Then, we rewrite Equations (35) as

$$\sum_{m=1}^{3} V_{jm}\, \widetilde{\mathcal{X}}_m^{\alpha\beta} = \mathcal{Y}_j^{\alpha\beta} \qquad \text{for } j = 1, 2, 3, \tag{36}$$

where $V_{jm} \equiv (\widetilde{E}_m)^{j-1}$ is the element of the Vandermonde matrix $V$, as in the case with unitarity (see Equation (15)). The simultaneous Equation (36) can be solved by inverting $V$, and we obtain

$$\widetilde{\mathcal{X}}_j^{\alpha\beta} = \sum_{m=1}^{3} (V^{-1})_{jm}\, \mathcal{Y}_m^{\alpha\beta}. \tag{37}$$

The factor $\mathrm{Im}\left[\widetilde{\mathcal{X}}_2^{e\mu}\widetilde{\mathcal{X}}_3^{e\mu*}\right]$ can be expressed in terms of $\widetilde{E}_j$ and $\mathcal{Y}_j^{e\mu}$:

$$\mathrm{Im}\left[\widetilde{\mathcal{X}}_2^{e\mu}\widetilde{\mathcal{X}}_3^{e\mu*}\right]$$

$$= \frac{-1}{\Delta\widetilde{E}_{21}\Delta\widetilde{E}_{32}} \frac{1}{\Delta\widetilde{E}_{31}\Delta\widetilde{E}_{32}}$$

$$\times \mathrm{Im}\left[\left\{\mathcal{Y}_3^{e\mu} - (\widetilde{E}_3 + \widetilde{E}_1)\mathcal{Y}_2^{e\mu} + \widetilde{E}_3\widetilde{E}_1\mathcal{Y}_1^{e\mu}\right\}\left\{\mathcal{Y}_3^{e\mu*} - (\widetilde{E}_1 + \widetilde{E}_2)\mathcal{Y}_2^{e\mu*} + \widetilde{E}_1\widetilde{E}_2\mathcal{Y}_1^{e\mu*}\right\}\right]$$

$$= \frac{1}{\Delta\widetilde{E}_{21}\Delta\widetilde{E}_{31}\Delta\widetilde{E}_{32}}\left(\widetilde{E}_1^2\,\mathrm{Im}\left[\mathcal{Y}_1^{e\mu}\mathcal{Y}_2^{e\mu*}\right] - \widetilde{E}_1\,\mathrm{Im}\left[\mathcal{Y}_1^{e\mu}\mathcal{Y}_3^{e\mu*}\right] + \mathrm{Im}\left[\mathcal{Y}_2^{e\mu}\mathcal{Y}_3^{e\mu*}\right]\right). \tag{38}$$

The quantities $\mathcal{Y}_j^{e\mu}$ $(j = 1, 2, 3)$ are calculated as follows:

$$\mathcal{Y}_1^{e\mu} = [NN^\dagger]_{\mu e} = [(\mathbf{1}+\eta)^2]_{\mu e} \simeq 2\,\eta_{\mu e}\,,$$

$$\mathcal{Y}_2^{e\mu} = \left[N\left\{\mathcal{E} + N^\dagger\mathcal{A}N - A_n\left(N^\dagger N - \mathbf{1}\right)\right\}N^\dagger\right]_{\mu e}$$

$$= \left[(\mathbf{1}+\eta)\left\{U\mathcal{E}U^{-1} + (\mathbf{1}+\eta)\mathcal{A}(\mathbf{1}+\eta) - A_n\left((\mathbf{1}+\eta)^2 - \mathbf{1}\right)\right\}(\mathbf{1}+\eta)\right]\Big|_{\mu e}$$

$$\simeq \left[U\mathcal{E}U^{-1} + \mathcal{A} + \{\eta, U\mathcal{E}U^{-1}\} + 2\{\eta, \mathcal{A}\} - 2A_n\eta\right]_{\mu e}$$

$$\simeq \Delta E_{31}X_3^{\mu e} + \Delta E_{21}X_2^{\mu e} + \Delta E_{31}\{\eta, X_3\}_{\mu e} + 2(A - A_n)\eta_{\mu e}\,,$$

$$\mathcal{Y}_3^{e\mu} = \left[ N \left\{ \mathcal{E} + N^\dagger \mathcal{A} N - A_n \left( N^\dagger N - \mathbf{1} \right) \right\}^2 N^\dagger \right]_{\mu e}$$

$$= \left[ (\mathbf{1} + \eta) \left\{ U \mathcal{E} U^{-1} + (\mathbf{1} + \eta) \mathcal{A} (\mathbf{1} + \eta) - A_n \left( (\mathbf{1} + \eta)^2 - \mathbf{1} \right) \right\}^2 (\mathbf{1} + \eta) \right] \Big|_{\mu e}$$

$$\simeq \left[ U \mathcal{E}^2 U^{-1} + \mathcal{A}^2 + \{ U \mathcal{E} U^{-1}, \mathcal{A} \} \right]_{\mu e} + \left[ \{ U \mathcal{E} U^{-1}, \{ \eta, \mathcal{A} \} \} + \{ \mathcal{A}, \{ \eta, \mathcal{A} \} \} \right]_{\mu e}$$

$$- \left[ 2 A_n \{ U \mathcal{E} U^{-1}, \eta \} + 2 A_n \{ \mathcal{A}, \eta \} \right]_{\mu e}$$

$$+ \left[ \{ \eta, U \mathcal{E}^2 U^{-1} \} + \{ \eta, \mathcal{A}^2 \} + \{ \eta, \{ U \mathcal{E} U^{-1}, \mathcal{A} \} \} \right]_{\mu e}$$

$$\simeq \Delta E_{31} (\Delta E_{31} + A) X_3^{\mu e} + A \, \Delta E_{21} X_2^{\mu e}$$

$$+ \Delta E_{31} \{ X_3, \{ \eta, \mathcal{A} \} \}_{\mu e} + \{ \mathcal{A}, \{ \eta, \mathcal{A} \} \}_{\mu e}$$

$$- 2 A_n \Delta E_{31} \{ X_3, \eta \}_{\mu e} - 2 A_n \{ \mathcal{A}, \eta \}_{\mu e}$$

$$+ \Delta E_{31}^2 \{ \eta, X_3 \}_{\mu e} + \{ \eta, \mathcal{A}^2 \}_{\mu e} + \Delta E_{31} \{ \eta, \{ X_3, \mathcal{A} \} \}_{\mu e}$$

$$= (\Delta E_{31} + A) \, \mathcal{Y}_2^{e\mu} - \Delta E_{31} \Delta E_{21} X_2^{\mu e}$$

$$- 2 (A - A_n) \Delta E_{31} \eta_{\mu e} - (A + 2 A_n) \Delta E_{31} \{ X_3, \eta \}_{\mu e}$$

$$+ \Delta E_{31} \{ X_3, \{ \eta, \mathcal{A} \} \}_{\mu e} + \Delta E_{31} \{ \eta, \{ X_3, \mathcal{A} \} \}_{\mu e} .$$

In the current scenario involving unitarity violation, we observe a nonvanishing contribution from $\mathcal{Y}_1^{\alpha\beta}$, necessitating knowledge of the explicit form of the energy eigenvalue $\widetilde{E}_1$. Given that $\mathcal{Y}_1^{e\mu}$ is of order $O(\eta_{\alpha\beta})$, to evaluate Equation (38) accurately to first order in both $\Delta E_{21} / \Delta E_{31}$ and $\eta_{\alpha\beta}$, we must calculate $\widetilde{E}_1$ solely to zeroth order in these parameters, i.e., assuming $\Delta E_{21} \to 0$ and $\eta_{\alpha\beta} \to 0$. Under these conditions, the characteristic equation of the $3 \times 3$ matrix (31) is defined by

$$0 = \det \left[ \mathbf{1} \, t - \left\{ \mathcal{E} + N^T \mathcal{A} N^* - A_n \left( N^T N^* - \mathbf{1} \right) \right\} \right]$$

$$\simeq \det \left[ \mathbf{1} \, t - \mathrm{diag}(0, 0, \Delta E_{31}) - U^{-1} \, \mathrm{diag}(A, 0, 0) \, U \right]$$

$$= t \left\{ t^2 - (\Delta E_{31} + A) \, t + A \Delta E_{31} \cos^2 \theta_{13} \right\}$$

$$= t \, (t - \lambda_+)(t - \lambda_-) ,$$

where $\lambda_\pm$ are the roots of the quadratic equation and are given by

$$\lambda_\pm \equiv \frac{\Delta E_{31} + A \pm \Delta \widetilde{E}_{31}}{2}$$

$$\Delta \widetilde{E}_{31} \equiv \sqrt{(\Delta E_{31} \cos 2\theta_{13} - A)^2 + (\Delta E_{31} \sin 2\theta_{13})^2} .$$

From this, we obtain the energy eigenvalues $\widetilde{E}_j$ ($j = 1, 2, 3$) to the leading order in $\Delta E_{21} / \Delta E_{31}$ and $\eta_{\alpha\beta}$:

$$\begin{pmatrix} \widetilde{E}_1 \\ \widetilde{E}_2 \\ \widetilde{E}_3 \end{pmatrix} \simeq \begin{pmatrix} \lambda_- \\ 0 \\ \lambda_+ \end{pmatrix}$$

The roots $\lambda_- = \widetilde{E}_1$ and $\lambda_+ = \widetilde{E}_3$ satisfy the quadratic equation

$$\lambda_\pm^2 - (\Delta E_{31} + A) \lambda_\pm + A \Delta E_{31} \cos^2 \theta_{13} = 0 .$$

Hence, the first two terms on the right-hand side of Equation (38) can be rewritten as

$$
\widetilde{E}_1^2 \operatorname{Im}\left[\mathcal{Y}_1^{e\mu} \mathcal{Y}_2^{e\mu*}\right] - \widetilde{E}_1 \operatorname{Im}\left[\mathcal{Y}_1^{e\mu} \mathcal{Y}_3^{e\mu*}\right]
$$

$$
= \operatorname{Im}\left[\mathcal{Y}_1^{e\mu}\left\{\widetilde{E}_1^2 \mathcal{Y}_2^{e\mu*} - \widetilde{E}_1 \mathcal{Y}_3^{e\mu*}\right\}\right]
$$

$$
\simeq \operatorname{Im}\left[\mathcal{Y}_1^{e\mu}\left\{\Delta E_{31} X_3^{\mu e*} \widetilde{E}_1^2 - \Delta E_{31}(\Delta E_{31} + A) X_3^{\mu e*} \widetilde{E}_1\right\}\right]
$$

$$
= \operatorname{Im}\left[\mathcal{Y}_1^{e\mu} A (\Delta E_{31})^2 X_3^{\mu e*} \cos^2 \theta_{13}\right]
$$

$$
= 2 A (\Delta E_{31})^2 \cos^2 \theta_{13} \operatorname{Im}\left[\eta_{\mu e} X_3^{\mu e*}\right],
$$

whereas the third term on the right-hand side of Equation (38) can be written as

$$
\operatorname{Im}\left[\mathcal{Y}_2^{e\mu} \mathcal{Y}_3^{e\mu*}\right]
$$

$$
= -\operatorname{Im}\left[\mathcal{Y}_2^{e\mu*}\left\{(\Delta E_{31} + A)\,\mathcal{Y}_2^{e\mu} - \Delta E_{31}\Delta E_{21} X_2^{\mu e}\right.\right.
$$
$$
\left.\left. -2(A - A_n)\Delta E_{31}\eta_{\mu e} - (A + 2A_n)\Delta E_{31}\{X_3, \eta\}_{\mu e}\right.\right.
$$
$$
\left.\left. +\Delta E_{31}\{X_3, \{\eta, \mathcal{A}\}\}_{\mu e} + \Delta E_{31}\{\eta, \{X_3, \mathcal{A}\}\}_{\mu e}\right\}\right]
$$

$$
= \Delta E_{31} \operatorname{Im}\left[X_3^{\mu e*}\left\{\Delta E_{31}\Delta E_{21} X_2^{\mu e}\right.\right.
$$
$$
\left.\left. +2(A - A_n)\Delta E_{31}\eta_{\mu e} + (A + 2A_n)\Delta E_{31}\{X_3, \eta\}_{\mu e}\right.\right.
$$
$$
\left.\left. -\Delta E_{31}\{X_3, \{\eta, \mathcal{A}\}\}_{\mu e} - \Delta E_{31}\{\eta, \{X_3, \mathcal{A}\}\}_{\mu e}\right\}\right].
$$

Thus, we obtain the expression for the factor $\operatorname{Im}\left[\widetilde{\mathcal{X}}_2^{e\mu} \widetilde{\mathcal{X}}_3^{e\mu*}\right]$:

$$
\operatorname{Im}\left[\widetilde{\mathcal{X}}_2^{e\mu} \widetilde{\mathcal{X}}_3^{e\mu*}\right]
$$

$$
= \frac{1}{\Delta \widetilde{E}_{21}\Delta \widetilde{E}_{31}\Delta \widetilde{E}_{32}}\left(\widetilde{E}_1^2 \operatorname{Im}\left[\mathcal{Y}_1^{e\mu} \mathcal{Y}_2^{e\mu*}\right] - \widetilde{E}_1 \operatorname{Im}\left[\mathcal{Y}_1^{e\mu} \mathcal{Y}_3^{e\mu*}\right] + \operatorname{Im}\left[\mathcal{Y}_2^{e\mu} \mathcal{Y}_3^{e\mu*}\right]\right)
$$

$$
\simeq \frac{\Delta E_{31}}{\Delta \widetilde{E}_{21}\Delta \widetilde{E}_{31}\Delta \widetilde{E}_{32}} \operatorname{Im}\left[X_3^{\mu e*}\left\{\Delta E_{21} X_2^{\mu e} + 4 A \eta_{\mu e} + 2 A_n(\{\eta, X_3\}_{\mu e} - \eta_{\mu e})\right\}\right] \tag{39}
$$

To complete the calculation of Equation (34), we need to estimate the two quantities:

$$
\operatorname{Im}\left[\eta_{\mu e} \widetilde{\mathcal{X}}_3^{e\mu*}\right]
$$

$$
= \frac{1}{\Delta \widetilde{E}_{31}\Delta \widetilde{E}_{32}} \operatorname{Im}\left[\eta_{\mu e}\left\{\widetilde{E}_1 \widetilde{E}_2 \mathcal{Y}_1^{e\mu*} - (\widetilde{E}_1 + \widetilde{E}_2)\mathcal{Y}_2^{e\mu*} + \mathcal{Y}_3^{e\mu*}\right\}\right]
$$

$$
\simeq \frac{1}{\Delta \widetilde{E}_{31}\Delta \widetilde{E}_{32}} \operatorname{Im}\left[\eta_{\mu e}\left(-\widetilde{E}_1 \mathcal{Y}_2^{e\mu*} + \mathcal{Y}_3^{e\mu*}\right)\right]
$$

$$
\simeq \frac{1}{\Delta \widetilde{E}_{31}\lambda_+} \operatorname{Im}\left[\eta_{\mu e}\left(-\lambda - \Delta E_{31} X_3^{\mu e*} + \Delta E_{31}(\Delta E_{31} + A) X_3^{\mu e*}\right)\right]
$$

$$
= \frac{1}{\Delta \widetilde{E}_{31}\lambda_+}\lambda_+ \Delta E_{31} \operatorname{Im}\left[\eta_{\mu e} X_3^{\mu e*}\right]
$$

$$
= \frac{\Delta E_{31}}{\Delta \widetilde{E}_{31}} \operatorname{Im}\left[\eta_{\mu e} X_3^{\mu e*}\right], \tag{40}
$$

$$\mathrm{Im}\left[\eta_{\mu e}\widetilde{\mathcal{X}}_2^{e\mu*}\right]$$

$$= \frac{-1}{\Delta\widetilde{E}_{21}\Delta\widetilde{E}_{32}}\mathrm{Im}\left[\eta_{\mu e}\left\{\widetilde{E}_3\widetilde{E}_1\mathcal{Y}_1^{e\mu*} - (\widetilde{E}_3 + \widetilde{E}_1)\mathcal{Y}_2^{e\mu*} + \mathcal{Y}_3^{e\mu*}\right\}\right]$$

$$\simeq \frac{-1}{\Delta\widetilde{E}_{21}\Delta\widetilde{E}_{32}}\mathrm{Im}\left[-(\widetilde{E}_3 + \widetilde{E}_1)\mathcal{Y}_2^{e\mu*} + \mathcal{Y}_3^{e\mu*}\right]$$

$$\simeq \frac{-1}{\Delta\widetilde{E}_{21}\Delta\widetilde{E}_{32}}\mathrm{Im}\left[-(\widetilde{E}_3 + \widetilde{E}_1)\Delta E_{31}X_3^{\mu e*} + \Delta E_{31}(\Delta E_{31} + A)X_3^{\mu e*}\right]$$

$$\simeq 0\,, \tag{41}$$

where terms of order $O((\Delta E_{21}/\Delta E_{31})^2)$, $O((\epsilon_{\alpha\beta})^2)$, and $O(\epsilon_{\alpha\beta}\Delta E_{21}/\Delta E_{31})$ have been neglected in Equations (39)–(41). Putting Equations (39)–(41) together, the final expression for T violation is given by

$$P(\nu_\mu \to \nu_e) - P(\nu_e \to \nu_\mu)$$

$$\simeq 16\frac{\Delta E_{31}}{A\,\Delta\widetilde{E}_{31}\,\cos^2\theta_{13}}\sin\left(\frac{\Delta\widetilde{E}_{31}L}{2}\right)\sin\left(\frac{\Delta\widetilde{E}_{21}L}{2}\right)\sin\left(\frac{\Delta\widetilde{E}_{32}L}{2}\right)$$

$$\times \mathrm{Im}\left[X_3^{\mu e*}\left\{\Delta E_{21}X_2^{\mu e} + 4\,A\,\eta_{\mu e} + 2A_n\left(\{\eta, X_3\}_{\mu e} - \eta_{\mu e}\right)\right\}\right]$$

$$+ 8\frac{\Delta E_{31}}{\Delta\widetilde{E}_{31}}\sin\left(\Delta\widetilde{E}_{31}L\right)\mathrm{Im}\left[\eta_{\mu e}X_3^{\mu e*}\right] \tag{42}$$

Due to the additional contribution proportional to $\sin(\Delta\widetilde{E}_{31}L)$, the energy dependence of Equation (42) in the scenario with unitarity violation differs from that in the scenarios with unitarity, such as the standard case (23) and the nonstandard interaction case (42). Therefore, if the contribution from unitarity violation is significant enough and the experimental sensitivity is sufficiently high, it may be possible to distinguish the unitarity violation scenario from both the standard and nonstandard interaction scenarios by examining the energy spectrum in T violation.

## 4. Conclusions

In this paper, we have derived the analytical expression for T violation in neutrino oscillations under three different scenarios: the standard three-flavor mixing framework, a scenario involving flavor-dependent nonstandard interactions, and a case with unitarity violation. In scenarios preserving unitarity, the T-violating component of the oscillation probability is proportional to $\sin(\frac{\Delta\widetilde{E}_{31}L}{2})\sin(\frac{\Delta\widetilde{E}_{21}L}{2})\sin(\frac{\Delta\widetilde{E}_{32}L}{2})$. However, in the case with unitarity violation, there is an additional contribution proportional to $\sin(\Delta\widetilde{E}_{31}L)$. Should future long-baseline experiments, such as $\mu$TRISTAN or other types of neutrino factories, achieve high sensitivity to T violation across a broad energy spectrum, it may become feasible to specifically probe unitarity in the $\nu_\mu \leftrightarrow \nu_e$ channel.

Moreover, we demonstrated that the coefficient of the term $\sin(\frac{\Delta\widetilde{E}_{31}L}{2})\sin(\frac{\Delta\widetilde{E}_{21}L}{2})$ $\sin(\frac{\Delta\widetilde{E}_{32}L}{2})(\Delta E_{31})^2(\Delta\widetilde{E}_{31}\Delta\widetilde{E}_{32}\Delta\widetilde{E}_{21})^{-1}$ varies depending on whether neutrino propagation follows the standard scheme or involves nonstandard interactions. In the standard scenario, this coefficient is proportional to $\Delta E_{21} = \Delta m_{21}^2/2E$. However, in the case with nonstandard interactions, there is an additional contribution that is energy-independent. Thus, it may be possible to observe the effects of nonstandard interactions by examining the energy dependence of T violation.

The purpose of this paper is to derive the analytical expression of T violation, and we did not quantitatively discuss the sensitivity of future experiments. The potential for T violation in neutrino oscillations deserves further study.

**Funding:** This research was partly supported by a Grant-in-Aid for Scientific Research of the Ministry of Education, Science and Culture, under Grant No. 21K03578.

**Data Availability Statement:** Data is contained within the article.

**Acknowledgments:** From 1989 to 1991, I was a postdoc at the University of North Carolina at Chapel Hill, and I would like to thank Frampton for giving me the opportunity to conduct research at UNC. I am delighted to celebrate Frampton's 80th birthday and wish him continued success and activity in the years to come.

**Conflicts of Interest:** The authors declare no conflicts of interest.

*Article*

# Flipped Quartification: Product Group Unification with Leptoquarks

James B. Dent [1,*], Thomas W. Kephart [2], Heinrich Päs [3] and Thomas J. Weiler [2]

1 Department of Physics, Sam Houston State University, Huntsville, TX 77341, USA
2 Department of Physics and Astronomy, Vanderbilt University, Nashville, TN 37235, USA; tom.kephart@gmail.com (T.W.K.); tom.weiler@vanderbilt.edu (T.J.W.)
3 Institut für Physik, Technische Universität Dortmund, D-44221 Dortmund, Germany; heinrich.paes@uni-dortmund.de
* Correspondence: jbdent@shsu.edu

**Abstract:** The quartification model is an $SU(3)^4$ extension with a bi-fundamental fermion sector of the well-known $SU(3)^3$ bi-fundamentalfication model. An alternative "flipped" version of the quartification model is obtained by rearrangement of the particle assignments. The flipped model has two standard (bi-fundamentalfication) families and one flipped quartification family. In contrast to traditional product group unification models, flipped quartification stands out by featuring leptoquarks and thus allows for new mechanisms to explain the generation of neutrino masses and possible hints of lepton-flavor non-universality.

**Keywords:** leptoquarks; beyond the standard model; quantification models; early Universe; thermodynamics; phenomenology

## 1. Introduction

The long-term goal of extending the standard model of particle physics is to develop a model that is more predictive than the standard model and to connect it with physics at higher energy scales. Many people have contributed to the progress toward this goal over the last 50 years, and the effort has continued up until this day (Paul Frampton has been a major contributor to this effort. This article is to acknowledge his work and celebrate his 80th birthday). These scales could be in descending order of energy, the Planck scale at $1.2 \times 10^{19}$ GeV, the string scale at about $10^{17}$ GeV, a grand unification scale around $10^{16}$ GeV, or some lower scale where proton decay can be avoided via a partial unification into a product gauge group. The latter two scales are typically set by vacuum expectation values (VEVs) of scalar fields, which give various non-standard model particles their masses but leave the SM fermions and gauge bosons massless. The SM particles themselves remain massless as the energy scale is lowered, until the Higgs scalar electroweak (EW) doubles develop a VEV at 246 GeV. The reason why the EW scale is at such a low energy relative to the Planck scale is one of the key puzzles of the SM, called the hierarchy problem, whose eventual resolution holds great promise for providing a deeper understanding of fundamental physics. Being able to predict other properties of the SM in a systematic way also provides us with the hope that the higher symmetry theory from which the SM descends can eventually be discovered.

Not all the information relevant to extending the SM will necessarily come from particle physics accelerators. The study of particle physics has long been assisted by astronomy, astrophysics, and cosmology. Cosmic rays in particular have played an important role in particle discoveries and searches. The highest energy cosmic rays, while scarce, are still the cause of the highest energy collisions of which we are aware. The origins and acceleration mechanism of these cosmic rays are still unknown, but these rare events are an important window of extreme energies and in turn a potential opportunity to understanding energy scales near unification.

In this paper, we will focus on a particular partial unification into the product gauge group $SU^4(3)$ called a quartification symmetry. Conformal field theories arise naturally as product gauge groups in compactifications of string theories on five dimensional orbifolds (for a review, see Lawrence et al. [1]). Such theories are potentially a way to connect physics at very high energies to physics at energies close to the EW scale. If such a scenario could be fully developed and if it had phenomenological relevance, then it would go a long way to filling out our understanding of a more complete theory of fundamental physics.

Before turning to our particular quartification model, we will first provide some technical remarks to help place it in context. Some phenomenological consequences will be discussed in the final section.

Product group unification schemes include trinification models, refs. [2–30] with gauge group $SU(3)_L \times SU(3)_C \times SU(3)_R$, and quartification models [14,31–38], where the gauge group is extended to $SU(3)_l \times SU(3)_L \times SU(3)_C \times SU(3)_R$, and in both classes of models, the fermions are accommodated in bi-fundamental representations. A generic feature of such product group unification schemes is the absence of leptoquarks, i.e., scalar or vector particles that allow transitions between quarks and leptons. In trinification models, leptons are defined by being bi-fundamental under the two $SU(3)$s that have no color, implying that there are no leptoquarks in such models. Likewise, in traditional quartification models, the $SU(3)$s are arranged in a way that particles have either $SU(3)_l$ or $SU(3)_C$ charges, preventing again the occurrence of leptoquarks. This property can be seen as both a blessing and a curse. On the positive side, the absence of transitions between quarks and leptons avoids the occurrence of various processes triggering fast proton decay, yet on the negative side, leptoquarks are attractive components of models for neutrino mass generation [39–41] and have been invoked to explain recent anomalies which suggested lepton-flavor non-universality [42–44] or both [45–47].

Here, we will concentrate on the phenomenology of a new class of quartification models obtained by "flipping" the $SU(3)_l$ and $SU(3)_R$ groups, which we call "flipped quartification". In contrast to traditional product-group unification schemes, flipped quartification allows for leptoquarks that are bi-fundamental under the $SU(3)_C$ and $SU(3)_l$ groups, albeit confined to the third generation, making them less likely of inducing fast proton decays. In addition, the model also singles out the $b$ quark as different from all the rest of the SM fermions in that, just above the electro-weak (EW) scale, the EW singlet $b_R$ can be in a nontrivial irreducible representation (irrep) of a new gauge group $SU(2)_l$, while all the other SM fermions are in $SU(2)_l$ singlets. This can happen when the $SU(2)_l$ symmetry breaks just above the EW scale where now the $b_R$ falls into its usual SM irrep, but with slightly different phenomenology due to nearby $SU(2)_l$ effects that the other SM fermions do not feel. This is a fairly conventional but interesting scheme for introducing new physics into the SM.

All quartification models contain an $SU(3)_l$ leptonic color sector to realize a manifest quark–lepton symmetry [48–50] and must contain at least three families to be phenomenologically viable, plus they contain the new fermions needed to symmetrize the quark and lepton particle content at high energies. Instead of fully quartified models, where all families are quartification families given by

$$3[(3\bar{3}11) + (13\bar{3}1) + (113\bar{3}) + (\bar{3}113)],\tag{1}$$

we will consider only hybrid models

$$n[(13\bar{3}1) + (113\bar{3}) + (1\bar{3}13)] + (3-n)[(3\bar{3}11) + (13\bar{3}1) + (113\bar{3}) + (\bar{3}113)].\tag{2}$$

where $n > 0$ families are trinification families and the the remaining $3 - n$ are quartification families. In particular, we concentrate on the $n = 2$ case [36]. One important thing to note here is that both the trinification and quartification family components of these models can be represented by quiver diagrams which are anomaly-free [51]. For models with only bi-fundamental fermions, there are no chiral gauge anomalies since for each **3** there is a **3̄** with equal and opposite charges. Furthermore, the descendent gauge groups are also guaranteed to be free of anomalies upon breaking the initial gauge symmetry with the 't Hooft matching conditions [52].

One can derive three family models with appropriate scalar content to permit gauge symmetry breaking to the SM and ultimately to $SU(3)_C \times U_{EM}(1)$ from orbifolded $AdS \otimes S^5$ (for a review, see [51]). In [53,54], two of us carried out a global search for $\Gamma = Z_n$ trinification models with three or more families, and in [36], quartification models of this type were derived from a $\Gamma = Z_8$ orbifolded $AdS \otimes S^5$. We leave the study of the UV completion of the present model for later work.

## 2. Flipped 2 + 1 Quartification Model

Under the original quartification gauge group $SU(3)_l \times SU(3)_L \times SU(3)_C \times SU(3)_R$, the representations of the two trinification plus one quartification family model (the 2 + 1 quartification model of reference [36]) were given by

$$2[(1\bar{3}\bar{3}1) + (11\bar{3}\bar{3}) + (1\bar{3}1\bar{3})] + [(\bar{3}311) + (1\bar{3}\bar{3}1) + (11\bar{3}\bar{3}) + (\bar{3}11\bar{3})] \tag{3}$$

We now "flip" the $R$ and $l$ designations such that

$$lLCR \to RLCl. \tag{4}$$

We are free to cyclically permute the groups and to reverse their order without changing the physics. Thus, we let

$$RLCl \to CLRl \tag{5}$$

which allows us to write our new 2 + 1 flipped quartification model in a form that conforms with the notation of earlier work. Symmetry breaking can easily be arranged with a single adjoint scalar VEV for each of $SU(3)_L$ and $SU(3)_l$ and a pair of adjoints for $SU(3)_R$ such that

$$SU(3)_L \qquad \to SU(2)_L \times U(1)_A \tag{6}$$

$$SU(3)_R \qquad \to U(1)_B \times U(1)_C \tag{7}$$

$$SU(3)_l \qquad \to SU(2)_l \times U(1)_D \tag{8}$$

where the charge operators $A$, $C$, and $D$ are of the form $diag(1, 1, -2)$ and $B$ is of the form $diag(1, -1, 0)$. Their weighting in forming weak hypercharge will be provided below.

To be more specific, the symmetry breaking from $SU(3)^4$ to $SU(3)_C \times SU(2)_L \times U(1)^4 \times SU(2)_l$ can be carried out with four adjoints $(1, 8, 1, 1)^H$, $(1, 1, 8, 1)^H$, $(1, 1, 8, 1)^H$ and $(1, 1, 1, 8)^H$ which break $SU(3)_L$ to $SU(2)_L \times U(1)$, $SU(3)_R$ to $U(1)^2$ and $SU(3)_l$ to $SU(2)_l \times U(1)$, respectively. The remaining $U(1)$s are broken by appropriately charged singlets of the respective groups. The standard model scalar doublet can come from an $(1, 3, 1, 1)^H$ irrep of $SU(3)_L$ to yield the standard model Higgs or in the present notation an $(1, 2, 1, 1)^H$. No light scalars are required beyond the SM Higgs.

Under the symmetry group $SU(3)_C \times SU(2)_L \times SU(2)_l \times U(1)_A \times U(1)_B \times U(1)_C \times U(1)_D$, the first two families decompose as in a standard trinification model,

$$(\bar{3}311) \to (321)_{-1000} + (311)_{2000} \tag{9}$$
$$(1\bar{3}\bar{3}1) \to (121)_{1-1-10} + (121)_{11-10} + (121)_{1020} + (111)_{-2-1-10} + (111)_{-21-10} + (111)_{-2020}$$
$$(\bar{3}131) \to (\bar{3}11)_{0110} + (\bar{3}11)_{0-110} + (\bar{3}11)_{00-20}$$

while the third family representations become

$$(3\bar{3}11) \rightarrow (321)_{-1000} + (311)_{2000} \tag{10}$$
$$(13\bar{3}1) \rightarrow (121)_{1-1-10} + (121)_{11-10} + (121)_{1020} + (111)_{-2-1-10} + (111)_{-21-10} + (111)_{-2020}$$
$$(113\bar{3}) \rightarrow (112)_{011-1} + (112)_{0-11-1} + (112)_{00-2-1} + (111)_{0112} + (111)_{0-112} + (111)_{00-22}$$
$$(\bar{3}113) \rightarrow (\bar{3}12)_{0001} + (\bar{3}11)_{000-2}.$$

Using the relation

$$Q = T_3 + Y \tag{11}$$

where $Q$ is the electric charge, $T_3$ is the third component of isospin, and $Y$ is the hypercharge, we can determine the hypercharge in terms of the $U(1)$ charges (designated by $A$, $B$, $C$, and $D$) as

$$Y = -\frac{1}{6}A + \frac{1}{2}B - \frac{1}{6}C + \frac{1}{3}D. \tag{12}$$

Charged singlets can be used to break $U(1)_A \times U(1)_B \times U(1)_C \times U(1)_D$ to the standard weak hypercharge $U(1)_Y$, resulting in

$$(3\bar{3}11) \rightarrow (321)_{\frac{1}{6}} + (311)_{-\frac{1}{3}} \tag{13}$$
$$(13\bar{3}1) \rightarrow (121)_{-\frac{1}{2}} + (121)_{\frac{1}{2}} + (121)_{\frac{1}{2}} + (111)_0 + (111)_1 + (111)_0$$
$$(\bar{3}131) \rightarrow (\bar{3}11)_{\frac{1}{3}} + (\bar{3}11)_{-\frac{2}{3}} + (\bar{3}11)_{\frac{1}{3}}$$

for the first two families, where as usual, each trinification family contains an SM family

$$Q_L^{1(2)} + d(s)_R + u(c)_R + l_L^{1(2)} + e(\mu)_R = (321)_{\frac{1}{6}} + (\bar{3}11)_{\frac{1}{3}} + (\bar{3}11)_{-\frac{2}{3}} + (121)_{\frac{1}{2}} + (111)_1 \tag{14}$$

plus the following vector-like states:

$$+(\bar{3}11)_{\frac{1}{3}} + (311)_{-\frac{1}{3}} + (121)_{-\frac{1}{2}} + (121)_{\frac{1}{2}} + (111)_0 + (111)_0. \tag{15}$$

The third family in Equation (10) becomes

$$(3\bar{3}11) \rightarrow (321)_{\frac{1}{6}} + (311)_{-\frac{1}{3}} \tag{16}$$
$$(13\bar{3}1) \rightarrow (121)_{-\frac{1}{2}} + (121)_{\frac{1}{2}} + (121)_{\frac{1}{2}} + (111)_0 + (111)_1 + (111)_0$$
$$(113\bar{3}) \rightarrow (112)_0 + (112)_{-1} + (112)_0 + (111)_1 + (111)_0 + (111)_1$$
$$(\bar{3}113) \rightarrow (\bar{3}12)_{\frac{1}{3}} + (\bar{3}11)_{-\frac{2}{3}}$$

which we rearrange in a more suggestive form

$$(321)_{\frac{1}{6}} + (\bar{3}11)_{-\frac{2}{3}} + (121)_{\frac{1}{2}} + (111)_1 \tag{17}$$
$$+(\bar{3}12)_{\frac{1}{3}} + [(112)_0 + (112)_0] + [(112)_{-1} + (111)_1 + (111)_1] + (111)_0$$
$$+(311)_{-\frac{1}{3}} + [(121)_{-\frac{1}{2}} + (121)_{\frac{1}{2}}] + (111)_0 + (111)_0.$$

The first line of Equation (17) contains an SM family except that $b_R$ is missing. The second line contains states in nontrivial $SU(2)_I$ irreps and their natural partners, and the last line contains the remaining states.

In order to complete the third SM family, a $(\bar{3}11)_{\frac{1}{3}}$ from the second line must be moved to the first line. To perform this, we can either (i) break $SU(2)_I \rightarrow 0$ at a scale $M_{ssb}$ or (ii) arrange to have the gauge coupling of $SU(2)_l$ run to large values, where at some scale $\Lambda_l$ this group becomes confining. We expect the lower bounds on $M_{ssb}$ and $\Lambda_l$ to be similar.

To complete the third family via spontaneous symmetry breaking, we introduce a scalar $SU(2)_l$ doublet $(1, 1, 2)_0$ whose VEV breaks $SU(2)_l$ completely so that $(\bar{3}12)_{\frac{1}{3}} \rightarrow (\bar{3}11)_{\frac{1}{3}} + (\bar{3}11)_{\frac{1}{3}}$. One of these two irreps can be identified with the $b_R$, hence completing the third family in the first line of Equation (17). The other we identify as the $b'_R$, which pairs with the $(\bar{3}11)_{-\frac{1}{3}}$ in the third line of Equation (17). The chargeless $SU(2)_l$ doublet leptonic states in the second line of Equation (17) also split into singlets, while the charge $-1$ doublet $SU(2)_l$ irreps split so that they can pair with the charge $+1$ singlet leptons in that line. Writing Equation (17) after the symmetry breaking, where we have moved half the split $(\bar{3}12)_{\frac{1}{3}}$ irrep into the first line and the other half into the third line gives

$$(321)_{\frac{1}{6}} + (\bar{3}11)_{\frac{1}{3}} + (\bar{3}11)_{-\frac{2}{3}} + (121)_{\frac{1}{2}} + (111)_1 \tag{18}$$
$$+ [(111)_0 + (111)_0 + (111)_0 + (111)_0] + [(111)_{-1} + (111)_{-1} + (111)_1 + (111)_1] + (111)_0$$
$$+ (\bar{3}11)_{\frac{1}{3}} + (\bar{3}11)_{-\frac{1}{3}} + [(121)_{-\frac{1}{2}} + (121)_{\frac{1}{2}}] + (111)_0 + (111)_0$$

The SSB has yielded a standard third family in the first line, states with identical charges to the extra trinification family in the third line, plus the new extra states of a quartification family in the second line. In the following, we concentrate on the properties of the $b$ quark.

Note that all three families have an extra $d'$ type quark in $(\bar{3}11)_{-\frac{1}{3}} + (\bar{3}11)_{\frac{1}{3}}$, which is typical of all trinification or $E_6$ models. For the first two families, these are in vector-like representations, so these particles can acquire mass at a high scale, and we will not discuss them further. However, in the third family, the $b'$ can not acquire a mass until $SU(2)_l$ is broken. Thus, the third family $b'$ is phenomenologically more interesting.

As we are completing the third family via spontaneous symmetry breaking, at some scale $M$, then the only chiral fermions below that scale are in the standard families. All the rest are vector-like; see Equation (18), and obtain masses around the scale $M_{ssb}$.

## 3. Phenomenological Implications

For spontaneous symmetry breaking of $SU(2)_l$, we find a phenomenology that is a straightforward extension of the SM: it contains the normal SM particle content in the first two families plus their trinification extension. The third quartified family contains a third normal family, its extended trinification content, plus the remaining extended quartification content composed of two $SU(2)_L$ singlet unit electric charged leptons and five Weyl neutrinos, some of which can be paired up after SSB.

*Extended $Z'$ bosons sector:* The gauge group of our $SU(3)^4$ flipped quartification model is rank 8, while the standard model is rank 4, so FQ has four additional uncharged $Z'$-like gauge bosons. Depending on how the spontaneous symmetry breaking proceeds, their masses can range from the initial $SU(3)^4$ breaking scale down to the current experimental limit on $Z'$ masses. The four $Z'$ masses can all be different within these bounds. We have yet to explore the full parameter space of allowed FQ models, so we are reluctant to give the full set of constraints on the $Z'$s yet, but we hope to come back to this interesting phenomenological question in future work.

*Leptoquarks, Hints for Lepton-Flavor-Non-Universality, and the Muon Anomaly:* A characteristic property of the "flipping" in the order of the quartification gauge group within the present construction is the likelihood of the presence of light leptoquarks. This could be realized by the scalar or vector representation that couple terms in the bi-fundamental fermions that are nontrivial in $SU(3)_C$ with those in $SU(2)_L$. (Bileptons and/or biquarks could also be present. For a full classification, see [55].) Leptoquarks have been a popular possibility to explain the recent $b$-physics anomalies pointing at lepton-flavor non-universality (see, e.g., [42,56]), though recent results from the LHCb collaboration are consistent with Standard Model predictions [57,58]. Regardless of these recent collider results, leptoquarks have a rich phenomenology that will continue to be explored in BSM scenarios of flavor physics and neutrino mass origins, to name a few (see, for example, ref. [59] for a review of the varieties of leptoquark phenomenology). For the most recent experimental results on leptoquarks, see the publications from ATLAS [60–62] and CMS [63–65].

There is another interesting leptoquark possibility in the flipped quartification model discussed above. Since the third family has $SU(3)_l$ quantum numbers, there is also the possibility of vector leptoquark contributions from this sector. Likewise, there are potential scalar $SU(3)_l$ leptoquarks if we were to add the appropriate scalar irreps.

An interesting result is Fermilab's recent confirmation of an anomalous result for the magnetic moment of the muon [66]. In [67], it had been shown that the anomalous magnetic moment of the muon could be explained by adding a vector-like doublet plus a scalar singlet to the particle content of the SM. In the present model the states $(112)_0 + (\overline{112})_0$ in the second line of Equation (17) can play the role of the vector-like doublet. See also [68].

Finally, while the model we have presented can be used to focus on *B* physics, other models in this class can be used to single out one or more right-handed charge $-\frac{1}{3}$ quarks. Then right-handed quarks are made to fall into flipped quartification families, while the remaining right-handed charge $-\frac{1}{3}$ quarks remain in trinification families. Future work can potentially lead to a whole class of models similar to flipped quartification where one or more fermions are singled out to differ from other normal family members, hence providing a rich and interesting BSM phenomenology.

Changing the model of particle physics to the FQ model has implications for astrophysics and cosmology. For instance, let us compare the thermodynamics of the early Universe for $SU(5)$ unification with that of the FQ model. When the $SU(5)$ gauge group breaks at a high scale, there is typically a first-order phase transition that produces magnetic monopoles and also causes inflation, after which the Universe evolves adiabatically until SM symmetry breaking. By contrast, the FQ model can undergo many phase transitions and have a much more complicated thermodynamics. Starting from $SU(3)^4$ and identifying one $SU(3)$ group as color, the other three can each break to $SU(2) \times U(1)$, and then two of the $SU(2)$s can break to $U(1)$s. At this stage, the gauge group is $SU(3) \times SU(2) \times U(1)^5$. Then the $U(1)^5$ part must break to $U(1)_Y$. These symmetry breakings can occur sequentially or some can happen concurrently. Some of these phase transitions can be first order, leading to particle production and entropy production. All the symmetry breakings that produce a $U(1)$ produce monopoles. All $U(1)$ breakings can produce cosmic strings. Monopole–antimonopole pairs can annihilate if they are at the end of a string. It is clear that the thermodynamics of the early Universe for FQ models can have a wide variety of implications for astro-particle physics and cosmology, all worthy of future study.

## 4. Summary

In this paper, we have discussed a novel class of quartification models with the curious feature that they—in contrast to traditional product group unification schemes—allow for the occurrence of leptoquarks and thus an interesting phenomenology for neutrino mass generation and other beyond-the-Standard-Model processes, such as lepton-flavor non-universality.

There are many FQ model building options to consider. The FQ model should be thought of as string-inspired, not string-derived. This allows us more leeway to explore potential phenomenologies; e.g., there are multiple ways to break the extra $U(1)$s to the required hypercharge $U(1)_Y$. Since it is a linear combination of $U(1)$s coming from different $SU(3)$s, this requires charged scalar representations living in multiple $SU(3)$s, e.g., bifundamental Higgses like $(1, 1, 3, \bar{3})_H$, etc. Such representations would arise naturally in, say, an $\mathcal{N} = 1$ SUSY $S^5/Z_4$ orbifolding.

From a purely phenomenological perspective, we could use any Higgs we like. Different choices lead to different symmetry breaking scenarios with different mass scales for the breaking. The breaking of $SU(3)^4$ to the non-abelian part of the standard model gauge group is easily accomplished with scalar octets, so that part of the phenomenology should be straightforward.

The $SU(3)^4$ scale is set by how much the gauge couplings need to run to get to their SM values. The initial values of the $SU(3)^4$ gauge couplings can be set by the fact that these $SU(3)$s can be in diagonal subgroups of some larger group, $SU(3)^p \times SU(3)^q \times SU(3)^r \times SU(3)^s$, where the ratios of $p/q/r/s$ set the initial values of these FQ $SU(3)$ couplings [69]. Consequently we can raise or lower the initial values by changing the string model orbifolding group, $\Gamma = Z_n$, where $n = p + q + r + s$.

This multitude of possibilities makes the FQ models a rich source of phenomenology, but it will take a dedicated effort to explore all the parameter space and to optimize the model with respect to new phenomenology and simple economical patterns of SSB. For this reason we have added an overview of phenomenological possibilities and SSBs in the final section of the manuscript but have not committed to a specific model. We are reluctant to make such a choice here before we feel comfortable with having accomplished all we can to select the best model. Hence, we believe this is better left to future work once we have fully explored all the options.

Within this model, we found that the third family of the Standard Model can be completed via spontaneous symmetry breaking of an unbroken $SU(2)_l$. Completion via spontaneous symmetry breaking leads to interesting leptoquarks and bileptons coupled only to the third family, which can potentially avoid proton decay but still extend standard model phenomenology. We leave the details to future work. Beyond the possibility of having leptoquarks, if the $SU(2)_l$ group becomes confining at a high scale, it leads to a possible composite $b$ quark. However, we have yet to build a successful phenomenology from this prospective and leave further considerations along these lines to future work.

**Author Contributions:** Writing—original draft, J.B.D., T.W.K., H.P. and T.J.W. All authors have read and agreed to the published version of the manuscript.

**Funding:** This research received no external funding.

**Institutional Review Board Statement:** Not applicable.

**Data Availability Statement:** The original contributions presented in the study are included in the article, further inquiries can be directed to the corresponding author.

**Acknowledgments:** We thank Gudrun Hiller for useful discussions about partial compositeness and flavor anomalies. This work was supported in part by the US Department of Energy under Grants DE-FG05-85ER40226 (JBD and TWK), DE-SC-0019235 (TWK), DE-SC-001198(TJW), and DE-FG03-91ER40833 (HP). TWK and HP thank the Aspen Center for Physics for hospitality, where this research was initiated some time ago. JBD acknowledges support from the National Science Foundation under Grant No. NSF PHY182080. This paper is dedicated to Paul Frampton's 80th birthday and to our coauthor, colleague, and friend Tom Weiler, who passed away in December of 2023 as we were finalizing this work.

## entropy

*Article*

# Status of Electromagnetically Accelerating Universe

## Paul H. Frampton

Dipartimento di Matematica e Fisica "Ennio De Giorgi", Università del Salento and INFN-Lecce, Via Arnesano, 73100 Lecce, Italy; paul.h.frampton@gmail.com

**Abstract:** To describe the dark side of the universe, we adopt a novel approach where dark energy is explained as an electrically charged majority of dark matter. Dark energy, as such, does not exist. The Friedmann equation at the present time coincides with that in a conventional approach, although the cosmological "constant" in the Electromagnetic Accelerating Universe (EAU) Model shares a time dependence with the matter component. Its equation of state is $\omega \equiv P/\rho \equiv -1$ within observational accuracy.

**Keywords:** accelerated expansion; dark matter; electromagnetism

## 1. Introduction to the EAU Model

Theoretical cosmology is at an exciting stage because about 95% of the energy in the Visible Universe remains incompletely understood. The 25% which is dark matter has constituents whose mass is unknown by over one hundred orders of magnitude. The 70% which is dark energy is, if anything, more mysterious. Although it can be parametrised by a cosmological constant with an equation of state $\omega = -1$, which provides an excellent phenomenological description, that is only a parametrisation and not a complete understanding.

In the present paper, we address the issues of dark matter and dark energy using a novel approach. We use only the classical theories of electrodynamics and general relativity. We shall not employ any knowledge of quantum mechanics or of theories describing short-range strong and weak interactions.

This paper may be regarded as a follow-up to our 2018 paper [1] entitled *On the Origin and Nature of Dark Matter* and we could have simply added *and Energy* to that title. We have, however, chosen *Status of Electromagnetic Accelerating Universe* because it more accurately characterises our present emphasis on the EAU model whose main idea is that electromagnetism dominates over gravitation in the explanation of the accelerating cosmological expansion. This idea takes us beyond the first paper [2] that applied general relativity to theoretical cosmology. This is not surprising, since in 1917, that author was obviously unaware of the fact [3,4] that is was discovered only in 1998 that the rate of cosmological expansion is accelerating.

The make up of this paper is that primordial black holes are discussed in Section 2, then primordial naked singularities are discussed in Section 3. Finally, in Section 4, there is a discussion.

## 2. Primordial Black Holes (PBHs)

Black holes may be classified into those which arise from the gravitational collapse of stars and others, which do not. We shall refer to all of the others as primordial. In general, PBHs with masses up to $10^5 M_\odot$ are expected to be formed during the first second after the Big Bang and arise from inhomogeneities and fluctuations of spacetime. The existence of PBHs was first proposed [5] by Novikov and Zeldovich and independently seven years later in the West by Carr and Hawking [6]. The idea that dark matter constituents are PBHs was first suggested by Chapline [7].

Shortly after the original presentation of general relativity [8–10], a metric describing a static black hole of mass M with zero charge and zero spin was discovered by Schwarzschild [11] in the form

$$ds^2 = -\left(1 - \frac{r_S}{r}\right)dt^2 + \left(1 - \frac{r_S}{r}\right)^{-1}dr^2 + r^2d\Omega^2 \tag{1}$$

Shortly thereafter, the Reissner–Nordstrom metric [12,13] for a static black hole with electric charge was found. It then took a surprising forty-five years until Kerr cleverly found a metric [14] of general relativity corresponding to such a solution with spin. We shall not discuss the case of non-zero spin in the present paper because, although we expect that all the objects we discuss do spin in nature, according to the calculations in [15], which use Kerr's generalisation, spin is an inessential complication in all of our subsequent considerations.

### 2.1. Primordial Intermediate Mass Black Holes (PIMBHs) as Galactic Dark Matter

Global fits to cosmological parameters have led to a consensus that about one quarter of the energy of the universe is in the form of electrically neutral dark matter. It seemed natural to propose [16] that in a galaxy like the Milky Way are between ten million and ten billion primordial black holes with masses between one hundred and one hundred thousand solar masses.

Black holes in this range of masses are naturally known as intermediate mass black holes (IMBHs) since they lie as an intermediate between the masses of stellar mass black holes and the masses of the supermassive black holes at galactic centers.

The existence of stellar mass black holes in nature was established sixty years ago in 1964 by the discovery in Cygnus X-1 of a black hole with a mass of about $15M_\odot$. Such X-ray binaries were studied in [17] and then in [18] and appear in the mass range between $5M_\odot$ and $100M_\odot$.

The existence of dark matter was first discovered by Zwicky [19,20] in 1933 in the Coma Clusters and its presence in individual galaxies was demonstrated convincingly by Rubin in the 1970s from the measurement of rotation curves, which demanded the existence of additional matter to what was luminous [21].

The PBH mass function is all important. Possible PBH masses extend upwards to many solar masses and without any obvious upper limit, far beyond what was thought possible in the twentieth century, when ignorance about PBHs with many solar masses probably prevented the MACHO [22] and EROS [23] collaborations from discovering a larger fraction of dark matter.

Black holes formed by gravitational collapse cannot satisfy $M_{BH} \ll M_\odot$ because stars powered by nuclear fusion cannot be far below $M = M_\odot$. This was contradicted by the studies in [5,6], which suggested that much lighter black holes can be produced in the earliest stages of the Big Bang.

Such PBHs are of special interest for several reasons. Firstly, they are the only type of black hole that can be so light, down to $10^{12}$ kg $\sim 10^{-18}M_\odot$, that Hawking radiation might conceivably be detected. Secondly, PBHs in the intermediate mass region $100M_\odot \leq M_{IMBH} \leq 10^5M_\odot$ can provide galactic dark matter.

The mechanism of PBH formation involves large fluctuations or inhomogeneities. Carr and Hawking [6] argued that we know there are fluctuations in the universe in order to seed structure formation and there must similarly be fluctuations in the early universe. Provided the radiation is compressed to a high density, meaning to a radius as small as its Schwarzschild radius, a PBH will form. Because the density in the early universe is extremely high, it is very likely that PBHs will be created. The two necessities are high density, which is guaranteed, and large inhomogeneities, which are possible.

During radiation domination,

$$a(t) \propto t^{1/2} \tag{2}$$

and

$$\rho_\gamma \propto a(t)^{-4} \propto t^{-2} \tag{3}$$

Ignoring factors $O(1)$ and bearing in mind that the radius of a black hole is

$$r_{BH} \sim \left( \frac{M_{BH}}{M_{Planck}^2} \right) \tag{4}$$

with

$$M_{Planck} \sim 10^{19} GeV \sim 10^{-8} kg \sim 10^{-38} M_\odot \tag{5}$$

and using the Planck density $\rho_{Planck}$

$$\rho_{Planck} \equiv (M_{Planck})^4 \sim (10^{-5}g)(10^{-33}cm)^{-3} = 10^{94} \rho_{H_2O} \tag{6}$$

the density of a general black hole $\rho_{BH}(M_{BH})$ is

$$\rho_{BH}(M_{BH}) \sim \left( \frac{M_{BH}}{r_{BH}^3} \right) = \rho_{Planck} \left( \frac{M_{Planck}}{M_{BH}} \right)^2 \sim 10^{94} \rho_{H_2O} \left( \frac{10^{-38} M_\odot}{M_{BH}} \right)^2 \tag{7}$$

which means that for a solar mass black hole

$$\rho_{BH}(M_\odot) \sim 10^{18} \rho_{H_2O} \tag{8}$$

while for a billion solar mass black hole

$$\rho_{BH}(10^9 M_\odot) \sim \rho_{H_2O}. \tag{9}$$

and above this mass, the density falls as $M_{BH}^{-2}$.

The mass of the PBH is derived by combining Equations (3) and (7). We see from these two equations that $M_{PBH}$ grows linearly with time, and using Planckian units or solar units, we find, respectively,

$$M_{PBH} \sim \left( \frac{t}{10^{-43} sec} \right) M_{Planck} \sim \left( \frac{t}{1 sec} \right) 10^5 M_\odot \tag{10}$$

which implies that if we insisted on PBH formation before the electroweak phase transition, $t < 10^{-12}s$, that

$$M_{PBH} < 10^{-7} M_\odot \tag{11}$$

Such an upper bound as that in Equation (11) explains why the MACHO searches at the turn of the twenty-first century [22,23], inspired by the clever suggestion of Paczynski [24], lacked motivation to pursue searching above $100 M_\odot$, because it was thought incorrectly at that time that PBHs were far too light. It was known correctly that the results of the gravitational collapse of normal stars, or even large early stars, are below $100 M_\odot$. Supermassive black holes with $M > 10^6 M_\odot$ such as $Sgr A^*$ in the Milky Way were beginning to be discovered in galactic centers but their origin was unclear and this will be discussed further in Section 2.2.

Using the mechanism for Hawking radiation provides the lifetime for a black hole evaporating *in vacuo* given by

$$\tau_{BH} \sim \left( \frac{M_{BH}}{M_\odot} \right)^3 \times 10^{64} years \tag{12}$$

so that to survive to the age $10^{10}$ years of the universe, there is a lower bound on $M_{PBH}$ to augment the upper bound in Equation (11), giving, as the full range of Carr–Hawking PBHs,

$$10^{-18}M_\odot < M_{PBH} < 10^{-7}M_\odot \tag{13}$$

The lowest mass possible for s surviving PBH in Equation (13) has the density $\rho \sim 10^{58}\rho_{H_2O}$. It is an object which has the physical size of a proton and the mass of Mount Everest.

The Hawking temperature $T_H(M_{BH})$ of a black hole is given by

$$T_H(M_{BH}) = 6 \times 10^{-8}K\left(\frac{M_\odot}{M_{BH}}\right) \tag{14}$$

which would be above the CMB temperature, and hence there would be outgoing radiation for all of the cases with $M_{BH} < 2 \times 10^{-8}M_\odot$. Hypothetically, if the dark matter halo were made entirely of the brightest possible (in terms of Hawking radiation) $10^{-18}M_\odot$ PBHs, the expected distance to the nearest PBH would be about $10^7$ km. Although the PBH temperature, according to Equation (14), is $\sim 6 \times 10^{10}K$, the inverse square law renders the intensity of Hawking radiation too small, by many orders of magnitude, to allow for detection by any foreseeable terrestrial apparatus.

The originally suggested mechanism produces PBHs with masses in the range up to $10^{-7}M_\odot$. We shall now discuss the formation of far more massive PBHs by a rather different mechanism. As already discussed, PBH formation requires very large inhomogeneities. Here, we shall illustrate how mathematically to produce inhomogeneities that are exponentially large.

In the simplest single-stage inflation, no exceptionally large-density perturbation is expected. Therefore, it is necessary to consider at least a two-stage hybrid inflation with respective fields called [25], inflaton, and waterfall. The idea then involves parametric resonance in that, after the first of the two stages of inflation, mutual couplings of the inflaton and waterfall fields cause both to oscillate arbitrarily wildly and produce perturbations which can grow exponentially. A second (waterfall) inflation then stretches the inhomogeneities further, thus enabling the production of PBHs with an arbitrarily high mass. This specific model may not describe nature but provides an existence theorem to confirm that arbitrarily large-mass PBHs can be produced mathematically. The resulting mass function is spiked, but it is possible that other PBH production mechanisms can produce a smoother mass function.

The full details of the model are presented in [26], where the inflaton and waterfall fields are denoted by $\sigma$ and $\psi$, respectively. Between the two stages of inflation, the $\sigma$ and $\psi$ fields oscillate, decaying into their own and mutual couplings. Specific modes of $\sigma$ and $\psi$ are amplified by parametric resonance. The resulting coupled equations for the two fields are of the Mathieu type with exponentially growing solutions. The numerical solution shows that the peak wave number $k_{peak}$ is approximately linear in $m_\sigma$. The resultant PBH mass, the horizon mass when the fluctuations re-enter the horizon, is approximately

$$M_{PBH} \sim 1.4 \times 10^{13}M_\odot \left(\frac{k_{peak}}{Mpc^{-1}}\right)^{-2} \tag{15}$$

Explicit plots were exhibited in [26] for the cases $M_{PBH} = 10^{-8}M_\odot, 10^{-7}M_\odot$ and $10^5 M_\odot$. At that time (2010), although not included in the paper, it was confirmed that parameters can always be chosen such that arbitrarily high-mass PBHs, at or even beyond the mass of the universe, may be produced. This is an important result to be borne in mind.

In the PBH production mechanism based on hybrid inflation with parametric resonance, the mass function is generally sharply spiked at a specific mass region. Such a peculiar mass function is not expected to be a general feature of PBH formation, only a property of this specific mechanism. But this specific mechanism readily demonstrates the possibility of the primordial formation of black holes with many solar masses. For com-pleteness, it should be pointed out that PBHs with masses up to $10^{-15} M_\odot$ were discussed even in the 1970s, for example, by Carr [27] and by Novikov, Polnarev, Starobinskii, and Zeldovich [28].

For dark matter in galaxies, PIMBHs are important, where the upper end must be truncated at $10^5 M_\odot$ to stay well away from galactic disk instability, first discussed by Ostriker et al [29]. They showed convincingly that an object with a mass one million solar masses out in the spiral arms of the Milky Way destabilizes the galactic disk to such an extent that the entire galaxy collapses.

The observations of rotation curves reveal that the dark matter in galaxies including the Milky Way fills out an approximately spherical halo somewhat larger in radius than the disk occupied by the luminous stars. Numerical simulations of structure formation suggest a profile of the dark matter of the NFW type [30]. Note that the NFW profile is independent of the mass of the dark matter constituent and the numerical calculations are restricted by the available computer size, for a system as large as a typical galaxy, to constituents which have many solar masses.

In our discussion a decade ago [16], we focused on galaxies like the Milky Way and restricted the mass range for dark matter constituents to lie within three orders of magnitude:

$$10^2 M_\odot < M < 10^5 M_\odot \tag{16}$$

We shall not repeat the lengthy entropy arguments in [16] here, just that the constituents were proposed to be primordial intermediate mass black holes, PIMBHs.

Assuming a total dark halo mass of $10^{12} M_\odot$, Equation (16) implies that the number of PIMBHs is between ten million ($10^7$) and ten billion ($10^{10}$). Assuming further that the dark halo has a radius $R$ of a hundred thousand ($10^5$) light years, the mean separation $\bar{L}$ of PIMBHs can then estimated by

$$\bar{L} \sim \left( \frac{R}{N^{1/3}} \right) \tag{17}$$

which translates approximately to

$$100 ly < \bar{L} < 1000 ly \tag{18}$$

which also provides a reasonable estimate of the distance to the nearest PIMBH from the Earth, which is very far outside the Solar System where the orbital radius of the outermost planet Neptune is $\sim 0.001$ ly.

To an outsider, it may be surprising that millions of intermediate mass black holes in the Milky Way have remained undetected. Ironically, they could have been detected more than two decades ago had the MACHO collaboration [22] persisted in its microlensing experiment at Mount Stromlo Observatory in Australia.

Dark matter was first discovered almost a century ago by Zwicky [19,20] in the Coma cluster, a large cluster at 99 Mpc containing over a thousand galaxies and with a total mass estimated at $6 \times 10^{14} M_\odot$ [31]. Convincing proof of the existence of cluster dark matter was provided by the Bullet cluster collision, where the distinct behaviours of the X-ray-emitting gas which did, and the dark matter which did not, was observable [32–34].

Since there is not the same disk stability limit [29] as for galaxies, the constituents of cluster dark matter can also involve PSMBHs up to much higher masses than those possible for the PIMBHs within galaxies.

The possible solution of the galactic dark matter problem cries out for experimental verification. Three methods have been discussed: wide binaries, distortion of the CMB, and microlensing. Of these, microlensing seems the most direct and promising. Microlensing experiments were carried out by the MACHO [22] and EROS [23] collaborations decades ago. At that time, it was believed that PBH masses were below $10^{-7} M_\odot$ by virtue of the Carr–Hawking mechanism. Heavier black holes could, it was then believed, arise only from the gravitational collapse of normal stars, or heavier early stars, and would have a mass below $100 M_\odot$.

For this reason, there was no motivation to suspect that there might be MACHOs which led to higher-duration microlensing events. The longevity, $\hat{t}$, of an event is

$$\hat{t} = 0.2 yrs \left( \frac{M_{PBH}}{M_\odot} \right)^{\frac{1}{2}} \tag{19}$$

which assumes a transit velocity of 200 km/s. Subsiting our extended PBH masses, one finds approximately $\hat{t} \sim 6, 20, 60$ years for $M_{PBH} \sim 10^3, 10^4, 10^5 M_\odot$, respectively, and searching for light curves with these higher values of $\hat{t}$ could be rewarding.

It is to be hoped that MACHO searches will soon resume at the Vera Rubin Observatory and focus on highest-longevity microlensing events. Is it possible that convincing observations showing only a fraction of a light curve could suffice? If so, only a fraction of the six years, for example, corresponding to PIMBHs with one thousand solar masses, could be enough to confirm the theory.

### 2.2. Primordial Supermassive Black Holes (PSMBHs) at Galactic Centers

Evidence for supermassive black holes at galactic centers arises from the observations of fast-moving stars around them and such stars being swallowed or torn apart by the strong gravitational field. The first discovered SMBH was Sgr $A^*$, at the core of the Milky Way, which was discovered in 1974 and has a mass $M_{Sgr A*} \sim 4.1 \times 10^6 M_\odot$. The SMBH at the core of the nearby Andromeda galaxy ($M31$) has a mass $M = 2 \times 10^8 M_\odot$, fifty times $M_{Sgr A*}$. The most massive core SMBH so far observed is for NGC4889, with a mass of $M \sim 2.1 \times 10^9 M_\odot$. Some galaxies contain two SMBHs in a binary, expected to be the result of a galaxy merger. Quasars contain black holes with even higher masses up to at least $4 \times 10^{10} M_\odot$.

A black hole with the mass of that of $Sgr A^*$ would disrupt the disk dynamics [29] were it out in the spiral arms, but at, or near to, the center of mass of the Milky Way, it is more stable. $Sgr A^*$ is far too massive to have been the result of a gravitational collapse, and if we take the view that all black holes either are the result of gravitational collapse or are primordial, then the galaxies' core SMBHs must be primordial. Nevertheless, it is probable that the PSMBHs are built up by merging and accretion from less massive PIMBH seeds.

## 3. Primordial Naked Singularities (PNSs)

Just as neutral black holes can be formed as PBHs in the early universe, it is natural to assume that objects can be formed based on the Reissner–Nordstrom metric [12,13]:

$$ds^2 = f(r)dt^2 - f(r)^{-1}dr^2 - r^2 d\theta^2 - r^2 \sin^2\theta d\phi^2 \tag{20}$$

where

$$f(r) \equiv \left( 1 - \frac{r_S}{r} + \frac{r_Q^2}{r^2} \right). \tag{21}$$

with

$$r_S = 2GM \quad r_Q = Q^2 G \tag{22}$$

The horizon(s) of the RN metric occur when

$$f(r) = 0 \tag{23}$$

which gives

$$r_\pm = \frac{1}{2} \left( r_S \pm \sqrt{r_S^2 - 4r_Q^2} \right) \tag{24}$$

It follows that for $2r_Q < r_S$, $Q^2 < M$, there are two horizons. On the other hand, when $2r_Q = r_S$, $Q^2 = M$, the RN black hole is named extremal and there is only one horizon. If $2r_Q > r_S$, $Q^2 > M$, the RN metric may be called super-extremal. In this case, there is no horizon at all and the $r = 0$ singularity becomes observable to a distant observer. This is called a naked singularity. With this last inequality, it is no longer a black hole, which, by definition, requires a horizon.

Consider two identical objects with mass M and charge Q, and then an electromagnetic repulsive force $F_{em} \propto k_e Q^2$ and a gravitational attraction $F_{grav} \propto GM^2$. Thus, for the electromagnetic repulsion to exceed the gravitational attraction, we need $Q^2 > GM^2/k_e$ and hence perhaps super-extremal Reissner–Nordstrom or naked singularities (NSs) (to anticipate NSs, we shall replace BH with NS for charged dark matter. If charges satisfy $Q^2 < M$, this replacement is unnecessary).

We cannot claim to understand the formation of PNSs. One idea hinted at in [35] is that extremely massive ones, charged PEMNSs, might begin life as electrically neutral PBHs. Then, during the dark ages, these selectively accrete electrons over protons. However this formation process evolves, it must be completed before the onset of accelerated expansion some 4 billlion years ago at cosmic time $t \sim 9.8$ Gy.

*Like-Sign-Charged Primordial Extremely Massive Naked Singularities (PEMNSs) and Accelerated Expansion: The EAU Model*

A novel EAU model was suggested in [36,37], where dark energy is replaced by charged dark matter in the form of PEMNSs or charged primordial extremely massive naked singularities (in [36,37] the PEMNSs were called PEMBHs). That discussion involved the new idea that, at the very largest cosmological distances, the dominant force is electromagnetism rather than gravitation. This differs from the assumption tacitly made in the first application of general relativity by Einstein [2].

The production mechanism for PBHs in general is not well understood, and for the PEMNSs, we shall make the assumption that they are formed before the accelerated expansion begins at $t = t_{DE} \sim 9.8$ Gy. For the expansion before $t_{DE}$, we shall assume that the $\Lambda CDM$ model is approximately accurate.

The subsequent expansion in the charged dark matter model will, in the future, depart markedly from the $\Lambda CDM$ case. We can regard this as advantageous because the future fate of the universe in the conventional picture does have certain unaesthetic features in terms of the extremely large size of the asymptotic extroverse.

In the $\Lambda CDM$ model, the introverse, or what is also called the visible universe, coincides with the extroverse at $t = t_{DE} \sim 9.8$ Gy with the common radius

$$R_{EV}(t_{DE}) = R_{IV}(t_{DE}) = 39 Gly. \tag{25}$$

The introverse expansion is limited by the speed of light and its radius increases from Equation (25) to 44 Gly at the present time $t = t_0$, but asymptotes only to

$$R_{IV}(t \to \infty) \to 58 Gly \tag{26}$$

The extroverse expansion is, by contrast, exponential and superluminal. Its radius increases from its value of 39 Gly in Equation (25) to 52 Gly at the present time $t = t_0$ and grows without limit. After only a trillion years, it attains an extremely large value, as follows:

$$R_{EV}(t = 1Ty) = 9.7 \times 10^{32} Gly. \tag{27}$$

This future for the $\Lambda CDM$ scenario seems distasteful because the introverse becomes of ever decreasing, and eventually vanishing, significance, relative to the extroverse.

One attempt at a possible formation mechanism of PEMNSs was provided in [35], where their common sign of electric charge, negative, arises from the preferential accretion of electrons relative to protons. This formation mechanism is not well understood (electrically neutral PEMBHS were first considered, with a different acronym, SLABs, in [38]). So, to create a cosmological model, we shall, for simplicity, assume that the PEMNSs are all formed before $t = t_{DE} \sim 9.8$ Gy and thereafter, the Friedmann equation, ignoring radiation, is

$$\left(\frac{\dot{a}}{a}\right)^2 = \frac{\Lambda(t)}{3} + \frac{8\pi G}{3}\rho_{matter} \tag{28}$$

where $\Lambda(t)$ is the cosmological "constant" generated by Coulomb repulsion between the PEMNSs. From Equation (28), in the $\Lambda CDM$ model with $a(t_0) = 1$ and constant $\Lambda(t) \equiv \Lambda_0$, we would predict that, in the distant future,

$$a(t \to \infty) \sim exp\left(\sqrt{\frac{\Lambda_0}{3}}(t - t_0)\right) \tag{29}$$

In the case of charged dark matter, with no dark energy, we must re-write Equation (28) as

$$\left(\frac{\dot{a}}{a}\right)^2 = \frac{8\pi G}{3}\rho_{cPEMNSs} + \frac{8\pi G}{3}\rho_{matter} \tag{30}$$

in which

$$\rho_{matter}(t) = \frac{\rho_{matter}(t_0)}{a(t)^3} \tag{31}$$

where matter includes normal matter and uncharged dark matter.

Of special interest to the present discussion is the expected future behaviour of the charged dark matter:

$$\rho_{PEMNSs}(t) = \frac{\rho_{PEMNSs}(t_0)}{a(t)^3} \tag{32}$$

so that the comparison of Equations (28) and (30) suggests that the cosmological constant is predicted to decrease from its present value. More specifically, we find that, asymptotically, the scale factor will behave as if matter-dominated and the cosmological constant will decrease at large future times as a power:

$$a(t \to \infty) \sim t^{\frac{2}{3}} \qquad \Lambda(t \to \infty) \sim t^{-2}. \tag{33}$$

so that a trillion years in the future, $\Lambda(t)$ will have decreased by some four orders of magnitude relative to $\Lambda(t_0)$. See Table 1.

**Table 1.** Cosmological "constant".

| Time | $\Lambda(t)$ |
|:---:|:---:|
| $t_0$ | $(2.0\ meV)^4$ |
| $t_0 + 10Gy$ | $(1.0\ meV)^4$ |
| $t_0 + 100Gy$ | $(700\ \mu eV)^4$ |
| $t_0 + 1Ty$ | $(230\ \mu eV)^4$ |
| $t_0 + 1Py$ | $(7.4\ \mu eV)^4$ |

In both the $\Lambda CDM$ model and the EAU model, the present time is an unusual one in cosmic history. In the former case, there is the present similarity between the the densities of dark matter and energy. In the latter case with charged dark matter, the present accelerated expansion is maximal and will disappear within a few more billion years.

In the EAU model, acceleration began about 4 Gy ago at $t_{DE} = 9.8Gy = t_0 - 4Gy$. This behaviour will disappear in a few more billion years. The value of the cosmological constant is predicted to fall like $a(t)^{-2}$ so that, when $t \sim \sqrt{2}t_0 \sim 19.5Gy \sim t_0 + 4.7Gy$, the value of $\Lambda(t)$ will be one half of its present value, $\Lambda(t_0)$. On the other hand, as discussed in [37], the equation of state associated with $\Lambda$ is accurately predicted to be $\omega = -1$, so close to that value that measuring the difference seems forever impracticable.

For charged dark matter, we now discuss the future time evolution of the introverse and extroverse. For the introverse, nothing changes from the $\Lambda CDM$, and after a trillion years, the introverse radius will be at its asymptotic value $R_{IV} = 58 Gly$, as stated in Equation (26). By contrast, the future for the extroverse is very different for charged dark matter than for the conventional $\Lambda CDM$ case. With the growth $a(t) \propto t^{\frac{2}{3}}$, we find that the radius of the extroverse at $t = 1$ Ty is

$$R_{EV}(t = 1 Ty) \sim 900 Gly. \tag{34}$$

This is in stark contrast to the extremely large value $9.7 \times 10^{32}$ Gly predicted by the $\Lambda CDM$ model, quoted in Equation (27) above. Equation (34) means that if there still exist scientific observers, their view of the distant universe will be quite similar to that of the present one and will include many billions of galaxies.

In the $\Lambda CDM$ case, such a hypothetical observational cosmologist, trillions of years in the future, could observe only the Milky Way and objects which are gravitationally bound to it, so that cosmology would become an extinct science.

The principal physics advantage of charged dark matter is that it avoids the idea of an unknown repulsive gravity inherent in "dark energy". Electromagnetism provides the only known long-range repulsion so it is more attractive to adopt it as the explanation for the accelerating universe. The secondary advantage of charged dark matter, that it provides a conducive environment for observational cosmology trillions of years into the future, is not by itself sufficient to choose this theory.

## 4. Discussion

Although this paper is essentially speculative, we are unaware of any fatal flaw. We have replaced the conventional make up for the slices of the universe's energy pie (5% normal matter; 25% dark matter; 70% dark energy) with a similar but crucially changed version (5% normal matter; 25% dark matter; 70% charged dark matter).

The term dark energy was coined by Turner [39] in 1998, shortly after the announcement of accelerated expansion [3,4]. An outsider familiar with $E = Mc^2$ might guess that dark energy and matter are equivalent. If our model is correct, they would be correct, although it has nothing to do with $E = mc^2$. Charged dark matter replaces dark energy, an ill-chosen name because it suggested that there exists an additional component in the Universe.

In April 2024, news [40] from the Dark Energy Spectroscopic Instrument (DESI) at Kitt Peak in Arizona, USA, gave a preliminary indication that the cosmological constant $\Lambda(t)$ is not constant but diminishing with time, as suggested by our Equation (33), and by our Table 1, thus providing possible support for the EAU model.

Other supporting evidence could appear in the foreseeable future from the James Webb Space Telescope (JWST), which might shed light on the formation of PBHs in the early universe, and also from the Vera C. Rubin Observatory in Chile, which will study long-duration microlensing light curves, which could provide evidence for the existence of PIMBHs inside the Milky Way.

It will be interesting to learn how these and other observations might support the idea that the observed cosmic acceleration is caused by charged dark matter.

**Funding:** This research received no external funding

**Institutional Review Board Statement:** Not applicable

**Data Availability Statement:** No new data were created or analyzed in this study. Data sharing is not applicable to this article.

**Acknowledgments:** We wish to thank the editors of *Entropy* magazine for coming up with the idea of this eightieth anniversary Festschrift, and to thank all of the other contributors for writing and submitting their stimulating papers.

**Conflicts of Interest:** The author declares no conflicts of interest

# String Invention, Viable 3-3-1 Model, Dark Matter Black Holes

**Holger B. Nielsen**

Niels Bohr Institute, 2200 Copenhagen, Denmark; hbech@nbi.ku.dk

**Abstract:** With our very limited memories, we provide a brief review of Paul Frampton's memories of the discovery of the Veneziano model, with this indeed being string theory, with Y. Nambu, and, secondly, his 3-3-1 theory. The latter is, indeed, a non-excluded replacement for the Standard Model with triangle anomalies being cancelled, as they must in a truly viable theory. It even needs (essentially) three as the family number! Moreover, primordial black holes as dark matter is mentioned. We end with a review of my own very speculative, utterly recent idea that for the purpose of the classical approximation, we could, using the functional integral as our rudimentary assumption taken over from quantum mechanics, obtain the equations of motion without the, in our opinion, very mysterious imaginary unit $i$, which usually occurs as a factor in the exponent of the functional integrand, which is this $i$ times the action. The functional integral without the mysterious $i$ leads to the prediction of some of the strongest features in cosmology, and also seems to argue for as few black holes as possible and for the cosmological constant being zero.

**Keywords:** string theory; invention of string; dual models; quantumfield theory; gauge theories; 3-3-1 model; anomalies; number of families; cosmology; dark matter; black holes; complex action; influence from future; cosmological constant; complex unit $i$

## 1. Introduction

Thinking about Paul, first, I do remember our nice time together in CERN when we were younger, getting nice dinners and visiting the mountains, once even with a couple of diplomat girls. However, first of all, we were working, and Paul was, in addition to working with me, etc., writing his book [1] on dual models with its several chapters. I had met Paul earlier than at CERN, I think in DESY. Additionally, once he even met my mother and uncle in Copenhagen, and I remember we were at a place in the far end of Nyhavn, and that was rather special in the way that I think my mother and uncle met exceedingly few of my colleagues.

In Geneva, we went to a lot of good restaurants together and there were many possibilities. During the same stay in CERN, I also worked with Lars Brink, and also, Colin Froggatt, with whom I am still working a lot, was there. In addition, Paul and I worked a bit together much later, and I have not given up the hope of us getting even more work together; in fact, I am going to Corfu soon, and Paul is to be there too.

*Plan of Paper*

In the next section, I review the very early stages of finding the string from the Veneziano model as conducted by Frampton and Nambu. In Section 3, we review the 3-3-1 model, which Paul created and which has been studied a lot, because it is a possible replacement for the Standard Model, which really could be true still. In Section 4, we look at which arguments one would speculate Nature may have used for finally selecting the model, 3-3-1 or the Standard Model. In Section 5, we move on to black holes, mainly with the idea that they should be dark matter as this is Paul's favorite dark matter model. Then, in Section 6, I refer to my own work, which attempts as a Random Dynamics project to derive quantum mechanics but instead discusses how one makes a classical theory from functional integral formulation with the, to me, mysterious **imaginary unit** $i$, in a fundamental theory such as quantum mechanics, **removed**. Section 7 provides a conclusion and birthday wishes.

## 2. Very Early String Theory

In the celebration of Nambu [2], Paul reveals how I, [3], and Lenny Susskind [4,5] should feel very lucky that Nambu [6,7] was a bit slow in publishing about the string from factorization of the dual model. (See also the book on the Birth of String Theory [8]).

Indeed, after a question by Nambu, Paul worked on factorizing the factor $(1-x)^{-2\alpha' p_2 \cdot p_3}$ in the Veneziano model to what would now be called string variables. The question was:

Can the $t$-dependence be factorized as

$$(1-x)^{-2\alpha' p_2 \cdot p_3} = F(p_2)G(p_3)? \tag{1}$$

where this is the factor in the Veneziano model notation

$$A(s,t) = \int_0^1 x^{-\alpha(s)-1}(1-x)^{-\alpha(t)-1}dx. \tag{2}$$

Paul carried out the factorization by writing

$$(1-x) = \exp(\ln(1-x)), \tag{3}$$

and expanding

$$\ln(1-x) = -\sum \frac{x^n}{n}. \tag{4}$$

He reached the factorization

$$F(p) = exp\left(i\sqrt{2}\alpha' p_\mu \sum_1^\infty \frac{a_\mu^{(n)} x^n}{\sqrt{n}}\right) \tag{5}$$

$$G(p) = exp\left(i\sqrt{2}\alpha' p_\mu \sum_1^\infty \frac{a_\mu^{(n)\dagger}}{\sqrt{n}}\right) \tag{6}$$

It is easy to check that the explicit solution is

$$(1-x)^{-2\alpha' p_3 \cdot p_2} = <0|F(p_2)G(p_3)|0>. \tag{7}$$

This was all very early, even compared to my own also unpublished version —even the first version, which was the same but without the word "almost" in the title. In fact, Knud Hansen, an experimentalist at the Niels Bohr Institute, commented on my first title without the "almost" so that I put in this word and then produced the title "An almost physical Interpretation of Dual Model" [3].

## 3. Extension of Standard Model That Could Work

The so-called 3-3-1 model [9,10], of which one can have some slightly different variations (see e.g., [11]), is a model carefully worked out to have **no anomalies** from the triangle diagrams for the fermions in the model. This model [10] is described as $SU(3)_c \times SU(3)_L \times U(1)_N$, or thinking with O'Raifeartaigh [12] on gauge **groups** rather than just the Lie algebra, $U(3)_c \times U(3)_L$, and the leptons have their left pair put together with the antiparticle of the right-handed singlet of the usual Standard Model, into a single antitriplet representation for the $SU(3)_L$ representation

$$\psi_{aL} \;=\; \begin{pmatrix} \nu_a \\ l_a' \\ l_a'^c \end{pmatrix} \sim (\mathbf{3}, 0) \tag{8}$$

For the quarks, we do not in the Standard Model have the $SU(2)$ singlet as for the leptons, and thus, instead, a new particle $J_a$ is introduced to complete the triplets under the $SU(3)_L$. The naive attempt would be to let all the three families of quarks we know also be represented as the lepton $SU(2)$-doublets as, e.g., the first family

$$Q_{1L} \;=\; \begin{pmatrix} u_1' \\ d_1' \\ J_1 \end{pmatrix} \sim (\mathbf{3}, 2/3), \tag{9}$$

but if we had all the quark families represented this way, then the anomaly from a triangle diagram (see Figure 1) with three external gauge particles from group $SU(3)_L$ would add up and the model would not be anomaly free.

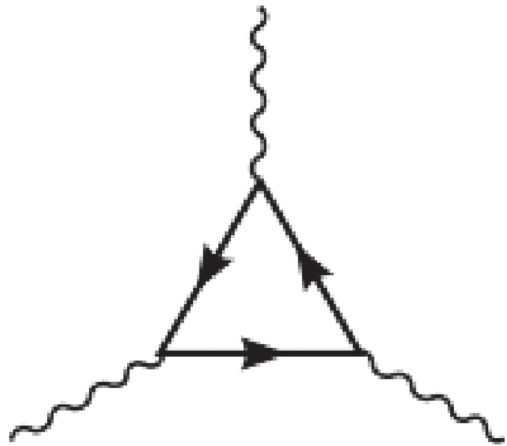

**Figure 1.** In this diagram, the three outgoing gauge boson (propagator) attachments denoting the gauge bosons from the gauge group $SU(3)_L$ replaced the weak $SU(2)$ and some of the $U(1)$ in the Standard Model, which is discussed in the text.

Instead, it must now be repaired by representing instead one of the quark families by the representation

$$Q_{aL} \;=\; \begin{pmatrix} J_a' \\ u_a' \\ d_a' \end{pmatrix} \sim (\mathbf{3}*, -1/3) \text{ for } \alpha = 2, 3. \tag{10}$$

Then, we can easily see that we can achieve a cancellation of the anomalies from the triangle diagram with three outgoing gauge particles from the $SU(3)_L$ group, with one extending the weak $SU(2)$ and, partly, the $U(1)$ of the Standard Model. We just have to know that the diagram anomaly is non-zero because it is obvious that the anomaly contribution is simply proportional to the number of left-handed fermions going around the triangle in the diagram. Thus, we must have arranged that there are equally as many left-handed particles going around with the representation $\mathbf{3}$ as with the conjugate $\mathbf{3}*$.

Since whatever choice we make with the different quark families, we always obtain them in triples because of the color representation, so they will always also participate with a multiple of three of the triangle diagram for three external $SU(3)_L$ gauge bosons, unless we choose higher than the sextet representations for them. Thus, under this attempt to keep close to the Standard Model and at least not postulate, e.g., $SU(3)_c$-octet-colored fermions, there must be lepton families in a multiple of three. With the just-sketched trick of having two quark families in the **3**∗, while there is only one in the **3**, we can indeed obtain the no anomaly conditions satisfied by the relations

$$
\begin{aligned}
(\text{``1 family''} * 3 + 3 \text{ leptons}) anomaly(\textbf{3} \; SU(3)_L\text{'s}) &\quad + \\
\text{``2 families''} * 3 * anomaly(\textbf{3*} \text{ of } SU(3)_L\text{'s}) &\quad \propto \\
3 + 3 - 2 * 3 &\quad = \quad 0.
\end{aligned}
\tag{11}
$$

Models of this type were made without any anomalies, so that they really could be true, and so close to the Standard Model, that they cannot be excluded yet. It is remarkable that it has its predictions much closer to the experiments than say grand unification theories, which only provide a number of coincidences of fine structure constants and proton decay, which are very remote in energy scale, while Paul's 3-3-1 model(s) have new particles much closer. (Actually there are limits to how far they can be put away). One test of 3-3-1 is the bilepton resonance in same sign leptons.

For the good reason of being, in this sense, closer to possible reality, it has already been well studied and deserved lots of citations; actually, if it really were true, it could be found very soon that it was indeed the case.

There are of course obvious dangers for the 3-3-1 model, namely that the relations between measured quantities, which come from the Standard Model as successfully predicted, do not quite come out in the 3-3-1 models. The question of whether the 3-3-1 model can manage to organize the successful predictions such as no flavor changing neutral currents (i.e., bounds for flavor changing neutral currents FCNC) and the $\rho$ parameter

$$
\rho \quad = \quad \frac{m_W^2}{m_Z^2 \cos^2 \theta_W}
\tag{12}
$$

is complicated because the model needs some scalar triplets under the $SU(3)_L$ in order to break the symmetry down to the Standard Model. In fact, one has three scalars that are triplets under the $SU(3)_L$ (the one connected to weak interactions), and one can make different models by leaving out one or the other of these scalars. It turns out, however, that there is a need to fit all the scales of the scalar triplets to adjust the 3-3-1 model to match the known conditions [11]. Thus, some simplified version(s) are excluded.

## 4. Some Day We Shall Explain Whether the 3-3-1 or the Old SM Is Right

The great thing is that the 3-3-1 model is closer than many other theories proposed, and then it should not take an unrealistically long time to find out if it is right or if the Standard Model is the right one. At that time, one would likely give some principles by word that point to the right model, in order to find out what the principles are behind the even deeper physics, which determines which model we should find in the LHC energy range. One might set out with some questions like:

- **Is the number of families determined?** The 3-3-1 model has the feature, via the disappearance of anomalies, needing (a multiple of) three families. Insisting on the asymptotic freedom of QCD requires exactly three families.
  This is of course a great victory for the 3-3-1 model, which, thus, does predict the number of families, and thus, it should support our belief in the model. However, is this something we should think about for the next level of theories so as to bring us onto the track of the next level of theories? Presumably, the next more fundamental level theory would not care if there some understanding was left of the number of families, or we would be left with that for the next level.
- **Small representations of the fermions** This is seemingly a good principle to apply for the details of both the 3-3-1 model and the Standard Model, since both (types of) models have the smallest non-trivial representations of the groups at hand.

One would have to make the concept of smallness of representations very detailed to make assuming such a detailed definition of smallness a good argument for why Nature should choose the one or the other (from a deeper physics point of view).

- **The gauge group** Philosophizing over the gauge group, you might propose thoughts like this:

  If in a deeper physics there are principles or mechanisms that favor making the gauge group at lower energies in some special simple group, then you might expect it to be better to use the same group several times and make the full group or Lie algebra a cross-product of several isomorphic groups, in the same way the 3-3-1 model has $SU(3)$ used twice. This saying would of course give a speculative argument in favor of the 3-3-1 model (relative to the Standard Model).

  Thus with such thinking in mind you would, if the Standard Model should finally turn out to be the right one, wonder very much why a gauge group was chosen with two different simple groups in it, in addition to the $U(1)$. I would say this mystery might have a possible explanation in the next item.

- **Which group has the strongest connection between the Lie algebras by rule(s) of allowed representation combinations?** I would say that it looks like Nature has chosen the Standard Model partly because it loves that the gauge **group** has obtained the center of the covering group divided out by a discrete group so as to connect the different Lie algebra cross-product factors. Having gauge **groups** is the O'Raifeartaigh way of looking at [12] the quantization rules connecting the allowed combinations of (irreducible) representations of the different Lie algebras.

  In fact, the Standard Model has a quantization rule for the $U(1)$ charge usually called $y/2$, which relates it to *both* the representation of the color $SU(3)_c$ and the weak $SU(2)$ requiring

  $$\text{"triality" for } SU(3)_c + s_W + y/2 \;=\; 0 \; (mod \; 1), \tag{13}$$

  which is of course the rule that ensures the quantization of the electric charge as it is believed in the Standard Model. Here, $s_W$ is the representation classifying number for the Standard Model weak $SU(2)$: i.e., the $(2s_W + 1)$-plet.

  You could not connect the given Lie algebras more by such a quantization rule than by this (13).

  In the 3-3-1 model, there is only such a quantization rule connection between the $U(1)_X$ and the color $SU(3)_c$. The $SU(3)_L$ is not connected this way. Thus, the 3-3-1 model is not quite as strongly connected as the Standard Model, which has such a quantization rule connection between all three Lie algebra cross-product factors $u(1)$, $su(2)$, and $su(3)$. In fact, the most intertangling quantization rule for the $U(1)$ charge $y/2$ is postulated for the Standard Model. Now, such a high degree of complication for the quantization rule can only be made provided the orders of the centers of the simple non-abelian groups, in the Standard Model, $SU(2)$ and $SU(3)$ are **incommensurable/mutually prime.** The two smallest incommensurable natural numbers that can be used in $SU(N)$-groups are actually 2 and 3.

  For example, if you have the Lie algebras as in 3-3-1, $u(1)$, $su(3)$, $su(3)$, you cannot make a non-trivial quantization rule connecting the $u(1)$ to *both* $su(3)$'s. (3 and 3 are, namely, not incommensurable).

  Thus, one could claim that the Standard Model would have to have the two different Lie algebras in order that one could then make such quantization rules for *both* non-abelian algebras.

- **Skewness** Once, I and Niels Brene [13–15], in our search for some characteristic properties of the Standard Model, proposed a concept of skewness that should basically mean that the (gauge) **group** (and here we really thought of the group in O'Raifeartaigh's sense [12] rather than just the Lie algebra) should have very few (outer) automorphisms compared to the rank. We found that the appropriately defined Standard model **group** $S(U(3) \times U(2))$ (which is defined as the group of $5 \times 5$ matrices com-posed from the $SU(2)$ and the $SU(3)$ matrices and imposing the condition given by the $S$, that the determinant of the $5 \times 5$ matrix should be 1). would be pointed out as **most skew**.

The 3-3-1 model group is $U(3)_c \times SU(3)_L$ because there is a rule connecting the $U(1)_X$ charges so that they have 1/3 modulus 1 for triplets of the color $SU(3)$, while the "$U(1)_X$ charge" is not connected similarly to the $SU(3)_L$ representation.

This **group** of the 3-3-1 model at least has the interesting sign of skewness, in that just the colored $SU(3)$ obtained a rule connecting it to the $U(1)_X$, while the other $SU(3)$ has no such connection to the $U(1)_X$. This, namely means that the obvious outer automorphism of the group $U(3)_c \times SU(3)_L$ consisting of permuting the two $SU(3)$-subgroups is prevented, so that at least this automorphism is not there.

However, you can make a complex conjugation of the $U(3)_c$ and of the $SU(3)_L$ separately, and thus, the outer automorphism group becomes $\mathbf{Z}_2 \times \mathbf{Z}_2$, while the Standard Model group only has an outer automorphism group $\mathbf{Z}_2$.

- **Holy number 3** The 3-3-1 model has "the holy number 3" we could fantasize first in the two SU(3) groups, and then from there, it is transferred with the anomaly avoidance to the number of families. Thus, it is really characterized by a "holy number".

  I do not see that you could say the same thing about the Standard Model.

- **Unification** In seeking theoretical stories that could be used to say a posteriori why Nature should have chosen one model or the other, the possibility of putting the model into a grand unification model is almost a must to be mentioned. I am afraid I shall not be able to study the possibilities of making a grand unification model extension of the 3-3-1 model before Paul is more than 80 years old, but we can hope to find out such possibilities when he gets to 90. It is not fitting in at all that $SU(5)$ is the starting point for most unifications for the Standard Model.

  It is, however, well known that the success for unification of the Standard Model without some helping complications like super-symmetry has not been so great again [16]. I have myself followed the attempt by Norma Mankoc Borstnik [17] and collaborators of having 10 extra dimensions in addition to the 4, which we see clearly, and $SO(10)$ is one of the studied unification extensions of the Standard Model.

  However, very recently I have been keen on accepting that the $SU(5)$ symmetry could be only approximate [18] and further gauge particles in it compared to the Standard Model should not exist in reality. Really, it is a lattice gauge theory I propose in which $SU(5)$ symmetry comes in the classical approximation but is broken by quantum corrections; you just have to multiply the quantum correction by a factor of 3 obtained by giving each family its own lattice.

## 5. Black Holes

For the purpose of the present article, I thought it best to read a little bit of Paul's papers, preferably about, e.g., dark matter.

Of course, I come in as one who definitely does not believe in dark matter being primordial black holes, in as far I have written with Colin Froggatt, who also was in CERN at the same time Paul and I were there, a long series of works about it being small (it could at first come from much bigger regions of an alternative vacuum, then contract, or still survive today) macroscopic objects; but then Paul's black holes sound very convincing!

After all, the black hole theory of dark matter is genuinely without new physics, whereas mine and Colin Froggatt's model [19–28] only formally is without new physics because it needs two phases of vacuum. Such two vacuum phases would a priori need new physics, unless we have such remarkably good luck that the calculations in QCD of the properties of vacuum for different values of the quark masses [29] should reveal a non-trivial behavior, and the experimental quark mass combination should just lie on the phase transition. At least one of the authors [30] looking for phase transitions represented the chance that we could have hope of really having no need for new physics in our model, by admitting with question marks on his phase plot that he did not know if the experimental quark combination was in the one phase or the other. Thus, it could miraculously be on the very border line. If so, there could be two phases existing in extensive regions of spacetime in the universe, e.g., inside and outside dark matter pearls [19].

Well, it must be admitted, that even if there was the appropriate phase transition between different phases that would still need the milder amount of new physics, there should be some principle, some law of nature, ensuring that the quark masses, say, would have just the right masses to be just on the phase border, so that more than one phase could be realized with essentially the same energy density. This hypothesis we have talked much about under the name "Multiple Point Critically Principle" [28,31–34] (MPP), and we once had the luck of predicting [28] the Higgs mass from assuming this hypothesis, before the Higgs was produced at LHC!

Actually, it turned out that what we thought was the main hypothesis in our MPP, namely that different phases of vacuum should have the same energy density or cosmological constant, Dvali and earlier Zeldovic had already proposed as a theorem [35–37]. Doubts concerning this theorem are discussed by C. Gross, Strumia, et al. [38].

*PBH Is a Possibility for Rather Heavy Dark Matter Particles*

Believing in Hawking radiation, the primordial black holes (PBHs) lighter than $10^{15}$ g $= 10^{12}$ kg would have radiated away or would be just about doing it today. This is a rather high mass compared to what is speculated in other models, and it has, of course, consequences for how high a number of incidences with the Earth we can have. We had, ourselves, a speculative model for dark matter once, in which the Tunguska event in which the trees were thrown down or even up by a big explosion in a 70 km large region should be due to the fall of a dark matter particle. This Tunguska particle was from the assumption that one fell on Earth every hundred years, and using the dark matter density 0.3 GeV/cm$^2$ in the solar system, having a mass of about $1.4 * 10^8$ kg. The PBHs have to be a few thousand times heavier, and thus, correspondingly, more seldomly hitting the Earth.

If so, it is of course excluded that the PBH component of the dark matter could be what is observed by the DAMA-LIBRA experiment [39].

Of course, this DAMA-LIBRA experiment [39] is still rather in contradiction (one found in 2021, e.g., an article with the title "Goodbye, DAMA/LIBRA: World's Most Controversial Dark Matter Experiment Fails Replication Test" by Ethan Siegel, alluding to the disagreeing Anais [40] experiment with another underground search for dark matter, which actually find nothing instead of confirming DAMA). Thus, one might attempt to declare DAMA as seeing something else or being mistaken somehow. If we shall uphold that DAMA saw many events of dark matter (whatever), it cannot be primordial black holes, so we rather must have several different components in the dark matter if there are also the PBHs. DAMA-Libra has achieved C.L. for the full exposure (2.86 t × yr) 13.7$\sigma$. However, the Anais experiment [40] sees no dark matter, i.e., a rate of $0.0003 \pm 0.0037$ cpd/kg/kev against DAMA-LIBRA of $0.0102 \pm 0.0008$ cpd/kg/keV for a range 2 keV to 6 keV. This rate means that DAMA-LIBRA sees a season-varying amplitude of counts per day per kg detector per bin of 1 keV energy in their scintillators of 0.0102 with uncertainty 8%. The Anais result is thus in contradiction with DAMA-LIBRA.

Froggatt and I propose that this contradiction comes about because DAMA-LIBRA is 1400 m down in the earth, while the other experiments such as Anais are typically higher up towards the earth's surface, so that the story could be made that the dark matter runs fast through the near surface region and does not have time to radiate electrons or X-rays before it gets stopped in the DAMA-LIBRA region. Maxim Khlopov's model [41] has it that the dark matter of his 0-helium should be stopped and make a nuclear interaction with nuclei inside the earth. This could also mean that it first became active deeper down at the DAMA depth. Such models have the chance of obtaining more counts for DAMA-LIBRA than for the higher up experiments. Most of the underground experiments use xenon as the scintillator liquid. The fact that it is a liquid could easily mean that dark matter of a type that should be stopped or move slowly to be observed would not be seen in the liquid xenon experiments. Alone, gravity might drive it too fast through the detector if it is fluid.

## 6. My Own Very Recent Work on "Kinetic Energy Being Unwanted"

With the bad excuse that my recent work [42] tends to predict that black holes should be kept only as a small fraction of the present energy density in the universe, it should be preferably inline with the relativistic contributions from the photons and the neutrinos, which have, respectively,

$$\Omega_\gamma h^2 = 2.480 \times 10^{-5} \tag{14}$$

$$\text{and } \Omega_\nu h^2 = 1.68 \times 10^{-5}, \tag{15}$$

namely, e.g., as [43] says, that about 0.04% of the energy density (=critical density) should be black holes. Already this is much compared to what many would have thought, but of course, if the dark matter really consisted of (primordial) black holes, then the $\Omega_{BH}$ would not be the here-estimated $4 \times 10^{-4}$ but up in the 0.2 region. My crazy idea is in the series of ideas [33,44–50] in which I speculated about a theory that seeks to combine physical equations of motion and initial states conditions, which we in cosmology make assumptions about similarly to how we make assumptions about the details of the laws of nature, the Lagrangian density, and the system of particles that exist. Thus, such a united theory should in principle tell, from the same formalism, about how the Universe started and even how it shall end and about the laws of nature, when you have put in the appropriate extra stuff, the Lagrangian, say. I long worked on that with Masao Ninomiya [44,45,51] and Keiichi Nagao [46–48,52,53], and we obtained the encouraging result that taking the Lagrangian or action to have complex coefficients in it **would not be seen in the effective equations of motions resulting**. (The very earliest thoughts on this future influence might have been with Colin Froggatt [50] and Don Bennett [33]). The only new information from such a complex action should be that it also makes sayings about the initial state conscience. In the newest idea, which I talked about in the Bled Workshop on "What comes beyond the Standard Models" [42], I want to say that the $i = \sqrt{-1}$ in quantum mechanics really is a bit strange; should such a fundamental theory as quantum mechanics really be based on the complex numbers?

Should Nature really in the so fundamental quantum mechanics take up these a priori just invented numbers, which Gerolamo Cardano around 1545 in his Ars Magna [54] was inventing, although his understanding was rudimentary? He even later described the complex numbers as being "as subtle as they are useless".

My recent work can be considered an attempt to ask: how needed really is this, is $i$ a priori not what you would expect Nature to choose?

Let us **take over from quantum mechanics** the idea of the functional integral, written almost symbolically

$$\text{Functional integral} = \int \exp(\frac{i}{\hbar} S[history]) \mathcal{D} history. \tag{16}$$

If you were thinking, as we have in mind here, on the development from the beginning of time $t$ to the end of time, you might without knowing better think that you should most simply obtain the expectation value for an operator $O$ at some moment of time $t$, i.e., the expectation value of the "Heisenberg operator" $O(t)$, by looking at the variable $O$ written in terms of the variables used in the functional integral taken at the time $t$ as constructed in terms of these variables at $t$. That is to say that with the introduction also of a normalization-denominator, one would naively propose to use as the expectation value

$$< O(t) > = \frac{\int_{with \ |i> \ and \ |f>} O(t) \exp(\frac{i}{\hbar} S[history]) \mathcal{D} history}{\int_{with \ |i> \ and \ |f>} \exp(\frac{i}{\hbar} S[history]) \mathcal{D} history}, \tag{17}$$

where I have also alluded to the fact that to make this expression meaningful, you need at the earliest time to give a boundary condition state $|i>$ and also one at the very final time $|f>$. This naive construction of the expectation value $<O(t)>$ does in general **not** give a really good expectation value in as far as it is typically **not even a real number**. Rather, this functional integral naive attempt to construct an expectation value in a seemingly not so bad way, is what is called **the weak value** [55] of the operator $O$ at time $t$.

The word *history* is here used to denote the general path in the thinkable development of the Universe from the beginning to the end, and $S$ is of course the action functional. Really, you can see that this weak value is more like a matrix element than a genuine expectation value, in as far as using the Heisenberg picture, the weak value becomes

$$< O(t) >_{weak\ value} \quad = \quad \frac{< f|O(t)|i >}{< f|i >} \text{ (Heisenberg picture).} \qquad (18)$$

We (Keiichi Nagao and I) made theorems about when it becomes real [56] for an operator or dynamical variable $O$, which really means the Hermitian of actually the slightly modified Hermiticity we required ("Q-hermiticity"). There is one case in which we can see easily that the weak value must be real, namely when the operator $O$ is Hermitean/real, and when a narrow bunch of paths *history*'s dominates the functional integral, i.e, what you can say would happen in a classical approximation, when the initial and final states $|i>$ and $|f>$ correspond to such a classical development. (Of course, if essentially only one history dominates, and $O(t)$ is the real function of the (real) variables used in the functional integral, you really just ask for the value of $O(t)$ on the classical path dominating and that is of course real).

In the classical approximation, one could indeed use this weak value to extract some classical paths.

It is well known that one of the uses of the functional integral formulation is that you can obtain the classical equations of motion by requiring the functional derivative $\delta$ of the integrand $\exp(\frac{i}{\hbar}S[history])$ to be zero. It even gives the classical equation of motion in the prequantum mechanics way as the requirement of the action being extremal.

However, if this classical use of the functional integral was the main application, we would **not need the** $i$!

Thus, if I say **I only care for the classical approximation**, my only quantum mechanics idea left over is **a functional integral of some sort**, then I can just **throw away the mysterious** $i$ **if I do not like it**. I would obtain a classical solution anyway, but now I would not obtain a lot of them as with the $i$; no, I would likely obtain only one or a **very little set of classical solutions** that would correspond to the truly maximal functional integrand. (Most other solutions would only survive as saddle points, or for a very special choice of the $< f|$ and $|i>$).

Note that the *history* that comes out as the winner that dominates in the case, when I leave out the $i$, still obeys the equations of motions, as if we had the $i$, because just multiplying equations of motion with $i$ or with any complex number ($\neq 0$) does not make any change at all. Thus, in this exercise of leaving out the $i$, one obtains the same equations of motion, but now of course there is only one or very few *histories* for which the integrand $\exp(\frac{i}{\hbar}S[history])$ is maximal and presumably dominates in the integral over all the other *histories*. This *history* is now selected by the functional integral formalism with the $i$ left out. We will, of course, assume that the *history* or *histories* with the very biggest integrand in the functional integral without the $i$ should now be the one (or a few) realized in the world. That is to say, what really happens in the world should be described by this maximal action solution. That is, so to speak, the model for the **initial conditions**, because by leaving out the $i$ the functional integral turned into a model, for the laws of nature, **uniting the equations of motion with the initial conditions**.

The reader might easily see how theories of the type with $i$ removed in functional–integral formulation leads to a prediction of the solution to the classical solution to be chosen, by simply noticing that the integrand in the functional integral is (exponentially) much larger for the $-S[hisoty]$ corresponding to a history that makes this $-S[hisory]$ larger than the one for which it is smaller.

### 6.1. Earlier Use of Turning the Phase and Throw an i Away

The idea of making an action or Lagrangian with the phase change like here is not quite new in the sense that you can consider Wick rotation [57] used in evaluating loop integrals as very similar. However, in the Wick rotation, it is purely a mathematical calculation method, while our idea is philosophically very different as far as we want to change a priori physics, by taking it that the fundamental physical action has indeed been turned in phase relative to the usual one.

In the attempts to study the holography of the Maldacena conjecture type, it is often used to compare under the name of correlation functions [58] what we below define as "weak values" [55] of say the conformal field theory CFT and an $Ads_5 \times S_5$ string theory. Here, when the question is about testing if two theories are equivalent, it is not so important if one puts an extra $i$ in or not, provided one does the same for both theories.

Moreover, Hawking in his theory for gravity allows himself to have the signature of the metric changed, a modification basically like that of our $i$ shift. When he now uses this formulation together with his and Hartle's no-boundary postulate [59], it is even philosophically the same game as ours. Thus, we shall consider Hartle and Hawking as the forerunners of the present work.

### 6.2. Could This Initial Condition Model Have Any Chance at All?

To achieve any idea about how such an "$i$-removed" might predict the chosen solution to the (classical) equations of motion (meaning choosing intimal conditions), we might think of a slightly simpler system, rather than the whole world with its field theories and gravity with black holes and other strange configurations and an energy concept that requires a special invention/or gauge choice, to make a system that might be nice to think about.

A still very general system that we could think about, and into which we might put a more primitive version of gravity, would be a, non-relativistic for simplicity, particle running in a potential in say a finite dimensional space, say dimension $N$. Such a particle in $N$ spatial dimensions really could be interpreted as many $N/3$ particles, namely by letting the $q_i$s be coordinates of the first three for the first particle, the next three for the second, ...

For such a system, we have an action

$$\text{action } S[q] = \int L(q, \dot{q}) dt \tag{19}$$

$$\text{where } L(q, \dot{q}) = T(\dot{q}) - V(q) \tag{20}$$

$$\text{and } T(\dot{q}) = \sum_{i=1}^{N} \frac{1}{2} m_i \dot{q}_i^2 \tag{21}$$

$$\text{and } V(q) = \text{"a potential function of } q\text{"}. \tag{22}$$

Here, we have let $q$ be a set of variables

$$q = (q_1, q_2, \ldots, q_{N-1}, q_N) = (\text{ordered set } q_i | i = 1, 2, \ldots, N).$$

We can imagine a complicated landscape with many peaks and valleys in the general potential $V(q)$. Then, we have to think of almost a "god" (a "god" in quotation marks) that has to figure out/calculate what classical solution will maximize this action $S[q]$.

If really we ask for such a maximal action, then the solution would be somewhat bad, because the kinetic energy is not bounded from above. Thus, if we seek a solution with as high a kinetic energy as possible, then the motions of the particles would be infinite or there should be some cut off, if it should make sense. In any case it would mean that the description with the particle variables we started from would not be good. In such a world, one would probably have used some different variables. However, instead of philosophizing here about what such divergently running variables could mean, if anything, let us just take the opposite sign for the action put into the functional integral integrand exponent.

We are allowed to think that the world without the *i* had replaced this *i* by $-1$ instead of just by 1, because the sign does **not** matter for the classical equations of motion. Thus, to avoid divergence of the kinetic energy (which might be postponed to a later work), we put the sign in the exponent in front of the action to be minus, so that the favorite solutions (the ones with high probability) have the highest potential energy and lowest kinetic one, as far as these can be combined.

This means we choose the opposite sign, namely to take the functional integral

$$\int \exp(-\frac{1}{\hbar}S[history])\mathcal{D}history \qquad (23)$$

$$\text{where } history \quad = \quad \text{an ordered function set } q : \text{time axis} \rightarrow \mathbf{R}^N.$$

Now, we make the "god" seek a solution, in which the world stands most of the time on top of the highest peaks in the landscape potential $V(q)$. (We assume to avoid problems an upper bound for the potential energy $V(q)$).

However, with this sign, and with just a tiny shaking, it may happen that the world/ particle slides down from the very peak, and now, the equations of motion will make it run down faster and faster from the peak, like a skier without the ability to break. (I remember a tour where I had come out together with Paul, but we were alone, when my ski made such a tour. Luckily for me I was not on the ski when it went down with high speed. As I was now walking in deep snow and slowly went down myself, people began to ask about if I had insurance. I began to fear the ski had hurt or killed somebody. However, shortly after, I saw the ski planted up down in the snow, but alas in two pieces. Oh dear. However, the ski was rented, and seemingly insured. The kind renter offered me a new ski for the rest of the day, but I thanked him no, and went on doing physics instead; it was enough for that day).

It is difficult to see that such a situation should not end in a total catastrophe in which the potential $V(q)$ which now comes into the exponent with a plus sign should go lower and lower and thus make a soon negligible contribution. Such a solution could of course not be the true winning dominant classical solution. What shall the "god" wishing a highest possible potential energy minus kinetic one do?

The solution to avoid the catastrophe preventing any hope of having the dominant *history* in the functional integral must be to arrange a way up to some neighboring hill-top as quickly as possible.

(Remember that we are just seeking the maximal "-action" solution, and that means, that the "god" that shall find it, also has the power also to arrange the future as good as "he" can by adjusting and fine-tuning the initial state).

### 6.3. Is This Scenario a Caricature of Cosmology?

We may optimistically interpret this to be a crude picture of what goes on in cosmology with regard to to the very strongest events:

- Standing on the highest peak as long as possible could be identified as a slow-rolling inflation: The inflation field stands on the highest place in the inflaton potential. It stands by making the uttermost effort to keep there as accurately as it can be arranged, and the physicists think it stayed too long to be believable and call it the **slow roll problem**, because it stayed so long.

  Finally, it could not avoid a bit of shaking, and at the end, the inflaton field rolled down. It is like putting a pen to stay just on the tip, just at the meta stable point, and then find it there for years. If nothing else would shake it, even a very weak quantum effect would do the job, and getting such a pen to stand straight up for a long time is not possible in practice.

- Next, it should run up a neighboring hill, and it would be better to be equally as high so that it can stop there again for a very long time.

  The way **up** is to let all the particles be shut away from each other, so that we have the gravitational potential growing, when particles go away from each other.

  The idea in the real world is that we can claim that the advice the "god" takes to come up on a high hill again quickly is to make a very strong Hubble–Lemaître expansion from at least the end of the inflation. (Really, there is already a Hubble–Lemaître expansion going on during the inflation, so it might be rather easy to achieve for "god" to just continue that; or rather, "he" arranged inflation in the inflation after his purpose also after inflation).

  Then, rather soon, most of the kinetic energy of the material/the particles should be converted into the potential energy from them being separated (think about a Newton gravity approximation). If this is carried out, then as the system approaches the next peak, the kinetic energy would be more and more suppressed and the system should be moving less and less.

  In the Universe as we know it in cosmology, the contributions to the energy density counted in the usual LFRW coordinate choice, which are expected to have much kinetic energy in them, should have been suppressed by the arrangement to find a **potential** peak.

  Now, the galaxies, etc., have run so far away that they have run most of their original speed off in the sense that the Hubble–Lemaître expansion has dropped very much compared to the original one, say at the end of inflation.

  Now, physicists believe that the Universe should continue to expand forever, but not terribly many years ago, one believed it could contract again some day; with possible doubts about what dark energy really is and the uncertainties in measuring it, we should rather say: only believing the strongest and most certain effects, it could well be that the universe would still slow down and approach a null Hubble–Lemaître constant. That would correspond to stopping on the hill-top. In any case, the expansion rate is minute today compared to what it was.

*6.4. Seeking FLRW Formulation*

In general relativity, even the concept of energy is coordinate choice dependent, and thus, we shall here, rather than making a full general relativity formulation, choose to consider a little subset of galaxies or just dust particles and study their kinetic energy, say in a frame of the center of mass of the little subset. If we take it that the scale factor $a$ in the usual Friedman–Lemaître–Robertson–Walker (FLRW) formulation is a true length scale, say the radius of the universe, and consider a subset of the galaxies of unit length extension (maybe the unit is Gparsec), then the time derivative $\dot{a}$ is the typical velocity in the essentially flat space frame for the little subset with its center of mass taken to be at rest. Now, we may look at the FLRW equations

$$\left(\frac{\dot{a}}{a}\right)^2 + \frac{kc^2}{a^2} - \frac{\Lambda c^2}{3} = \frac{\kappa c^4}{3}\rho \tag{24}$$

$$2\frac{\ddot{a}}{a} + \left(\frac{\dot{a}}{a}\right)^2 + \frac{kc^2}{a^2} - \Lambda c^2 = -\kappa c^2 p. \tag{25}$$

The goal of the theory (the "god") without the $i$ is to avoid kinetic energy and keep the energy as potential energy as much as possible, and that will mean concentrating on the little subset of galaxies to keep the time derivative of the scale parameter $\dot{a}$ as small as we can weighted with time. Thus, it is most important to keep the velocity $\dot{a}$ small in the long time intervals.

Now, it is well known and easy to see from the FLRW Equation (24) assuming a dust model (matter dominance), otherwise the velocity of a galaxy makes is not important, that the following cases can happen:

- **If** the parameters, such as the $\Lambda$, are so large and positive that the Universe, as it seems empirically for the time being, goes into a $\Lambda$-dominated **DeSitter universe in the long run**, then

$$\frac{\dot{a}}{a} \quad \rightarrow \quad costant \tag{26}$$

$$a \quad \rightarrow \quad \infty \tag{27}$$

$$\text{so that } \dot{a} \quad \rightarrow \quad \infty \tag{28}$$

and our formal kinetic energy goes to infinity. Thus, this case would **not be chosen** in our model.

- **If** on the other hand the $\Lambda$ is so negative that the **Universe recontracts** at some time, then, with the matter dominance ansatz

$$\rho \quad \propto \quad \frac{1}{a^3} \tag{29}$$

$$\text{when } a \text{ gets small } \dot{a} \quad \propto \quad \frac{1}{\sqrt{a}} \tag{30}$$

$$\text{and thus } \dot{a} \quad gets \quad \text{huge.} \tag{31}$$

Thus, this **should** also **not be chosen,** (if one wants to minimize kinetic energy).

- **If** the Hubble–Lemaitre constant $\frac{\dot{a}}{a}$ keeps from both growing up and from turning down, then there is a chance that the velocity $\dot{a}$ could be kept small. However, the most favorable and thus our model prediction would be to have

$$\dot{a} \quad \rightarrow \quad 0 \tag{32}$$

$$\text{even though } a \quad \rightarrow \quad \text{large or } \infty \tag{33}$$

$$\text{but then } \Lambda \quad = \quad 0. \tag{34}$$

This is the case we must have to avoid the kinetic energy in the long run.

Thus, indeed, the **prediction is that** $\Lambda = 0$**.**

If we require there to be a very small velocity $\dot{a}$ over a long time once the $a$ goes through very large values, so that the density $\rho$ has gone almost to zero, then, in fact, both the $\Lambda$ term and the space curvature term $k/a^2$ have to be zero. Thus, we see that our "throwing away the *i*" model indeed predicts both that the universe is flat, i.e., $k/a^2 \sim 0$, and that the cosmological constant $\Lambda$ be "zero".

We can certainly see that to the very first approximation, namely if we compare to energy densities in the reheating era, the cosmological constant $\Lambda$ is indeed minute, and thus, the prediction that it should be zero is very good. However, in today's best fit, we know that the cosmological constant is not quite zero. However, there is still so many ways of making alternative speculations by replacing $\lambda$ with something else like domain walls or quintessence running $\Lambda$, so that it is absolutely not quite excluded, and that the cosmological constant could be avoided. Even experimental uncertainties, that might be needed to repair the various tensions, could also make the zero $\Lambda$ become a possibility.

### 6.5. Field versus Particle Kinetic Energy, a Little Problem?

Even though the above description of the losing and quick recovery of the potential energy as being very crudely arranged in our picture of cosmology sounds by words ok, there is actually a mistake in it, in that the rushing down the peak describing the slow roll inflation is a description in terms of the **fields**, whereas the description of the Hubble–Lemaître expansion as climbing up a gravitational potential is a description in terms of **particles**.

If we consider the fields even classically as the most fundamental description, then we should define kinetic and potential energy in terms of fields concerning both situations considered.

To obtain an idea about how to translate between kinetic and potential energy concepts defined for fields versus for particles, let us consider, e.g., in [60], the expression for the energy momentum tensor of a real scalar field $\phi(x)$ (Klein–Gordon equation field)

$$T_{\mu\nu} = \phi_{,\mu}\,\phi_{,\nu} - \frac{1}{2}g_{\mu\nu}g^{\rho\tau}\phi_{,\rho}\,\phi_{,\tau} - \frac{1}{2}m^2\phi^2 g_{\mu\nu} \tag{35}$$

so that energy density $T_{00} = \phi_{,0}\,\phi_{,0} - \frac{1}{2}g_{00}g^{00}\phi_{,0}\,\phi_{,0} - \frac{1}{2}g_{00}\phi_{,i}\,\phi_{,i} - \frac{1}{2}m^2\phi^2 g_{00}$

$$= \frac{1}{2}\phi_{,0}\,\phi_{,0} + \frac{1}{2}(1+2\varphi(x))\phi_{,i}\,\phi_{,i} + \frac{1}{2}(1+2\varphi(x))m^2\phi^2$$

$$= \frac{1}{2}\phi_{,0}\,\phi_{,0} + \frac{1}{2}(1+2\varphi(x))(\phi_{,i}\,\phi_{,i} + m^2\phi^2) \tag{36}$$

where we assumed $g_{ii} = \eta_{ii} = 1$ for the spatial indices, and $i = 1, 2, 3$; and in flat space, $g_{00} = -1$. However, in a Newton gravity approximate situation

$$g_{00}(x) = -1 - \frac{2\varphi(x)}{c^2} \tag{37}$$

$$= -1 - 2\varphi(x) \tag{38}$$

where $\varphi(x)$ is the gravitational potential.

It is obvious that you could look naively at Equation (36) and see the first term $\frac{1}{2}\phi_{,0}\,\phi_{,0}$ is in the field terminology the kinetic term, while the second term $\frac{1}{2}(1+2\varphi(x))(\phi_{,i}\,\phi_{,i} + m^2\phi^2)$ is the potential one, because the first term consists of the time derivatives $\phi_{,0}$ of the Klein–Gordon field $\phi$, while the second term has only the $\phi$ field itself, without the **time** derivative. There is a gradient term with spatial derivatives $\frac{1}{2}(1+2\varphi(x))\phi_{,i}\,\phi_{,i}$, but that must principally be counted as potential. We can see that interaction with the Newtonian gravitational field $\varphi(x)$ is solely in the potential part.

It looks promising for the potential energy for the particle language being identical to a part of the field-wise potential energy, but I do not trust that. We must investigate/think about what really happens when such a Klein–Gordon field describes a particle coming with significant velocity and run up a slowly varying Newtonian gravitational potential $\varphi(x)$. That the particle is running with significance (but we can for simplicity still think that it is non-relativistic, not to be nervous about using Newton approximation) means that there is at first a gradient part of the energy, but that is counted as potential field-wise. However, as long as the particle has a rather well-determined momentum in spite of being localized relative to the very little variation in space of the gravitational potential, then the vibrations of the Klein–Gordon field will be like a harmonic oscillator and there will be necessarily equally as much potential and kinetic energy provided when one counts the zero for the potential energy at the minimum in the approximate harmonic oscillator potential. This means that the two terms we pointed out as kinetic and potential energy, respectively, will be approximately equally as big when integrated over the relatively large region to which the particle is located. This approximate equality is to be understood with the normalization, in that both terms would be zero if the Klein–Gordon field $\phi$ were zero. This means that, some time later, when the particle has climbed up a hill and come to a region where the Newtonian gravitational potential $\varphi(x)$ is higher, this normalization of the potential is different.

Now of course what happens under the climbing up of the particle is that it slows down, and that means that the gradient term will be smaller when it has come higher up.

Both the kinetic and potential term will have sunk, counted in the local normalization, meaning that they would be zero when the Klein–Gordon field is zero.

Thus, we see that, indeed, by climbing up, the kinetic energy term in the field-wise sense will have fallen. The potential term will also have fallen if one uses the local normalization, but using the same normalization it must have increased because the total energy should be conserved.

Thus, after all we came to, even when strictly using the definition of kinetic and potential energy separation by means of the fields (i.e., the Klein–Gordon field $\phi$), then we find that the climbing of the hill indeed gives a conversion of the kinetic energy to potential energy as the naive thinking in the particle definition of the the kinetic and potential energies.

Thus, indeed, the making of the Hubble–Lemaitre expansion to make the kinetic energy into potential energy is a good idea for the "god" to perform, even when we take the field definition of these concepts of various energies as the fundamental one.

However, it should be noted that defining kinetic energy and potential energy by means of fields or by means of particles is **not** quite the same:

We calculated here, as is easily seen, the energy of the particle in excess of the gravitational potential to be **half potential and half kinetic.**

In the particle definition, however, the energy due to the motion of the particle is counted as **purely kinetic**. The Einstein mass energy is a bit more doubtful for the particle and should presumably be counted as potential.

### 6.6. What to Think about Black Holes for the "God" ?

It sounds obvious that if the "god", in quotation marks, is so eager to make kinetic energy into potential energy, so that he makes/arranges the material in the Universe to Hubble–Lemaitre expand dramatically to achieve that, he would consider making black holes, primordial or later, as a kind of catastrophe, since it would undo "his" great efforts with the expansion. Thus, "he" should put the density of black holes as low as is possible for "him" within "his" competence of arranging details in initial conditions.

However, even ordinary matter and dark matter of a different nature than black holes should be kept down in amounts, because, as I claimed above, half of the Einstein mass energy that should be for a Klein–Gordon field particle could be counted as kinetic and half as potential.

Thus, even having ordinary or non-black-hole dark matter seems not to be wanted.

One can find comfort in the idea of the determination having high potential energy by telling that in the first 70,000 years or so we had a radiation-dominated universe, so that these massive particles came in after a major work of climbing up the next hill is already over in the first approximation.

Thus, one could still claim that the first and strongest Hubble–Lemaitre expansion was performed without any problem of thinking about the masses of the surviving particles. It was, namely, radiation dominance.

### 6.7. Our Dark Matter Model from the Point of View of the "God" in Quotation Marks

The "god" in quotation marks sat and read Paul Frampton's work that dark matter could be black holes, and thus, at the time of the humans, there should be about 24 percent, after the mass/energy of the density, that should be dark matter. "He" got rather sad by reading that and said to himself: "This is terrible when I now have made so much effort to expand the universe so as to make the (gravitational) potential energy so positive as possible, then to spoil it by making such a lot of black holes. That theory shall never be right!" However, then he thought a bit and added, "Well, there are many proposals that are not going to be true" This encouraged "him" a bit, but then he thought on, "But I am also in a theory that is proposed, but that proposed theory is even worse than that of the dark matter being black holes, so I am very likely not true, I do not exist! It is very much more likely". Then, "he" became very depressed: "It is to be so much dark matter and in addition I do not even exist". He now became so depressed that he almost lost all courage for life. He almost thought it was better not to exist at all.

A little later (whatever that means for a timeless "god"), he sat and googled quite without interest; "he" really was seriously depressed.

The "god" in quotation marks was so sad that he was googling almost at random and among other things, "Columbia-plot" [61] (see Figure 2) came about and he saw that lattice QCD calculations led to there being a phase transition, of second order, but anyway, as a function of the light quark masses of the vacuum.

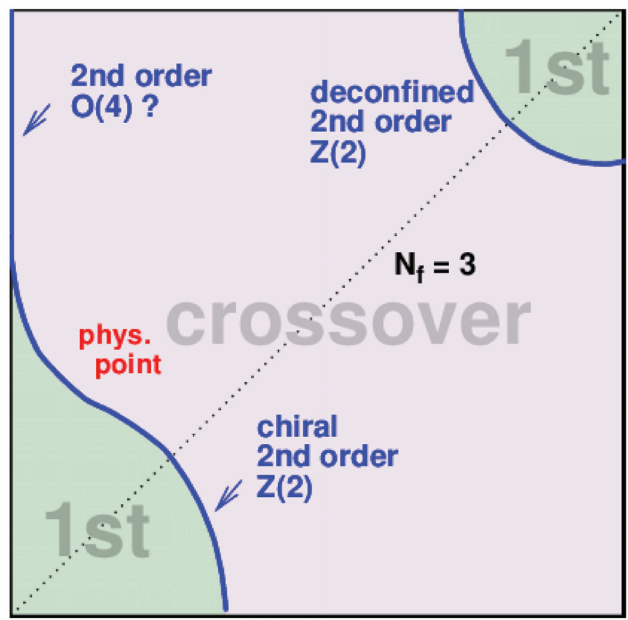

**Figure 2.** "Columbia-plot": on the abscissa is taken to be common mass of the up and down quarks, and on the ordinate, the strange quark mass. By simulation, one has for each combination of quark masses looked for whether increasing the temperature leads to a phase transition of first order or only a so-called crossover, which means that there is no genuine phase transition as function of temperature, but possibly a rather steep variation of an order parameters. It is the phase transition separating curve, which separates the lower left corner, the zero quark mass point, from the middle of the diagram, which is the phase transition, that "god" wants to use to make the Universe have two sorts of vacua. The quark mass axes goes from 0 to ∞.

Now he got the idea that he could put the quark masses just on the phase transition curve. It was not too far from where he would have thought of putting them anyway. Then, he could obtain different phases of vacuum that were realized in Nature in different places and times, and he could hope for some surface, a domain wall between the phases, and of course, there would be a bit of potential difference between the two phases for the nucleons. He got the idea of using such potential difference to catch the nucleons as in a jail and let the skin between the phases contract itself around the caught nucleons and squeeze them so hard that he could catch very many in a small region. Then, he would have very heavy pearls of jails for nuclei and could make use of that in two ways:

**First:** The pearls would pick up electrons of course and become almost neutral, but in any case, relative to their size, hugely heavy and thus function as dark matter.

**Secondly:** He could get rid of most nucleons and thus most matter, and thus of most stars that could finally even develop into neutron stars or even worse to black holes. It would be wonderful: he would only have some dark matter that would not be able to clump severely into clumps where the gravitational potential would become decreased so that potential energy was converted into kinetic energy. He was so glad for this idea, because if he could not get rid of the baryons totally, this was the best he could do: to catch and keep them from interacting and losing their potential energy into kinetic energy. The clumping would be much less than for ordinary matter that was not jailed. Thus, no serious clumps could be formed and decrease the potential energy of gravity, and nothing of the bad stuff such as stars and black holes should come out of it. He had an alternative to the dark matter being black holes, and he got rid of the ordinary matter that had a tendency to make stars and later even black holes. Hurrah, he became very happy.

However, suddenly, he heard some explosions, it seemed! What was that? He looked and saw that inside his small jails for the nucleons, fusion bombs had appeared. The nucleons had first formed helium under the high pressure inside the pearls, and then, in an explosive way, the helium had also combined with carbon. The latter happened so explosively that a lot of nucleons were expelled out from his so smartly invented jails because of the high temperature from the fusion of the helium to carbon. About one-fifth of the nucleons in the jails had escaped due to the explosions, and he had got a lot of freely running nucleons back. That was a disappointment for him, but at least most of the nucleons stayed in the jails.

However, he was still much more happy now that the terrible scenario of the 24 percent of black holes had been avoided, and he contemplated: It was ME who adjusted the quark masses so that we obtained two phases with the same energy density, so that they could be in balance and coexist without the one immediately decaying to the other one. Thus, if HE had done such a thing, should he not then exist? In spite of the little bad luck in that the nucleons had made this fusion bomb and some of them had escaped, the situation was still much better than with a lot of black holes, and a thread of his very existence. Now, HE at least believed in himself; whether the others would believe in him or not, he now did himself.

### 6.8. The Ratio of Dark to Ordinary Matter

The explosion of helium fusion to carbon (or some other heavy elements, but it would have been carbon most likely) was one of the first points studied by Colin Froggatt and me concerning dark matter. In [62], we wrote in 2005: "Before the further internal fusion process took place, the main content of the balls was in the form of $^4He$ nuclei. Now the nucleons in a $^4He$ nucleus have a binding energy of 7.1 MeV in normal matter in our phase, while a typical "heavy" nucleus has a binding energy of 8.5 MeV for each nucleon. Let us, for simplicity, assume that the ratio of these two binding energies per nucleon is the same in the alternative phase and use the normal binding energies in our estimate below. Thus we take the energy released by the fusion of the helium into heavier nuclei to be 8.5 MeV − 7.1 MeV = 1.4 MeV per nucleon. Now we can calculate what fraction of the nucleons, counted as a priori initially sitting in the heavy nuclei, can be released by this 1.4 MeV per nucleon. Since they were bound inside the nuclei by 8.5 MeV relative to the energy they would have outside, the fraction released should be $(1.4 \text{ MeV})/(8.5 \text{ MeV}) = 0.165 = 1/6$. So we predict that the normal baryonic matter should make up 1/6 of the total amount of matter, dark as well as normal baryonic. According to astrophysical fits, giving 23% dark matter and 4% normal baryonic matter relative to the critical density, the amount of normal baryonic matter relative to the total matter is $4\%/(23\% + 4\%) = 4/27 = 0.15$".

Not so many dark matter models give the ratio of the dark to ordinary to be of order unity directly; it more comes out as a miracle that the order of magnitude is the same.

## 7. Conclusions and Birthday Wishes

We have talked about the fact that it was very close, and that it would have been Paul and Nambu who would have been the known inventors of string theory and not myself and Susskind. (Both combinations in addition to Nambu).

Additionally, we talked about a genuinely still not excluded replacement for the Standard Model, the 3-3-1 model(s) (there are possibilities for some variations of the model), a model in which one, in order to cancel the (triangle) anomalies, has to have the number of families being a (multiple of) the number colors in QCD. This is of course a remarkably good prediction. There are three families and three colors.

An important sign of the 3-3-1 model is that in the spectrum of leptons of the same sign one should find resonances signaling decaying gauge particles.

We also mentioned that if dark matter is indeed primordial black holes, they have to be rather heavy compared to many alternative pictures, but most severely, they would be so heavy that there would be so long between them hitting earth that the DAMA-LIBRA experiment, which in spite of being in contradiction seemingly with even very similar experiments as the Anais experiment, with very high statistics for having seen dark matter in the underground, should not see so much as they saw.

At the end, I sneaked in my own crazy theory of the last weeks of little "god" in quotation marks seeking to govern the world so as to put most of the energy as **potential energy**, and that this could be interpreted to mean that just after the "reheating" time, when the inflation in which all energy was indeed potential ended, it was organized that a huge expansion quickly could convert the kinetic energy back to potential energy.

In fact, we argued that such a "god" disfavoring kinetic energy would make the cosmological constant $\Lambda$ and curvature term in the FLRW–cosmological equations preferably zero.

If there really was such an organization that liked to arrange the kinetic energy to quickly come back to being potential, e.g., by Hubble–Lemaitre expansion, then of course the production of black holes in which matter falls into the black hole in a strong gravitational potential and thus sees its potential energy converted into kinetic energy, would be seen as a bad thing to do. Thus, this "god" would only accept black holes to the extent it would be almost unavoidable.

*Congratulation*

Let me first give thanks for the very nice times we have had together and for the many discussions, etc. Then, the best wishes for that the "god", if "he" exists, has planned a great future for Paul, and if he does not exist, that Paul may have a very lucky and successful future anyway, in the latter case, even with dark matter being primordial black holes. Good luck with the birthday!

**Funding:** This research received no external funding.

**Institutional Review Board Statement:** Not applicable.

**Data Availability Statement:** The original contributions presented in the study are included in the article, further inquiries can be directed to the corresponding author.

**Conflicts of Interest:** The authors declare no conflicts of interest.

*Review*

# Hunting for Bileptons at Hadron Colliders

**Gennaro Corcella**

INFN, Laboratori Nazionali di Frascati, Via E. Fermi 54, 00044 Frascati, Italy; gennaro.corcella@lnf.infn.it

**Abstract:** I review possible signals at hadron colliders of bileptons, namely doubly charged vectors or scalars with lepton number $L = \pm 2$, as predicted by a 331 model, based on a $SU(3)_c \times SU(3)_L \times U(1)_X$ symmetry. In particular, I account for a version of the 331 model wherein the embedding of the hypercharge is obtained with the addition of three exotic quarks and vector bileptons. Furthermore, a sextet of $SU(3)_L$, necessary to provide masses to leptons, yields an extra scalar sector, including a doubly charged Higgs, i.e., scalar bileptons. As bileptons are mostly produced in pairs at hadron colliders, their main signal is provided by two same-sign lepton pairs at high invariant mass. Nevertheless, they can also decay according to non-leptonic modes, such as a TeV-scale heavy quark, charged 4/3 or 5/3, plus a Standard Model quark. I explore both leptonic and non-leptonic decays and the sensitivity to the processes of the present and future hadron colliders.

**Keywords:** BSM phenomenology; bileptons; hadron colliders

## 1. Introduction

The Standard Model (SM) of electroweak and strong interactions is a complete theory, but it exhibits several drawbacks, such as the hierarchy problem in the Higgs sector, neutrino masses, or Dark Matter, which call for a theory with a more general gauge structure and possibly new particles. As well-motivated SM extensions, such as supersymmetry or extra dimensions, have provided no visible signal at the LHC thus far, it is mandatory to explore alternative scenarios. In this paper, I review the work carried out in the last few years [1–3] in the framework of the $SU(3)_L \times SU(3)_C \times U(1)_X$ model [4–6], also known as the 331 model, and its possible signals at the Large Hadron Collider (LHC) and at a future 100 TeV hadron collider (FCC-*hh*).

Among its main features, this model predicts the existence of bileptons, i.e., gauge bosons ($Y^{--}, Y^{++}$) of charge $Q = \pm 2$ and lepton number $L = \pm 2$, which is why one often refers to it as a bilepton model. Furthermore, in the specific formulation of [4], one is capable of explaining the asymmetry of the third quark family, i.e., top and bottom quarks, with respect to the other two, while the existence of three families, i.e., $N_f = N_C = 3$, $N_C$ being the number of colours, is a consequence of the requirement of an anomaly-free theory (see also the detailed discussion in [7]).

As will be detailed later on, the scenario that will be investigated, besides the vectors $Y^{++(--)}$, predicts a number of new particles, which may possibly be within reach of the LHC or a future hadron collider, such as FCC-*hh*. Among those, one has heavy quarks with charge 5/3, usually labelled $T$, or charge 4/3, i.e., $D$ or $S$, which typically have a mass of the order of a few TeVs (see the analysis in [3]). Moreover, a complete description of the model requires the inclusion of a Higgs sector, which is a sextet of $SU(3)_L$ and is needed to provide mass to the leptons. In the Higgs sector, the prediction of new doubly charged scalars is particularly relevant. Such Higgs-like bosons with charge $\pm 2$ have been intensively searched by the experimental collaborations in different new physics models, setting mass bounds between 900 and 1100 GeV [8,9] at $\sqrt{s} = 13$ TeV and integrated luminosities $\mathcal{L} = 139$ fb$^{-1}$ and 12.9 fb$^{-1}$, respectively. As far as I know, no specific search for vector bileptons has been undertaken so far.

On the other hand, as will be detailed in the following, Refs. [1,3] investigated the phenomenology of vector bileptons, decaying into leptonic or non-leptonic final states, while Ref. [2] explored both vector and scalar bileptons, concentrating on final states with same-sign lepton pairs. All such papers published results for reference points that are not yet excluded by the experimental searches, with a bilepton mass just below the exclusion range, in order to maximize the production cross section.

The plan of this contribution is the following. In Section 2, I shall review the main ingredients of the 331 model, in the version proposed in [4]. In Section 3, I shall critically present the phenomenological results contained in Refs. [1–3]. In Section 4, some concluding remarks will be presented.

## 2. Theoretical Framework

Following Ref. [4], the gauge structure of the bilepton model is $SU(3)_c \times SU(3)_L \times U(1)_X$, with the fermions (quarks) in the fundamental of $SU(3)_c$ arranged into triplets of $SU(3)_L$. As anticipated in the Introduction, the third quark family (top and bottom) is treated asymmetrically with respect to the first two families in the electroweak $SU(3)_L$. In detail, as for the first two families, one has

$$Q_1 = \begin{pmatrix} u_L \\ d_L \\ D_L \end{pmatrix}, \quad Q_2 = \begin{pmatrix} c_L \\ s_L \\ S_L \end{pmatrix}, \quad Q_{1,2} \in (3,3,-1/3) \tag{1}$$

under $SU(3)_c \times SU(3)_L \times U(1)_X$, while, for the third one, it is

$$Q_3 = \begin{pmatrix} b_L \\ t_L \\ T_L \end{pmatrix}, \quad Q_3 \in (3,\bar{3},2/3). \tag{2}$$

In the above formulation, $D$, $S$, and $T$ are quarks, with charge 4/3 ($D$ and $S$) or 5/3 ($T$). In the following, I will explore scenarios wherein such quarks are either within or outside the reach of present and future hadron colliders.

The right-handed quarks ($q$), as happens in the SM, are singlets even under $SU(3)_L$. Their representations are the following:

$$(d_R, s_R, b_R) \in (\bar{3}, 1, 1/3) \tag{3}$$

$$(u_R, c_R, t_R) \in (\bar{3}, 1, -2/3) \tag{4}$$

$$(D_R, S_R) \in (\bar{3}, 1, 4/3) \tag{5}$$

$$T_R \in (\bar{3}, 1, -5/3). \tag{6}$$

One can notice that adding such new particles to the Standard Model states is not enough to cancel the $SU(3)_L$ anomalies [4,7]. Therefore, one has to introduce new leptonic states in three $\bar{3}$ representation. As a result, the three lepton families, unlike the quarks, are arranged in a 'democratic' manner as triplets of $SU(3)_L$:

$$l = \begin{pmatrix} l_L \\ \nu_l \\ \bar{l}_R \end{pmatrix}, \quad l \in (1,\bar{3},0), \quad l = e, \ \mu, \ \tau. \tag{7}$$

As discussed in [4,7], these assignments of quarks and leptons lead to the cancellation of the anomaly of $SU(3)_L$, while the $SU(3)_C$ one is cancelled as happens in the SM, i.e., through a complete balance between left-handed colour triplets and right-handed anti-triplets in the quark sector.

The electroweak symmetry breaking of this 331 model occurs through scalar fields $\rho$, $\eta$, and $\chi$, which are arranged as triplets of $SU(3)_L$:

$$\rho = \begin{pmatrix} \rho^{++} \\ \rho^+ \\ \rho^0 \end{pmatrix} \in (1,3,1), \quad \eta = \begin{pmatrix} \eta^+ \\ \eta^0 \\ \eta^- \end{pmatrix} \in (1,3,0), \quad \chi = \begin{pmatrix} \chi^0 \\ \chi^- \\ \chi^{--} \end{pmatrix} \in (1,3,-1). \quad (8)$$

The breaking of $SU(3)_L \times U(1)_X \rightarrow U(1)_{em}$ is achieved in two steps. First, the vacuum expectation value (vev) of the neutral component of $\rho$ provides mass to novel gauge bosons $Z'$, $Y^{++}$, and $Y^+$ and heavy quarks $D$, $S$, and $T$. In this first step, the original gauge group $SU(3)_L \times U(1)_X$ breaks into $SU(2)_L \times U(1)_Y$. In the second step, it is $\chi^0$ and $\eta^0$ that receive a vev, and one has the usual breaking from $SU(2)_L \times U(1)_Y$ to $U(1)_{em}$.

In detail, the scalar potential reads as follows:

$$\begin{aligned} V = {}& m_1\, \rho^*\rho + m_2\, \eta^*\eta + m_3\, \chi^*\chi \\ & + \lambda_1(\rho^*\rho)^2 + \lambda_2(\eta^*\eta)^2 + \lambda_3(\chi^*\chi)^2 \\ & + \lambda_{12}\rho^*\rho\,\eta^*\eta + \lambda_{13}\rho^*\rho\,\chi^*\chi + \lambda_{23}\eta^*\eta\,\chi^*\chi \\ & + \zeta_{12}\rho^*\eta\,\eta^*\rho + \zeta_{13}\rho^*\chi\,\chi^*\rho + \zeta_{23}\eta^*\chi\,\chi^*\eta \\ & + \sqrt{2}f_{\rho\eta\chi}\rho\,\eta\,\chi. \end{aligned} \quad (9)$$

The neutral component of each triplet acquires a vev and can be expanded as

$$\rho^0 = \frac{1}{\sqrt{2}}v_\rho + \frac{1}{\sqrt{2}}\left(\mathrm{Re}\,\rho^0 + i\mathrm{Im}\rho^0\right) \quad (10)$$

$$\eta^0 = \frac{1}{\sqrt{2}}v_\eta + \frac{1}{\sqrt{2}}\left(\mathrm{Re}\,\eta^0 + i\mathrm{Im}\eta^0\right) \quad (11)$$

$$\chi^0 = \frac{1}{\sqrt{2}}v_\chi + \frac{1}{\sqrt{2}}\left(\mathrm{Re}\,\chi^0 + i\mathrm{Im}\,\chi^0\right). \quad (12)$$

As detailed in [1], one first determines the potential minimization conditions and then, after spontaneous symmetry breaking, the gauge and the mass eigenstates of $\rho$, $\eta$, and $\chi$. The explicit expression of the mass matrices of the scalar sector, both neutral and charged, are provided in [1], and we do not report them here for the sake of brevity.

As anticipated in the Introduction and discussed in detail in [2], it is necessary to add to the scalar sector a $SU(3)_L$ sextet in order to provide masses to leptons. This implies that the particle spectrum of the bilepton model includes doubly charged Higgs bosons ($H^{\pm\pm}$) capable of decaying into same-sign lepton pairs. In other words, decays like $H^{\pm\pm} \rightarrow l^\pm l^\pm$ would be evidence of the presence of sextet representation of $SU(3)_L$.

Still on decays of doubly charged scalars, as pointed out in [2], in principle, as for the Standard Model Higgs, one should have amplitudes proportional to the Yukawa coupling, hence to the masses of the final-state particles. However, for the sake of generality and putting vector and scalar bileptons on the same footing, following [2], I shall consider a scenario where the branching ratios of doubly charged Higgs bosons are not proportional to the mass, but, referring, e.g., to decays into same-sign lepton pairs, one has $\mathrm{BR}(Y^{\pm\pm} \rightarrow l^\pm l^\pm) \simeq \mathrm{BR}(H^{\pm\pm} \rightarrow l^\pm l^\pm)$, $Y^{\pm\pm}$ being vector bileptons.

After electroweak symmetry breaking, one ends up with a rich Higgs sector. In detail, we have 5 scalar Higgs bosons, one of them being the Standard Model with mass about 125 GeV and 4 neutral pseudoscalar Higgs bosons, out of which 2 are the Goldstones of the $Z$ and $Z'$ massive vector bosons. Furthermore, one has 6 charged Higgses, 2 of which are the charged Goldstones and 3 are doubly charged Higgses, 1 of which is a Goldstone boson.

As the main goal of this investigation is the phenomenology of doubly charged vectors and scalars, we point out that Ref. [2] contains a thorough discussion of the vertices where pairs $Y^{\pm\pm}Y^{\pm\pm}$ or $H^{\pm\pm}H^{\pm\pm}$ are involved. We do not report the formulas in the present contribution for brevity and refer to [2] for such couplings.

## 3. Phenomenology at the LHC and Future Colliders

### 3.1. Leptonic Decays of Bileptons

In this section, I present the main results contained in [1–3] regarding the phenomenology of doubly charged scalars and vectors at the LHC (13 or 14 TeV) and future colliders, namely FCC-*hh*. Typical contributions to bilepton production in hadron collisions are presented in Figure 1: an initial-state $q\bar{q}$ pair annihilates and a $B^{++}B^{--}$ pair, $B$ being a doubly charged vector or scalar, is produced. As can be seen, bilepton-pair production can be mediated by the exchange of, e.g., a neutral Higgs or a vector ($Z$, $Z'$, or photon) in the *s*-channel, or a heavy TeV-scale quark, charged 5/3, in the *t*-channel. Ref. [1] also discusses the production of bilepton pairs in association with jets and presents some typical diagrams for such processes as well.

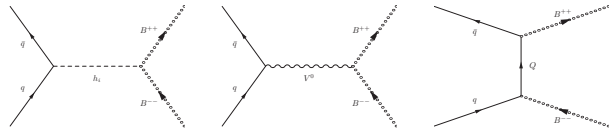

**Figure 1.** Characteristic diagrams for the production of bilepton pairs in hadron collisions.

In detail, Refs. [1,2] account for decays of bileptons into same-sign lepton pairs, say $Y^{++} \to \mu^+\mu^+$, while [3] deals with non-leptonic decays, i.e., decays into a light (SM) quark (antiquark) and a heavy Tev-scale antiquark (quark), e.g., $Y^{++} \to T\bar{b}$, $\bar{D}u$. The results are presented for a few benchmarks, determined in such a way as to be not yet excluded by the experimental searches although capable of yielding a remarkable cross section and number of events.

In order to determine the benchmarks and scan the parameter space, we had to implement the bilepton model in the SARAH 4.9.3 code [10]. In particular, Refs. [1,2] carry out a phenomenological investigation for a bilepton mass about 880 GeV, just below the experimental exclusion limit. In Ref. [2], where the phenomenologies of vector and scalar bileptons are compared, one sets the masses to the same value:

$$M_{Y^{++}} \simeq M_{H^{++}} \simeq 878.3\,\text{GeV},\tag{13}$$

while exotic Higgs bosons, $Z'$, and heavy quarks are assumed to have masses well above 1 TeV; hence, they are too heavy to contribute to any bilepton phenomenology (see [2] for their actual values in the reference points). One can then explore the process

$$pp \to Y^{++}Y^{--}(H^{++}H^{--}) \to (l^+l^+)(l^-l^-),\tag{14}$$

setting the following cuts on final-state lepton transverse momentum ($p_T$), rapidity ($\eta$), and invariant opening angle:

$$p_{T,l} > 20\,\text{GeV}, |\eta_l| < 2.5, \Delta R_{ll} > 0.1.\tag{15}$$

In [1,2], one assumes democratic leptonic branching ratios of bileptons, namely BR($Y^{++} \to l^+l^+$) $\simeq$ BR($H^{++} \to l^+l^+$) $\simeq 1/3$, for any lepton flavour ($e$, $\mu$ or $\tau$).

After the cuts are applied, the leading-order cross sections of processes in Equation (14), computed by means of MadGraph 2.6.1 [11] at $\sqrt{s} = 13$ TeV, read as follows:

$$\sigma(pp \to YY \to 4l) \simeq 4.3\,\text{fb}\ ;\ \sigma(pp \to HH \to 4l) \simeq 0.3\,\text{fb}.\tag{16}$$

At 14 TeV, one has instead $\sigma(pp \to YY \to 4l) \simeq 6.0$ fb and $\sigma(pp \to HH \to 4l) \simeq 0.4$ fb. As discussed in [2], the higher cross section in the case of vector-pair production can be explained in terms of the bilepton helicity. In the case of doubly charged Higgs production, only the amplitudes where the intermediate vectors ($\gamma$, $Z$, $Z'$) have helicity zero contribute, while, in case of doubly charged vectors, all helicities 0 and $\pm 1$ play a role. For processes mediated by scalars, $Y^{++}$ and $Y^{--}$ can still rearrange their helicities in a few different ways to achieve angular-momentum conservation and a total vanishing helicity in the centre-of-mass frame. Similar results are also found in [12], where the authors investigated vector and scalar bilepton pairs at hadron colliders at parton level in the LO approximation.

As for backgrounds, as pointed out in [1,2], the main one is due to same-sign lepton-pair production mediated by a $Z$-boson pair, i.e.,

$$pp \to ZZ \to (l^+ l^-)(l^+ l^-), \tag{17}$$

while processes mediated by neutral Higgs pairs are negligible due to the tiny coupling of the Higgs with leptons. After setting the cuts, the LO cross section of the process (17) is provided by $\sigma(pp \to ZZ \to 4l) \simeq 6.1$ fb at 13 TeV and 6.6 fb at 14 TeV. For an integrated luminosity $\mathcal{L} = 300$ fb$^{-1}$, at 13 TeV, one has $N(YY) \simeq 1302$ lepton pairs mediated by doubly charged vectors, while scalars yield $N(HH) \simeq 120$ and the $ZZ$ background $N(ZZ) \simeq 1836$ events. At 14 TeV and $\mathcal{L} = 3000$ fb$^{-1}$, such numbers read $N(YY) \simeq 17880$, $N(HH) \simeq 1260$, and $N(ZZ) \simeq 19740$. For $S$ signal and $B$ background events, one can define a significance (in units of standard deviations)

$$s = \frac{S}{\sqrt{B + \sigma_B^2}}, \tag{18}$$

where $\sigma_B$ is the systematic error on $B$ gauged about $\sigma_B \simeq 0.1B$ in [2]. Following [13], the denominator of the significance (18) sums in quadrature the intrinsic statistical fluctuation of the background $\sqrt{B}$ and the uncertainty in the background $\sigma_B$, obtaining $s = S/\sqrt{\sqrt{B^2} + \sigma_B^2}$. One can find a significance $s \simeq 6.9$ for vector pairs at 13 TeV and $\mathcal{L} = 300$ fb$^{-1}$ and $s = 0.6$ for scalars, which clearly means that only doubly charged vector bileptons may possibly be visible at 13 TeV. At 14 TeV and high integrated luminosity, one has $s \simeq 9$ for $Y^{++}Y^{--}$ and $s \simeq 0.64$ for $H^{++}H^{--}$ production. Reference [2] explores several distributions of relevant leptonic observables, yielded by vector and scalar bileptons, as well as $ZZ$ background. For the sake of conciseness, we present in Figure 2 only those referring to the hardest-lepton transverse momentum $p_{T,1}$ and the same-sign lepton invariant mass.

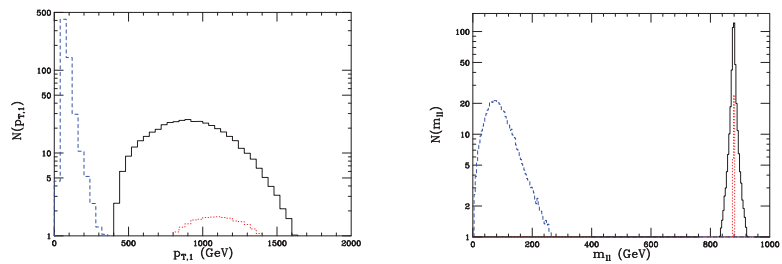

**Figure 2.** Distributions of the transverse momentum of the hardest lepton (**left**) and of the same-sign lepton invariant mass (**right**). The solid histograms are the spectra yielded by vector bileptons, the dots correspond to scalar doubly charged Higgs bosons, and the blue dashes to the $ZZ$ Standard Model background.

As for the transverse momentum $p_{T,1}$, the $ZZ$ distribution is sharp and peaked at low $p_T$, while those yielded by the $HH$ and $YY$ bileptons are much broader and peak at about 1 TeV. In fact, the background $Z$ bosons are much lighter than bileptons and decay into different-sign lepton pairs, while $Y^{\pm\pm}$ and $H^{\pm\pm}$ decay into same-sign electrons and muons. Furthermore, for every value of $p_T$, the $HH$ spectrum is well below the $YY$ one.

Regarding the same-sign lepton invariant mass $m_{ll}$, as expected, the 331 signal peaks at $m_{ll} \simeq 900$ GeV, while the $Z$-background distribution is instead a broad spectrum, significant up to about 350 GeV and maximum around 70 GeV. The signal spectra are rather narrow: the authors of [2] quoted $Y^{++}$ and $H^{++}$ widths about 7 GeV and 400 MeV, hence much smaller than their masses.

Before concluding this subsection, one can then point out that, as should have been expected due to the obtained significances, the distributions in Figure 2, as well as those published in [2], seem to show that discriminating the 331 signal from the background should be feasible, with doubly charged vectors dominating over scalars.

### 3.2. Non-Leptonic Decays of Bileptons

While possible decays into same-sign lepton pairs would be the 'smoking gun' for bilepton discovery at the LHC, depending on the mass spectrum, it is also possible that vector and scalar bileptons could well decay into non-leptonic final states, such as a TeV-scale heavy quark and a light quark. This was in fact the main purpose of the exploration in [3], which I shall summarize hereafter.

Unlike Refs. [1,2], the more recent work in [3] took advantage of the results of Ref. [14], where the authors, by using renormalization group arguments, provided the estimate $m_Y = (1.29 \pm 0.06)$ TeV for the bilepton mass. Making use of this finding, Ref. [3] concentrated on doubly charged vectors $Y^{\pm\pm}$ and considered two benchmark cases: one scenario with all heavy quarks $D$, $S$, and $T$ lighter than $Y^{\pm\pm}$ and another one where only the mass of $D$ is lower than $m_Y$, while $S$ and $T$ are heavier. More precisely, the first benchmark, labelled BM I in [3], sets all TeV-scale quark masses to 1 TeV, i.e.,

$$m_D = m_S = m_T = 1 \text{ TeV}, \tag{19}$$

while, in the second one, i.e., BM II, one has the following mass values:

$$m_D = 1.2 \text{ TeV}, \ m_S = 1.5 \text{ TeV}, \ m_T = 1.5 \text{ TeV}. \tag{20}$$

Both BM I and BM II are consistent with a light SM-like Higgs boson with mass about 125 GeV; all the other BSM particles have masses much above 1 TeV; therefore, they are not relevant for bilepton phenomenology.

Unlike Refs. [1,2], wherein bileptons could only decay leptonically, in the benchmark points of [3], one has substantial branching fractions into both leptonic and hadronic final states. In detail, one has

$$\text{BR}(Y^{++} \to l^+ l^+) \simeq 20.6\% \text{ (BM I)}, \ 32.5\%\text{(BM II)}, \tag{21}$$

for each lepton flavour $l = e, \mu, \tau$, and

$$\text{BR}(Y^{++} \to u\bar{D}, c\bar{S}, T\bar{b}) \simeq 12.7\% \text{ (BM I)}, \ \text{BR}(Y^{++} \to u\bar{D}) \simeq 2.5\% \text{ (BM II)}. \tag{22}$$

The total bilepton widths instead read as follows:

$$\Gamma(Y^{\pm\pm}) \simeq 17.9 \text{ GeV (BM I)}; \ \Gamma(Y^{\pm\pm}) \simeq 11.4 \text{ GeV (BM II)}. \tag{23}$$

The larger width in BM I is clearly due to the fact that decays into final states with all three heavy quarks $D$, $S$, and $T$ are permitted.

Before presenting some numerical results, one should also explore the phenomenology of TeV-scale quark decays. In BM I, the heavy quarks exhibit three-body decays into a Standard Model quark and a same-sign lepton pair or a lepton–neutrino pair, through a virtual bilepton, with the following branching fractions:

$$\text{BR}(D(S) \to u(c)l^-l^-) \simeq \text{BR}(D(S) \to d(s)l^-\nu_l) \simeq 16.7\% \text{ (BM I).} \tag{24}$$

In BM II, $S$ and $T$ are heavier than singly and doubly charged bileptons and can therefore decay into final states with a real $Y^\pm$ or $Y^{\pm\pm}$. While the $D$ rates are the same as in BM I, i.e., Equation (24), $S$ can decay into real bileptons as follows:

$$\text{BR}(S \to cY^{--}) \simeq 50.5\%, \ \text{BR}(S \to sY^-) \simeq 49.5\% \text{ (BM II).} \tag{25}$$

As for $T$, charged 5/3, its decay rates are

$$\text{BR}(T \to bl^+l^+) \simeq 19.4\%, \ \text{BR}(T \to tl^+\bar{\nu}_l) \simeq 13.9\% \text{ (BM I);} \tag{26}$$

$$\text{BR}(T \to bY^{++}) \simeq 64.6\%, \ \text{BR}(T \to tY^+) \simeq 35.4\% \text{ (BM II).} \tag{27}$$

The total decay widths are provided by

$$\Gamma(D) \simeq \Gamma(S) \simeq 3.4 \times 10^{-3} \text{ GeV}, \ \Gamma(T) \simeq 3.0 \times 10^{-3} \text{ GeV (BM I);} \tag{28}$$

$$\Gamma(D) \simeq 1.3 \times 10^{-2} \text{ GeV}, \ \Gamma(S) \simeq 1.5 \text{ GeV}, \ \Gamma(T) \simeq 1.1 \text{ GeV (BM II).} \tag{29}$$

In other words, in BM I, all TeV-scale quarks have a pretty small width of the order $\mathcal{O}(10^{-3} \text{ GeV})$; in BM II, $D$ is still quite narrow, having a width $\mathcal{O}(10^{-2} \text{ GeV})$, while the widths of $S$ and $T$ are of the order of 1 GeV since they are capable of decaying into states with real bileptons.

The production cross sections of bilepton pairs at the LHC (13 and 14 TeV) and FCC-*hh* are provided by

$$\sigma(pp \to Y^{++}Y^{--}) \simeq 0.75 \text{ fb (LHC, 13 TeV),} \tag{30}$$

$$\sigma(pp \to Y^{++}Y^{--}) \simeq 1.12 \text{ fb (LHC, 14 TeV),} \tag{31}$$

$$\sigma(pp \to Y^{++}Y^{--}) \simeq 393.89 \text{ fb (FCC}-hh\text{),} \tag{32}$$

with the FCC-*hh* cross sections about 500 and 350 times larger than the LHC ones.

Following [3], in BM I, I shall account for primary decays of $Y^{\pm\pm}$ into quarks $T$, which further decay into a bottom quark and a same-sign muon pair, hence a final state with four *b*-flavoured jets and two same-sign muon pairs:

$$pp \to Y^{++}Y^{--} \to (T\bar{b})(\bar{T}b) \to (b\bar{b}\mu^+\mu^+)(b\bar{b}\mu^-\mu^-) \text{ (BM I).} \tag{33}$$

In reference point BM II, I shall instead explore primary decays into quarks $D$ and final states with four *u*-quark initiated light jets accompanied by four muons (4*u*4*μ*):

$$pp \to Y^{++}Y^{--} \to (\bar{D}u)(D\bar{u}) \to (u\bar{u}\mu^+\mu^+)(u\bar{u}\mu^-\mu^-) \text{ (BM II).} \tag{34}$$

In Ref. [3], a few representative diagrams of processes (33) and (34) are presented as well.

A first rough estimation of the predicted number of events at the LHC and FCC-*hh* can be obtained by multiplying the inclusive cross sections in Equation (30) by the relevant branching ratios, assuming a perfect tagging efficiency and no cut on final-state jets and leptons. At the LHC, one obtains

$$\sigma(pp \to YY \to 4b4\mu) \simeq 4.55 \times 10^{-4} \text{ fb (LHC, 13 TeV, BM I),} \tag{35}$$

$$\sigma(pp \to YY \to 4b4\mu) \simeq 6.80 \times 10^{-4} \text{ fb (LHC, 14 TeV, BM I),} \tag{36}$$

$$\sigma(pp \to YY \to 4u4\mu) \simeq 1.31 \times 10^{-5} \text{ fb (LHC, 13 TeV, BM II),} \tag{37}$$

$$\sigma(pp \to YY \to 4u4\mu) \simeq 2.03 \times 10^{-5} \text{ fb (LHC, 14 TeV, BM II).} \tag{38}$$

Such cross sections are too small to see any event at 300 fb$^{-1}$ and at 3000 fb$^{-1}$ (HL-LHC), even before imposing any acceptance cut. Therefore, the investigation in [3] discarded the LHC environment and the analysis was concentrated on FCC-*hh*, where the cross sections are remarkable:

$$\sigma(pp \to YY \to 4b4\mu) \simeq 0.24 \text{ fb (FCC} - hh, \text{ BM I)}, \tag{39}$$

$$\sigma(pp \to YY \to 4u4\mu) \simeq 6.87 \times 10^{-3} \text{ fb (FCC} - hh, \text{ BM II)}. \tag{40}$$

The scenario BM I at FCC-*hh* yields a few hundreds events; BM II is less promising but still worthwhile to investigate.

As for the backgrounds, one considers, above all, four *b* quarks and two *Z* bosons decaying into muon pairs (background $b_1$),

$$pp \to b b \bar{b} \bar{b} Z Z \to b b \bar{b} \bar{b} \mu^+ \mu^- \mu^+ \mu^-, \tag{41}$$

and four top quarks with the subsequent *W*s decaying into muons and requiring, as in [2], a small missing energy due to the muon neutrinos (background $b_2$):

$$pp \to t t \bar{t} \bar{t} \to (bW^+)(bW^+)(\bar{b}W^-)(\bar{b}W^-) \to b b \bar{b} \bar{b} \mu^+ \mu^+ \mu^- \mu^- \nu_\mu \nu_\mu \bar{\nu}_\mu \bar{\nu}_\mu. \tag{42}$$

As discussed in [3], the simulation of (42) accounted for electroweak corrections as well since, as pointed out in [15], at both LO and NLO, they can contribute up to 10% of the total cross section.

Reference [3] also considered the following backgrounds with four light jets and two *Z* bosons and with two light jets, two *b*-jets, and two *Z*s:

$$pp \to jjjjZZ \to jjjj\mu^+ \mu^- \mu^+ \mu^-, pp \to jjb\bar{b}ZZ \to jjb\bar{b}\mu^+ \mu^- \mu^+ \mu^-. \tag{43}$$

In Equation (43), *j* is either a light-quark or gluon-initiated jet, mistagged as a *b*-jet.

I cluster the final states of hadrons in four jets according to the $k_T$ algorithm and apply the following acceptance cuts on jets and muons:

$$p_{T,j} > 30 \text{ GeV}, \ p_{T,\mu} > 20 \text{ GeV}, \ |\eta_j| < 4.5, |\eta_\mu| < 2.5,$$
$$\Delta R_{jj} > 0.4, \ \Delta R_{\mu\mu} > 0.1, \ \Delta R_{j\mu} > 0.4. \tag{44}$$

The cuts in (44) correspond to a conservative choice of the so-called 'overlap removal' algorithm used at the LHC to discriminate lepton and jet tracks at the LHC [16]. As for the four-top background (42), Ref. [3] sets the additional cut MET < 200 GeV on the missing transverse energy due to the neutrinos in the final state. In [3], the MET cut was consistently set even on neutrinos coming from hadron decays.

In principle, one should account for the *b*-tagging efficiency, as well as the probability of mistagging a light jet as a *b*-jet. Such efficiencies depend on the jet rapidity, transverse momentum, and flavour; however, for an explorative analysis, like the one in [3], one can implement such effects in a flat manner, i.e., independently of the jet kinematics and of the flavour of the light jets. The *b*-tagging efficiency ($\epsilon_b$) and the mistag rate ($\epsilon_j$, with $j = u, d, s, c$) are then set to the following values, as in [17]:

$$\epsilon_b = 0.82 \ , \ \epsilon_j = 0.05. \tag{45}$$

After setting such cuts, the signal (*s*) cross section of process (33) amounts to $\sigma(4b4\mu)_s \simeq 6.24 \times 10^{-2}$ fb, leading to $N(4b4\mu)_s \simeq 90$ events at FCC-*hh* for an integrated luminosity $\mathcal{L} = 3000$ fb$^{-1}$ and after setting all cuts and *b*-tagging efficiency. As for the backgrounds (41)–(43), one obtains $\sigma(4b4\mu)_{b_1} \simeq 1.28 \times 10^{-2}$ fb, $\sigma(4b4\mu + \text{MET})_{b_2} \simeq 3.34 \times 10^{-2}$ fb, $\sigma(4j4\mu)_{b_3} \simeq 4.43$ fb, $\sigma(2b2j4\mu)_{b_4} \simeq 1.34$ fb. Including also *b*-tagging and mistag efficiencies and rounding to the nearest ten, one computes the following number of background events at FCC-*hh*: $N(4b4\mu)_{b_1} \simeq 20$, $N(4b4\mu + \text{MET})_{b_2} \simeq 50$. The backgrounds $b_3$ and $b_4$ yield too few events to be significant. Regarding BM II and the decay chain (34), the cross section is about $\sigma(4j4\mu)_s \simeq 1.88 \times 10^{-3}$ fb at FCC-*hh*. As a result, considering that some extra suppression is due to the efficiency of jet/lepton tagging, the BM II reference point was eventually discarded in [3]. Several observables were presented in [3] for the purpose of the benchmark BM I; as for leptonic decays, in Figure 3, the hardest muon transverse momentum ($p_{T,1}$) and the invariant mass of same-sign muons ($M_{\mu\mu}$) are plotted for both signal and background. The background $p_{T,1}$ spectra are substantial only at low transverse momenta, peaking about $p_{T,1} \simeq 100$ GeV and rapidly vanishing at large $p_T$, while the signal ones are broad and substantial up to $p_{T,1} \simeq 2$ TeV. Above a few hundred GeV, the signal greatly dominates over the background; this was expected since same-sign muons are related to the decay of a TeV-scale resonance. Regarding $M_{\mu\mu}$, unlike the backgrounds, whose spectra peak at low values and are negligible above 500 GeV, the signal yields a broad invariant-mass spectrum, shifted towards large values and exhibiting a maximum about 700 GeV.

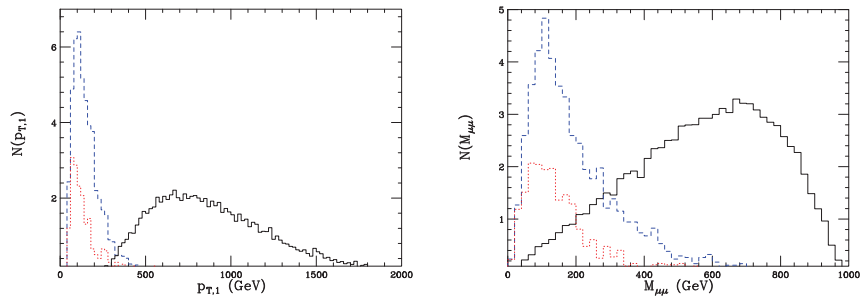

**Figure 3.** Distributions of the transverse momentum of the hardest muons (**left**) and of the same-sign muon invariant mass (**right**). The solid histograms are the signals, the dashes correspond to four tops, and the dots to the *bbZZ* background.

## 4. Discussion

I reviewed recent phenomenological work carried out in Refs. [1–3] within the $SU(3)_C \times SU(3)_L \times U(1)_X$, i.e., 331 model, which has, above all, appropriate features to explain the number of quark and lepton families and the asymmetry between the third and first two quark families. I worked in the framework of [4] and explored the possibility to discover doubly charged bileptons, i.e., doubly charged vectors or scalars with lepton number $\pm 2$, at the LHC and at a future 100 TeV collider FCC-*hh*. I first considered the production of bilepton pairs and decays into same-sign leptons, and then I accounted for non-leptonic decays too. In both cases, a few benchmarks, not yet excluded by the experimental searches, were determined to maximize the cross section at the LHC. Regarding the leptonic decays, it was found that a discovery of bileptons is feasible, with a possible signal due to vector bileptons dominating over the scalars, because of helicity arguments. Decays into non-leptonic final states are more cumbersome since bileptons decay into a heavy TeV-scale quark and a light Standard Model quark: the cross section of the resulting decay chain is too small at the LHC, even in the high-luminosity phase, to provide any signal. Nevertheless, non-leptonic decays of bileptons are expected to be visible at a future FCC-*hh*.

Before concluding, I wish to stress that, while the work presented here deals with the primary production of bileptons, it is certainly a very interesting scenario regarding TeV-scale quarks $T$, $S$, and $D$, which are heavier than $Y^{++}$, so they can be produced in processes like $pp \to T\bar{T}$ and decay according to, e.g., $T \to Y^{++}b$. A study of heavy-quark production and decays, along the lines of [18] but specific to the 331 model, is currently in progress [19].

In summary, as the most-studied models of new physics have provided no visible signal yet, exploring alternative scenarios is compelling. The bilepton model is certainly an appealing framework from the theoretical viewpoint, and, as summarized in this contribution, it features a rich phenomenology that may be a first indication of new physics in the next LHC run as well as HL-LHC and FCC-*hh*. It is then hopeful and desirable that the experimental collaborations use the results presented here and join the effort to search for bileptons at present and future colliders.

**Funding:** This research received no external funding

**Data Availability Statement:** The results presented in this paper are openly available in Refs. [1–3]. The release of the computing codes used to obtain such results is under way. For the time being, the codes are available upon request from the author.

**Acknowledgments:** It has been a great honour collaborating with Paul Frampton, who introduced me to the 331 model and gave me countless hints on physics beyond the Standard Model. I also acknowledge Antonio Costantini and Claudio Corianò, coauthors of Refs. [1–3].

**Conflicts of Interest:** The author declares no conflicts of interest.

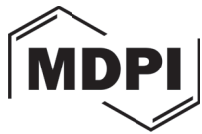

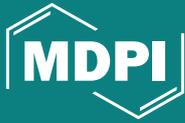